\documentclass[journal]{IEEEtran}
\usepackage{amsmath}
\usepackage{amssymb}
\usepackage{amsfonts}
\usepackage{amsthm}
\usepackage{graphicx}
\usepackage{mathtools}
\usepackage{balance}
\usepackage{bm}
\usepackage{colortbl}
\usepackage{lipsum}      
\usepackage{changepage}
\usepackage{caption}
\usepackage{color}
\usepackage{pgf,tikz}
\usepackage{balance}
\usepackage{mathtools}
\usepackage{bbm}
\usepackage{array}
\usepackage{relsize}
\usepackage{cite}
\usepackage{amsthm}
\usepackage{verbatim}
\usepackage{epstopdf}
\usepackage{array}
\usepackage{url}
\usepackage{stfloats}
\usepackage[hidelinks]{hyperref}
\usepackage{url}
\usepackage[linesnumbered,ruled]{algorithm2e}
\usepackage{algpseudocode}
\newtheorem{theorem}{Theorem}
\newtheorem{corollary}{Corollary}
\newtheorem{proposition}{Proposition}
\newtheorem{lemma}{Lemma}
\newtheorem{remark}{Remark}

\newtheorem{definition}{Definition}

\newtheorem{problem}{Problem}

\newcommand{\eqdef}{\mathrel{\mathop:}=}
\newcommand*{\QEDclosed}{\hfill\ensuremath{\blacksquare}}

\DeclareMathOperator*{\argmin}{arg\,min}

\usepackage{tikz,amsmath}
\usetikzlibrary{arrows, decorations.markings}
\usetikzlibrary{positioning,fit,calc} % used for the efficient working of the positioning system  
\tikzset{block/.style={draw, text width=1.5cm, minimum height=1cm, align=center,node distance = 0.5cm},   
	input/.style    = {coordinate}, % Input
	output/.style   = {coordinate}, % Output  
	line/.style={-latex,node distance = 0.25cm}   % the lesser the width the greater will be the diagram window  
}
\tikzset{block2/.style={draw, text width=0.9cm, minimum height=1.2cm, align=center,node distance = 0.9cm},   
	input/.style    = {coordinate}, % Input
	output/.style   = {coordinate}, % Output  
	line/.style={-latex,node distance = 0.25cm}   % the lesser the width the greater will be the diagram window  
}

\begin{document}

\title{Tutorial-Cum-Survey on Semantic and Goal- Oriented Communication: Research Landscape, Challenges, and Future Directions}
% Alternative title - Probabilistic Impossibility: A Deep ReLU Neural Networks That Provably Fail to Learn an Optimum Detector

\author{Tilahun~M.~Getu,~\IEEEmembership{Member,~IEEE},~Georges~Kaddoum,~\IEEEmembership{Senior Member,~IEEE},~and~Mehdi~Bennis,~\IEEEmembership{Fellow,~IEEE}
	 % <-this % stops a space
	\thanks{\IEEEcompsocthanksitem T. M. Getu is with the Electrical Engineering Department, \'Ecole de Technologie Sup\'erieure (ETS), Montr\'eal, QC H3C 1K3, Canada. He was with the Communications Technology Laboratory, National Institute of Standards and Technology (NIST), Gaithersburg, MD 20899, USA and the Electrical Engineering Department, ETS, Montr\'eal, QC H3C 1K3, Canada (e-mail: tilahun-melkamu.getu.1@ ens.etsmtl.ca).}

	\thanks{\IEEEcompsocthanksitem G. Kaddoum is with the Electrical Engineering Department, \'Ecole de Technologie Sup\'erieure (ETS), Montr\'eal, QC H3C 1K3, Canada, and the Cyber Security Systems and Applied AI Research Center, Lebanese American University, Beirut, Lebanon (e-mail: georges.kaddoum@etsmtl.ca).}

	\thanks{\IEEEcompsocthanksitem M. Bennis is with the Centre for Wireless Communications, University of Oulu, 90570 Oulu, Finland (e-mail: mehdi.bennis@oulu.fi).}

	\thanks{\IEEEcompsocthanksitem This research was supported by the U.S. Department of Commerce and its agency NIST.}

}

\IEEEtitleabstractindextext{%
\begin{abstract}
Amid the global rollout of fifth-generation (5G) wireless communication system services, researchers in academia, industry, and national laboratories have been developing proposals and roadmaps for the sixth generation (6G). Despite the many 6G proposals and roadmaps put forward, the materialization of 6G as presently envisaged is fraught with many fundamental interdisciplinary, multidisciplinary, and transdisciplinary (IMT) challenges. To alleviate some of these challenges, semantic communication (SemCom) and goal-oriented SemCom (effectiveness-level SemCom) have emerged as promising technological enablers of 6G. SemCom and goal-oriented SemCom are designed to transmit only semantically-relevant information and hence help to minimize power usage, bandwidth consumption, and transmission delay. Consequently, SemCom and goal-oriented SemCom embody a paradigm shift that can change the status quo that wireless connectivity is an opaque data pipe carrying messages whose context-dependent meaning and effectiveness have been ignored. On the other hand, 6G is critical for the materialization of major SemCom use cases (e.g., machine-to-machine SemCom) and major goal-oriented SemCom use cases (e.g., autonomous transportation). The paradigms of \textit{6G for (goal-oriented) SemCom} and \textit{(goal-oriented) SemCom for 6G} call for the tighter integration and marriage of 6G, SemCom, and goal-oriented SemCom. To facilitate this integration and marriage of 6G, SemCom, and goal-oriented SemCom, this comprehensive tutorial-cum-survey paper first explains the fundamentals of semantics and semantic information, semantic representation, theories of semantic information, and definitions of semantic entropy. It then builds on this understanding and details the state-of-the-art research landscape of SemCom and goal-oriented SemCom in terms of their respective algorithmic, theoretical, and realization research frontiers. This paper also exposes the fundamental and major challenges of SemCom and goal-oriented SemCom, and proposes novel future research directions for them in terms of their aforementioned research frontiers. By presenting novel future research directions for SemCom and goal-oriented SemCom along with their corresponding fundamental and major challenges, this tutorial-cum-survey article duly stimulates major streams of research on SemCom and goal-oriented SemCom theory, algorithm, and implementation for 6G and beyond.        
\end{abstract}

\begin{IEEEkeywords}
6G, SemCom, goal-oriented SemCom, semantics and semantic information, research landscape of SemCom, fundamental challenges of SemCom, future directions of SemCom, research landscape of goal-oriented SemCom, fundamental challenges of goal-oriented SemCom, future directions of goal-oriented SemCom.
\end{IEEEkeywords}
}

\maketitle

\IEEEdisplaynontitleabstractindextext
\IEEEpeerreviewmaketitle

\section{Introduction}
\label{sec: introduction}
\subsection{Motivation}
\label{subsec: motiv}
In tandem with the worldwide rollout of fifth-generation (5G) wireless communication system services, researchers in academia, industry, and national laboratories have been developing visions \cite{Saad_6G_Vision_20,Letaief_Edge_AI_Vision'22,Alwis_Survey_GG_Networks'21,Alsabah_6G_Wireless_Commun_Network'21,You_Towards_6G'21,6G_mailbox_theory'21,Road_towards_6G'21,Roadmap_6G_Privacy_and_Security'21,6G_Ecosystem'21,Survey_6G_Networks'21,Akyildiz_6G_and_Beyond'20,Dang_Alouini_6G_20,Lu_6G_survey_20,Yaacoub_PIEEE_20,zhao2019survey_IRS_19,Chowdhury_6G_2020,Viswanathan_6G_2020,Bariah_6G_2020,Tataria_6G_Wire_Systems'21,Uusitalo_6G_Hexa-X'21,Fettweis_6G_Per_Tactile_Internet'21,De_Lima_6G'21,Khan_6G_20,Rappaport_wires_commun_above_100GHz,Chen_6G_2020,Shaping_Future_6G_Networks'22} of the forthcoming wireless communication technology colloquially known as the sixth generation (6G). 6G as it is envisioned nowadays is driven by a variety of anticipated applications as diverse as: 
\begin{itemize}
	\item Multi-sensory extended reality (XR) applications, connected robotic and autonomous systems, wireless brain--computer interactions, and blockchain and distributed ledger technologies \cite{Saad_6G_Vision_20}.
	
	\item Haptic communication, massive internet of things (IoT) \cite{IoT_Connectivity_in_6G'21}, and massive IoT-integrated smart cities, and automation and manufacturing \cite{Spec_stu_6G_19}.
	
	\item Accurate indoor positioning, new communication terminals, high-quality communication services on board aircraft, worldwide connectivity and integrated networking, communications that support industry verticals \cite{RWH_6G_VTM_19}, holographic and tactile communications, and human bond communications \cite{Dang_Alouini_6G_20}.
	
	\item Industrial IoT \cite{Gui_6G_20}, internet of robots \cite{Chen_6G_2020}, flying vehicles \cite{Bariah_6G_2020}, and wireless data centers \cite{Bariah_6G_2020,Wireless_DCN_18}.
	
	\item Smart Grid 2.0, Industry 5.0, personalized body area networks, Healthcare 5.0; internet of industrial smart things, and internet of healthcare \cite{Alwis_Survey_GG_Networks'21}.
	
	\item The internet of no things (the Metaverse) \cite{InoT_Maier'20,Edge_Enabled_Metaverse'22}.
\end{itemize} 
These applications are motivated by numerous trends and use cases that drive 6G.

The trends and use cases that have been proffered to date are as heterogeneous as: 
\begin{itemize}
	\item The emergence of smart surfaces and environments, the massive availability of small data, from self-organizing networks to self-sustaining networks, the convergence of 3CLS (communications, computing, control, localization, and sensing), and the end of the smartphone era \cite{Saad_6G_Vision_20}.
	
	\item The use of mobile edge, cloud, and fog computing, intelligent distributed computing and data analytics, and dynamic infrastructure \cite{MMMLA18}.
	
	\item Artificial intelligence (AI)-enabled autonomous wireless networks and the convergence of intelligent sensing, communication, computing, caching, and control \cite{Zhang_6G_networks_2019}.
	
	\item Multi-sensory holographic teleportation, real-time remote healthcare, autonomous cyber-physical systems, intelligent industrial automation, high-performance precision agriculture, space connectivity, and smart infrastructure and environments \cite{Akyildiz_6G_and_Beyond'20}.
	
	\item Digital twinning \cite{Viswanathan_6G_2020,Digital_Twin'20,Digital_Twin_Survey'19,Digital_Twin_in_Industry'19,Digital_Twin_Enabling_Techs'20}.
	
	\item Knowledge systems, ubiquitous universal computing, man-machine interfaces, and the three dimensions of data, energy, and computing \cite{Viswanathan_6G_2020}.
	
	\item Space exploration, travel by air and sea, maglev transportation, intelligent driving, the internet of vehicles, and 6C (capturing, communication, caching, cognition, computing, and controlling) functions \cite{Gui_6G_20}. 
	
	\item Ubiquitous super 3D connectivity \cite{Chowdhury_6G_2020}; 
	
	\item The 3C (user-centralized, content-centralized, and data-centralized) paradigm \cite{Chen_6G_2020}.
	
	\item Intelligent vehicle-to-everything \cite{Tang_SVT_6G_2019}, collaborative robots (CoBots) \cite{Spec_stu_6G_19}.
	
	\item Huge scientific data applications, application-aware data burst forwarding, emergency and disaster rescue, and socialized internet of things \cite{ITU-T_additional_representative_use_cases}.
	
	\item Globalized ubiquitous connectivity, enhanced on-board connectivity, and pervasive intelligence \cite{Road_towards_6G'21}.
	
	\item Increasing elderly population and gadget-free communication \cite{Alwis_Survey_GG_Networks'21}.
		
	\item Intelligent Internet of medical things \cite{6G_Vision_Int_Healthcare'20}.
\end{itemize}
These trends and use cases that drive 6G are believed to be achievable with the materialization of numerous 6G technology enablers.   

The 6G technology enablers -- that are envisaged for 6G broadband access quantified by KPI (key performance indicator) impact on system capacity, system latency, and system management \cite{Akyildiz_6G_and_Beyond'20} -- can be categorized as infrastructure-level enablers, spectrum-level enablers, and algorithm/protocol-level enablers \cite{Scoring_Terabit_per_second_goal'20,6G_BBC_20_White_Paper}. The algorithm/protocol-level enablers put forward by various researchers are edge AI \cite{Letaief_Edge_AI_Vision'22,Saad_6G_Vision_20,Xiao_Edge_Intel_20,Zhou_Edge_AI'19,Park_Commun_Efficient_Dis_Learning'21}; semantic communications \cite{ECSSCD_6G_19,Strinati_Beyond_Shannon'20,SemCom_for_6G_Future_Internet'22}; pervasive AI \cite{Spec_stu_6G_19}; orbital angular momentum (OAM) multiplexing \cite{Zhang_6G_networks_2019}; ubiquitous sensing \cite{MMMLA18}; ultra-low-latency communications \cite{MMMLA18}; network harmonization and interoperability, intelligent proactive caching and mobile edge computing (MEC), multi-objective optimization and routing optimization, massive IoT and big data analytics, configurable multi-antenna systems, and intelligent cognitive radio and self-sustaining wireless networks \cite{SSSMM19}; blockchain-based spectrum sensing and molecular communications \cite{Zhang_6G_networks_2019}; intelligent radio, AI-enabled closed-loop optimization, intelligent wireless communication, and hardware-aware communications \cite{Letaief_6G_Roadmap_2019}; symbiotic radio and super IoT \cite{Zhang_6G_visions_China_Com_2019}; AI and photonics-based cognitive radio \cite{Zong_6G_Technologies_19}; multi-mode multi-domain joint transmission and intelligent transmission \cite{Yang_6G_Wirels_commun_19}; autonomous wireless systems with AI \cite{Gacanin_Auto_Wireless_with_AI_19}; model-aided wireless AI \cite{AZa_WND_19,Zappone_Mod_Aid_Wireless_AI_19}; data-oriented transmission \cite{Yang_Data_Oriented_trans_2019}; device-centric wireless communications and demand-driven opportunistic networking \cite{Coll-Perales_5G_and_beyond_19}; delta-orthogonal multiple access \cite{Al-Eryani_D-OMA_19}; coded caching and rate splitting \cite{6G_BBC_20_White_Paper}; ambient backscatter communication \cite{Akyildiz_6G_and_Beyond'20}; disaggregation and virtualization \cite{Giordani_To_6GNs_20}; intelligent device-to-device communication \cite{Zhang_D2D_20}; index modulation \cite{basar2019reconfigurable}; self-driving networks \cite{Akyildiz_6G_20,Self_driving_nets_18}; AI/machine learning (ML)-driven air interface design and optimization, and networking with the sixth sense \cite{Viswanathan_6G_2020}; the seamless integration of wireless information and energy transfer \cite{Chowdhury_6G_2020}; tactile internet, multi-access edge computing \cite{Bariah_6G_2020}; and intelligent internet of intelligent things \cite{Peltonen_2020_6G}. A number of widely recognized spectrum-level enablers of 6G have also been proposed: above 6 GHz for 6G (from small cells to tiny cells) and transceivers with integrated surfaces \cite{Saad_6G_Vision_20}; the holistic management of communication, computation, caching, and control resources \cite{ECSSCD_6G_19}; superfast wireless broadband connectivity \cite{MMMLA18}; multi-band ultrafast-speed transmission \cite{Yang_6G_Wirels_commun_19}; an all-spectrum reconfigurable front end for dynamic spectrum access \cite{Akyildiz_6G_and_Beyond'20}; and optical wireless communications \cite{Scoring_Terabit_per_second_goal'20}. These spectrum-level enablers are complemented -- in view of the advent of diverse 6G services\footnote{The 6G service classes foreseen thus far are: mobile broadband reliable low-latency communication, massive ultra-reliable low-latency communication (URLLC), human-centric services, and multi-purpose 3CLS and energy services \cite{Saad_6G_Vision_20}; holographic communications, high-precision manufacturing, sustainable development and smart environments, and battery-free communication \cite{ECSSCD_6G_19}; computation-oriented communication, contextually agile enhanced mobile broadband (eMBB) communication, and event-defined URLLC \cite{Letaief_6G_Roadmap_2019}; ubiquitous mobile broadband, ultra-high-speed low-latency communication, and ultra-high data density \cite{Zong_6G_Technologies_19}; secure wireless computing for private data \cite{Gui_6G_20}; network-as-an intelligent-service \cite{Khan_6G_20}; and digital replica \cite{Samsung_6G}.} -- by many 6G infrastructure-level enablers.

The 6G infrastructure-level enablers that have been contemplated to date are communication with reconfigurable intelligent surface (RIS) \cite{Saad_6G_Vision_20}; integrated terrestrial, airborne, and satellite networks \cite{Saad_6G_Vision_20}; energy transfer and harvesting \cite{Saad_6G_Vision_20}; three-dimensional (3D) coverage \cite{ECSSCD_6G_19}; cell-free networks, metamaterials-based antennas, fluid antennas, software-defined materials, programmable metasurfaces, and wireless power transfer and energy harvesting \cite{Spec_stu_6G_19}; tiny-cell communication and cell-free communications \cite{SSSMM19}; supermassive multiple-input multiple-output (MIMO), large intelligent surfaces, and holographic beamforming \cite{Zhang_6G_networks_2019}; satellite-assisted IoT communications \cite{Zhang_6G_visions_China_Com_2019}; holographic radio and photodiode-coupled antenna arrays \cite{Zong_6G_Technologies_19}; multi-purpose converged, full-spectral, and all-photonic radio access networks (RANs) \cite{Zong_6G_Technologies_19}; hyperspectral space-terrestrial integration networks \cite{Zong_6G_Technologies_19}; super flexible integrated networks \cite{Yang_6G_Wirels_commun_19}; airplane-aided integrated networking \cite{Huang_Integ_netw_for_6G_wirels_19}; extremely large aperture arrays, holographic massive MIMO, six-dimensional positioning, large-scale MIMO radar, and intelligent massive MIMO \cite{Bjrnson2019Massive_MIMO}; integrated access and backhaul networks \cite{6G_BBC_20_White_Paper,Scoring_Terabit_per_second_goal'20}; internet of space things \cite{Akyildiz_6G_and_Beyond'20}; antenna-on-glass \cite{Viswanathan_6G_2020}; and zero-touch networks \cite{Alwis_Survey_GG_Networks'21}. In light of these many infrastructure-level enablers and the aforementioned spectrum-level and algorithm/protocol-level enablers, realizing 6G as it is currently imagined needs both an evolutionary and a revolutionary paradigm shift \cite{Saad_6G_Vision_20}.

A paradigm shift has to address the following fundamental challenges related to 6G \cite{Getu_Metrics_of_SemCom_and_GO_SemCom_2023}:
\begin{itemize}
	
\item Managing ultra-heterogeneity.

\item Guaranteeing an ultra-high data rate for most users.

\item Ensuring ultra-reliability and low latency for most users.

\item Taming ultra-high complexity in 6G networks.

\item Incorporating various KPIs in the design.

\item Addressing ultra-high mobility.

\item Being highly energy efficient.

\item Supporting energy-efficient AI.

\item Accommodating users' needs or perspectives \cite{KDHB_18}.

\item Ensuring security, privacy, and trust. 

\item Attaining \textit{full intelligence and autonomy}.

\item Coping with the inevitable technological uncertainty associated with 6G technology enablers \cite{Latvaaho2019KeyDAC}.

\end{itemize}
Addressing the fundamental challenges listed amounts to overcoming numerous interdisciplinary, multidisciplinary, and transdisciplinary (IMT) challenges that are intertwined with several technological challenges. To alleviate these challenges, 6G systems and networks should be holistically designed to minimize power usage, bandwidth consumption, and transmission delay by reducing to a minimum the transmission of information that is \textit{semantically redundant or irrelevant}. Semantic-centric information transmission calls for the efficient transmission of \textit{semantics} (meaning) by a semantic transmitter followed by faithful recovery by a semantic receiver. This communication paradigm is now widely recognized as semantic communication (SemCom).\footnote{Throughout this tutorial-cum-survey paper, the acronym SemCom stands for \textit{wireless SemCom}. When we discuss SemCom in the optical and quantum domains, we will explicitly refer to it as \textit{optical SemCom} and \textit{quantum SemCom}, respectively. For further information about all these types of SemCom, the reader is referred to the survey in \cite{Getu_Metrics_of_SemCom_and_GO_SemCom_2023}.} 

SemCom has the potential to change the status quo perception that wireless connectivity is an opaque data pipe carrying messages whose context-dependent \textit{meaning} and \textit{effectiveness} (or goal) have been ignored \cite{Seo_Sem_Naive_Com_with_Contextual_Reasoning'21}, as the designers of traditional communication systems have viewed it \cite{Kountouris_Semantics_EmpoweredCF'21}. In stark contrast to traditional communication systems that aim to offer a high data rate and a low symbol (bit) error rate, SemCom focuses on extracting the meaning of information transmitted by a source and interpreting the semantic information at the destination \cite{Luo_SemCom_Overview'22}. SemCom's chief objective is to convey the intended meaning, which depends on not only the physical content of the message, but also the sender's personality, intention, and other human-oriented factors that could reflect the real quality of experience (QoE) (i.e., the subjective experience) of human users \cite{Shi_From_SemCom_to_Sematic-aware_Networking'20}. SemCom's fundamental goal is therefore to ensure the successful delivery of the transmitted information's representative meaning \cite{Zhang_Wisdom_Evolutionary_6G'21}. Toward a meaning-centric communication system design, SemCom can be designed to focus on conveying the \textit{interpretation} of transmitted messages instead of an exact or approximate reproduction of them to achieve a meaning-centric communication system \cite{Bao_Towards_Theory_SemCom'11}.

In addition to minimizing the divergence between the meaning of the transmitted messages and the meaning inferred from the recovered messages \cite{Tong_FL_ASC'21}, SemCom aims to transmit only the semantic information that is relevant to the communication goal, thereby significantly reducing data traffic \cite{Xie_DL-based_SemCom'21}. To this end, it transmits fewer data than traditional communication techniques \cite{Tong_FL_ASC'21}. In contrast to those techniques focusing on mere data reconstruction, SemCom takes its inspiration from human-to-human communication, whose goal is understanding and delivering the meaning behind a message \cite{LLL_Reasoning_Based_SemCom'22}. To deliver the meaning behind a message, SemCom is designed as a system that attempts to communicate the true meaning of a message rather than ensuring the exact replication of the information transmitted by a source \cite{Lu_RL-powered_SemCom'21}. What matters in the design of SemCom is the source data's semantic content rather than the source’s average probabilistic information \cite{Strinati_Beyond_Shannon'20}. Via a semantic content transmission relevant only for accurate interpretation at the destination, SemCom can promote the effective utilization of available network capacity by transmitting semantic content relevant only for accurate interpretation at the destination  \cite{Kalfa_Toward_GO_Semantic_Signal_Processing'21}. It is possible to make effective use of a network's capacity by avoiding the bit-by-bit reconstruction of the transmitted information at the receiver. To this end, SemCom uses a semantic encoder to incorporate the purpose of transmission, simplify the data to be transmitted, and eliminate the transmission of redundant information \cite{Zhou_WiCom_Letters'22}. More specifically, it embodies the \textquotedblleft provisioning of the right and significant piece of information to the right point of computation (or actuation) at the right point in time'' \cite{SemCom_Net_Systems'21}. This design philosophy is of paramount importance for networked systems and intelligence tasks.   

When it comes to intelligence tasks such as speech recognition and speech transmission, SemCom extracts and transmits only task-related features -- and discards the irrelevant ones -- thereby considerably reducing bandwidth consumption \cite{Weng_SemCom_Speech_Recognition'21,Weng_SemCom_Sys_Speech_Trans'21}. The significant reduction in bandwidth consumption that SemCom achieves represents a paradigm shift from \textquotedblleft how to transmit'' to \textquotedblleft what to transmit'' \cite{Iyer_Sem_Com_4_Int_Wireless_Networking'22}. When it comes to \textquotedblleft how to transmit,'' it is worth noting that conventional wireless communication systems have steadily approached the Shannon limit \cite{Strinati_Beyond_Shannon'20}. Consequently, a breakthrough must be made to be able to support the unparalleled proliferation of mobile devices, the insatiable desire for high data rates, and the emergence of new and highly heterogeneous 6G use cases and applications \cite{Cognitive_SemCom_Systems'22}. To these ends, SemCom is emerging as a promising communication paradigm for the design, analysis, and optimization of 6G networks and 6G systems, and a possibly revolutionary one at that. This necessitates our underneath discussion on why SemCom is for 6G.

\subsection{Why SemCom for 6G?}
\label{subsec: Why_SemCom}
\begin{figure*}[t!]
	\centering
	\includegraphics[scale=0.52]{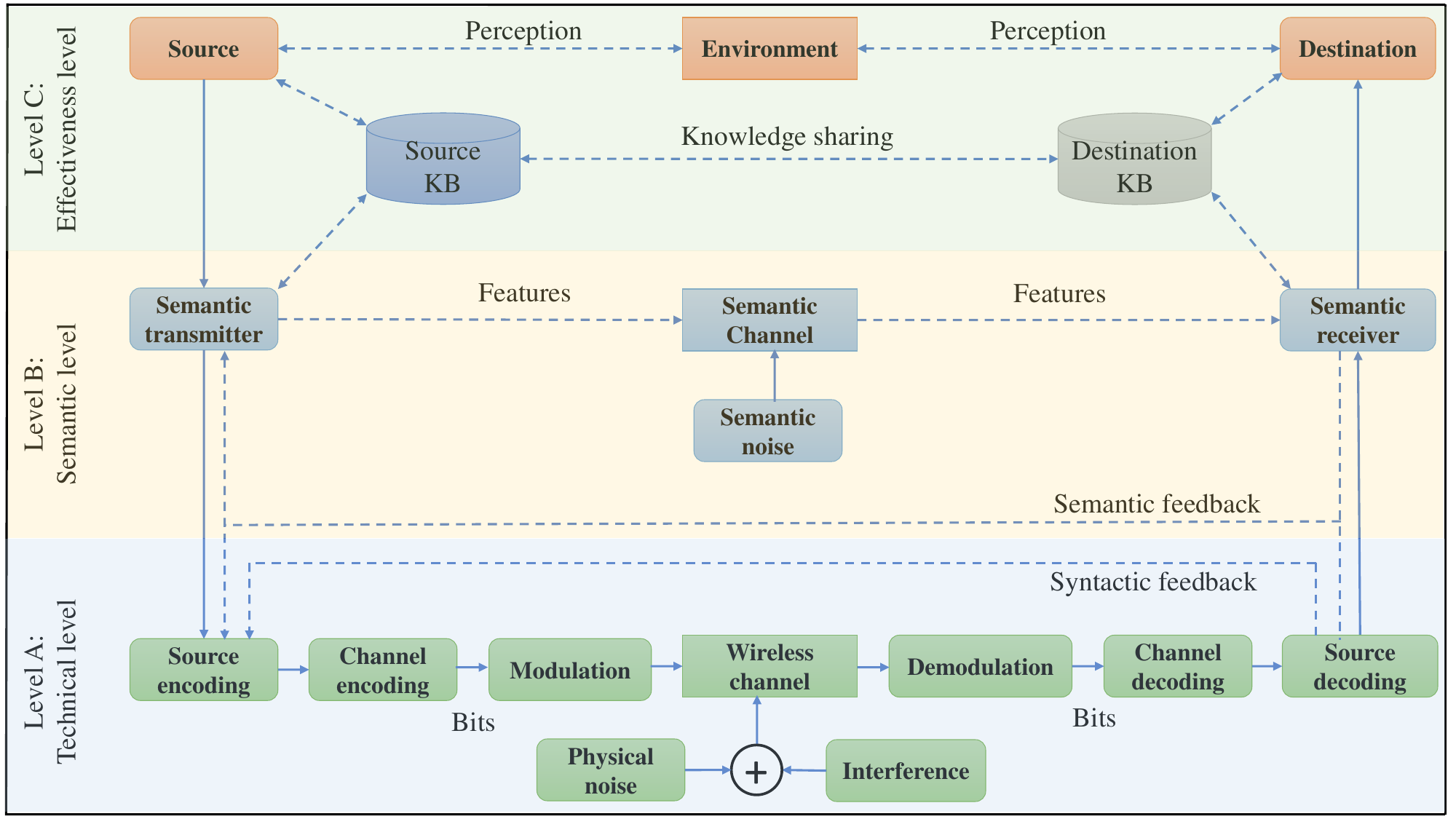}  \\ [4mm]
	\caption{Weaver's three levels of communication \cite[FIGURE 1]{Luo_SemCom_Overview'22} -- KB: knowledge base.}
	\label{fig: Three_lavels_of_communication}  
\end{figure*}

Although SemCom is a promising paradigm shift for the design, analysis, and optimization of 6G networks and 6G systems, it is a classic communication paradigm that was first proposed around 1950. At that time, the notion and relevance of SemCom were pointed out by Weaver \cite[Ch. 1]{Shannon_Weaver_Math_Theory_Commun'49}, even though Shannon deliberately ignored\footnote{\textquotedblleft  Frequently the messages have meaning; that is they refer to or are correlated according to some system with certain physical or conceptual entities. These semantic aspects of communication are irrelevant to the engineering problem'' \cite[p. 1]{Shannon_Math_Theory_of_Communication'1948}.} the semantic aspects of communication in his classic masterpiece \cite{Shannon_Math_Theory_of_Communication'1948}. Weaver envisioned communication using semantics and outlined three hierarchical levels of communication (see Fig. \ref{fig: Three_lavels_of_communication}) that fundamentally differentiate the broad subject of communication \cite[p. 4]{Shannon_Weaver_Math_Theory_Commun'49}:  
\begin{itemize}
	\item \textquotedblleft \textbf{Level A}. How accurately can the symbols of communication be transmitted? (The technical problem). 
	\item \textquoteleft \textbf{Level B}. How precisely do the transmitted symbols convey the desired meaning? (The semantic problem). 
	\item \textquoteleft \textbf{Level C}. How effectively does the received meaning affect conduct in the desired way? (The effectiveness problem).'' 
\end{itemize}

\textit{The technical problem} focuses on the accurate transmission of a message's symbols (bits) and is driven by Shannon information theory (see Fig. \ref{fig: Three_lavels_of_communication}), which is a classical information transmission theory. This theory has steered the design of generations of communication systems and resulted in wireless connectivity being viewed as an opaque data pipe carrying messages \cite{Kountouris_Semantics_EmpoweredCF'21} -- whose context-dependent meaning and effectiveness have been ignored  -- and the sender and receiver are considered agents without intelligence \cite{Shi_to_Semantic_Fidelity'21}. In this old paradigm, a huge amount of semantically irrelevant and redundant data are transmitted, and an enormous amount of communication resources such as transmission power and bandwidth are consumed to do so \cite{Shi_to_Semantic_Fidelity'21}. Aside from the bandwidth and power expenditure, acquiring, processing, and sending an excessive amount of distributed real-time data -- that is likely to be useless to the end users or outdated by the time they reach them -- will produce communication bottlenecks, and increase latency and safety issues in emerging cyber-physical and autonomous networked systems \cite{Kountouris_Semantics_EmpoweredCF'21}. Accordingly, communication system designers may need to look beyond the technical problem for a paradigm where communication in itself is a means to achieving specific goals rather than an end goal \cite{Kountouris_Semantics_EmpoweredCF'21,Juba_Universal_SemCom'08}. To achieve particular goals, how precisely the transmitted symbols convey the desired meaning (i.e., \textit{the semantic problem} in Fig. \ref{fig: Three_lavels_of_communication}) needs to be factored in. To this end, it is useful to design an \textit{understand-first-and-then-transmit system} that can accomplish joint optimization \cite{Shi_to_Semantic_Fidelity'21}.   

\begin{figure*}[htb!]
	\centering
	\includegraphics[scale=0.52]{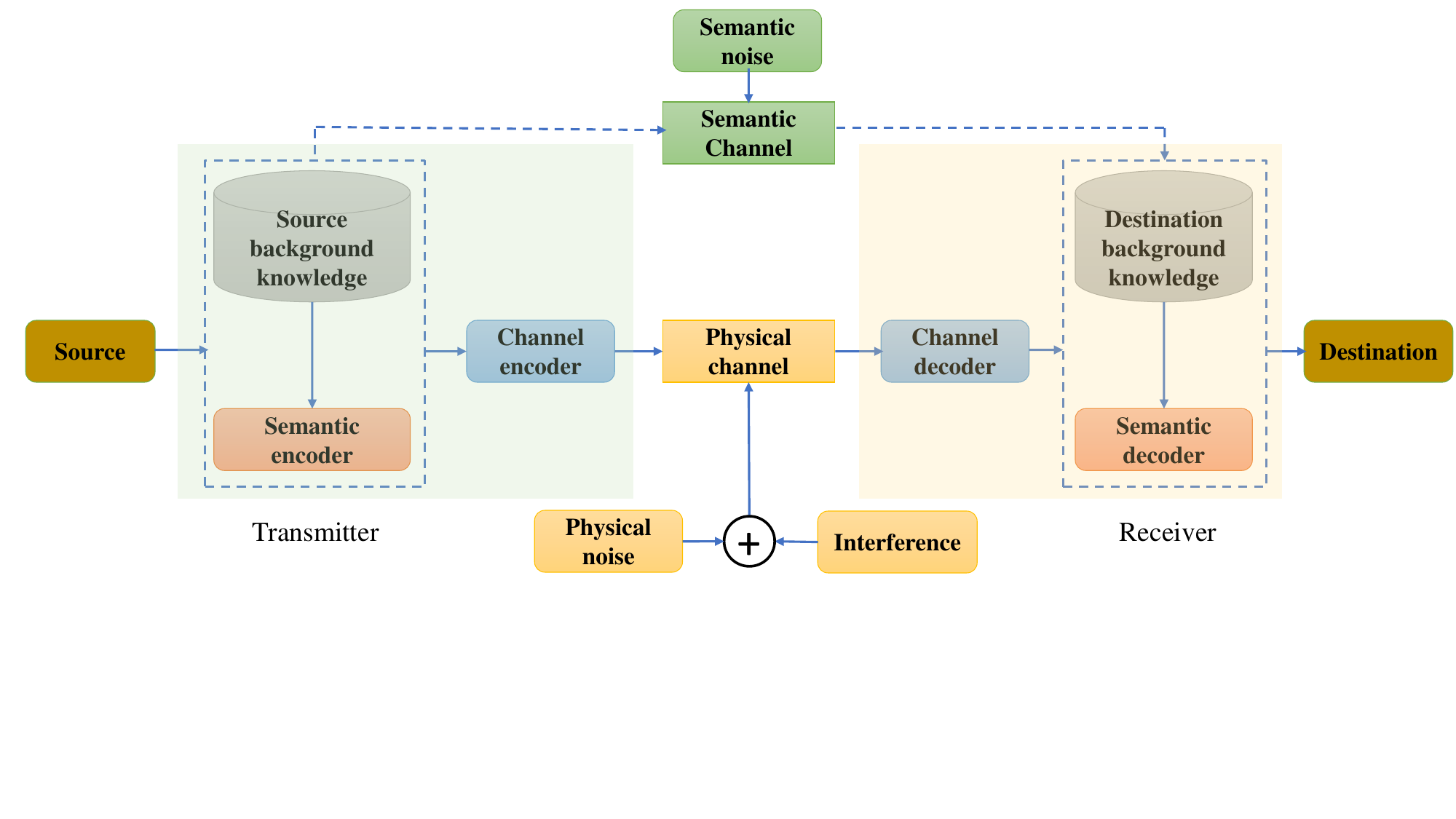}  \vspace{-2.5cm} 
	\caption{The main components in a SemCom system -- modified from \cite[Fig. 3]{Qin_Sem_Com_Principles_Apps'22}.}
	\label{fig: SemCom_main_components}
\end{figure*}

SemCom interprets information at its semantic level -- rather than by bit sequence \cite{Bar_Carnap_Theory_SemInfo'1954} -- and seeks the meaning behind the transmitted symbols (bits) \cite{Luo_SemCom_Overview'22}. For example, when a transmitter dispatches \textquotedblleft I have just brought a yellow banana,'' a receiver may receive a message like \textquotedblleft I have brought a banana'' \cite{Lu_RL-powered_SemCom'21}. Although this message is not exactly what was conveyed by the transmitter, one can still understand the overall idea. However, if a receiver receives a message like \textquotedblleft I have just brought a yellow banner'' \cite{Lu_RL-powered_SemCom'21} (which has a lower bit-level error rate), the meaning behind this received message is quite different from the transmitted message's encoded meaning, which is naturally undesirable. Semantic-level communication, a.k.a. SemCom, is crucial for meaning-level-accurate transmission and reception. Fig. \ref{fig: SemCom_main_components} schematizes SemCom between a semantic transmitter and a semantic receiver over a virtual semantic channel using features that encode semantic information. To encode semantic information at the transmitter, a system designer would exploit the source's background knowledge base (KB). Despite some fundamental challenges, as shown in Fig. \ref{fig: SemCom_main_components}, the source KB is shared with the destination KB, which is then used when the decoded message is interpreted by the semantic receiver. The semantic receiver's interpretation can be drastically affected by semantic noise, which causes semantic information to be misunderstood and semantic decoding errors to occur. This will generate a misleading between the transmitter's intended meaning and the reconstructed meaning of the receiver \cite{Hu_Robust_SemCom_with_Masked_VQ-VAE'22}. Semantic noise happens naturally in SemCom over a semantic channel (see Fig. \ref{fig: SemCom_main_components}) and can be caused by various factors:
\begin{itemize}
	\item Multiple possible interpretations of the recovered symbols at the technical level due to ambiguity in some words, sentences, or symbols used in the dispatched message \cite{Luo_SemCom_Overview'22,Qin_Sem_Com_Principles_Apps'22}.  
	
	\item Semantic ambiguity\footnote{Semantic ambiguity can arise from \textit{dialect} and \textit{polysemy}. Polysemy involves an instance of a word (or phrase) being employed to convey two or more different meanings in different contexts \cite{Shi_to_Semantic_Fidelity'21}.  Dialect, on the other hand, concerns the variant of a language that is understood mainly by a specific group of speakers of a given language \cite{Shi_to_Semantic_Fidelity'21}.} -- which is common in natural language processing (NLP) \cite{Qin_Sem_Com_Principles_Apps'22} -- in the reconstructed symbols when multiple sets of data with different meanings are represented by the same semantic symbol \cite{Shi_to_Semantic_Fidelity'21}. 
	
	\item Adversarial examples (possibly created with adversarial noise \cite{Goodfellow_Adversarial_Examples'14}) that can detrimentally mislead a semantic decoder in its interpretation (see Fig. \ref{fig: SemCom_main_components}), especially in deep learning (DL)-based SemCom \cite{Qin_Sem_Com_Principles_Apps'22,Hu_Robust_SemCom'22,Hu_Robust_SemCom_with_Masked_VQ-VAE'22}.
	
	\item Interference (a jamming signal) emitted by a malicious attacker that is received by the antenna transmitting the signal of interest \cite{Hu_Robust_SemCom_with_Masked_VQ-VAE'22}.
	
	\item A disturbance in the estimated \textit{syntactic information}\footnote{Since it is specific in nature, syntactic information can be produced directly through a subject’s sensing function \cite{hong_Theory_Semantic_Info'17,Zhong_Theory_Sem_Information_Book_Chapter}.} -- resulting in symbol/bit-level errors -- due to considerable physical noise, signal fading, and/or wireless interference \cite{Qin_Sem_Com_Principles_Apps'22,Shi_to_Semantic_Fidelity'21,Luo_SemCom_Overview'22}.

	\item A mismatch between the source KB and the destination KB (even in the absence of syntactic errors) \cite{SemCom_for_6G_Future_Internet'22}.
\end{itemize}

Substantial semantic noise can impede SemCom's faithfulness and applications, and cause semantic errors. In the SemCom model in \cite[Fig. 2]{Bao_Towards_Theory_SemCom'11}, a semantic error occurs if the message to be sent is \textquotedblleft true'' at the source with respect to (w.r.t.) the transmitter's world model, background KB, and inference procedure, but \textquotedblleft false'' at the destination w.r.t. the receiver's world model, background KB, and inference procedure \cite{Bao_Towards_Theory_SemCom'11}. Semantic errors\footnote{Identifying those semantic errors that are caused by physical noise, interference, or wireless channel fading, and those that are the consequence of a mismatch between the source KB and destination KB is fundamentally challenging.} can occur due to symbol/bit-level errors during transmission as a result of physical noise, interference, or wireless channel fading \cite{Luo_SemCom_Overview'22}, or be caused by a mismatch between the source KB and destination KB \cite{Luo_SemCom_Overview'22,Shi_to_Semantic_Fidelity'21}. A recurring mismatch between the source KB and destination KB can be corrected each time using semantic feedback (see Fig. \ref{fig: Three_lavels_of_communication}) for robust SemCom. A system designer may design a SemCom system with semantic feedback and syntactic feedback (see Fig. \ref{fig: Three_lavels_of_communication}) to ensure SemCom is robust. At the technical level, syntactic feedback can be incorporated between source decoding and source encoding. This feedback can improve the quality of decoded bits/symbols, which will in turn ensure minimal noise (i.e., semantic noise) at the semantic level. At the semantic level, semantic feedback can be incorporated between a semantic transmitter and a semantic receiver -- more specifically, between a semantic encoder and a semantic decoder (see Fig. \ref{fig: SemCom_main_components}) -- to alleviate the aforementioned background KB mismatch. Such a mismatch can happen in practice since the source and destination are always adding knowledge to their KBs and can be exploited to convey semantically-secure messages \cite{Choi_Unified_Approach'22,Choi_Unified_View_SemInfo'22}. Semantically-secure messages are difficult to decode (interpret) without the intended receiver’s KB \cite{Choi_Unified_Approach'22,Choi_Unified_View_SemInfo'22}, which paves the way for a \textit{secure-by-design} paradigm shift in 6G and beyond. 

When it comes to 6G communication and networking, secure \textit{human-to-human} (H2H), \textit{human-to-machine} (H2M), and \textit{machine-to-machine} (M2M) systems can be designed using SemCom \cite{Qiao_What_is_SemCom'21}. SemCom also has the potential to be a key enabler of 6G edge intelligence with efficient communication and computation overheads -- despite uncertain wireless environments and limited resources -- while overcoming the challenges faced by 6G communication networks \cite{Yang_SemCom_meets_Edge_Intelligence'22,SemCom_for_6G_Future_Internet'22}. SemCom therefore empowers 6G with the possibility of designing a variety of 6G systems that can benefit greatly from a system design that incorporates not only a semantic component but also an effectiveness component (see Fig. \ref{fig: Three_lavels_of_communication}). Concerning the latter, a communication system designer seeks to ensure the received meaning in affecting conduct in the desired way \cite[p. 4]{Shannon_Weaver_Math_Theory_Commun'49}. This design paradigm has inspired various 6G use cases based on goal-oriented SemCom, as discussed below.  

\subsection{Why Goal-Oriented SemCom for 6G?}
\label{subsec: Why_6G_for_GO_SemCom}  
SemCom deals with the transmission of complex data structures (e.g., features, patterns, and data lying on manifolds) or, in general, abstract concepts \cite{Zhang_Goal-Oriented_Commun'22}. SemCom in which the effectiveness of semantic transmission is explicitly defined and focused on can be qualified as a goal-oriented SemCom \cite{Zhang_Goal-Oriented_Commun'22}. Thus, SemCom is a broader concept than goal-oriented SemCom,\footnote{Throughout this tutorial-cum-survey paper, task-oriented communication and goal-oriented communication are detailed under the heading goal-oriented SemCom. Nevertheless, the authors of \cite{Sem_Empowered_Commun'22} underline that goal communication is much broader than SemCom, which they classify -- per Weaver’s vision -- as \textit{semantic-level SemCom} and \textit{effectiveness-level SemCom}. While the former emphasizes semantic transmission for data reduction and the delivery of the meaning behind the transmitted content, the latter focuses on effectively employing semantic information -- at a suitable time -- for successful task execution \cite{Sem_Empowered_Commun'22}. Moreover, we shall also underscore the view in the wireless communication research community that SemCom and goal-oriented SemCom are more or less different terminologies for the same thing.} since the semantics of information are not necessarily linked to a system's overarching goal \cite{Zhang_Goal-Oriented_Commun'22}. Per this view, goal-oriented SemCom is a subset of SemCom that takes a pragmatic approach to SemCom where the receiver is interested in the significance (semantics) of the source's transmitted message and the message's effectiveness in accomplishing a certain goal \cite{Zhang_Goal-Oriented_Commun'22}. To this end, goal-oriented SemCom is aimed at extracting and transmitting only task-relevant information so that the transmitted signal is substantially compressed, communication efficiency is improved, and low end-to-end latency is achieved \cite{Xie_Robust_IB'22}. Goal-oriented SemCom is therefore very useful for 6G since communication is not an end but a means to achieve specific goals \cite{Kountouris_Semantics_EmpoweredCF'21}.

A wide variety of 6G use cases such as autonomous transportation, consumer robotics, environmental monitoring, telehealth, smart factories, and networked control systems (NCSs) require ultra-low latency, very high reliability, and ultra-large transmission bandwidth. These stringent requirements can be met by transmitting only the information that is semantically relevant for the effective performance of the desired action. Consequently, goal-oriented SemCom is also quite crucial for designing and realizing 6G w.r.t. minimizing power usage, bandwidth consumption, and transmission delay while aiming to effectively achieve one or more goals. Therefore, the aforementioned 6G use cases are also promising of M2M goal-oriented SemCom, which will be vital for the design and realization of 6G like H2H goal-oriented SemCom and H2M goal-oriented SemCom.    

On the other hand, ongoing developments in SemCom, goal-oriented SemCom, and 6G are mutually reinforcing \cite{SemCom_for_6G_Future_Internet'22}. This justifies the need for the following discussion on why 6G is crucial for SemCom and goal-oriented SemCom.

\subsection{Why 6G for SemCom and Goal-Oriented SemCom?}
\label{subsec: Why_6G_for_SemCom}
Ongoing 6G developments are core enablers -- and hence huge opportunities -- for further development of SemCom and goal-oriented SemCom. SemCom and goal-oriented SemCom can hugely benefit from the emergence of AI-native networks, ubiquitous connectivity, and trustworthiness-native networks \cite{SemCom_for_6G_Future_Internet'22}. AI-native networks \cite{Shen_AI-assisted_Net'20,NeuroRAN_6G_21,Native_AI_6G_Nets'21} are emerging in 6G because of the following driving trends:
\begin{itemize}
	\item The migration of data processing from the network core to the network edge \cite{Wang_DL_Edge_Computing'20,Dustdar_Edge_AI'20,Zhou_Edge_AI'19}. 
	
	\item A fundamental change from cloud AI to distributed AI \cite{Native_AI_6G_Nets'21}.
	
	\item An emerging paradigm shift from connection-oriented communication to task-oriented communication \cite{Native_AI_6G_Nets'21} and computation-oriented communications \cite{Letaief_6G_Roadmap_2019}.
\end{itemize} 
These communication- and computation-oriented paradigm shifts enable ubiquitous connectivity, which will in turn promote the development of 6G wireless systems that are based on SemCom and goal-oriented SemCom. uch systems can be designed and optimized to materialize a global ubiquitous connectivity along the maturation of the following 6G technology enablers \cite{SemCom_for_6G_Future_Internet'22}:
\begin{itemize}
	\item Space-air-ground integrated network \cite{Liu_SAGIN_18,Kato_Opt_SAGN_19,Huang_Integ_netw_for_6G_wirels_19}.
	
	\item (sub-)Terahertz (THz) communications \cite{KMS_Hug_B5G_fast_netws_19,Rappaport_wires_commun_above_100GHz,Chaccour_THz_VR_2020,Chaccour_Fellowshop_Com_Sensing'22} (despite its major multi-faceted challenges \cite{Tekbiyik_PhyCom_2019}).
	
	\item Supermassive (ultra-massive) MIMO \cite{Bjrnson2019Massive_MIMO,Ammar_User_Centric_CF_M_MIMO'22}.
\end{itemize}
Moreover, the design of trustworthiness-native networks is going to be at the forefront of 6G research \cite{Roadmap_6G_Privacy_and_Security'21,Letaief_Edge_AI_Vision'22}, which will in turn facilitate the realization of SemCom and goal-oriented SemCom through secure-by-design 6G networks. 

\begin{figure*}[htb!]
	\centering
	\includegraphics[scale=0.8]{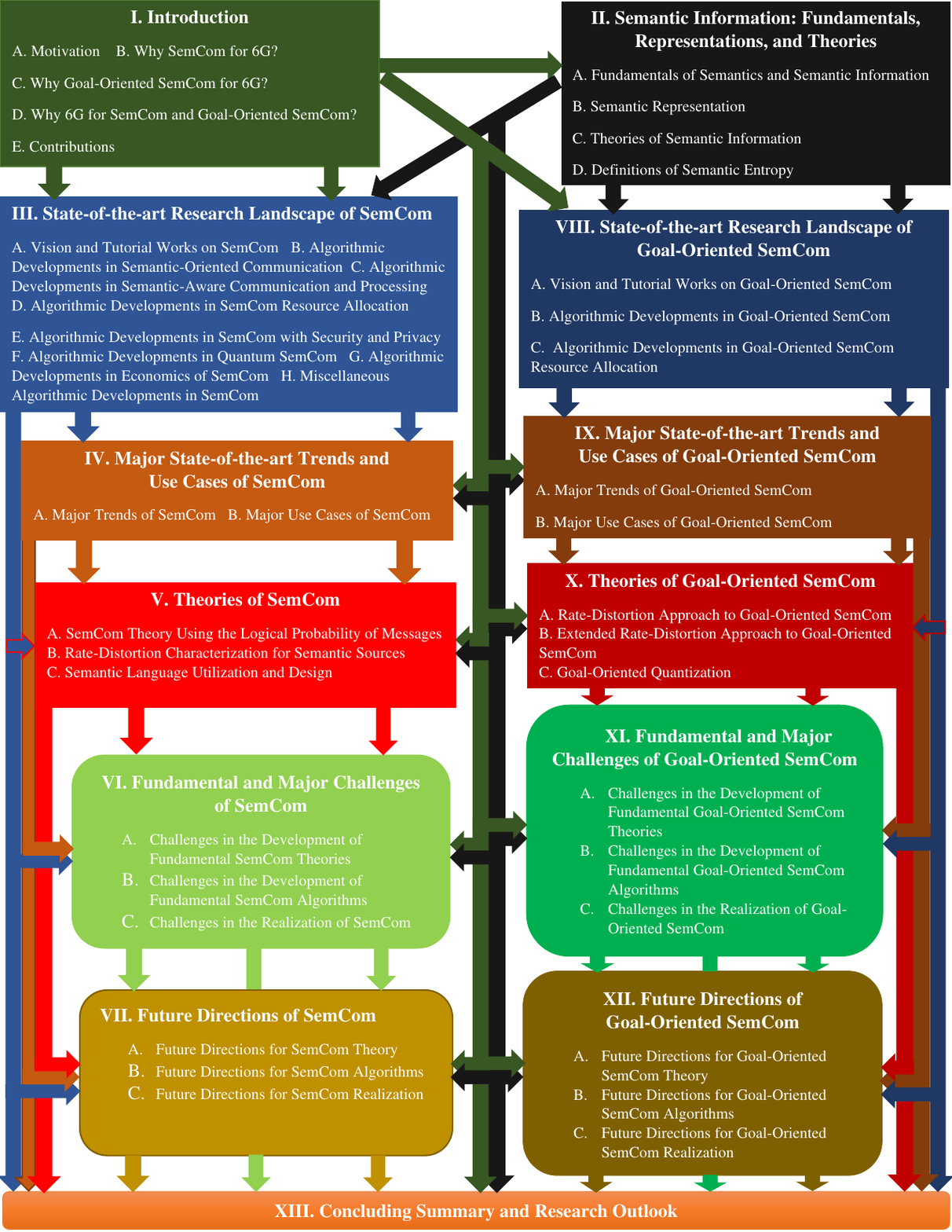}  \vspace{4mm} 
	\caption{Concept map, structure, and organization of this tutorial-cum-survey paper.}
	\label{fig: Concept_map_SemCom_20230409}
\end{figure*}

Apart from secure-by-design 6G networks, the realization of many SemCom use cases (e.g., H2H SemCom, H2M SemCom, M2M SemCom, and KG-based SemCom) and major goal-oriented SemCom use cases (e.g., autonomous transportation, consumer robotics, environmental monitoring, telehealth, smart factories, and NCSs) require a possibly autonomous and integrated 6G network be designed and realized. Therefore, the paradigm of \textit{6G for SemCom and goal-oriented SemCom} and the previously discussed paradigms of \textit{SemCom and goal-oriented SemCom for 6G} necessitate the tighter integration and marriage of 6G, SemCom, and goal-oriented SemCom. Facilitating the tighter integration and marriage of 6G, SemCom, and goal-oriented SemCom, this comprehensive\footnote{This tutorial-cum-survey paper presents almost everything about SemCom and goal-oriented SemCom except their performance assessment metrics. These metrics are comprehensively surveyed by the first three authors in \cite{Getu_Metrics_of_SemCom_and_GO_SemCom_2023}.} tutorial-cum-survey paper delivers the contributions enumerated in Section \ref{subsec: contributions}. To put these contributions in perspective, we compare and contrast them with those of prior survey and tutorial papers in Tables \ref{table: Scope_of_tutorial_cum_survey} and \ref{table_add: Scope_of_tutorial_cum_survey}. Meanwhile, the concept map, structure, and organization of this article are depicted in Fig. \ref{fig: Concept_map_SemCom_20230409}.

\begin{table*}
	\centering
	\begin{tabular}{| l | l | l | l | l | l| }
		\hline
		Themes of SemCom and goal-oriented SemCom & Scope of & Scope of    & Scope of  & Scope of    & Scope of   \\   
		&  Ref. \cite{SemCom_for_6G_Future_Internet'22} & Ref. \cite{Luo_SemCom_Overview'22}  & Scope of \cite{Zhang_Wisdom_Evolutionary_6G'21}   & Ref. \cite{Qin_Sem_Com_Principles_Apps'22}  & Ref. \cite{Qiao_What_is_SemCom'21}   \\   \hline  \hline
		Fundamentals of semantics and semantic information,   & Partially  & --   & Partially & --  & --  \\ 
		semantic representations, theories of semantic     &  &   &  &   &   \\ 
		information, and definitions of semantic entropy &  &  &   &  &   \\ \hline
		State-of-the-art research landscape of SemCom  & Partially  & Partially  & Partially  & Partially & Partially  \\ \hline
		Major state-of-the-art trends and use cases of SemCom & Partially & Partially  & --  & Partially    & Almost completely   \\ \hline
		State-of-the-art theories of SemCom & Partially &  -- & -- & Partially  & --  \\ \hline
		Fundamental and major challenges (in theory, algorithm,  & Partially & Partially  & --   & Partially   & --  \\ 
		 and realization) of SemCom  &  &  &   &  &    \\ \hline
		Future directions (in theory, algorithm, and & Partially  & Partially & --   & Partially & --    \\ 
		 realization) of SemCom &  &  &   &  &    \\ \hline
		State-of-the-art research landscape of   & Partially & -- & -- & Partially  & --  \\ 
		goal-oriented SemCom &  &  &  &   &   \\ \hline
		Major state-of-the-art trends and use cases of  & Partially & --  & --   & Partially & --   \\
		goal-oriented SemCom &  &  &  &  &   \\ \hline
		State-of-the-art theories of goal-oriented SemCom & Partially & -- & --   & -- &    \\  \hline
		Fundamental and major challenges (in theory, algorithm,  & Partially  &--  & --   & --  & -- \\    
		and realization) of goal-oriented SemCom &  &  &   &  &  \\    \hline
		Future directions (in theory, algorithm,   & Partially  & -- & -- & --  &  -- \\
		and realization) of goal-oriented SemCom  &  &  &   &  &    \\  \hline    
	\end{tabular}  \\ [3mm]
\begin{tabular}{| l | l | l | l | l | l| }
	\hline  
	Themes of SemCom and goal-oriented SemCom & Scope of & Scope of    & Scope of  & Scope of    & Scope of   \\   
	&  Ref. \cite{Zhang_Goal-Oriented_Commun'22} & Ref. \cite{Sem_Empowered_Commun'22}  & Ref. \cite{Gunduz_Beyond_Transmitting_Bits'22}   & Ref. \cite{Chaccour_Building_NG_SemCom_Networks'22}  & Ref. \cite{Engineering_SemCom'22}   \\   \hline  \hline
	Fundamentals of semantics and semantic information,   &--  & Partially  & Partially &  Partially & Almost completely   \\ 
	semantic representations, theories of semantic     &  &   &  &   &   \\ 
	information, and definitions of semantic entropy &  &  &   &  &   \\ \hline
	State-of-the-art research landscape of SemCom  & -- &  Partially & Partially  & Partially &  Partially \\ \hline
	Major state-of-the-art trends and use cases of SemCom &-- & --  & -- & --   &  --  \\ \hline
	State-of-the-art theories of SemCom & -- & --  & Partially  & Partially  & --  \\ \hline
	Fundamental and major challenges (in theory, algorithm,  & -- & Partially & --   & Partially    &  Partially \\ 
	and realization) of SemCom  &  &  &   &  &    \\ \hline
	Future directions (in theory, algorithm, and & -- & Partially  & --   & Partially  & --   \\ 
	realization) of SemCom &   &  &   &  &    \\ \hline
	State-of-the-art research landscape of   & Partially  & Partially & Partially & --   & --   \\ 
	goal-oriented SemCom &  &  &  &   &   \\ \hline
	Major state-of-the-art trends and use cases of  & -- & -- & --  & --  &  --  \\
	goal-oriented SemCom &  &  &   &  &   \\ \hline
	State-of-the-art theories of goal-oriented SemCom & -- & -- & Partially  & --  & --   \\  \hline
	Fundamental and major challenges (in theory, algorithm,  & --  & Partially  & --  & -- & -- \\    
	and realization) of goal-oriented SemCom &  &  &   &  &  \\    \hline
	Future directions (in theory, algorithm,   & Partially  & Partially & --  & --  & --  \\
	and realization) of goal-oriented SemCom  &  &  &   &  &    \\  \hline    
\end{tabular}   \\ [3mm]
\caption{Scope of this tutorial-cum-survey paper w.r.t. the scope of prior survey and tutorial papers on SemCom and/or goal-oriented SemCom -- Ref.: reference; \textquotedblleft --'' means the specific reference didn't discuss the theme listed on a given row.}
\label{table: Scope_of_tutorial_cum_survey}
\end{table*}
\begin{table*}
	\centering
\begin{tabular}{| l | l | l | }
	\hline
	Themes of SemCom and goal-oriented SemCom & Scope of  & \hspace{6mm}  \textbf{Scope of this}    \\   
	&  Ref. \cite{Goal-oriented_SemCom_Thesis} & \textbf{tutorial-cum-survey paper}      \\   \hline  \hline
	Fundamentals of semantics and semantic information,   & -- & \textbf{Comprehensively}     \\ 
	semantic representations, theories of semantic     &  &    \\ 
	information, and definitions of semantic entropy &  &   \\ \hline
	State-of-the-art research landscape of SemCom  & --  &  \textbf{Comprehensively}  \\ \hline
	Major state-of-the-art trends and use cases of SemCom & -- &  \textbf{Completely}  \\ \hline
	State-of-the-art theories of SemCom & --  & \textbf{Completely}   \\ \hline
	Fundamental and major challenges (in theory, algorithm,  & -- & \textbf{Completely}    \\ 
	and realization) of SemCom  &  &     \\ \hline
	Future directions (in theory, algorithm, and & -- &   \textbf{Completely} \\ 
	realization) of SemCom &  &  \\ \hline
	State-of-the-art research landscape of   & Partially & \textbf{Comprehensively} \\ 
	goal-oriented SemCom &   &   \\ \hline
	Major state-of-the-art trends and use cases of  & Partially  & \textbf{Completely} \\
	goal-oriented SemCom &  &    \\ \hline
	State-of-the-art theories of goal-oriented SemCom & Partially & \textbf{Completely}  \\  \hline
	Fundamental and major challenges (in theory, algorithm,  & -- &  \textbf{Completely}  \\    
	and realization) of goal-oriented SemCom &  &   \\    \hline
	Future directions (in theory, algorithm,   & -- &  \textbf{Completely}    \\
	and realization) of goal-oriented SemCom  &  &    \\  \hline    
\end{tabular}  
\caption{Scope of this tutorial-cum-survey paper w.r.t. the scope of prior survey and tutorial papers on SemCom and/or goal-oriented SemCom -- Ref.: reference; \textquotedblleft --'' means the specific reference didn't discuss the theme listed on a given row.}
\label{table_add: Scope_of_tutorial_cum_survey}
\end{table*}

%\newpage

\newpage

\subsection{Contributions}
\label{subsec: contributions}
The contributions of this holistically comprehensive -- in depth and breadth -- tutorial-cum-survey paper are enumerated below. 
\begin{enumerate}
	\item This paper explains the fundamentals of semantics and semantic information, semantic representations, theories of semantic information, and definitions of semantic entropy.
	
	\item This paper details the state-of-the-art research landscape of SemCom.
	
	\item This paper presents the major state-of-the-art trends and use cases of SemCom. 
	
	\item This paper discusses the state-of-the-art theories of SemCom. 
	
	\item This paper uncovers the fundamental and major challenges (in theory, algorithm, and realization) of SemCom.
	
	\item This paper offers novel future research directions (in theory, algorithm, and realization) of SemCom.

	\item This paper documents the state-of-the-art research landscape of goal-oriented SemCom.
	
	\item This paper provides the major state-of-the-art trends and use cases of goal-oriented SemCom.
	
	\item This paper discusses the state-of-the-art theories of goal-oriented SemCom. 
	
	\item This paper exposes the fundamental and major challenges (in theory, algorithm, and realization) of goal-oriented SemCom.
	
	\item This paper provides novel future research directions (in theory, algorithm, and realization) of goal-oriented SemCom.      
\end{enumerate}

\textit{Notation}: scalars, vectors, and matrices are represented by italic letters, bold lowercase letters, and bold uppercase letters, respectively. Sets, quantization regions, quantizers, and KBs are denoted by calligraphic letters. $\mathbb{N}$, $\mathbb{R}$, $\mathbb{R}_+$, $\mathbb{R}^n$, $\mathbb{R}_+^n$, and $\mathbb{R}^{m\times n}$ denote the set of natural numbers, real numbers, non-negative real numbers, $n$-dimensional vectors of real numbers, $n$-dimensional vectors of non-negative real numbers, and $m\times n$ matrices of real numbers, respectively. $\eqdef$ denotes an equality by definition. For $n, k\in\mathbb{N}$, $[n]\eqdef \{1, 2, \ldots, n\}$ and $\mathbb{N}_{\geq k} \eqdef \{k, k+1, k+2, \ldots\}$. For a matrix $\bm{A} \in \mathbb{R}^{m \times n}$, its element in the $i$-th row and the $j$-th column is denoted by $(\bm{A})_{i,j}$ for all $i\in[m]$ and $j\in[n]$. The symbols $\sim$, $\models$, $|\cdot|$, $\| \cdot \|$, $(\cdot)^T$, and $\mathcal{O}(\cdot)$ denote distributed as, the propositional satisfaction relation, absolute value, the Euclidean norm, transpose, and the Landau notation, respectively. The notation $\min(\cdot)$ (or $\min\{ \cdot\}$), $\max(\cdot)$ (or $\max\{ \cdot\}$), $\mathbb{E}\{\cdot\}$, $\mathbb{E}_X\{\cdot\}$, and $\mathbb{P}(\cdot)$ stand for minimum, maximum, expectation, expectation w.r.t. the random variable (RV) $X$, and probability, respectively. $\mathbb{P}(A|B)$ represents the probability of event $A$ conditioned on event $B$.
% 3CLS (Communications, computing, control, localization, and sensing)
\begin{table}
	\centering
	\begin{tabular}{ | l | l |  }
		\hline
		Abbreviation & Definition    \\ \hline \hline 
		3C  & User-centralized, content centralized, and  \\ 
		    &  data-centralized   \\ \hline
		3CLS & Communications, computing, control,   \\ 
		& localization, and sensing  \\ \hline
		3D  &  Three-dimensional \\ \hline
		5G  & Fifth-generation   \\ \hline
		5GNR & 5G new radio  \\ \hline
		6C  & Capturing, communication, caching, cognition, \\ 
		 &  computing, and controlling  \\ \hline
		6G  & Sixth-generation   \\ \hline 
		ADJSCC	&  Attention DL-based JSCC \\ \hline
	 	AE & Autoencoder  \\ \hline
	 	AF & Amplify-and-forward \\ \hline
		AI & Artificial intelligence  \\ \hline
		AR &  Augmented reality \\ \hline
		ASC & Adaptable semantic compression  \\ \hline
	 	AVs & Autonomous vehicles   \\ \hline
	 	BA algorithm & Blahut–Arimoto algorithm   \\ \hline
		BCP & Bar-Hillel-Carnap paradox    \\ \hline
		BERT & Bidirectional encoder representations from    \\ 
		&   transformers \\ \hline
		 Bi-LSTM & Bidirectional long short-term memory	 \\ \hline
		BitCom & Bit communication  \\ \hline
		CB-TBMA & Compressed IB-TBMA   \\ \hline
		CCCA & Cache-computing coordination algorithm   \\ \hline
		CDRL & Collaborative deep RL  \\ \hline
		CE & Cross entropy  \\ \hline
		CKG & Cross-modal KG  \\ \hline
		CNN(s) & Convolutional neural network(s)  \\ \hline
		 CoBots & Collaborative robots  \\ \hline
		 CP-DQN & Content popularity-based DQN  \\ \hline
		  CRI & Channel rate information  \\ \hline
		  CSED & Combined semantic encoding and decoding \\ \hline
		  CSI &  Channel state information \\ \hline
		  CTSF & Communication toward semantic fidelity      \\\hline
		  CU  & Centralized unit  \\ \hline
		  D2D & Device-to-device    \\ \hline
		 DF & Decode-and-forward   \\  \hline
		 DII &  Data importance information \\ \hline
		 DL	& Deep learning   \\ \hline
		DNN(s)&   Deep neural network(s)  \\ \hline
		 DQN &Deep Q-network   \\ \hline
		 DRL &  Deep RL \\ \hline
		 DTI & Data type information   \\ \hline
		 DT-JSCC & Discrete task-oriented JSCC  \\ \hline
		 DIB &  Distributed IB  \\ \hline
		 EEIM &  Evolutionary energetic information model \\ \hline 
		 eMBB & Enhanced mobile broadband \\ \hline
		 ESC &  Emergent semantic communication \\ \hline
		FFTs & Fast Fourier transforms   \\ \hline
		FL & Federated learning   \\ \hline  
		FPS & Frames per second  \\ \hline
		F-user & Far user  \\ \hline
		GAN &  Generative adversarial network \\ \hline
		GAXNet & Graph attention exchange network \\ \hline
		GD	& Gradient descent  \\ \hline
		GFlowNets & Generative flow networks     \\ \hline
		GIB & Graph IB  \\ \hline
		GOQ & Goal-oriented quantization	 \\ \hline
		H2H & Human-to-human \\ \hline
		H2M &  Human-to-machine   \\ \hline
		HARQ & Hybrid automatic repeat request  \\ \hline
		HTC & Human-type communication   \\ \hline
		IB & Information bottleneck   \\ \hline
		IC &  Incentive compatibility \\ \hline
		IE-SC & Intelligent and efficient semantic communication  \\ \hline
		IMT & Interdisciplinary, multidisciplinary, and \\ 
		& transdisciplinary  \\ \hline
	\end{tabular}   \\ [2mm]
	\caption{List of abbreviations and acronyms I.}
	\label{table: abbreviations_and_acronyms_I}
\end{table}

\begin{table}
	\centering
	\begin{tabular}{ | l | l |  }
		\hline
		Abbreviation & Definition    \\ \hline \hline 
	IoT & Internet of things  \\ \hline
	IoV &  Internet of vehicles \\ \hline
	IR & Impulse radio \\ \hline
	IR$^2$ SemCom & Interference-resistant and robust SemCom  \\ \hline
	iSemCom &  Intelligent SemCom \\ \hline
	iSemCom-HetNet & An iSemCom-enabled heterogeneous network \\ \hline 
	IS-JSCC & iterative semantic JSCC   \\ \hline
	ISS &  Image-to-graph semantic similarity  \\ \hline
	ISSC & Image segmentation semantic communication    \\ \hline
	JSC & Joint source and channel  \\ \hline
	JSCC & Joint source-channel coding  \\ \hline
	JSemC & Joint semantic-channel  \\ \hline
	JSNC & Joint semantics-noise coding  \\ \hline
	KB & Knowledge base    \\ \hline
	KG(s) & Knowledge graph(s)  \\ \hline
	KL & Kullback–Leibler \\ \hline
	KPI	&  Key performance indicator \\ \hline
	LDPC & Low-density parity check codes  \\ \hline
	L-MMSE & Linear minimum MSE    \\ \hline
	LSTM & Long short-term memory   \\  \hline
		M2M & Machine-to-machine   \\ \hline
		MA-POMDP & Multi-agent partially observable Markov decision     \\ 
		&  process   \\ \hline
		MARL & Multi-agent RL  \\ \hline
		MEC & Mobile edge computing \\ \hline
		MIMO & Multiple-input multiple-output    \\ \hline
		ML & Machine learning   \\ \hline
		MSE & Mean squared error  \\ \hline
		MSP & Mobile service provider\\ \hline
		MTC & Machine-type communication   \\ \hline
       NCSs & Networked control systems    \\ \hline
       NeSy AI & Neuro-symbolic AI  \\ \hline
	   NeuroComm & Neuromorphic wireless cognition	 \\ \hline
		NLP & Natural language processing  \\ \hline
		NOMA &  Non-orthogonal multiple access \\ \hline
		NPMs & DNN-based protocol models  \\ \hline
		NTSCC & Nonlinear transform source-channel coding  \\ \hline
		N-user &  Near user \\ \hline
		OAM  &  Orbital angular momentum \\ \hline
		O-DU & Open distributed unit   \\ \hline
		O-RAN & Open RAN  \\ \hline
		O-RU &  Open radio unit    \\  \hline
		OSI & Open system interconnection  \\ \hline
		PAI  &  Partial algorithm information \\ \hline
		PCM &   Pulse code modulation \\ \hline
		PCV & Point cloud video  \\ \hline
		PDF & Probability density function \\ \hline
	 	PHY & Physical layer  \\ \hline
	 	PMF & Probability mass function \\ \hline
	 	QAM & Quadrature amplitude modulation   \\ \hline
	 	OFDM & orthogonal frequency division multiplexing  \\ \hline
	 	QKD & Quantum key distribution  \\ \hline
	 	QML & Quantum ML  \\ \hline
		QoE & Quality of experience  \\ \hline
	 	QoS  &  Quality of service \\ \hline
		RANs & Radio access networks   \\ \hline 
		RB & Resource block \\ \hline
		Ref. & Reference   \\ \hline
		RHS &   Right-hand side \\ \hline
		RIB &  Robust IB   \\   \hline
		RIS & Reconfigurable intelligent surface   \\ \hline 
		RL & Reinforcement learning    \\ \hline
		RL-ASC & RL-based adaptive semantic coding   \\ \hline
		ROI & Region-of-interest  \\ \hline
		RS &  Reed Solomon \\ \hline
		RSUs & Roadside units   \\ \hline
		RV & Random variable   \\ \hline
		S-AI layer & Semantic-empowered application-intent layer   \\ \hline
		SC & Semantic coding \\ \hline
		SC-AIT & SemCom paradigm with AI tasks   \\ \hline 
	\end{tabular}   \\ [2mm]
	\caption{List of abbreviations and acronyms II.}
	\label{table: abbreviations_and_acronyms_II}
\end{table}

\begin{table}
	\centering
	\begin{tabular}{ | l | l |  }
		\hline
		Abbreviation & Definition    \\ \hline \hline 
		SCT & Semantic coded transmission   \\ \hline
		SE & Semantic extraction    \\ \hline
		Seb & Semantic base  \\ \hline
		SEED & Semantic/effectiveness encoded data  \\ \hline
		SemCom & Semantic communication    \\ \hline
		SemComNet & SemCom-enabled network   \\ \hline
		Seq2Seq-SemCom & Sequence-to-sequence SemCom    \\ \hline
		SF & Semantic forward   \\ \hline
		SFV & Semantic feature vector  \\ \hline
		SINR & signal-to-interference-plus-noise ratio  \\ \hline
		SI plane &  Semantic intelligence plane \\ \hline
		S-IF & Semantic information flow  \\ \hline
		SIT & Semantic information theory   \\ \hline
		SNC & Semantic native communication  \\ \hline
		SNN & Spiking neural network  \\ \hline
		SNR & Signal-to-noise ratio    \\ \hline  
		S-NP layer & Semantic-empowered network protocol layer    \\ \hline
		S-PB layer & Semantic-empowered physical-bearing layer \\ \hline
		SPM & Semantic protocol model  \\ \hline
		S-SE &  Spectral efficiency \\ \hline
		STM & System throughput in message \\ \hline
		SVC & Semantic video conferencing  \\ \hline
		T5 & Text-to-text transfer Transformer \\ \hline
		TBMA & Type-based multiple access \\ \hline
		TechnicCom & Technical communication    \\ \hline
		THz & Terahertz   \\ \hline
		TSSI & Theory of strongly semantic information   \\ \hline
		UA & User association \\ \hline
		UAV & Unmanned aerial vehicle  \\ \hline
		URLLC & Ultra-reliable low-latency communication \\  \hline
		UT & Universal Transformer  \\ \hline
		VIB & Variational IB   \\ \hline
		VQA & Visual question answering  \\ \hline
		VR &  Virtual reality \\ \hline
		VSRAA-SM & Video semantics-based resource allocation   \\ 
		& algorithm for spectrum multiplexing scenarios  \\ \hline
		 w.r.t. & With respect to  \\ \hline
		WITT & Wireless image transmission transformer  \\ \hline
		XR & Extended reality   \\ \hline
	\end{tabular}   \\ [2mm]
	\caption{List of abbreviations and acronyms III.}
	\label{table: abbreviations_and_acronyms_III}
\end{table}

\newpage

\section{Semantic Information: Fundamentals, Representations, and Theories}
\label{sec: semantics_information_fundamentals_representations_and_theories}
SemCom and goal-oriented SemCom revolves around the transmission of semantic information. One must therefore understand semantic information and its theories before they can design rigorous SemCom and goal-oriented SemCom systems. To this end, we hereinafter present the fundamentals of semantics and semantic information, semantic representation, theories of semantic information, and definitions of semantic entropy. We begin with the fundamentals of semantics and semantic information.

\subsection{Fundamentals of Semantics and Semantic Information}
\label{subsec: fundamentals_of_semantics_and_semantics_information}
The word \textit{semantics} originates from  natural or formal languages and the concept of compositionality \cite[p. 125]{Tong_Zhu__6G'21}. The concept of compositionality asserts that the meaning of a sentence is decided by three ingredients: the composition rule (syntax) of a sentence; the context of a sentence; and the meaning (semantics) of each component of the sentence \cite[p. 125]{Tong_Zhu__6G'21}. Semantics has been studied for centuries \cite{Gunduz_Beyond_Transmitting_Bits'22} in a variety of disciplines such as linguistics \cite{Daniel_NLP'00}, philosophy \cite{Stanford_Encyclopedia_of_Philosophy'22,Floridi_Phil_Conceptions_of_Information'09}, cognitive sciences \cite{Samsonovich_Toward_Sematic_General_Theory_of_Everything'10}, neuroscience \cite{Brain_Inside_Out_19}, biology \cite{Wolfgang_SemInfo_in_Natrure'15}, and robotics \cite{Garg_Semantics_for_Robotic_Mapping'20}. Because semantics is used to mean something different in each discipline, it is a highly complex as well as controversial topic, to the extent that it would be very difficult to provide a concise definition for it that would be widely accepted \cite{Gunduz_Beyond_Transmitting_Bits'22}. In short and in general, however, semantics can be defined as the study of meaning and is closely connected to \textit{semiotics}, the study of \textit{signs} \cite{Gunduz_Beyond_Transmitting_Bits'22}. As for signs, all communication systems are built upon signs \cite{Gunduz_Beyond_Transmitting_Bits'22}. A system of signs and rules amounts broadly to a language, and a language's rules applied to signs are categorized into \textit{syntax}, \textit{semantics}, and \textit{pragmatics} \cite{Theory_of_Signs_Morris_Book'38}:
\begin{itemize}
	\item Syntax studies signs and their relationships to one another, and is concerned only with signs and their relationships \cite{Gunduz_Beyond_Transmitting_Bits'22}. 
	
	\item Semantics aims to understand the relationships between signs and the objects to which they apply (i.e., the \textit{designata}). It is built upon syntax and studies signs and their relationship to the world \cite{Gunduz_Beyond_Transmitting_Bits'22}. According to Chomsky \cite{Syntactic_Structures_Chomsky'57}\footnote{\textquotedblleft Through his now famous sentence \textquoteleft Colorless green ideas sleep furiously,' Chomsky argued that it is possible to construct grammatically consistent but semantically meaningless phrases, hence the separation between syntax and semantics'' \cite{Gunduz_Beyond_Transmitting_Bits'22}.}, syntax is independent of semantics \cite{Gunduz_Beyond_Transmitting_Bits'22}.
	
	\item Pragmatics considers the context of communication and studies the signs and their relationships with users \cite{Gunduz_Beyond_Transmitting_Bits'22}. Accordingly, pragmatics takes into consideration all the personal and psychological factors -- apparent in human communications -- as well as the impact of a sign on the designata \cite{Gunduz_Beyond_Transmitting_Bits'22}. 
\end{itemize}

\begin{figure}[t!]
	\centering
	\includegraphics[scale=0.28]{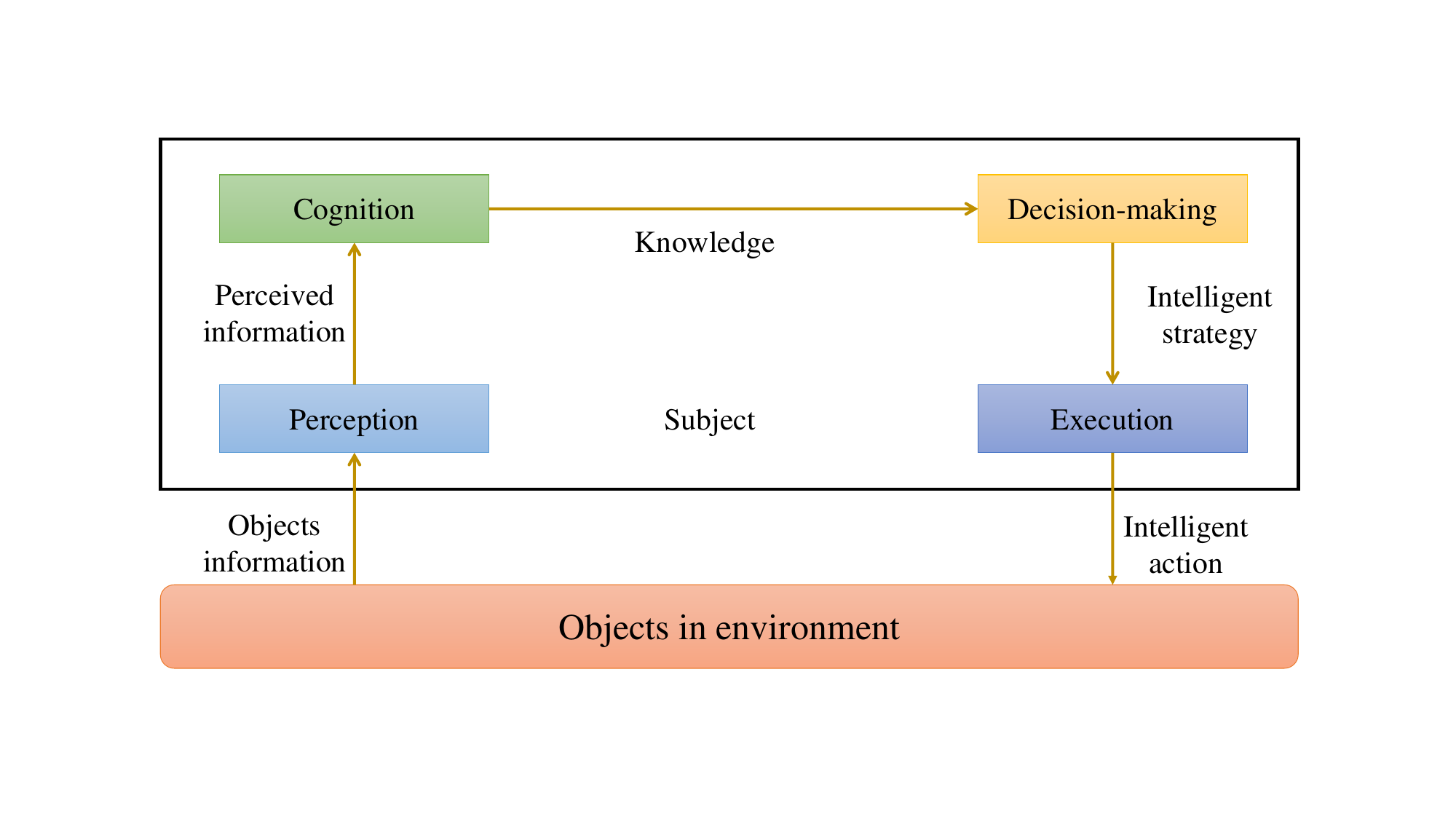}  \vspace{-1.0cm} 
	\caption{Model of information ecosystem \cite[Fig.1]{hong_Theory_Semantic_Info'17}.}
	\label{fig: Model_of_Information_Ecosystem}
\end{figure}

Because semantics is built upon syntax and studies signs and their relationship to the world \cite{Gunduz_Beyond_Transmitting_Bits'22}, the fundamental notion of semantic information hinges upon the information ecosystem, which is a complete process of \textit{information-knowledge-intelligence conversion} \cite{hong_Theory_Semantic_Info'17,Zhong_Theory_Sem_Information_Book_Chapter}, as shown in Fig. \ref{fig: Model_of_Information_Ecosystem}. As seen in Fig. \ref{fig: Model_of_Information_Ecosystem}, the bottom part represents the object in the environment that generates the object information, whilst the top part of it portrays the subject interacting with the object through the following processes \cite{hong_Theory_Semantic_Info'17}: the object information is transformed -- via perception -- to perceived information, which constitutes knowledge (cognition) that is deployed in an intelligent strategy (decision-making), which leads to the taking of intelligent action (execution) on the object and completes the basic information ecosystem.\footnote{Per the information ecosystem model schematized in Fig. \ref{fig: Model_of_Information_Ecosystem}, perception leads to action in line with the contemporary \textit{outside-in} neuroscientific framework \cite{Bear_MF_Neuroscience'16,Kandel_Principles_of_Neural_Science'21}. In this framework (the perception-action framework), our knowledge is considered to emerge from the perceptual associations of cause-and-effect relationships and inductive reasoning \cite[p. 32]{Brain_Inside_Out_19} -- ascertaining the view that \textit{the brain is an organ for perceptual representation}. This perceptual representation-centered view would imply the presumption of a hidden \textquotedblleft homunculus'' that would decide whether to respond or not \cite[p. 32]{Brain_Inside_Out_19}. Contrarily, the emerging \textit{inside-out} framework (the action-perception framework) -- which was first advocated by the author of \cite{Brain_Inside_Out_19} -- considers action primarily as a source of knowledge \cite[p. 32]{Brain_Inside_Out_19}: action corroborates the meaning and importance of sensory signals by giving a second opinion \cite[p. 32]{Brain_Inside_Out_19}.} In light of this ecosystem, the theory of semantic information must fulfill the constraints or the requirements dictated by the \textit{ecological process of information} \cite{hong_Theory_Semantic_Info'17} as depicted in Fig. \ref{fig: Model_of_Information_Ecosystem}. Per Fig. \ref{fig: Model_of_Information_Ecosystem}, only those types of object information that are participating in the subject-object interaction process -- among all the types that exist -- are viewed as meaningful \cite{hong_Theory_Semantic_Info'17}, hence the basis of semantic information. Accordingly, the following definitions \cite{hong_Theory_Semantic_Info'17} on \textit{object information} (or \textit{ontological information}) and \textit{perceived information} (or \textit{epistemological information}) ensue.
\begin{definition}[\textbf{Object information \cite[Definition 1]{hong_Theory_Semantic_Info'17}}]
\label{object_info}
\textquotedblleft The object information concerning an object is defined as the set of states at which the object may stay and the pattern with which the states vary presented by the object itself.'' 
\end{definition}

\begin{definition}[\textbf{Perceived information \cite[Definition 2]{hong_Theory_Semantic_Info'17}}]
\label{perceived_info}
\textquotedblleft The perceived information a subject possesses about an object is defined as the trinity of the form (named the syntactic information), the meaning (the semantic information), and the utility (the pragmatic information), all of which are perceived by the subject from the object information.''
\end{definition}

If we compare Definitions \ref{object_info} and \ref{perceived_info}, the object information originates from the real world, while the perceived information is the outcome of the object information being perceived by a subject \cite{hong_Theory_Semantic_Info'17}. Consequently, the perceived information can comprise more intentions than the object information \cite{hong_Theory_Semantic_Info'17}. Per Definition \ref{perceived_info}, meanwhile, semantic information is the meaning a subject perceives from the object information, which matches the notion of semantics in semiotics \cite{hong_Theory_Semantic_Info'17}. This leads us to the following definition of \textit{comprehensive information} \cite{hong_Theory_Semantic_Info'17}.
\begin{definition}[\textbf{Comprehensive information \cite[Definition 3]{hong_Theory_Semantic_Info'17}}]
\label{Comprehensive_info}
Comprehensive information is the trinity of syntactic, semantic, and pragmatic information.
\end{definition}
\noindent Comprehensive information calls for more information research looking into the mutual relationships among syntactic, semantic, and pragmatic information -- contrary to their separate studies as approached predominantly in semiotics -- as defined below \cite{hong_Theory_Semantic_Info'17}.
\begin{definition}[\textbf{Mutual relation among syntactic, semantic, and pragmatic information \cite[Definition 4]{hong_Theory_Semantic_Info'17}}]
\label{Mutual_relation}
\textquotedblleft The syntactic information is specific in nature and can directly be produced through subject’s sensing function while the pragmatic information is also specific in nature and can directly be produced through subject’s experiencing. However, the semantic information is abstract in nature and thus cannot be produced via subject’s sensing organs and experiencing directly. The semantic information can only be produced based on both syntactic and pragmatic information just produced already, that is, by mapping the joint of syntactic and pragmatic information into the semantic information space and then naming it.''
\end{definition}

\begin{figure}[htb!]
	%\centering
	\includegraphics[scale=0.27]{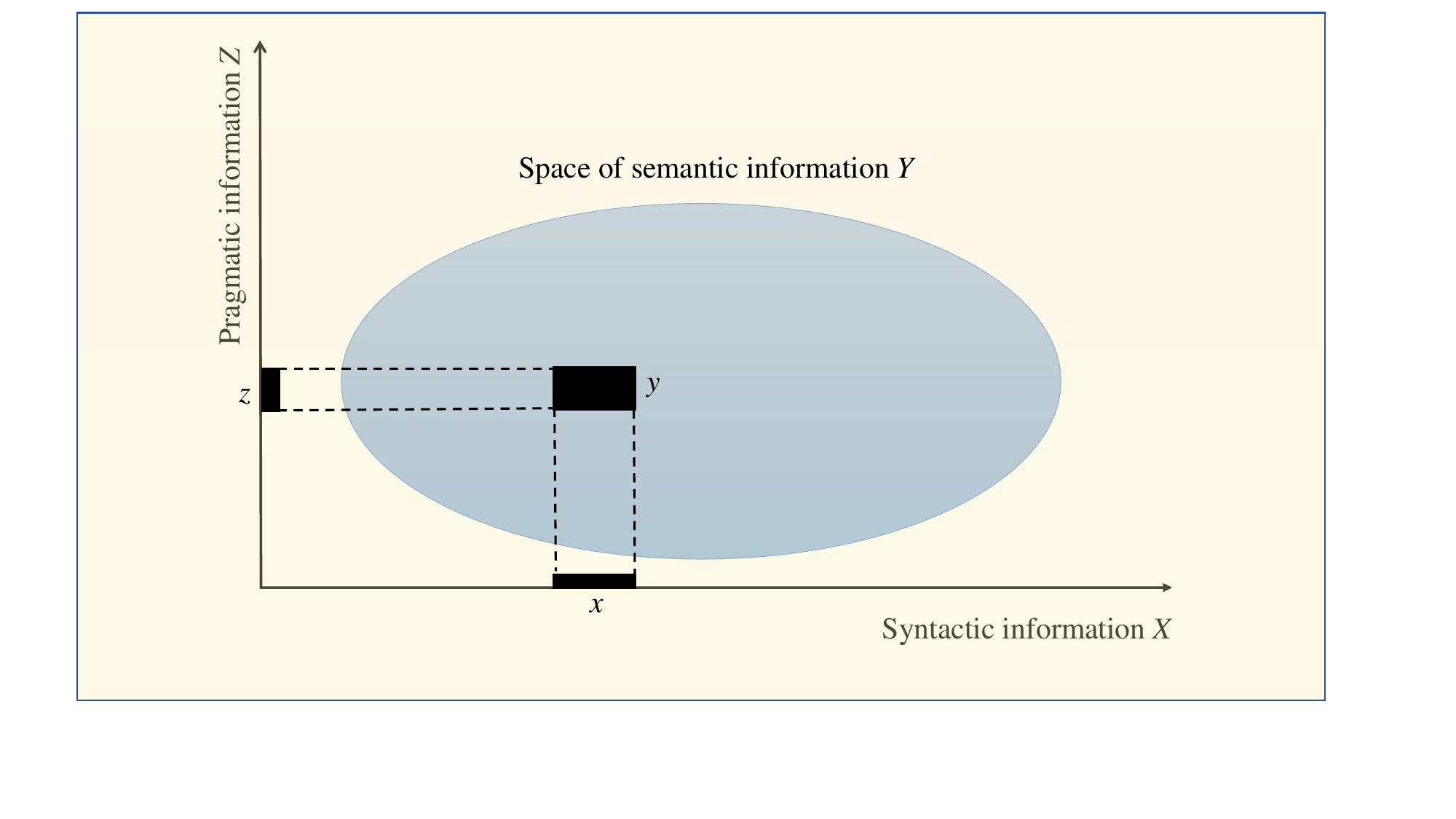}  \vspace{-0.5cm} 
	\caption{Relation of semantic information to syntactic information and pragmatic information \cite[Fig.2]{hong_Theory_Semantic_Info'17}, \cite[Fig. 7]{SemCom_for_6G_Future_Internet'22}.}
	\label{fig: SemInfo_wrt_syntactic_and_pragmatic_information}
\end{figure}

As asserted in Definition \ref{Mutual_relation} and illustrated in Fig. \ref{fig: SemInfo_wrt_syntactic_and_pragmatic_information}, semantic information is obtained by jointly mapping the syntactic information and the pragmatic information into the semantic information space. To clarify, let $X$, $Y$, and $Z$ denote syntactic, semantic, and pragmatic information, respectively. If the instance of $X$ is $x$, then the instance of $Y$, which is the semantic information $y$, will be different for different instances of $Z$ denoted by $z$ \cite{hong_Theory_Semantic_Info'17}:
\begin{itemize}
	\item If $\{x=$apple, $z=$ nutritious$\}$, then $y=$ fruit.
	
	\item If $\{x=$apple, $z=$ used for information processing$\}$, then $y=$ iPad.
	
	\item If $\{x=$ apple, $z=$ information processor in pocket$\}$, then $y=$ iPhone.
\end{itemize}
Semantic information has more importance than syntactic and pragmatic information, and can serve as the \textit{legal representative} of the perceived information \cite{hong_Theory_Semantic_Info'17}. Furthermore, any theory of semantic information should take into account the information ecosystem (as in Fig. \ref{fig: Model_of_Information_Ecosystem}) as a \textit{big picture} \cite{hong_Theory_Semantic_Info'17}. 

In light of the information ecosystem depicted in Fig. \ref{fig: Model_of_Information_Ecosystem}, the author of \cite{Wolfgang_SemInfo_in_Natrure'15} proposes an energy-based perspective of semantic information by assuming that syntax, semantics, and pragmatics are structural features of information in biological evolution. More specifically, the author of \cite{Wolfgang_SemInfo_in_Natrure'15} argues that semantic information is an exclusive feature of biological evolution by proposing an information model named the \textit{evolutionary energetic information model} (EEIM). EEIM takes into account the transformation of energy into information from the outside to the inside of a biological organism: \textquotedblleft Energy is transformed into information via senses and the application of syntactic and semantic rules (by accompanied pragmatics) evaluating energy on its usability as information'' \cite{Wolfgang_SemInfo_in_Natrure'15}. Some of the attributes of EEIM are \cite[Table 1]{Wolfgang_SemInfo_in_Natrure'15}:
\begin{itemize}
	\item \textbf{On evolution} -- \textquotedblleft Information is exclusively being \textquoteleft produced' by evolution.'' 
	
	\item \textbf{On energy} -- \textquotedblleft Information is a quality of informational energy''; \textquotedblleft Informational energy provides semantic information to accordingly prepared receiver instances''; \textquotedblleft Not all energy is informational.''
	
	\item \textbf{On semiotics} -- \textquotedblleft Information is bound to syntax, semantics and pragmatics''; \textquotedblleft Information requires a sending instance''; \textquotedblleft Information requires a biological receiver able to interpret incoming information.''\footnote{In light of the perspective that semantic information concerns meaning as \textit{interpreted} by our brains, the following are fundamental questions worth pondering: How does the human brain interpret information? How does the human brain give meaning to information (or something)? From a systems neuroscience standpoint, (observer-independent) semantic information/memory is generally learned after frequent encounters with the same thing or event, which contrasts the one-trial \textquotedblleft acquisition'' of (observer-dependent) \textit{episodic memory}  concerning a unique event \cite[p. 126]{Brain_Inside_Out_19}.}  
\end{itemize}

Apart from the aforementioned perspective of semantic information being driven by the information ecosystem and biological evolution, the authors of \cite{Kountouris_Semantics_EmpoweredCF'21} advocate for evaluating and extracting the semantic value of data at a macroscopic scale, a mesoscopic scale, and a microscopic scale that correspond to the source level, the link level, and the system level, respectively. At the microscopic scale or source level, semantics connote the relative importance of different events, outcomes, or observations from a stochastic process or source of information \cite{Kountouris_Semantics_EmpoweredCF'21}. At the mesoscopic scale or link level, the semantics of information concern a composite nonlinear multivariate function comprising the vector of information attributes, which can be either objective (innate) or subjective (contextual) \cite{Kountouris_Semantics_EmpoweredCF'21}. At the macroscopic scale or system level, the semantics of information alludes to the effective distortion and timing mismatch -- quantified end-to-end -- between information generated at a point/region in space-time and its reconstructed/estimated version at another point/region in space-time, while considering all sources of variability and latency \cite{Kountouris_Semantics_EmpoweredCF'21}.\footnote{Latency may be due to sensing latency and accuracy, data gathering, transmission latency, decoding, processing, etc. \cite{Kountouris_Semantics_EmpoweredCF'21}.}  

After the semantic value of data is extracted at the macroscopic, mesoscopic, and microscopic scales, it has to be represented semantically. Semantic representation can be viewed as a method of efficient compression, which relieves the burden of processing, storage, and transmission \cite{Zhang_a_New_Paradigm'22}. Semantic information can then be directly utilized for design, analysis, learning, and other intelligent tasks in networked intelligence while benefiting from being compact and informative as a result of semantic representation \cite{Zhang_a_New_Paradigm'22}, as discussed below.

\subsection{Semantic Representation} 
\label{subsec: semantic_representation}
While semantic content is the \textquotedblleft meaningful'' part of the data, semantic representation is the \textquotedblleft minimal way to represent this meaning'' \cite{Chaccour_Building_NG_SemCom_Networks'22}. Accordingly, semantic representations of a SemCom language must satisfy three fundamental notions: minimalism, generalizability, and efficiency \cite{Chaccour_Building_NG_SemCom_Networks'22}. The semantic representation of data can be achieved using knowledge graphs (KGs) \cite{Knowledge_Graphs_Survey'22}; NLP \cite{Torfi_NLP_Survey'20,Trands_in_DL-Based_NLP'18,Daniel_NLP'00}; deep neural networks (DNNs) \cite{Hanin_2019_Univ_Approx,DL_Approx_Theory'21,Barron2018_DL_Approximation'18}; \textit{toposes} \cite{Topos_and_Stacks'21,Heng_Math_for_Future_Comp_Commun'22}; causal representation learning \cite{Toward_causal_learning_21,Causal_ML'22}; and quantum corollas \cite{Tetlow_Toward_Semantic_Info_Theory'22}. While the last way is a quantum (quantum mechanical) method, the first five are classical mechanisms. These mechanisms -- along with their pros and cons -- are discussed below, beginning with KG.

\subsubsection{Knowledge Graph (KG)}
one way in which semantic information can be represented (being embedded into) is by using  KGs \cite{Gunduz_Beyond_Transmitting_Bits'22}. A KG is composed of three major components: nodes, edges, and labels \cite{Gunduz_Beyond_Transmitting_Bits'22}. A node can be any object, place, or person, and an edge defines the relationship between two nodes \cite{Gunduz_Beyond_Transmitting_Bits'22}. KG embedding amounts to embedding the components of a KG, including its entities and relationships into continuous vector spaces to simplify manipulation while preserving the KG's inherent structure \cite{Wang_KG_Embedding'17}. KG embedding is a crucial technology for solving problems in KGs \cite{Gunduz_Beyond_Transmitting_Bits'22}. State-of-the-art KG techniques can be categorized into translational distance models that employ distance-based scoring and semantic matching models that employ similarity-based scoring functions \cite{Gunduz_Beyond_Transmitting_Bits'22}. Meanwhile, KG-based semantics is deployed in data integration, recommendation systems, and real-time ranking \cite{Gunduz_Beyond_Transmitting_Bits'22}, and has been exploited in recently proposed text SemCom techniques \cite{Cognitive_SemCom_Systems'22,Jiang_Reliable_SemCom'22,Wang_SemCom_Per_Optimization'21,Wang_JSAC_Performance_Optimi_SemCom'22}. Despite KGs having such broad applicability, they have inherent downsides, since they can merely represent simplified causal graphs and hence are limited to the \textit{expressivity} of a graph \cite{Chaccour_Building_NG_SemCom_Networks'22}. KGs would thus fail to characterize highly complex tasks despite their causal structure \cite{Chaccour_Building_NG_SemCom_Networks'22}. DNNs are crucial in overcoming this limitation, as discussed below. 

\subsubsection{DNNs}
pertaining to deep networks' inherent ability in universal function approximation \cite{Hanin_2019_Univ_Approx,DL_Approx_Theory'21,Barron2018_DL_Approximation'18,Pinkus99approximationtheory,Barron_ANN_Approx_94,Getu_error_bounds_21,Daubechies_NL_Approximation_and_DReLU_NNs'19}, DNNs can be effectively employed for semantic representation. As a result, many DL-enabled SemCom techniques (e.g., \cite{Xie_DL-based_SemCom'21,Peng_Robust_DL-Based_SemCom'22,Weng_DL-enabled_SemCom'22,DL_Enabled_SemCom_with_Task_unaware_TX'22,Weng_SemCom_Sys_Speech_Trans'21,Weng_SemCom_Speech_Recognition'21,SemCom_for_speech_signals'20,Huang_Toward_SemCom'23,Xu_Wireless_Image_Transmission'22,Eirina_JSCC'19}) have been developed using an end-to-end trained joint source/channel encoder and a joint source/channel decoder that have been designed using state-of-the-art deep networks such as \textit{Transformers} \cite{Transformer_paper_NIPS2017,Universal_transformers'19,Liu_Swin_Transformer'21}, DNNs \cite{JS_DL_15,TW_DL_PWC_17,DG_JSAC_ML_in_z_air_19}, and convolutional neural networks (CNNs) \cite{Gu_CNN_Advances_2017}. In spite of the fact that DNNs are mature and easy to reparameterize; integrate fairly easily with existing AI/ML models; and have been widely adopted for semantic representation and therefore the design of SemCom systems, they are:
\begin{itemize}
	\item ineffective at reasoning\footnote{Since DNNs are not knowledge-driven networks, their reasoning capacity is limited by the statistical nature of the data \cite{Chaccour_Building_NG_SemCom_Networks'22}. Consequently, DNNs are widely believed to be poor at solving commonsense inference tasks \cite{DL_Appraisal_normal_Marcus'18}.} the cause, context, or effect of an event or source of data \cite{Chaccour_Building_NG_SemCom_Networks'22};
	
	\item unable to capture the complexity of data or learn a proper representation if the data are not purely statistical \cite{Chaccour_Building_NG_SemCom_Networks'22}.
\end{itemize}
Accordingly, DNNs face a major challenge when it comes to accurate semantic representation due to their limited contextual information reasoning capabilities that are not amenable to statistical relationships \cite{Chaccour_Building_NG_SemCom_Networks'22}. This calls for another semantic representation technique known as NLP, as discussed below.  

\subsubsection{NLP}
the NLP methodology for semantic representation centers on using a human language to describe the semantic information contained in raw data \cite{Chaccour_Building_NG_SemCom_Networks'22}. The semantic representation of data has been facilitated by advancements in DL-driven NLP \cite{Torfi_NLP_Survey'20,Trands_in_DL-Based_NLP'18}. These advancements have inspired the development of a number of text SemCom techniques \cite{Xie_DL-based_SemCom'21,Zhou_WiCom_Letters'22,Xie_Task-Oriented_MU-SemCom'22} in which natural languages can be used to describe the data \cite{Chaccour_Building_NG_SemCom_Networks'22}. Thus, semantic representation using NLP has the following advantages: 
\begin{itemize}
	
\item being readily understandable and decodable for design purposes \cite{Chaccour_Building_NG_SemCom_Networks'22}, and 

\item being appropriate for specific data structures that greatly depend on text data \cite{Chaccour_Building_NG_SemCom_Networks'22}.
\end{itemize}
Nevertheless, semantic representation by NLP is restricted by syntax, pragmatics, and wording \cite{Chaccour_Building_NG_SemCom_Networks'22}, which translates to the following major challenge: converting the bit-pipeline problem to a word-pipeline problem \cite{Chaccour_Building_NG_SemCom_Networks'22}. In alleviating this challenge, toposes are vital, as presented below.

\subsubsection{Toposes}
originated from \textit{homological algebra} and \textit{algebraic topology}, toposes transform every data structure by a family of objects in a well-defined topos \cite{Topos_and_Stacks'21,Heng_Math_for_Future_Comp_Commun'22}. Accordingly, semantic representation using toposes translates current data structures to well-defined \textit{morphisms} that make it possible to extract the unobserved semantic information \cite{Chaccour_Building_NG_SemCom_Networks'22}. To this end, employing toposes for semantic representation has two benefits:
\begin{itemize}
	\item toposes are capable of disentangling unobserved contextual patterns \cite{Chaccour_Building_NG_SemCom_Networks'22}, and
	
	\item toposes can reason beyond statistical boundaries \cite{Chaccour_Building_NG_SemCom_Networks'22}.
		
\end{itemize}
Despite these crucial advantages, semantic representation using toposes has its drawbacks because several topos concepts remain mathematically intractable and cumbersome to characterize \cite{Chaccour_Building_NG_SemCom_Networks'22}. Consequently, deploying toposes for semantic representation faces these major challenges in connection with SemCom: 
\begin{itemize}
	\item toposes cannot readily be unified with coexisting AI frameworks \cite{Chaccour_Building_NG_SemCom_Networks'22}, and
	
	\item toposes can be inherently quite computationally complex when handling raw data at the SemCom transmitter \cite{Chaccour_Building_NG_SemCom_Networks'22}.
\end{itemize}
 These challenges can make toposes unattractive and non-scalable for the overall design of an end-to-end communication system \cite{Chaccour_Building_NG_SemCom_Networks'22}. In mitigating this challenge in part, causal representation learning is useful, as discussed below.

\subsubsection{Causal Representation Learning}
semantic representation via causal representation learning \cite{Toward_causal_learning_21,Causal_ML'22} aims to learn a minimalist representation that can partly expose the unknown causal structure of the data \cite{Chaccour_Building_NG_SemCom_Networks'22}. The exposed data structure can reveal the semantic content elements of the data and their relationships -- hence the context -- while assuring high generalizability and minimalism \cite{Chaccour_Building_NG_SemCom_Networks'22}. Apart from its minimalism and generalizability, causal representation learning for semantic representation has the following benefits: 
\begin{itemize}
	\item it can leverage \textit{interventions} and \textit{counterfactuals} \cite{Pearl_Causality'09,Causality_Pearl_Primer'2016,Elements_Causal_Inference'17} to understand the structure of the data beyond \textit{associative logic}, and 
	
	\item the context of transmission is implicitly characterized by the causes of the semantic content elements \cite{Chaccour_Building_NG_SemCom_Networks'22}.
\end{itemize}
Accordingly, semantic representation via causal representation learning is restricted by the need to pose suitable interventions or counterfactuals at the apprentice \cite{Chaccour_Building_NG_SemCom_Networks'22}. In this vein, a major fundamental challenge of causal representation learning is embedding a structural causal model within a DNN that can characterize both statistical and causal properties \cite{Chaccour_Building_NG_SemCom_Networks'22}.

Apart from semantic representation, the fundamental performance quantification of both SemCom and goal-oriented SemCom systems/algorithms necessitate a fundamental theory of semantic information. To this end, we discuss next some of the existing theories of semantic information.

\subsection{Theories of Semantic Information}
\label{subsec: theories_semantics_information}
The first theory of semantic information was proposed by Carnap and Bar-Hillel in the early 1950s and was based on logical probabilities \cite{Tech_Report_Theory_of_Sem_Info'52,Bar_Carnap_Theory_SemInfo'1954}. They used logical probabilities (as opposed to the statistical probabilities used in the Shannon information theory) over the content of a sentence to quantify the amount of information in a sentence in a given language \cite{Bao_Towards_Theory_SemCom'11,Bao_Towards_Theory_of_SemCom'11}. In their theory, information is perceived as a set of excluded possibilities \cite{Langel_dissertation_09}, and a sentence's logical probability is measured by the likelihood that it would be true in all possible situations \cite{Bao_Towards_Theory_of_SemCom'11}. To this end, Carnap and Bar-Hillel's semantic information theory (SIT) asserts that \textquotedblleft A and B'' provides more information than \textquotedblleft A'' or \textquotedblleft B'' (since \textquotedblleft A and B'' is less likely to be true), \textquotedblleft A'' provides more information than \textquotedblleft A or B'', and a tautology (which is always true) provides no information \cite{Bao_Towards_Theory_of_SemCom'11}. This SIT is considered a model-theoretical approach to assign probabilistic values to logical sentences and thereby affirm a close relationship between the quantity of information in a sentence and the set of its models \cite{Bao_Towards_Theory_of_SemCom'11}. Thus, a consistent sentence that has fewer models comprises more information \cite{Bao_Towards_Theory_of_SemCom'11}, in line with Nilsson's probabilistic logic \cite{NILSSON_ProbLogic'86}. 

The above SIT does not consider the qualification of the information content as \textit{truthful} \cite{Floridi_Theory_Strongly_SemInfo'04,Stanford_Encyclopedia_of_Philosophy'22}. However, Floridi developed a \textit{theory of strongly semantic information} (TSSI)\footnote{We note that the author of \cite{Adriaans_Critical_Anakysis'10} attacked -- fiercely and rather subjectively -- Floridi and his TSSI while seeking to defend modern information theory as follows: \textquotedblleft I will defend the view that notions that are associated with truth, knowledge, and meaning all can adequately be reconstructed in the context of modern information theory and that consequently there is no need to introduce a concept of semantic information'' \cite{Adriaans_Critical_Anakysis'10}.} to capture truthfulness by defining semantic-factual information in terms of its data space as well-formed, meaningful, and truthful data \cite{Floridi_Theory_Strongly_SemInfo'04,Floridi_Phil_Conceptions_of_Information'09,Stanford_Encyclopedia_of_Philosophy'22}. TSSI is aimed at solving the \textit{Bar-Hillel-Carnap paradox} (BCP) \cite{Floridi_Theory_Strongly_SemInfo'04,Floridi_Phil_Conceptions_of_Information'09} in Carnap and Bar-Hillel's SIT, wherein contradictions provide an infinite amount of information \cite{Bao_Towards_Theory_of_SemCom'11}. TSSI resolves this paradox by working from the basic idea that the informativeness of a statement is measured by a positive/negative degree of semantic distance (or deviation) from \textquotedblleft truth'' \cite{Floridi_Theory_Strongly_SemInfo'04,Floridi_Phil_Conceptions_of_Information'09}. This makes TSSI completely different from Carnap and Bar-Hillel's SIT, which specifies informativeness as a function over all situations \cite{Bao_Towards_Theory_of_SemCom'11}.

Despite Floridi's attempt to resolve the BCP with his TSSI aimed at capturing truthfulness, D’Alfonso argues \cite{Simon_Quantification_SemInfo'11} that TSSI is incomplete in regard to quantifying all possible statements and that there exist propositional sentences that cannot be assessed using the TSSI approach \cite{Floridi_Theory_Strongly_SemInfo'04,Bao_Towards_Theory_of_SemCom'11}. As a result, D’Alfonso took inspiration from the existing works on \textit{truthlikeness} \cite{Niiniluoto_Truthlikeness'87,Oddie_Likeness_to_Truth'86} (the degree of being similar to the truth) and put forward the \textit{value aggregate method} \cite[Section 4]{Simon_Quantification_SemInfo'11} that seeks to capture both inaccuracy and vacuity using the formal models of truthlikeness. In doing so, D’Alfonso attempts to extend information quantification to the semantic concept of quantity of misinformation, where semantic information and semantic misinformation are defined true semantic content and false semantic content, respectively\cite{SemCom_for_6G_Future_Internet'22}. Apart from this theory and the other aforementioned SITs, other semantic information modeling approaches/theories include semantic information G theory \cite{Lu_SemInfo_G_Theory'19}, the theory of the semantics of questions and the pragmatics of answers \cite{Groenendijk_Stokhof_dissertation'84}, semantic information analysis from the vantage point of thermodynamics \cite{Kolchinsky_SemInfo_AA_NE_Stat_Physics'18}, the algebraic theory of semantic information \cite{Langel_dissertation_09}, the conceptual space theory of semantics \cite{Conceptual_Spaces_Book'00,The_Geometry_of_Meaning_Book'14}, causal semantics \cite{Elements_Causal_Inference'17}, information algebra \cite{Kohlas_Schmid_Algebraic_ToI'14}, the theory of information flow \cite{Barwise_Seligman_Book'97}, universal semantic communication \cite{Juba_Universal_SemCom_I'08,Juba_Universal_SemCom'08}, semantic coding \cite{Willems_semantic_compaction'05}, SIT via organized complexity \cite{Okamoto_Unified_Paradigm'16}, and SIT via quantum corollas \cite{Tetlow_Toward_Semantic_Info_Theory'22}.

Despite the numerous theories of semantic information that exist, as highlighted above, existing theories are fundamentally incomplete. From the fundamental standpoint of neuroscience and cognitive science, semantics can be defined in (and is closely related to) the context of subjective experience, i.e., \textquotedblleft meaningful'' parallels \textquotedblleft meaningful to a subject/person'' \cite{Samsonovich_Toward_Sematic_General_Theory_of_Everything'10}. In this vein, the essence of semantics becomes synonymous with the notion of potential experience \cite{Samsonovich_Toward_Sematic_General_Theory_of_Everything'10}. This calls for a \textit{semantic general theory of everything} \cite{Samsonovich_Toward_Sematic_General_Theory_of_Everything'10}.\footnote{\textquotedblleft In conclusion, to develop a unified semantic theory applicable to subjective experience and objective physical reality, we need to better understand semantics itself, from a conceptual to a computational level. This sort of knowledge can be extracted from natural language and all documents, viewed as a collective product of all human minds of all generations'' \cite{Samsonovich_Toward_Sematic_General_Theory_of_Everything'10}.} However, the development of such a fundamental general theory is fraught with numerous fundamental challenges. Without being bogged down by the fundamental challenges, however, current research progress on SemCom and goal-oriented SemCom can be guided by a rigorous definition of semantic entropy. Accordingly, we move on to the existing definitions of semantic entropy.

\subsection{Definitions of Semantic Entropy}
\label{subsec: Defns_semantic_entropy}
Around 1952, Carnap and Bar-Hillel introduced \cite{Tech_Report_Theory_of_Sem_Info'52} the concept of semantic entropy of a sentence -- within a given language -- which is defined as \cite[eq. (1)]{Gunduz_Beyond_Transmitting_Bits'22}
\begin{equation}
\label{H_s_e_CB}
H(s,e) \eqdef -\log c(s,e),
\end{equation}
where $c(s,e)$ is the degree of confirmation of sentence $s$ on the evidence $e$, which is written as \cite[eq. (2)]{Gunduz_Beyond_Transmitting_Bits'22}
\begin{equation}
\label{c_s_e_defn}
c(s,e) \eqdef \frac{m(e,s)}{m(e)},
\end{equation}
where $m(e, s)$ and $m(e)$ denote the logical probability of $s$ on $e$ and of $e$ \cite{Gunduz_Beyond_Transmitting_Bits'22}, respectively.

Regarding KB being useful to correctly infer the transmitted message -- even when the receiver is unable to directly decode a semantic message -- given a set of logical relationships \cite{Sem_Empowered_Commun'22}, the authors of \cite{Choi_Unified_View_SemInfo'22} introduce the notion of \textit{knowledge entropy}. The knowledge entropy of a KB is defined as the \textit{uncertainty of the answers it computes} \cite{Sem_Empowered_Commun'22}. Thus, knowledge entropy equates to the average semantic entropy of query (message) $x$ calculable from the KB $\mathcal{K}$ \cite[eq. 1]{Choi_Unified_View_SemInfo'22}, \cite[eq. (7)]{Sem_Empowered_Commun'22}:
\begin{equation}
\label{KB_entropy}
H(\mathcal{K})  \eqdef   \frac{1}{|\mathcal{K}|} \sum_{x \in \mathcal{K} }   H(x),
\end{equation}
where $H(\mathcal{K})$ is the knowledge entropy of $\mathcal{K}$, $|\mathcal{K}|$ is the size of the KB, and $H(x)$ is the semantic entropy of each message $x$ and is defined as \cite[eq. (4)]{Sem_Empowered_Commun'22}
\begin{equation}
\label{Semantic_entropy_of_x}
H(x) \eqdef - \big[ m(x) \log (m(x)) + \big( 1-m(x)\big) \log \big(1-m(x) \big)    \big],
\end{equation}
where $m(x)$ is the logical probability pertaining to the probability of being true in a given world model \cite{Sem_Empowered_Commun'22}.

Grounded on a language comprehension model -- contrary to the probabilistic structure of a language -- in the context of the structure of the world, the authors of \cite{Venhuizen_semantic_entropy'19} derive semantic entropy that is given by \cite[eq. (3)]{Gunduz_Beyond_Transmitting_Bits'22}
\begin{equation}
\label{Sem_entropy_language_comprehension_model}
H(\bm{v}_t) \eqdef - \sum_{\bm{v}_M \in \mathcal{V}_{\mathcal{M}}} \mathbb{P}(\bm{v}_M | \bm{v}_t) \log \mathbb{P}(\bm{v}_M | \bm{v}_t), 
\end{equation}
where $\mathcal{M}$ denotes the set of models that reflects the probabilistic structure of the world and $\mathcal{V}_{\mathcal{M}} \eqdef  \{ \bm{v}_M | \bm{v}_M(i) = 1 \hspace{2mm} \textnormal{if and only if} \hspace{2mm} M_i= M  \hspace{2mm} \textnormal{and} \hspace{2mm} M \hspace{2mm} \textnormal{is a unique model in} \hspace{2mm} M\}$ \cite{Gunduz_Beyond_Transmitting_Bits'22}. The comprehension-centric semantic entropy model expressed by (\ref{Sem_entropy_language_comprehension_model}) quantifies uncertainty w.r.t. the whole meaning space and relies on both linguistic experience as well as world knowledge \cite{Gunduz_Beyond_Transmitting_Bits'22}.

Aside from the above semantic entropy definitions -- expressed in (\ref{H_s_e_CB}) and (\ref{Sem_entropy_language_comprehension_model}) -- that pertain to the language system, semantic entropy concerning intelligent tasks has also been studied in \cite{Melamed_semantic_entropy'97}. The author of \cite{Melamed_semantic_entropy'97} proffers an information-theoretic method for measuring semantic entropy in translation tasks by employing translational distributions of words in parallel text corpora \cite{Gunduz_Beyond_Transmitting_Bits'22}. According to this method, the semantic entropy of each word $w$ is defined as \cite[eq. (4)]{Gunduz_Beyond_Transmitting_Bits'22}
\begin{multline}
\label{Melamed_semantic_entropy}
H(w) \eqdef H(T|w)+ N(w)= -\sum_{t\in T} \mathbb{P}(t|w) \log \mathbb{P}(t|w) + \\  \mathbb{P}(NULL|w) \log F(w),
\end{multline}
where $T$ denotes the set of target words, $H(T|w)$ represents the translational inconsistency of a source word $w$, $N(w)$ stands for the contribution of null links of $w$, and $F(w)$ is the frequency of $w$ \cite{Gunduz_Beyond_Transmitting_Bits'22}. In addition, the authors of \cite{Liu_Semantic_Entropy'20} define semantic entropy for classification tasks by considering the membership degree in axiomatic fuzzy set theory \cite{Liu_Semantic_Entropy'20}. In accordance with their framework, the authors of \cite{Liu_Semantic_Entropy'20} first obtain the matching degree regarding the characterization of the semantic entropy of the data samples in class $C_j$ on semantic concept $\varsigma$ as \cite[eq. (5)]{Gunduz_Beyond_Transmitting_Bits'22}, \cite[eq. (12)]{Liu_Semantic_Entropy'20}
\begin{equation}
\label{membership_degree_defn}
D_j(\varsigma) \eqdef \frac{\sum_{x \in \mathcal{X}_{C_j} } \mu_{\varsigma}(x)}{\sum_{x \in \mathcal{X}} \mu_{\varsigma} (x) },
\end{equation}
where $\mathcal{X}_{C_j}$ denotes the set of data for class $C_j$ w.r.t. all $j \in \{1, 2, \ldots , m\}$ and $\mathcal{X}$ stands for the data set of all classes \cite{Gunduz_Beyond_Transmitting_Bits'22}. Using (\ref{membership_degree_defn}), the semantic entropy of class $C_j$ on $\varsigma$ is defined as \cite[eq. (6)]{Gunduz_Beyond_Transmitting_Bits'22}, \cite[eq. (13)]{Liu_Semantic_Entropy'20}  
\begin{equation}
\label{Sem_entropy_on_macthing_degree_defn}
H_{C_j} (\varsigma) \eqdef  -D_j(\varsigma)  \log_2 D_j(\varsigma). 
\end{equation}
Using (\ref{Sem_entropy_on_macthing_degree_defn}), the semantic entropy of concept $\varsigma$ on $\mathcal{X}$ is defined as \cite[eq. (7)]{Gunduz_Beyond_Transmitting_Bits'22}, \cite[eq. (14)]{Liu_Semantic_Entropy'20}
\begin{equation}
\label{semantic_entropy_of_varsigma_defn}
H(\varsigma)   \eqdef  \sum_{j=1}^m H_{C_j} (\varsigma).
\end{equation}
The uncertainty in designing the classifier is minimized by the definitions in (\ref{membership_degree_defn})-(\ref{semantic_entropy_of_varsigma_defn}), which can be used to obtain the optimal semantic description for each class \cite{Gunduz_Beyond_Transmitting_Bits'22}.

Whereas the above-mentioned definitions apply mainly to a single task, the authors of \cite{Chattopadhyay_Quantifying_Task'21} investigate an information-theoretic framework to quantify the semantic information of any source for any task. Regardless of the task at hand, the authors of \cite{Chattopadhyay_Quantifying_Task'21} define semantic entropy as the minimum number of semantic queries about data $X$ whose answers are sufficient to predict task $Y$ \cite{Gunduz_Beyond_Transmitting_Bits'22}. Mathematically, the semantic information quantification of the work in \cite{Chattopadhyay_Quantifying_Task'21} is given by \cite[eq. (8)]{Gunduz_Beyond_Transmitting_Bits'22}, \cite[eq. (2)]{Chattopadhyay_Quantifying_Task'21},
\begin{equation}
\begin{aligned}
\label{Sem_entropy_Chattopadhyay_defn}
\quad &H_Q(X;Y)   \eqdef \displaystyle\min_{E} \mathbb{E}_X\big\{ \big| Code^E_Q(X) \big|\big\}  \\
\quad   & \textrm{s.t.} \hspace{4mm}\mathbb{P}(y|Code^E_Q(x))  = \mathbb{P}(y|x), \hspace{3mm} \forall x,y,
\end{aligned}
\end{equation}
where $E$ denotes the semantic encoder and $Code^E_Q(x)$ represents the query vector extracted from $X$ with $E$ \cite{Gunduz_Beyond_Transmitting_Bits'22}. As seen in (\ref{Sem_entropy_Chattopadhyay_defn}), one needs to find the optimal semantic encoder (that encodes $X$ into the minimal representation that can faithfully predict the task \cite{Gunduz_Beyond_Transmitting_Bits'22}) to be able to obtain the semantic entropy.

Apart from the previously outlined techniques for measuring semantic entropy, several new methods -- such as semantic information pursuit and variational inference -- are emerging and they need to be further investigated \cite{Gunduz_Beyond_Transmitting_Bits'22}. To summarize, all of the aforementioned definitions except the last one are task-oriented, whereas the last one that can be applied to different tasks \cite{Gunduz_Beyond_Transmitting_Bits'22}. However, designing the corresponding optimal semantic encoder is as challenging as obtaining semantic entropy \cite{Gunduz_Beyond_Transmitting_Bits'22}. Therefore, there is no unifying definition for semantic entropy: existing definitions lack the operational relevance of the Shannon entropy in many engineering problems \cite{Gunduz_Beyond_Transmitting_Bits'22}.

In what follows, we discuss extensively the state-of-the-art research landscape of SemCom.   

\section{State-of-the-Art Research Landscape of SemCom}
\label{sec: SemCom_research_landscape}
\begin{figure*}[htb!]
	\centering
	\includegraphics[scale=0.52]{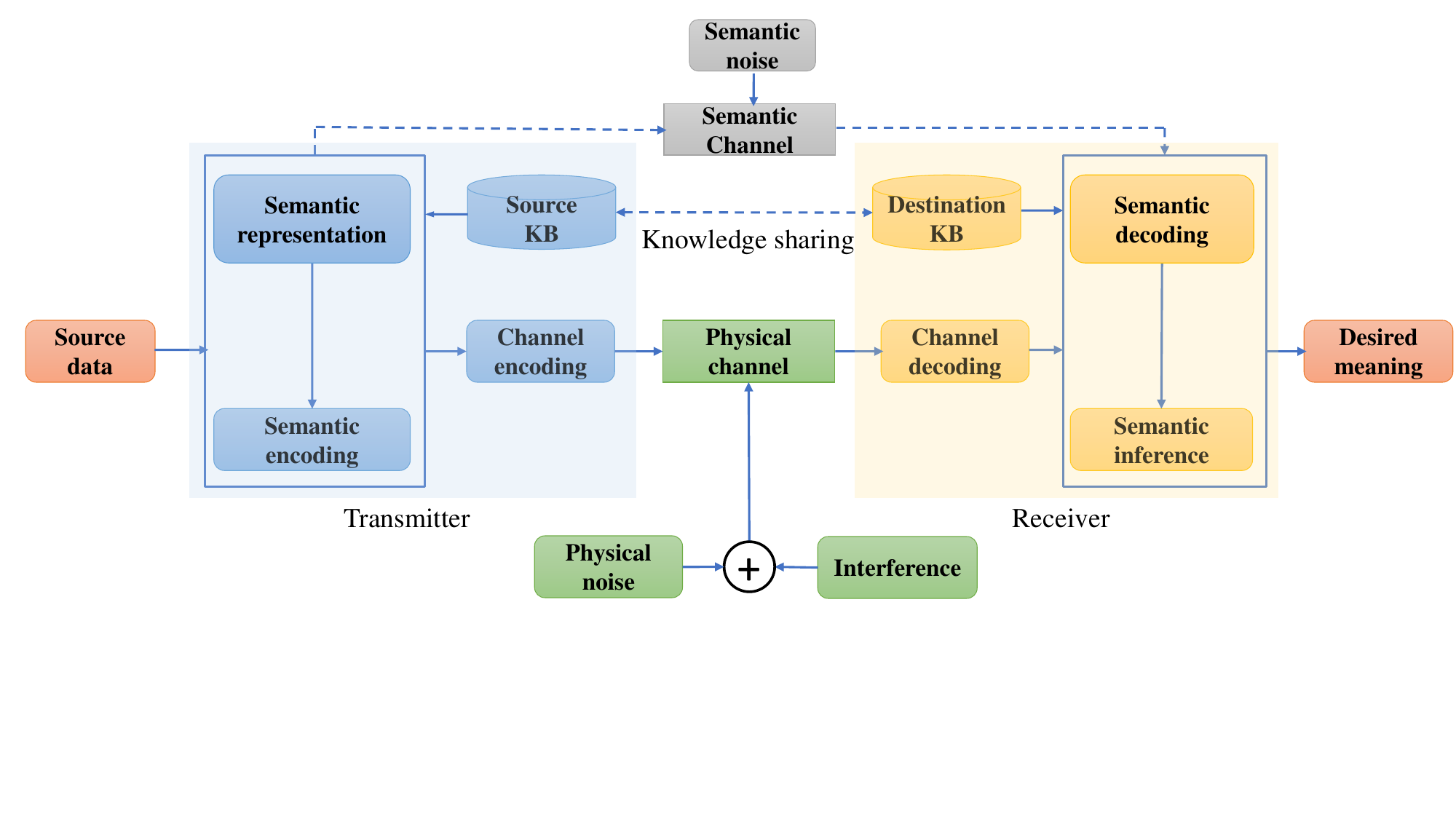}  \vspace{-2.1cm} 
	\caption{System model for semantic-oriented communications -- modified from \cite[Fig. 6(b)]{SemCom_for_6G_Future_Internet'22}.}
	\label{fig: SemCom_oriented_system_model}
\end{figure*}

SemCom aims to convey a desired meaning. A desired meaning can be communicated through a SemCom transceiver as shown in Fig. \ref{fig: SemCom_oriented_system_model}. In Fig. \ref{fig: SemCom_oriented_system_model}, a semantic representation inspired by \cite[Fig. 2]{Shi_to_Semantic_Fidelity'21} is used to convert the source data to a semantic modality \cite{Shi_to_Semantic_Fidelity'21} that will be encoded semantically by a semantic encoder (w.r.t. a source KB). The semantic encoder's function in various state-of-the-art studies -- such as those on DL-based SemCom (e.g., \cite{Xie_Lite_distributed_SemCom'21,Xie_DL-based_SemCom'21,Hu_Robust_SemCom'22,Weng_SemCom_Sys_Speech_Trans'21,Weng_SemCom_Speech_Recognition'21,Xie_TO_MU_SemCom'21}) -- jointly encompasses semantic representation and semantic encoding as schematized in Fig. \ref{fig: SemCom_oriented_system_model}. This figure also shows a receiver's decoding in reference to the desired meaning using the in tandem operation -- w.r.t. the destination KB -- of the semantic decoding and semantic inference blocks. In a DL-based SemCom (e.g., \cite{Xie_Lite_distributed_SemCom'21,Xie_DL-based_SemCom'21,Hu_Robust_SemCom'22,Weng_SemCom_Sys_Speech_Trans'21,Weng_SemCom_Speech_Recognition'21,Xie_TO_MU_SemCom'21}), these blocks' semantic inference and semantic decoding tasks are combined and performed by a semantic decoder (see Fig. \ref{fig: SemCom_main_components}). Per Fig. \ref{fig: SemCom_oriented_system_model}, semantic decoding and semantic inference can suffer greatly from semantic noise (as discussed in Section \ref{subsec: Why_SemCom}) when there is a mismatch between the source KB and destination KB. The destination KB needs to be shared with the source KB in real time for effective SemCom, similar to productive human conversation requiring common knowledge of the communicating parties' languages and cultures \cite{SemCom_for_6G_Future_Internet'22}. This knowledge sharing facilitates KB-assisted semantic extraction (SE), which is a SemCom technique for joint semantic encoding and decoding.   

The joint semantic encoding and decoding process is regarded as SE \cite{SemCom_for_6G_Future_Internet'22}. SE is a core component of a semantic transceiver, and four major state-of-the-art SE techniques exist \cite[Fig. 10]{SemCom_for_6G_Future_Internet'22}, \cite[Fig. 3]{Yang_SemCom_meets_Edge_Intelligence'22}: DL-based SE, reinforcement learning (RL)-based SE, KB-assisted SE, and semantic-native SE, which we discuss below. Apart from these SE techniques, there are also some specialized SE approaches for specific semantic-aware communication scenarios \cite{SemCom_for_6G_Future_Internet'22}.

We begin with DL-based SE, which leverages advancements in DL \cite{Lecun_DL_Nature_15,IGYAC16,PSSSYYTTRMMSS19,QFQH_18,CZPHH19,LWXPJLM18} and NLP \cite{Trands_in_DL-Based_NLP'18}. DL-based SE aims to enhance the SemCom system's robustness in low signal-to-noise ratio (SNR) regimes by modeling the semantic (channel) encoder and the semantic (channel) decoder -- at the transmitter and receiver, respectively -- as two separate learnable sections that are linked through a random channel \cite{SemCom_for_6G_Future_Internet'22,Yang_SemCom_meets_Edge_Intelligence'22}. The random channel -- which is often modeled by a generative adversarial network (GAN) \cite{GANs_NIPS2014,Creswell_GANs_18,Wang2019GenerativeAN} or an untrainable layer -- is trained using the DNN-based semantic (channel) decoder and the DNN-based semantic (channel) encoder in an end-to-end manner with differentiable loss functions such as cross entropy (CE) and mean squared error (MSE). MSE- and CE-mediated end-to-end training treats the joint semantic encoding and decoding process as a \textquotedblleft black box'' \cite{Castelvecchi_Black_Box_16}, which affirms a \textit{fundamental lack of interpretability}. The lack of interpretability makes the effectiveness of DL-based SE hard to quantify \cite{SemCom_for_6G_Future_Internet'22}. As highlighted so far, DL-based SE considers only the semantic coding problem, without any semantic understanding \cite{SemCom_for_6G_Future_Internet'22}. This challenge can be alleviated in part by using RL-based SE. 

As a first RL-based SE scheme, the authors of \cite{Lu_RL-powered_SemCom'21,Rethinking_modern_com_Lu_2022} propose to integrate RL into an end-to-end text SemCom system, wherein the encoder and decoder are viewed as an agent that interacts with sentences that are considered to be in an external environment \cite{SemCom_for_6G_Future_Internet'22}. For this RL-based SE technique, simulation results demonstrate it outperforms DL-based SE for non-differential semantic metric optimization. Nonetheless, RL-based SE suffers from inflexibility when it comes to SE for variable goal-oriented SemCom \cite{SemCom_for_6G_Future_Internet'22}. This type of SemCom's effectiveness can be improved by using KB-assisted SE \cite{SemCom_for_6G_Future_Internet'22}.

KB-assisted SE incorporates the KB into the encoder and decoder in an end-to-end manner with synchronized KBs at both ends to efficiently extract semantic information for scenarios with multiple communication tasks \cite{Yang_SemCom_with_AI_Tasks'21,SemCom_for_6G_Future_Internet'22,Yang_SemCom_meets_Edge_Intelligence'22}. The KB-assisted SE technique can achieve goal-based SE with retraining \cite{SemCom_for_6G_Future_Internet'22} using a typical KB, which comprises a computational ontology, facts, rules, and constraints \cite{SemCom_for_6G_Future_Internet'22}. Meanwhile, a KB in Semcom is of source information, goals corresponding to desired tasks, and methods of reasoning that can be understood, recognized, and learned by all parties involved in communication \cite{Yang_SemCom_meets_Edge_Intelligence'22}. Moreover, KB-assisted SE has also addressed goal-based SE without retraining \cite{SemCom_for_6G_Future_Internet'22}. However, it lacks self-adaptability for the possible evolution of a communication goal \cite{SemCom_for_6G_Future_Internet'22}. In this vein, it would be especially challenging to build a general KB that can capture the complex and diverse relationships that exist between semantic information and tasks/goals \cite{SemCom_for_6G_Future_Internet'22}. This limitation is partly addressed by semantic-native SE.

\begin{figure*}[t!]
	\centering
	\includegraphics[scale=0.50]{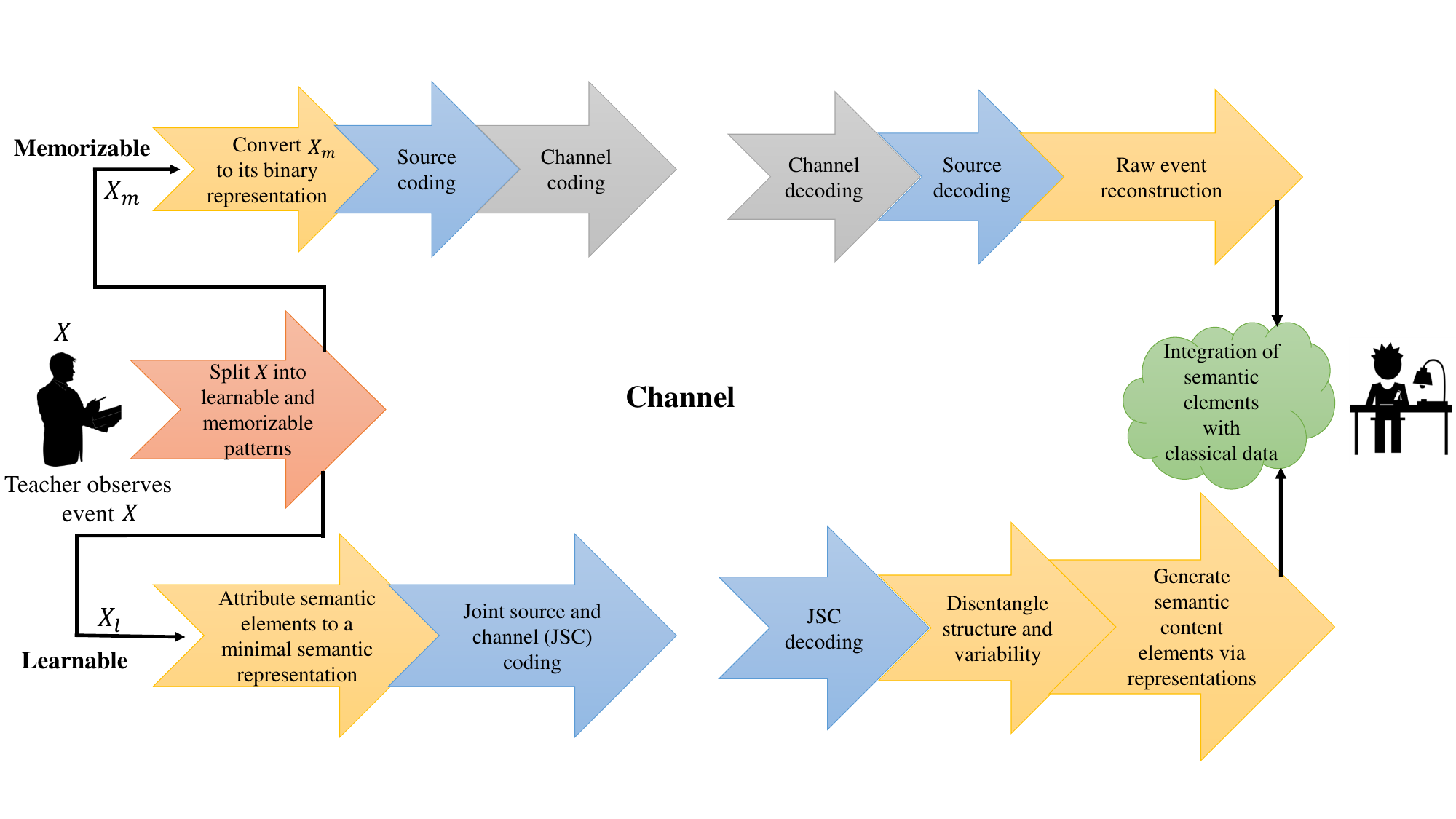}  \vspace{-0.35cm} 
	\caption{SemCom with memorizable and learnable data patterns -- modified from \cite[Fig. 8]{Chaccour_Building_NG_SemCom_Networks'22}.}
	\label{fig: SemCom_with_LMPs}
\end{figure*}

Semantic-native SE underpins the fact that the aforementioned SE techniques are effective for communication systems with unchanging semantics, whereas semantics often vary over time in real-world scenarios \cite{SemCom_for_6G_Future_Internet'22}. To address this challenge and give transceivers contextual reasoning ability, the authors of \cite{Seo_Sem_Naive_Com_with_Contextual_Reasoning'21} introduce \textit{System 1} and \textit{System 2} (inspired by the book in \cite{Kahneman_Thinking_Fast_and_Slow'2011}) semantic native communication (SNC). In System 1 SNC, a speaker conceptualizes and symbolizes an entity of interest (e.g., an abstract idea, a physical phenomenon, or an object) as a semantic representation -- that is decodable as the intended entity by its listener -- to be communicated to a target listener \cite{Seo_Sem_Naive_Com_with_Contextual_Reasoning'21}. When System 1 SNC is infused with contextual reasoning such that the speaker locally and iteratively communicates with a virtual agent built on a listener’s unique way of coding its semantics, it would follow that System 2 SNC can allow the speaker to extract its listener-tailored effective semantics \cite{Seo_Sem_Naive_Com_with_Contextual_Reasoning'21}. Despite System 2 SNC having effective semantics, System 1 and 2 SNC have a limitation: semantic-native SE through SNC is still a theoretical model that is difficult to generalize in practice \cite{SemCom_for_6G_Future_Internet'22}. 

\begin{figure*}[t!]
	\centering  \vspace{-0.6cm}
	\includegraphics[scale=0.40]{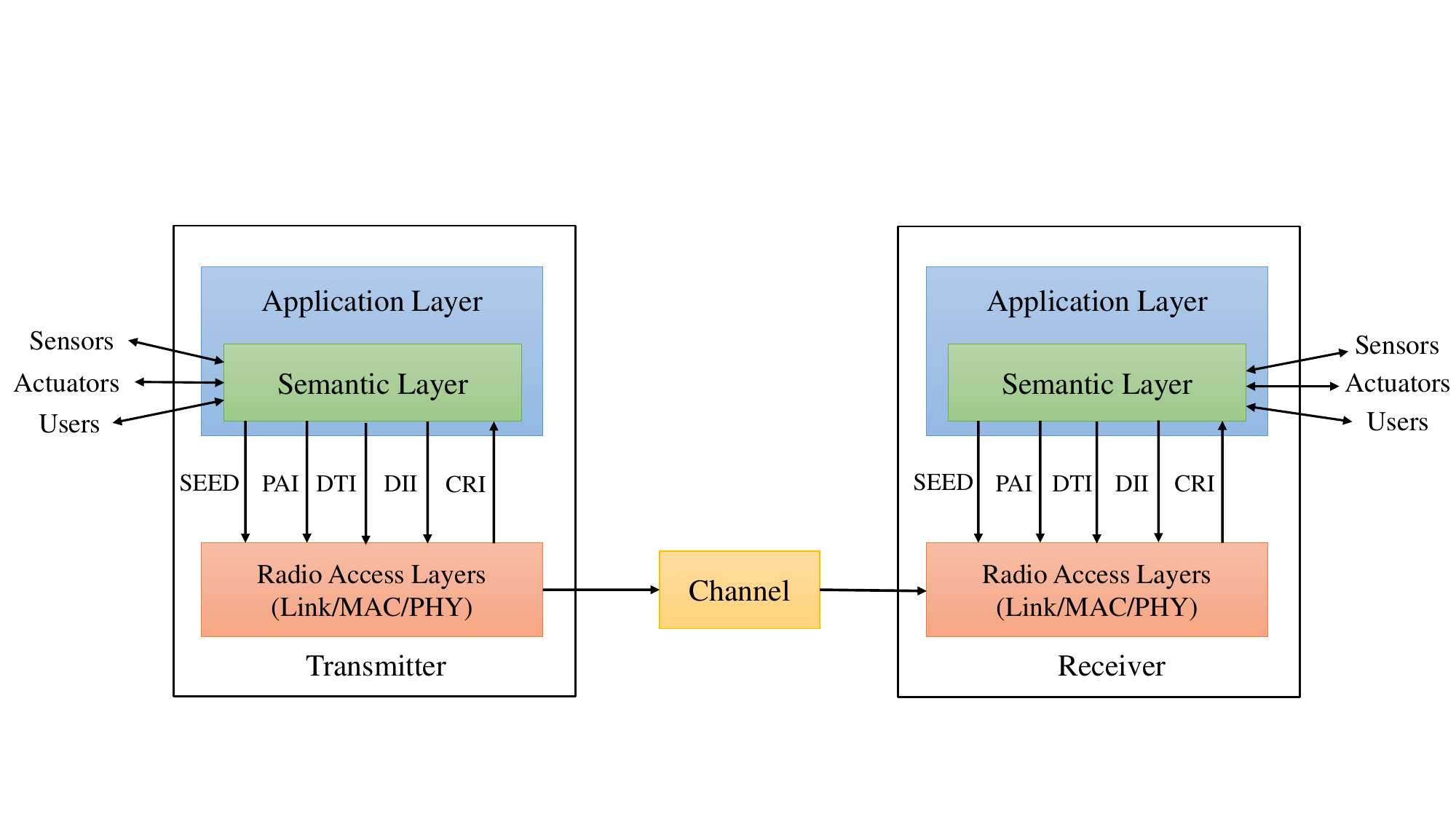}  \vspace{-0.75cm} 
	\caption{A layered architecture of a SemCom system -- \cite[Figure 2]{Qiao_What_is_SemCom'21}: SEED: semantic/effectiveness encoded data; PAI: partial algorithm information; DTI: data type information; DII: data importance information; and CRI: channel rate information.}
	\label{fig: Layered_SemCom_Architecture}
\end{figure*}

Considering the wide variety of SemCom techniques and trends that exist, the authors of \cite{Chaccour_Building_NG_SemCom_Networks'22} emphatically clarify\footnote{This paper does not necessarily make distinctions regarding the design philosophies of SemCom and goal-oriented SemCom in order to incorporate a variety of views on SemCom and goal-oriented SemCom.} what SemCom is and is not:
\begin{itemize}
	\item SemCom is not data compression. 
	
	\item SemCom is not only an \textquotedblleft AI for wireless'' concept.
	
	\item SemCom is not only goal-oriented communication.
	
	\item SemCom is not only application-aware communication.
\end{itemize} 
In light of these clarifications, the authors of \cite{Chaccour_Building_NG_SemCom_Networks'22,haccour_Disentangling_Learnable'22} propose a SemCom system with memorizable and learnable data patterns, which is schematized in Fig. \ref{fig: SemCom_with_LMPs}. Per Fig. \ref{fig: SemCom_with_LMPs}, the teacher first observes an event $X$ that is subsequently split into memorizable and learnable patterns. The learnable component $X_l$ and the memorizable component $X_m$ are then transformed into a binary representation and a minimal semantic representation, respectively. Thereafter, the underneath cascaded signal processing follow: 
\begin{itemize}
	\item The binary representation of the memorizable part is transmitted via a physical channel after it is transformed by the source encoder followed by the channel encoder. The received signal is then transformed via the channel decoder followed by the source decoder. Thereafter, a raw event reconstruction occurs as shown in Fig. \ref{fig: SemCom_with_LMPs} \cite{Chaccour_Building_NG_SemCom_Networks'22}. 
	
	\item The semantically represented learnable component is transformed by joint source and channel (JSC) coding prior to its transmission via the physical channel. Once transformed by the physical channel, the received signal is passed through the JSC decoder, whose output is employed to determine the materialization of the structure and the variability of the task, as shown in Fig. \ref{fig: SemCom_with_LMPs}. Then, the semantic content is generated via representation \cite{Chaccour_Building_NG_SemCom_Networks'22}.
\end{itemize}

Once the mentioned cascaded signal processing is completed, the end-to-end SemCom system is concluded by integrating the reconstructed (classical) memorizable component and (semantic) learned content into the recovered content from the teacher \cite{Chaccour_Building_NG_SemCom_Networks'22}, as depicted in Fig. \ref{fig: SemCom_with_LMPs}. The \textit{total achievable capacity} $C_T$ of the SemCom system depicted in Fig. \ref{fig: SemCom_with_LMPs} is given by \cite[eq. (16)]{Chaccour_Building_NG_SemCom_Networks'22}
\begin{equation}
\label{Total_achievable_capacity_defn}
C_T \eqdef C_C + C_R = W \log_2 (1+\gamma) + \Omega \log_2 (1+\eta_{b,d}), 
\end{equation}
where $C_C \eqdef W \log_2 (1+\gamma)$ is the Shannon capacity for $W$ and $\gamma$ being the bandwidth and the signal-to-interference-plus-noise ratio (SINR), respectively, and $C_R \eqdef \Omega \log_2 (1+\eta_{b,d})$ is the \textit{reasoning capacity} \cite[Proposition 3]{Chaccour_Building_NG_SemCom_Networks'22} for $\Omega$ and $\eta_{b,d}$ being the maximum computing capability of the server deployed to represent/generate the semantic representation and the \textit{communication symmetry index} \cite[Proposition 2]{Chaccour_Building_NG_SemCom_Networks'22} per second, respectively \cite{Chaccour_Building_NG_SemCom_Networks'22}.

Aside from the above-discussed works that attempt to set out the concept of SemCom, some other state-of-the-art works proffer useful SemCom architectures. One such architecture, the layered SemCom architecture that is proposed by the authors of \cite{Qiao_What_is_SemCom'21} is shown in Fig. \ref{fig: Layered_SemCom_Architecture}. The authors of \cite{Qiao_What_is_SemCom'21} propose a semantic \textit{open system interconnection} (OSI) model that is built on the conventional OSI protocol stack and has a semantic layer added as a sub-layer of the application layer (see Fig. \ref{fig: Layered_SemCom_Architecture}). Due to its sub-layer position, the semantic layer interfaces with sensors, actuators, and users, and has access to algorithms and the content of data in a specific application \cite{Qiao_What_is_SemCom'21}. It, therefore, executes semantic/effectiveness encoding/decoding and sends (see Fig. \ref{fig: Layered_SemCom_Architecture}) lower radio access layers semantic/effectiveness encoded data (SEED) such as data importance information (DII), partial algorithm information (PAI), and data type information (DTI) over a control channel \cite{Qiao_What_is_SemCom'21}. The semantic layer also receives (per Fig. \ref{fig: Layered_SemCom_Architecture}) channel rate information (CRI) from the radio access layers over a control channel. This CRI is employed to mitigate semantic noise for semantic symbol error correction or control computing in the application layer \cite{Qiao_What_is_SemCom'21}.

\begin{figure*}[t!]
	\centering
	\includegraphics[scale=0.54]{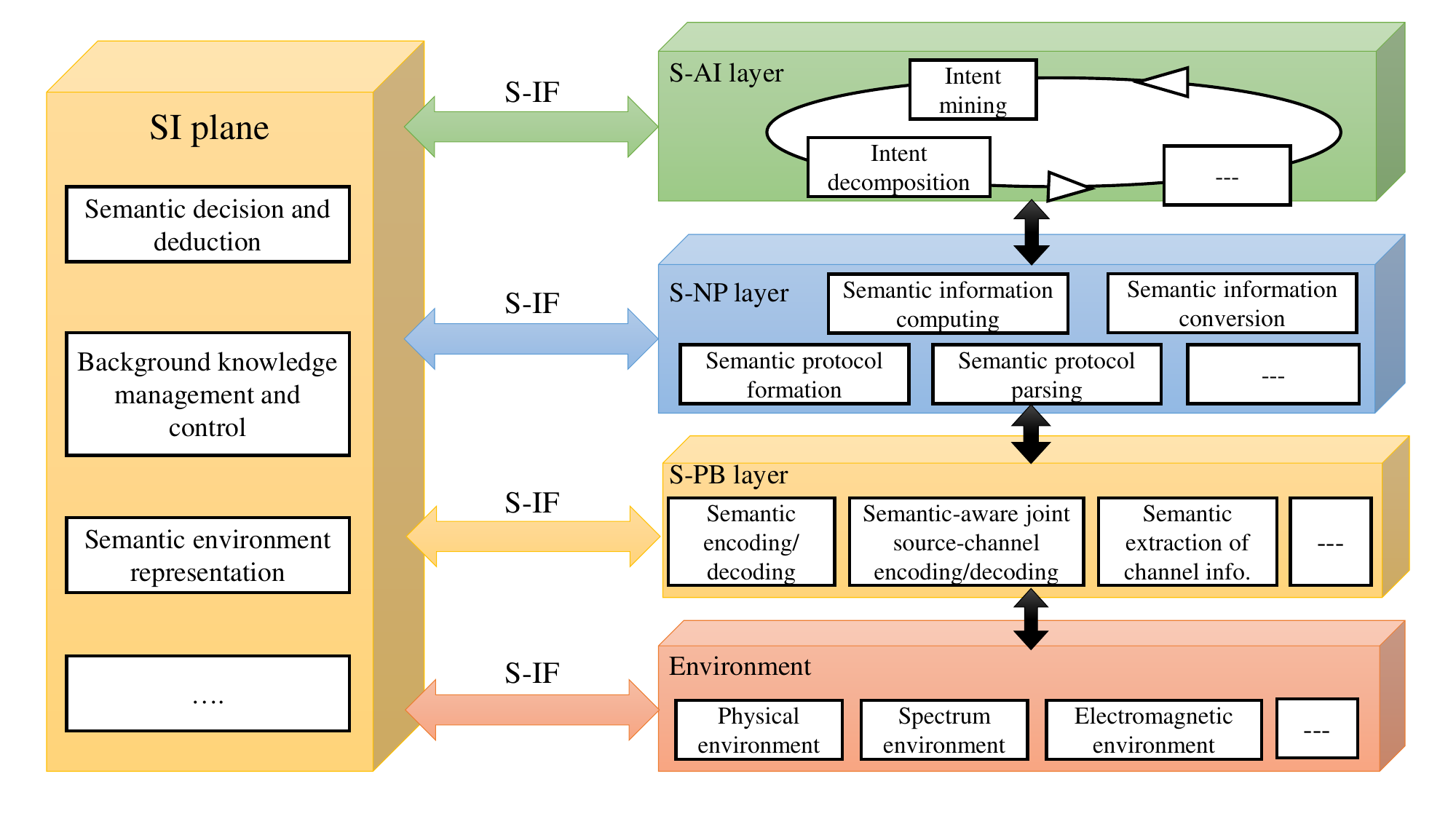}  \vspace{-0.5cm} 
	\caption{The intelligent and efficient semantic communication (IE-SC) network architecture \cite[Fig. 3]{Zhang_Wisdom_Evolutionary_6G'21} -- SI plane: semantic intelligence plane; S-PB layer: semantic-empowered physical-bearing layer; S-NP layer: semantic-empowered network protocol layer; S-AI layer: semantic-empowered application-intent layer; S-IF: semantic information flow; and info.: information.}
	\label{fig: IE-SC_Architecture_20221103}
\end{figure*}

\begin{figure}[t!]
 	%\centering
 	\hspace{-0.4cm}\includegraphics[scale=0.40]{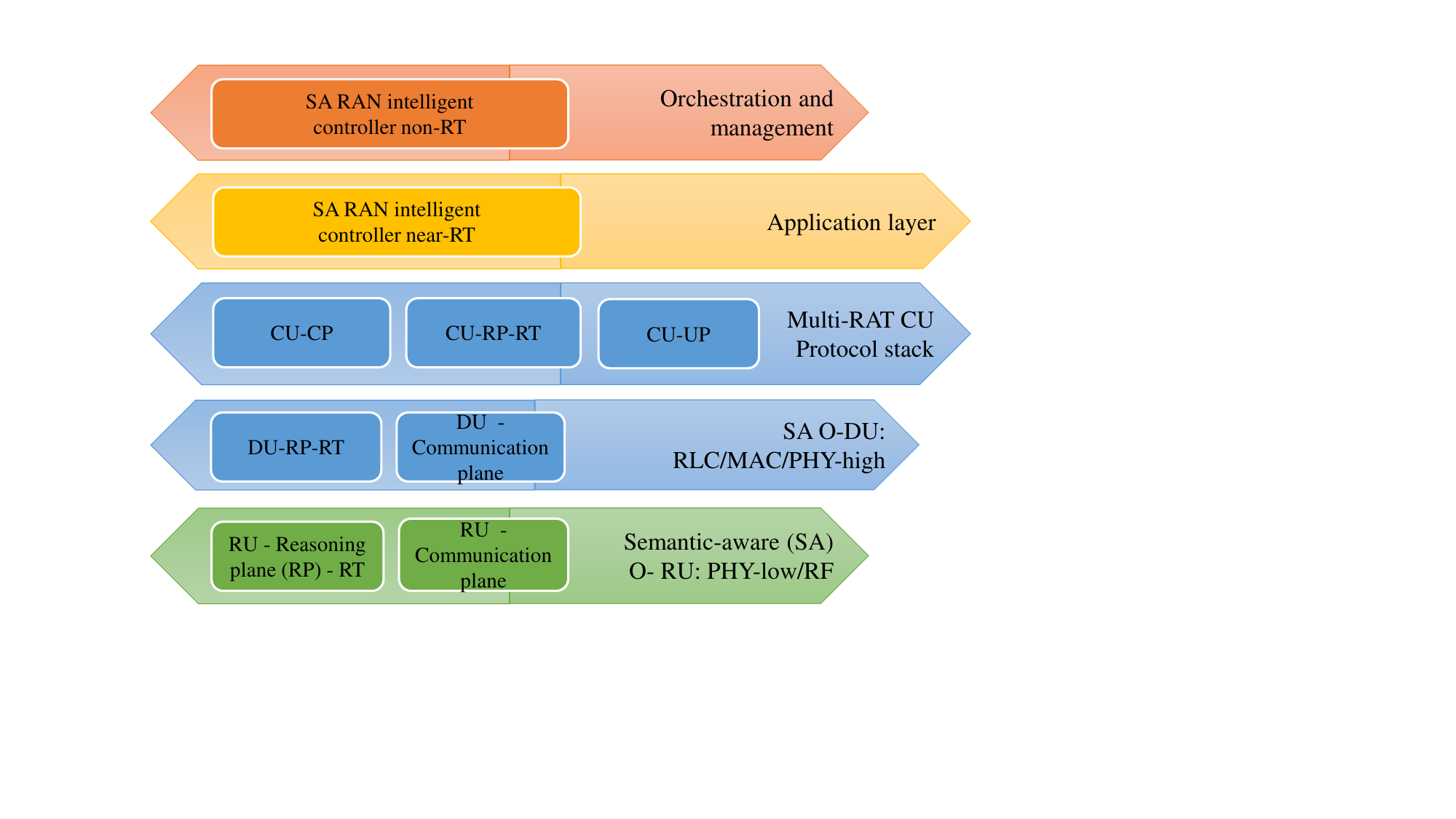}  \vspace{-1.5cm} 
 	\caption{O-RAN architecture for SemCom-enabled 6G and beyond -- modified from \cite[Figure 13]{Chaccour_Building_NG_SemCom_Networks'22}: SA: semantic-aware; RT: real-time; RP: reasoning plane; CU: centralized unit; RU: radio unit; DU: distributed unit; O-RU: open radio unit; O-DU: open distributed unit.}
 	\label{fig: AI-native_O-RAN_architecture_20221223}
 \end{figure}

The authors of \cite{Zhang_Wisdom_Evolutionary_6G'21} put forward the intelligent and efficient semantic communication (IE-SC) network architecture, which is shown in Fig. \ref{fig: IE-SC_Architecture_20221103}. This architecture comprises a semantic intelligence plane (SI plane), a semantic-empowered physical-bearing layer (S-PB layer), a semantic-empowered network protocol layer (S-NP layer), a semantic-empowered application-intent layer (S-AI layer), and a semantic information flow (S-IF) \cite{Zhang_Wisdom_Evolutionary_6G'21}. Despite this architecture's promises of intelligent and efficient SemCom, it is radically different from previous architectures and may not be interoperable with the state-of-the-art OSI model of 5G networks and their evolution. Meanwhile, the authors of \cite{Zhang_Wisdom_Evolutionary_6G'21} advance the traditional human-machine-thing architecture with their \textit{human–machine–thing–genie} architecture (to orchestrate the physical and digital worlds) wherein \textit{genie} is envisioned to be the main entity of the digital world that is used as an AI-empowered \textit{super intelligent} agent for physical communication objects. Furthermore, the authors of \cite{Chaccour_Building_NG_SemCom_Networks'22} propose the open-RAN (O-RAN) architecture for SemCom-enabled 6G and beyond, which is shown in Fig. \ref{fig: AI-native_O-RAN_architecture_20221223}. In this architecture that incorporates the introduction of the reasoning plane, real-time AI-oriented blocks are incorporated in the open radio unit (O-RU), the open distributed unit (O-DU), and the centralized unit (CU) \cite{Chaccour_Building_NG_SemCom_Networks'22}.    

 
 We now move on to the state-of-the-art vision and tutorial works on SemCom.

\subsection{Vision and Tutorial Works on SemCom}
\label{subsec: vision_and_tutorial_works_SemCom} 
In this section, we present the existing vision and tutorial works on SemCom. We begin with the vision works.     

\subsubsection{Vision Works on SemCom}
\label{subsubsec: vision_works_SemCom}
the authors of \cite{SemCom_Game'18} first introduce SemCom for text transmission (text SemCom) by formulating the text SemCom problem as a static Bayesian game and a dynamic game. The authors of \cite{Qiao_What_is_SemCom'21} explain the principles of SemCom; introduce H2H SemCom, H2M SemCom, and M2M SemCom systems as well as techniques for different areas in which H2M and M2M SemCom can be applied; and detail an approach for designing SemCom systems based on KGs \cite{Knowledge_Graphs_Survey'22}. The authors of \cite{Yang_SemCom_meets_Edge_Intelligence'22} envision edge-driven SemCom and SemCom-driven edge toward the efficient \textit{intelligentization} of future networks and discuss the corresponding open research issues. The authors of \cite{Strinati_Beyond_Shannon'20} propose their vision of 6G wireless networks, wherein SemCom and goal-oriented SemCom are key technologies that derive a crucial paradigm shift from Shannon's information-theoretic principles. Departing from these principles while not necessarily resorting to bandwidth or power, the authors of \cite{Strinati_Beyond_Shannon'20} advocate that increased effectiveness and reliability can be achieved by identifying the information that would be necessary to make a receiver extract precisely the intended meaning or to actuate the right procedures to accomplish a predefined goal efficiently. The authors of \cite{Zhang_Wisdom_Evolutionary_6G'21} put forward a systematic design for SemCom networks in the context of ubiquitous 6G networks that is based on an intelligent and efficient semantic communication network architecture.

The authors of \cite{Wang_Transformer_empowered_6G'22}  propose \textit{Transformer}-based solutions for several massive MIMO and SemCom problems, demonstrate Transformer-based architectures' superiority over other architectures, and discuss key challenges as well as open issues affecting Transformer-based solutions. The authors of \cite{Beck_SemCom_Info_Bottleneck_View'22} propose an information-theoretic framework wherein the semantic context is explicitly introduced as a hidden RV in the communication system design by recasting SemCom's system design problem as an \textit{information bottleneck} (IB) \cite{IB_applications_Goldfeld'20,IB_method_Tishby'00} optimization problem. In light of Weaver's three levels of communication (see Fig. \ref{fig: Three_lavels_of_communication}), the authors of \cite{Phopovski_SE_Filtering'19} introduce the concept of a \textit{semantic-effectiveness} plane for effective filtering and control for post-5G wireless connectivity. The authors of \cite{Shi_to_Semantic_Fidelity'21} put forward an \textit{understand-first-and-then-transmit} SemCom framework. The authors of \cite{Shi_From_SemCom_to_Sematic-aware_Networking'20} propose a federated edge AI-based architecture to support resource-efficient semantic-aware networking. 

The authors of \cite{Commun_Beyond_Transmitting_Bits'22} introduce a unified framework for semantics-guided source and channel coding by proposing an end-to-end SemCom system named \textit{semantic coded transmission}. The authors of \cite{Dong_Semantic_Cognitive_Intell'22} envision a new intelligence paradigm named \textit{edge semantic cognitive intelligence} for 6G networks that is emerging at the confluence of edge intelligence and SemCom. In view of SNC, the authors of \cite{Zhao_Semantic-Native_Communication'22} introduce a visionary\footnote{The crux of the vision in \cite{Zhao_Semantic-Native_Communication'22} is that information is not scalar as in\linebreak Shannon's case but topological space much like the stored information/knowledge in our brains.} SemCom work that is inspired by a topological space perspective and wherein higher-order data semantics live in a \textit{simplicial complex}. Based on this perspective, a transmitter first maps its data into a $k$-order simplicial complex and then learns its high-order correlations \cite{Zhao_Semantic-Native_Communication'22}. A simplicial autoencoder (AE) CNN is then used to encode this simplicial structure and its features into semantic embeddings in latent space for transmission \cite{Zhao_Semantic-Native_Communication'22}. Following the transmission and propagation of this SemCom signal, the receiver decodes the simplicial Laplacians from the received embeddings using a bilinear decoder and then infers the missing (or distorted) data using a simplicial convolutional decoder \cite{Zhao_Semantic-Native_Communication'22}. In summary, the transmitter and receiver collaboratively train a simplicial CNN AE to accomplish a SemCom task \cite{Zhao_Semantic-Native_Communication'22}. 

To address the lack of interpretability evident in the DNN-based protocol models (NPMs), the authors of \cite{Seo_SemCom_Protocols'22} proffer a semantic protocol model (SPM) that is constructed by converting an NPM into an interpretable symbolic graph written in the probabilistic logic programming language known widely as \textit{ProbLog} \cite{Raedt_ProbLogAP'07}. The authors of \cite{Seo_SemCom_Protocols'22} substantiate that the proposed SPM closely approximates an NPM while utilizing only 0.02\% memory. The authors of \cite{Pokhrel_UBT_SemCom'22} propose a new SemCom approach to 6G networks by proffering and assessing a hashing-based SemCom framework. In this framework, the authors' design of SE and domain adaptation epitomizes the \textit{joint optimization of information gathering, dissemination, and decision-making} over 6G networks based on SemCom \cite{Pokhrel_UBT_SemCom'22}. The authors of \cite{Chaccour_Building_NG_SemCom_Networks'22} disseminate a holistic vision of an end-to-end SemCom network that is rooted in overarching concepts of AI, causal reasoning, communication theory, networking, information theory, transfer learning, and minimum description length theory.

We now continue with existing tutorial works on SemCom.

\subsubsection{Tutorial Works on SemCom}
\label{subsubsec: tutorial_works_SemCom}
the authors of \cite{Qin_Sem_Com_Principles_Apps'22} provide an overview of SemCom theory, frameworks, and DL-enabled system design. The authors of \cite{Luo_SemCom_Overview'22} provide an overview of recent works on SemCom, summarize the open issues, and highlight the corresponding challenges in theoretical research and practical implementation. The authors of \cite{Rethinking_modern_com_Lu_2022} provide an overview of existing semantics-aware communication techniques (and their underlying shortcomings), rethink the design of semantics-aware communications systems and highlight some related concerns such as implementation cost; and establish a joint semantics-noise coding solution for the semantic coding problem and an RL-based similarity-targeted SemCom technique for both differentiable and non-differentiable semantic similarity metrics. The authors of \cite{Niu_Towards_SemCom'22} present a brief tutorial on SemCom and its information-theoretic perspectives. 

The authors of \cite{SemCom_for_6G_Future_Internet'22} offer a comprehensive review of the fundamentals, applications, and challenges of SemCom. The authors of \cite{Luo_SemCom_Overview'22} provide an overview of DL-based SemCom, its open issues, and corresponding future research directions. The authors of \cite{Engineering_SemCom'22} put forward a survey aimed at providing a clear picture of state-of-the-art SemCom developments. The authors of \cite{Gunduz_Beyond_Transmitting_Bits'22} offer a tutorial for communication theorists and practitioners that provides an introduction to contemporary tools and advancements in SemCom. The authors of \cite{Zhang_a_New_Paradigm'22} offer a brief discussion on the functionalities of SemCom-enabled networked intelligence and their respective open issues, and review related works. The authors of \cite{Jiang_Wireless_Semantic_Transmission'22} discuss recent developments in state-of-the-art SemCom that exploit the conventional modules in wireless systems. The authors of \cite{Sem_Empowered_Commun'22} disseminate a tutorial-cum-survey that aims to provide a comprehensive understanding of state-of-the-art developments in SemCom -- generally, semantics-empowered communication -- and its applications.  

Apart from the aforementioned vision and tutorial works on SemCom, the expanding body of state-of-the-art works on SemCom for the transmission of text, image, video, and multi-modal data encompass numerous SemCom techniques and trends such as cognitive SemCom \cite{Cognitive_SemCom_Systems'22}, implicit SemCom \cite{Xiao_Reasoning_on_the_Air'22}, adaptive SemCom \cite{Dai_Adaptive_SemCom'22}, context-based SemCom \cite{zhang_Context-Based_SemCom'22,Liu_Context-Based_SemCom'22}, digital SemCom \cite{Learning_Based_Digital_SemCom'22,Fu_VQ_SemCom'22}, cross-modal SemCom \cite{Li_Cross-Modal_SemCom'22}, sequence-to-sequence SemCom (Seq2Seq-SemCom) \cite{Lee_EQ2SEQ-SC'22}, SemCom with conceptual spaces \cite{SemCom_with_Conceptual_Spaces'22}, inverse SemCom \cite{Du_RIS-aided_Encoding'22}, one-to-many SemCom \cite{Hu_One-to-Many_SemCom'22}, quantum key distribution (QKD)-secured SemCom \cite{Kaewpuang_Cooperative_Resource_Management'22}, encrypted SemCom \cite{Luo_Encrypted_SemCom'22}, and quantum SemCom \cite{Chehimi_Quantum_SemCom'22}.  

We now continue with state-of-the-art algorithmic developments in semantic-oriented communication.

\subsection{Algorithmic Developments in Semantic-Oriented Communication}
\label{subsec: semantic_oriented_communs}
This section discusses the state-of-the-art SemCom techniques for text transmission, audio transmission or recognition, image transmission or recognition, video transmission, and multi-modal signal transmission. We begin with the techniques for text transmission.

\subsubsection{SemCom for Text Transmission} 
\label{subsubsec: SemCom_for_Text_Transmission}
the authors of \cite{SemCom_Game'18} first introduce SemCom for text transmission (text SemCom) by formulating the text SemCom problem as a static Bayesian game and a dynamic game. In accordance with these games, the authors integrate semantic inference and physical layer (PHY) communications to optimize the entire transceiver. However, the text SemCom scheme in \cite{SemCom_Game'18} quantifies semantic error at the word level as opposed to the sentence level. The authors of \cite{Farsad_DL_JSCC'18} represent the channel by a \textit{dropout layer} \cite{Dropput_Hinton_JMLR'14} and put forward a text SemCom scheme made up of an encoder and a decoder that are implemented by a stacked bidirectional long short-term memory (Bi-LSTM) network \cite{Bi_LSTM_Alex'05} and a stacked long short-term memory (LSTM) network \cite{Greff_LSTM_Search_Space_Odyssey'17}, respectively.   

The authors of \cite{Xie_DL-based_SemCom'21} develop a DL-based text SemCom system (a Transformer-based system \cite{Peters_BERT_Paper'18} as in Fig. \ref{fig: DeepSC_Original_Model_20221103}) -- named \textit{DeepSC} -- that performs joint semantic-channel coding to produce a superior performance gain than existing techniques in which the source and channel are coded separately in low SNR regimes. The authors of \cite{Xie_Lite_distributed_SemCom'21} build on the DeepSC system and propose a lite distributed text SemCom system -- dubbed \textit{L-DeepSC} -- for IoT networks that considers the participating devices' limited power and computing capabilities. The authors of \cite{Zhou_WiCom_Letters'22} took inspiration from L-DeepSC and DeepSC, and put forward a \textit{Universal Transformer} (UT)-based text SemCom system that incorporates an adaptive circulation mechanism in the UT. This UT-based text SemCom technique offers a small performance improvement over DeepSC for low SNR regimes. However, the fidelity of DeepSC, UT-based text SemCom, and L-DeepSC can be destroyed by considerable semantic noise. To combat literal semantic noise and adversarial semantic noise, the authors of \cite{Peng_Robust_DL-Based_SemCom'22} develop a DL-enabled robust SemCom system named \textit{R-DeepSC}. R-DeepSC improves system robustness in various wireless environments and outperforms DeepSC when the corpus is erroneous \cite{Peng_Robust_DL-Based_SemCom'22}. The aforementioned SemCom techniques that employ end-to-end DNNs do not generalize well under varying channel conditions \cite{Yao_Semantic_Coding'22}, however. To address this challenge, the authors of \cite{Yao_Semantic_Coding'22} develop a semi-neural framework with an iterative joint source-channel coding (JSCC) architecture, named \textit{iterative semantic JSCC} (IS-JSCC).

Considering that most semantic metrics are non-differentiable, the authors of \cite{Lu_RL-powered_SemCom'21} introduce an RL-based optimization paradigm that is a self-critic policy gradient approach for possibly large-scale and complex text semantic transmission. In this technique, the authors handle the non-differentiable semantic channel optimization problem by training the decoupled semantic transceiver using self-critic stochastic iterative updating \cite{Lu_RL-powered_SemCom'21}. This semantic transceiver -- whose encoder and decoder are made up of Bi-LSTM and LSTM, respectively -- is named \textit{SemanitcRL-JSCC} and improves the recovery of semantically meaningful sentences and the handling of semantic noise \cite{Lu_RL-powered_SemCom'21}. In the spirit of the work in \cite{Lu_RL-powered_SemCom'21}, the authors of \cite{Rethinking_modern_com_Lu_2022} also propose an RL-powered text SemCom paradigm and introduce a joint semantics-noise coding (JSNC) technique. The authors of \cite{Luo_SemCom_with_relay'21} introduce SemCom over wireless relay channels and develop an AE-based text SemCom scheme for wireless relay channels with a \textit{semantic forward} (SF) protocol. The purpose of the SF protocol is to enable direct SemCom when the source node and the destination node have different background KBs \cite{Luo_SemCom_with_relay'21}. To this end, the relay node can cooperatively use the background KB of the source and that of the destination to forward semantic information between the source and the sink \cite{Luo_SemCom_with_relay'21}.

All the above-discussed text SemCom techniques assume fixed codeword length and are possibly inefficient as well as inflexible when it comes to handling varying sentence length \cite{Jiang_Deep_Source-Channel_coding'22}. The authors of \cite{Jiang_Deep_Source-Channel_coding'22} address this limitation by exploiting hybrid automatic repeat request (HARQ) and Reed Solomon (RS) channel coding, and combining them with semantic coding (SC) to propose a text SemCom technique named \textit{SC-RS-HARQ}. SC-RS-HARQ benefits from the performance gains of SC and the reliability of the RS channel coding and HARQ \cite{Jiang_Deep_Source-Channel_coding'22}. The authors of \cite{Jiang_Deep_Source-Channel_coding'22} also put forward an end-to-end text SemCom architecture made up of a Transformer and a fully connected DNN -- dubbed SCHARQ -- that has been demonstrated to considerably reduce the  number of bits required for sentence semantic transmission and the sentence error rate \cite{Jiang_Deep_Source-Channel_coding'22}. All the previously highlighted text SemCom techniques overlook the contextual correlation among sentences at the transmitter and fail to take historical text into consideration in the decoding process \cite{Liu_Context-Based_SemCom'22}. Considering historical text in the decoding process and contextual correlation among sentences at the transmitter, the authors of \cite{Liu_Context-Based_SemCom'22} proffer a context-based text SemCom technique. This technique has been demonstrated to outperform DeepSC in low SNR regimes \cite{Liu_Context-Based_SemCom'22}. The authors of \cite{Lee_EQ2SEQ-SC'22} propose a computationally efficient -- in extracting semantic information -- text SemCom technique named seq2seq-SemCom. In seq2seq-SemCom, the pre-trained encoder-decoder transformers are integrated with end-to-end SemCom systems, and the channel encoder and decoder are composed of 5G new radio (5GNR)-compliant modules \cite{Lee_EQ2SEQ-SC'22}. The seq2seq-SemCom technique works with all (general) text corpora -- unlike DeepSC, which is dependent on specific datasets -- and outperforms DeepSC in terms of semantic similarity as demonstrated via link-level simulations that closely resemble actual 5G system \cite{Lee_EQ2SEQ-SC'22}.
 
We now move on to discuss state-of-the-art SemCom techniques for audio transmission or recognition.  

\subsubsection{SemCom for Audio Transmission or Recognition} 
\label{subsubsec: SemCom_for_Audio_Transmission_or_Recognition}
SemCom has a variety of applications in semantic-aware (semantic-empowered) speech/audio transmission and recognition systems. We refer to such systems as \textit{audio SemCom} and discuss audio SemCom techniques below. To begin with, the authors of \cite{Weng_SemCom_Sys_Speech_Trans'21,SemCom_for_speech_signals'20} propose a DL-enabled audio SemCom system named \textit{DeepSC-S} that improves transmission efficiency by transmitting only semantic information. This audio SemCom scheme adopts a joint semantic encoder/decoder and channel encoder/decoder for efficient learning and speech feature extraction, and for mitigating wireless channel distortion. The authors of \cite{Weng_SemCom_Speech_Recognition'21} put forward another DL-enabled audio SemCom system -- dubbed \textit{DeepSC-SR} -- for speech recognition by exploiting a joint semantic encoder/decoder and channel encoder/decoder for learning and extracting speech features while mitigating wireless channel impairments.

The authors of \cite{Han_Semantic-aware_Speech2Text_Transmission'22} propose a semantic-aware speech-to-text transmission audio SemCom system with a soft alignment module that extracts only the text-related semantic features and a redundancy removal module that drops the semantically redundant content. This technique reduces semantic redundancy and outperforms DeepSC-SR \cite{Han_Semantic-aware_Speech2Text_Transmission'22}. Extending the work in \cite{Han_Semantic-aware_Speech2Text_Transmission'22}, the authors of \cite{Han_Semantic-preserved_Com_System'22} proffer a DL-based speech-to-text transmission and a speech-to-speech transmission audio SemCom systems that also deploy a soft alignment module and a redundancy removal module. Apart from audio SemCom for speech transmission or speech recognition, the authors of \cite{Weng_DL-enabled_SemCom'22} develop a DL-enabled audio SemCom system -- termed \textit{DeepSC-ST} -- for speech recognition and synthesis. DeepSC-ST recovers the text transcription by using text-related semantic features and reconstructs the speech sample sequence through a joint semantic-channel coding scheme deployed to learn and extract semantic features as well as mitigate channel impacts \cite{Weng_DL-enabled_SemCom'22}. The work in \cite{Weng_DL-enabled_SemCom'22}, meanwhile, demonstrates that DeepSC-ST outperforms traditional communication systems for speech recognition and speech synthesis tasks, especially in low SNR regimes.

The authors of \cite{Tong_FL_ASC'21} exploit federated learning (FL) for audio SemCom and investigate a \textit{wav2vec} \cite{Schneider_wav2vec_paper'19}-based autoencoder made of CNNs \cite{J_Xu_CNN_17,ICSM_19} that comprises an audio SemCom system that can effectively encode, transmit, and decode audio semantic information while reducing communication overhead. This audio SemCom's AE is trained using FL \cite{Amiri_FederatedLO'20,DG_JSAC_ML_in_z_air_19,Li_Fed_Learning'20} to improve the accuracy of semantic information extraction and substantially reduces transmission error compared with existing audio coding schemes that are based on pulse code modulation (PCM), low-density parity check codes (LDPC), and 64-QAM (quadrature amplitude modulation) \cite{Tong_FL_ASC'21}.

We now move on to discuss SemCom techniques for image transmission or recognition.

\subsubsection{SemCom for Image Transmission or Recognition} 
\label{subsubsec: SemCom_for_Image_Transmission_or_Recognition} 
hereinafter, we term SemCom techniques/systems for image transmission or recognition \textit{image SemComs}. Among the first image SemCom works on image transmission over wireless channels, the authors of \cite{Eirina_JSCC'19} investigate an image SemCom architecture named \textit{deep JSCC} whose encoder and decoder functions are parameterized by CNNs and trained jointly on the same dataset to minimize the average MSE of the reconstructed image. The joint training enables deep JSCC to not suffer from the \textit{cliff effect} as demonstrated in \cite{Eirina_JSCC'19}, while offering a graceful performance degradation as the channel SNR varies w.r.t. the SNR presumed during training. Building on deep JSCC, meanwhile, the authors of \cite{Kurka_Deep_JSCC-f'20} introduce an AE-based JSCC scheme -- named \textit{DeepJSCC-f} -- that uses the channel output feedback. As demonstrated in \cite{Kurka_Deep_JSCC-f'20}, DeepJSCC-f significantly improves end-to-end reconstruction quality for fixed-length transmission and average delay for variable-length transmission. Furthermore, the authors of \cite{Bandwidth_Agile_Image_Transmission'21} build on DeepJSCC-f \cite{Kurka_Deep_JSCC-f'20} and deep JSCC \cite{Eirina_JSCC'19} by exploring the use of DL-based methods for progressive image transmission over wireless channels and introduce \textit{DeepJSCC-l}. DeepJSCC-l is a group of DL-based JSCC algorithms made up of CNN-based AEs that are able to encode and decode images over multiple channels while supporting flexible bandwidth-adaptive transmission \cite{Bandwidth_Agile_Image_Transmission'21}.

The aforementioned JSCC schemes presume that any complex value can be transmitted over a wireless channel \cite{Tung_DeepJSCC-Q'22}. However, this can create compatibility problems for hardware/protocols that can accept only certain sets of channel inputs (e.g., as prescribed by digital modulation) \cite{Tung_DeepJSCC-Q'22}. To overcome this limitation, the authors of \cite{Tung_DeepJSCC-Q'22} develop \textit{DeepJSCC-Q}, which is an end-to-end-trained JSCC scheme for wireless image transmission using a finite channel input alphabet. DeepJSCC-Q has been demonstrated to perform comparably to prior JSCC schemes that permit complex value channel input whenever high modulation orders are available \cite{Tung_DeepJSCC-Q'22}. The authors of \cite{Zhang_Wireless_Information_Transmission_of_Image'22} took inspiration from the aforementioned DL-based wireless image transmission technique and developed a DL-based image SemCom system -- dubbed \textit{MLSC-image} -- that is trained in an end to end manner for wireless image transmission. MLSC-image incorporates a multi-level semantic feature extractor that extracts both high-level semantic information (e.g., text semantics and segmentation semantics) and low-level semantic information (e.g., local spatial details of images) \cite{Zhang_Wireless_Information_Transmission_of_Image'22}. The semantic features are combined and then encoded by a joint semantic-channel (JSemC) encoder into symbols to be transmitted over the physical channel, whose output is fed to a JSemC decoder and then to an image reconstruction module \cite[Fig. 1]{Zhang_Wireless_Information_Transmission_of_Image'22}. The JSemC encoder and decoder enable MLSC-image to outperform deep JSCC \cite{Eirina_JSCC'19} in low compression ratio regimes (though it performs worse than deep JSCC in high compression ratio regimes) \cite{Zhang_Wireless_Information_Transmission_of_Image'22}. Moreover, the aforementioned JSCC schemes adapt the compression ratio in source coding and the channel coding rate dynamically in accordance with the channel SNR \cite{Xu_Wireless_Image_Transmission'22}. Instead of a resource allocation strategy, the authors of \cite{Xu_Wireless_Image_Transmission'22} deploy channel-wise soft attention to scale features according to the SNR in their proposed JSCC technique named attention DL-based JSCC (ADJSCC) \cite[Fig. 5]{Xu_Wireless_Image_Transmission'22}. ADJSCC's adaptability, robustness, and versatility have been demonstrated by simulations \cite{Xu_Wireless_Image_Transmission'22}.

The afore-discussed image SemCom works revolve around JSCC. The authors of \cite{Pan_IM-SemCom'22} stray away from JSCC and devised an \textit{image segmentation semantic communication} (ISSC) system to manage the transmission of the massive amount of visual data perceived by vehicles' visual sensors over the internet of vehicles (IoV) \cite{Tang_SVT_6G_2019}. For IoV applications, the proposed ISSC system efficiently transmits the semantic features of images that are extracted using a \textit{Swin Transformer} \cite{Liu_Swin_Transformer'21}-based multi-scale semantic feature extractor that can broaden the receptive area of an image \cite{Pan_IM-SemCom'22}. The ISSC system's encoder and decoder are jointly designed and end-to-end-trained to globally optimize the model parameters \cite{Pan_IM-SemCom'22}. To this end, the ISSC system has been demonstrated to perform better than traditional coding schemes in the low SNR regimes \cite{Pan_IM-SemCom'22}. The authors of \cite{Yang_WITT'22} propose another Swin Transformer-based image SemCom scheme \cite{Liu_Swin_Transformer'21} -- dubbed wireless image transmission transformer (WITT) -- that is optimized for image transmission while considering the wireless channel's effect. It has been proven to outperform a CNN-based deep JSCC scheme as well as classical separation-based schemes \cite{Yang_WITT'22}.

The highlighted state-of-the-art JSCC techniques do not integrate any hyperpriors as side information though this is a promising concept that is widely deployed in modern image codecs \cite{Dai_NLT_SCC'22}. To address this limitation, the authors of \cite{Dai_NLT_SCC'22} devised a joint source-channel coding architecture that unifies the concept of nonlinear transform coding \cite{NTC_Balle'21} and deep JSCC and is named nonlinear transform source-channel coding (NTSCC). NTSCC is a class of high-efficiency deep JSCC techniques that can adapt to the source distribution under the nonlinear transform \cite{Dai_NLT_SCC'22}. The authors of \cite{Lee_Joint_Transmission_Recognition_for_IoTs'19} put forward a DNN-constructed joint transmission-recognition scheme that, unlike the previously discussed JSCC-based image SemCom techniques, makes IoT devices transmit data effectively to a server for image recognition. This DL-enabled joint transmission-recognition technique was shown to outperform a JPEG-compressed scheme and a compressed sensing-based scheme under analog transmission and digital transmission at all SNRs on CIFAR-10 image database \cite{Lee_Joint_Transmission_Recognition_for_IoTs'19}. Furthermore, the authors of \cite{Hu_Robust_SemCom'22} propose a framework that introduces a robust design to image SemCom for an end-to-end robust image SemCom system that can withstand semantic noise through \textit{adversarial training} that incorporates samples with semantic noise in the training dataset \cite{Hu_Robust_SemCom'22}. Concerning this robust framework, simulation results confirm that it can considerably improve the robustness of image SemCom systems against (image) semantic noise \cite{Hu_Robust_SemCom'22}.    

The authors of \cite{Huang_Toward_SemCom'23} develop an image SemCom technique that is both bandwidth sensitive and richer in semantics than the aforementioned image SemCom techniques by devising an RL-based adaptive semantic coding (RL-ASC) approach that encodes images beyond the pixel level. In this technique, a convolutional semantic encoder is used to extract semantic information that is in turn encoded by adaptive quantization in accordance with an RL-based semantic bit allocation model \cite{Huang_Toward_SemCom'23}. On the receiver side, meanwhile, the authors of \cite{Huang_Toward_SemCom'23} design a generative semantic decoder that exploits the attention model to fuse the local and global features and is deployed to reconstruct the transmitted semantic concepts \cite{Huang_Toward_SemCom'23}. Furthermore, the authors of \cite{Huang_Toward_SemCom'23} corroborate that their proposed RL-ASC approach can promote multiple vision tasks for SemCom scenarios. Most of the aforementioned JSCC schemes are optimized using traditional semantic metrics such as \textit{peak signal-to-noise ratio} and \textit{multi-scale structural similarity} and hardly account for human visual perception in SemCom \cite{Wang_Perceptual_Learned'22}. To overcome this limitation, the authors of \cite{Wang_Perceptual_Learned'22} devise a deep JSCC architecture that merges an encoder, a wireless channel, a decoder/generator, and a discriminator that are learned jointly w.r.t. both perceptual and adversarial losses. This deep JSCC technique has been shown to produce results that are more visually pleasing to humans than those of state-of-the-art image-coded transmission techniques and the aforementioned deep JSCC schemes \cite{Wang_Perceptual_Learned'22}.

We now move on to discuss SemCom techniques for video transmission.  

\subsubsection{SemCom for Video Transmission} 
\label{subsubsec: SemCom_for_Video_Transmission}
SemCom techniques/systems for video transmission are termed \textit{video SemComs}, and we present below state-of-the-art video SemCom techniques.

To overcome conventional video compression's limitation that it reduces the resolution under limited bandwidth, the authors of \cite{Jiang_Wireless_SemCom'22} study semantic video conferencing (SVC), which maintains high resolution by transmitting key points to represent motion for scenarios in which the video background is almost static and the speakers do not change frequently. For this type of scenario, the authors of \cite{Jiang_Wireless_SemCom'22} develop SVC techniques that exploit HARQ and channel state information (CSI) feedback. These techniques considerably improve transmission efficiency \cite{Jiang_Wireless_SemCom'22}. The authors of \cite{Wang_Wireless_Deep_Video_Transmission} developed another video SemCom technique that also benefits from semantic information: a video SemCom framework that exploits nonlinear transform and conditional coding architecture to adaptively extract semantic features across video frames that are transmitted through a group of  variable-length (learned) deep JSCC codecs and a wireless channel. This video SemCom technique performs significantly better than traditional wireless video coded transmission schemes \cite{Wang_Wireless_Deep_Video_Transmission}.

The authors of \cite{Tung_DeepWiVe'21} devised an end-to-end JSCC video transmission scheme -- named \textit{DeepWiVe} -- that leverages DNNs to directly map video signals to channel symbols. DeepWiVe combines video compression, channel coding, and modulation steps in a single neural transform and is capable of dynamic bandwidth allocation and residual estimation without the need for distortion feedback \cite{Tung_DeepWiVe'21}. Furthermore, DeepWiVe overcomes the cliff effect, achieves a graceful degradation in channel quality, and produces superior video quality in highly bandwidth-constrained scenarios compared to both H.264 and H.265 \cite{Tung_DeepWiVe'21}. Moreover, following the popularity of point cloud video (PCV) and PCV streaming, the authors of \cite{Huang_IS-SemCom'22} developed an interest-aware SemCom scheme for immersive point cloud video streaming. This video SemCom technique aims to tackle the challenges associated with real-time PCV streaming on resource-constrained devices by using a region-of-interest (ROI) selection module (i.e., a two-stage efficient ROI selection method that considerably reduces the data volume), a lightweight decoder network, and an intelligent scheduler (for adaptive online  PCV streaming) \cite{Huang_IS-SemCom'22}. This video SemCom technique has been demonstrated to outperform an AI-driven technique by at least 10 frames per second (FPS) \cite{Huang_IS-SemCom'22}.

We now proceed to discuss SemCom techniques for multi-modal signal transmission.

\subsubsection{SemCom for Multi-Modal Signal Transmission} 
\label{subsubsec: SemCom_for_Multimodal_Signal_Transmission}
the previously highlighted text SemCom, audio SemCom, image SemCom, and video SemCom techniques focus on the efficient semantic transmission of text, audio, image, and video signal, respectively. However, many driving applications and services of 6G (see Sec. \ref{subsec: motiv}) aim at offering immersive experiences with low latency and high reliability by transmitting multi-modal signals \cite{Li_Cross-Modal_SemCom'22,Zhou_Cross-Modal_Collaborative_Commun'20}. This requires that multi-modal signals be transmitted efficiently, which can be accomplished by using an emerging SemCom technique named \textit{cross-modal SemCom} \cite{Li_Cross-Modal_SemCom'22}. The cross-modal SemCom paradigm is inspired by cross-modal communication \cite{Zhou_Cross-Modal_Collaborative_Commun'20} and comprises three modules \cite{Li_Cross-Modal_SemCom'22}:
\begin{itemize}
	
	\item \textit{Cross-modal semantic encoder}: it accepts video, audio, and haptic signals as inputs and produces explicit semantics as well as implicit semantics -- while reducing encoding polysemy -- for transmission \cite{Li_Cross-Modal_SemCom'22}. 
	
	\item \textit{Cross-modal semantic decoder}: it ensures the multi-modal source signals and the multi-modal recovered signals are consistent at the bit and semantic levels, while reducing decoding ambiguity \cite{Li_Cross-Modal_SemCom'22}.
	
	\item \textit{Cross-modal KG (CKG)}: it provides essential background knowledge and signal patches for cross-modal semantic encoding and decoding \cite{Li_Cross-Modal_SemCom'22}.  
	
\end{itemize}

We now move on to discuss algorithmic developments in semantic-aware communication and processing.

\subsection{Algorithmic Developments in Semantic-Aware Communication and Processing}
\label{subsec: semantic_aware_commuications}
A communication system that exploits semantic information in its design is termed hereinafter \textit{semantic-aware communication}. In \cite{Yun_Attention_based_RL'21}, the authors study the problem of air-to-ground URLLC for a moving ground user and put forward a multi-agent deep RL (DRL) framework, dubbed graph attention exchange network (GAXNet). In GAXNet, each unmanned aerial vehicle (UAV) locally constructs an attention graph that measures its level of attention to its neighboring UAVs using semantic representation encoding while sharing the attention weights with other UAVs using SemCom to minimize any attention mismatch between them. This semantic-aware scheme has been demonstrated \cite{Yun_Attention_based_RL'21} to attain lower latency with higher reliability than the state-of-the-art QMIX scheme (see \cite{Rashid_QMIX'18}). Also in the context of collaborative deep RL (CDRL), the authors of \cite{Semantics_Aware_CDRL'21} propose a semantic-aware CDRL framework that enables knowledge to be efficiently transferred among heterogeneous agents that are distributed across a resource-constrained wireless cellular network and have semantically related tasks. They, therefore, introduce a new heterogeneous federated DRL algorithm \cite{Semantics_Aware_CDRL'21} for selecting the best subset of semantically-associated DRL agents for collaboration. This CDRL algorithm has been shown to offer an 83\% improvement in maximum reward compared with baseline methods \cite{Semantics_Aware_CDRL'21}.

Apart from CDRL algorithms, SemCom has also inspired the development of semantic-aware algorithms for the Metaverse \cite{Edge_Enabled_Metaverse'22}. When it comes to metaverse applications, the authors of \cite{Wang_Semantic-Aware_Sensing'22} developed a semantic-aware transmission framework for transforming and transmitting sensing data from the physical world to a mobile service provider (MSP) in the Metaverse. In this framework, the authors of \cite{Wang_Semantic-Aware_Sensing'22} put forward a semantic-aware sensing data transmission scheme and establish a contest theory-based incentive mechanism. In their transmission scheme, they proffer a semantic encoding algorithm for data sensing that considerably reduces the amount of data needed, storage costs, and transmission costs, while ensuring the MSP performs well in the Metaverse. The contest-based incentive mechanism, on the other hand, aims to boost the data uploading frequency of all transmitters by setting rewards and to support the MSP in enhancing its quality of service (QoS) \cite{Wang_Semantic-Aware_Sensing'22}. When it comes to SemCom-aided virtual transportation networks in the Metaverse, the authors of \cite{Ng_Stochastic_Resource_Allocation'22} attempt to address the resource allocation problem marked by stochastic user demand by proposing a stochastic semantic transmission scheme that is based on a two-stage stochastic integer programming. This scheme has been demonstrated to minimize the transmission costs of virtual service providers whilst taking into account users' demand uncertainty \cite{Ng_Stochastic_Resource_Allocation'22}.

The authors of \cite{Sun_Semantic-driven_Computation'22} put forward a semantic-driven computation offloading and resource allocation scheme for UAV-assisted vehicle data collection and edge-cloud collaboration to complete intelligent tasks. They propose a CNN segmentation scheme for its offloading decisions and a multi-agent deep Q-network (DQN) algorithm for its resource allocation. Simulation results confirm their semantic-driven scheme improves the multi-objective optimization of latency, energy consumption, and task performance \cite{Sun_Semantic-driven_Computation'22}. Furthermore, in regard to using \textit{type-based multiple access} (TBMA) as a semantic-aware multiple access protocol for remote inference, the authors of \cite{Zhu_Semantics-Aware_Remote_Est'22} devised an IB-inspired design principle for TBMA named IB-TBMA. In their IB-TBMA protocol, the shared codebook is jointly optimized with channel statistics that are based strictly on data and a decoder that is based on artificial neural networks \cite{Zhu_Semantics-Aware_Remote_Est'22}. The authors of \cite{Zhu_Semantics-Aware_Remote_Est'22} also propose the \textit{compressed IB-TBMA} (CB-TBMA) protocol, which enhances IB-TBMA by making it possible to reduce the number of codewords (via an IB-inspired clustering phase). The authors of \cite{Zhu_Semantics-Aware_Remote_Est'22} show -- with numerical results -- the power of joint codebook and neural decoder design in CB-TBMA and IB-TBMA while demonstrating the benefits of codebook compression.

We now continue with a discussion of state-of-the-art algorithmic developments in SemCom resource allocation.  

\subsection{Algorithmic Developments in SemCom Resource Allocation}
\label{subsec: Resource_allocation_SemCom}
The authors of \cite{Wang_SemCom_Per_Optimization'21,Wang_JSAC_Performance_Optimi_SemCom'22} propose a performance optimization framework for semantic-driven wireless networks in a text SemCom system. The authors of \cite{Wang_JSAC_Performance_Optimi_SemCom'22} formulate an optimization problem for such networks that jointly considers wireless resource constraints, transmission delay requirements, and SemCom performance and whose goal is to maximize the total metric of semantic similarity -- a semantic metric that was introduced by the authors of \cite{Wang_SemCom_Per_Optimization'21,Wang_JSAC_Performance_Optimi_SemCom'22} -- by optimizing resource block allocation for the transmission of partial semantic information (modeled by a KG). To solve this problem, the authors of \cite{Wang_SemCom_Per_Optimization'21,Wang_JSAC_Performance_Optimi_SemCom'22} develop an \textit{attention proximal policy optimization algorithm} that has been shown \cite{Wang_JSAC_Performance_Optimi_SemCom'22} to considerably reduce the amount of data needed to be transmitted.

After defining the semantic spectral efficiency (S-SE) metric to quantify the communication efficiency of a text SemCom system named DeepSC \cite{Xie_DL-based_SemCom'21}, the authors of \cite{Resource_allocation_text_SemCom'22} formulate a semantic-aware resource allocation problem as an optimization problem that aims to maximize the overall S-SE of all users and report on its optimal solution, the validity and feasibility of which are demonstrated \cite{Resource_allocation_text_SemCom'22} using simulation results. Meanwhile, the authors of \cite{Yan_QoE_Aware_Resource_Allocation_SemCom'22} study a semantic-aware resource allocation problem in a multi-cell multi-task network in the context of semantic-aware resource allocation. They formulate a QoE maximization problem for the network that constrains the number of semantic symbols transmitted, channel assignment, and power allocation, and provides a \textit{matching theory}-based solution whose effectiveness is demonstrated \cite{Yan_QoE_Aware_Resource_Allocation_SemCom'22}.

In the broader context of intelligent SemCom (iSemCom) -- more specifically, SemCom enabled by AI models \cite{Xia_Resource_Management_SemCom'22} -- and an iSemCom-enabled heterogeneous network (iSemCom-HetNet), the authors of \cite{Xia_Resource_Management_SemCom'22} investigate user association (UA) and bandwidth allocation problems for an iSemCom-HetNet by introducing auxiliary KBs into the system model and then developing a new performance metric termed \textit{system throughput in message} (STM). The authors of \cite{Xia_Resource_Management_SemCom'22} approach the joint optimization of UA and BA via STM maximization subject to KB matching and wireless bandwidth constraints, and propose a two-stage solution whose superiority and reliability over two baseline algorithms are demonstrated \cite{Xia_Resource_Management_SemCom'22}. The authors of \cite{Xia_Wireless_SemCom'22} built on the work in \cite{Xia_Resource_Management_SemCom'22} and introduce two general SemCom-enabled network (SemComNet) scenarios that are based on all possible knowledge-matching states between mobile users and base stations: \textit{perfect knowledge matching-based SemComNet} and \textit{imperfect knowledge matching-based SemComNet}. They delineate the semantic channel capacity model for the perfect matching scenario mathematically.

Apart from resource allocation algorithms in text SemCom, there have also been algorithmic developments in resource allocation for video SemCom. The authors of \cite{Zhu_Video_Semantics-Based_Resource_Allocation'21} develop a video semantics-based resource allocation algorithm for spectrum multiplexing scenarios (VSRAA-SM). VSRAA-SM can be used to optimize vehicle-to-infrastructure semantic understanding tasks and vehicle-to-vehicle information transmission tasks \cite{Zhu_Video_Semantics-Based_Resource_Allocation'21}. Furthermore, the authors of \cite{Zhang_Opt_in_Image_SemCom'23} devise an image SemCom framework that enables a set of servers to collaboratively transmit images to their respective users using SemCom techniques. In their framework, all servers
 must jointly decide -- subject to limited wireless resource constraints -- which semantic information to transmit and which corresponding resource block (RB) allocation scheme o use \cite{Zhang_Opt_in_Image_SemCom'23}. The authors of \cite{Zhang_Opt_in_Image_SemCom'23} formulate this problem as an optimization problem whose target is minimizing the average transmission latency while satisfying the image-to-graph semantic similarity (ISS) requirement -- ISS is the authors' proposed semantic metric. To solve this problem, the authors develop a value decomposition-based entropy-maximized multi-agent RL algorithm. This algorithm is shown to considerably reduce transmission latency and improve convergence speed compared with traditional multi-agent RL algorithms \cite{Zhang_Opt_in_Image_SemCom'23}.

We now proceed to discuss state-of-the-art algorithmic developments in SemCom with regard to security and privacy.

\subsection{Algorithmic Developments in SemCom with Security and Privacy}
\label{subsec: Alg_Developments_in_SemCom_with_Security_and_Privacy}
Security and privacy must be carefully considered in the design of networks in 6G and beyond. To this end, some state-of-the-art SemCom works \cite{Luo_Encrypted_SemCom'22,Kaewpuang_Cooperative_Resource_Management'22} have developed a SemCom system with security and privacy features.

The authors of \cite{Luo_Encrypted_SemCom'22} proffer an encrypted SemCom system \cite[Fig. 1]{Luo_Encrypted_SemCom'22} that provides two modes of semantic transmission -- encrypted and unencrypted -- without the need to change the semantic encoder or decoder. To realize this system, the authors designed the structure of the secret key, encryptor, and decryptor for SemCom, so they can be embedded in a shared SemCom model. To make the encrypted SemCom system universal and confidential, they put forward an adversarial encryption training scheme that ensures the accuracy of SemCom -- in encrypted as well as unencrypted mode -- while preventing attackers from eavesdropping on the {transmitted semantic information. The results of simulations in which this adversarial training scheme was used to demonstrate that the encrypted SemCom system can considerably enhance the privacy protection capability of a SemCom system \cite{Luo_Encrypted_SemCom'22}.

The considerable potential that SemCom represents for 6G and beyond is justified by the fact that many SemCom techniques outperform their traditional counterparts, especially in low SNR regimes. Nonetheless, this attribute of SemCom can be a security risk because an eavesdropper can easily decode semantic information received over a very noisy channel \cite{Zhang_Wireless_Image_Transmission'22}. The authors of \cite{Zhang_Wireless_Image_Transmission'22} therefore put forward a SemCom framework that takes into consideration both semantic decoding efficiency and its risk of privacy leakage. This framework -- named \textit{SecureMSE} -- employs a loss function that flexibly regulates the efficiency-privacy tradeoff \cite{Zhang_Wireless_Image_Transmission'22}. Computer experiments demonstrate SecureMSE's effectiveness and robustness when it comes to addressing this tradeoff \cite{Zhang_Wireless_Image_Transmission'22}.

The authors of \cite{Sagduyu_Is_SemCom_Secure'22} study an AE-based SemCom system that is enabled by DL and deep networks such as DNNs and its ability to convey information from a source to a destination while preserving the semantic information. The authors demonstrate that the use of DNNs makes the SemCom system vulnerable to adversarial attacks in which the attacker attempts to manipulate the deep network inputs. More specifically, they substantiate that:
\begin{itemize}
	\item Adversarial attacks can be launched in multiple domains: $1)$ a computer vision attack that injects a malicious perturbation into the input image at the source, or $2)$ a wireless attack that sends a perturbation signal that is received by the decoder while superimposed on the transmitted signal \cite{Sagduyu_Is_SemCom_Secure'22}. 
	
	\item Both a computer vision attack and a wireless attack are effective individually \cite{Sagduyu_Is_SemCom_Secure'22}. When these attacks are combined (i.e., a multi-domain adversarial attack), they are even more effective in reducing SemCom performance \cite{Sagduyu_Is_SemCom_Secure'22}. 
	
	\item Multi-domain adversarial attacks can not only increase the reconstruction loss but also make the SemCom system so unreliable that the recovered information cannot preserve the semantics of the transmitted message \cite{Sagduyu_Is_SemCom_Secure'22}.
\end{itemize}

In a similar spirit as the work in \cite{Sagduyu_Is_SemCom_Secure'22}, the authors of \cite{Sagduyu_TO-Commun'22} present novel attack vectors that are based on backdoor and adversarial attacks, demonstrate empirically that goal-oriented communications are also vulnerable to stealth manipulations by smart adversaries. Consequently, the authors underscore the need for novel security mechanisms that can promote the safe adoption of task-oriented communications in 6G and beyond.

We now continue to discuss state-of-the-art algorithmic developments in quantum SemCom.

\subsection{Algorithmic Developments in Quantum SemCom}
\label{subsec: Quantum_SemCom}
The authors of \cite{Kaewpuang_Cooperative_Resource_Management'22} stray from encryption-based that is secured SemCom and propose a SemCom system secured by QKD (see \cite{Gisin_QCommun'07,Van_Meter_QNetworking'14,Cao_QKD_Networks_Survey'22}) and in which edge devices need to meet security requirements in QKD nodes and the QKD service providers must supply QKD resources to minimize the deployment cost. The authors develop a decision-making scheme for QKD service providers in a QKD-secured SemCom system by proposing resource allocation methods for the secure transmission of edge devices' semantic information, cost management, and cooperation among QKD service providers. When it comes to resource allocation, the authors implement two-stage stochastic programming to resolve the QKD service providers' solutions so that edge devices can transmit their semantic information that employs resources in the resource pool.

The authors of \cite{Chehimi_Quantum_SemCom'22} were inspired by rapid developments in SemCom \cite{Sem_Empowered_Commun'22,Chaccour_Building_NG_SemCom_Networks'22,Qin_Sem_Com_Principles_Apps'22,SemCom_for_6G_Future_Internet'22}; ML \cite{Jordan_ML_Science'15,MUWCM19,Ghahramani'2015_Probabilistic_ML}; quantum ML (QML) \cite{JBPPNWS_17,SSSMM19,NW_QDL_2019}; quantum computing \cite{Nielsen_Chuang_QC'10,Scherer_Math_for_QC'19,Preskill2018quantumcomputingin}; quantum communication \cite{Gisin_QCommun'07,Imre_Advanced_QC'12,Cariolaro2015QuantumC}; and quantum networking \cite{Van_Meter_QNetworking'14,Bassoli_QCNs'2021,Djordjevic_QC_QN_and_QS'22} and introduce a quantum semantic communications framework for developing reasoning-based future communication systems with quantum semantic representations that exhibit minimalism, efficiency, and accuracy. This framework employs quantum embedding and high-dimensional Hilbert spaces to extract the meaning of classical data \cite{Chehimi_Quantum_SemCom'22}. An unsupervised QML technique named quantum clustering is exploited for minimalistic and efficient contextual information extraction and accurate characterization of the semantics of the message to be sent \cite{Chehimi_Quantum_SemCom'22}. The quantum semantic representations -- that are constructed -- are then transmitted using quantum communication links \cite{Chehimi_Quantum_SemCom'22} established through \textit{quantum entanglement}\footnote{Quantum entanglement is a remarkable quantum mechanical phenomenon that the states of two or more quantum subsystems are correlated in a manner that is not possible in classical systems \cite{Van_Meter_QNetworking'14}. It is a unique quantum mechanical resource driving numerous applications of quantum computation, quantum information, quantum communication, and quantum networking \cite{Van_Meter_QNetworking'14,Nielsen_Chuang_QC'10}.} \cite{Gisin_QCommun'07,Imre_Advanced_QC'12,Cariolaro2015QuantumC}. 

We now move on to discuss state-of-the-art algorithmic developments in the economics of SemCom.

\subsection{Algorithmic Developments in Economics of SemCom}
\label{subsec: Economics_SemCom}
The authors of \cite{Liew_SemCom_ICASSP'22} develop an energy allocation framework for wireless-powered SemCom-based IoT. More particularly, they derive the valuation of energy based on SemCom performance metrics and maximize the wireless power transmitters' revenue using a DL-based auction while maintaining the desired properties of \textit{incentive compatibility} (IC) and \textit{individual rationality} \cite{Liew_SemCom_ICASSP'22}. Building on the work in \cite{Liew_SemCom_ICASSP'22}, the authors of \cite{Liew_Economics_of_SemCom_Auction_Approach'22} put forward incentive mechanisms -- i.e., a hierarchical trading system -- for \textit{semantic model trading} and \textit{semantic information trading}. Regarding the former, the proposed mechanism \cite{Liew_Economics_of_SemCom_Auction_Approach'22} helps to maximize the semantic model providers' revenue from semantic model trading and incentivizes them to take part in SemCom system development. As for semantic information trading, the authors' auction approach promotes trading between multiple semantic information sellers and buyers while ensuring individual rationality, IC, and a balanced budget.

Further development in the economics of SemCom is provided by the authors of \cite{Luong_Edge_Computing_for_SemCom'22}, who put forward a DL-based auction for edge computing trading in a SemCom-enabled Metaverse system. Their proposed auction aims to maximize the edge computing providers' revenue and attain individual rationality as well as IC \cite{Luong_Edge_Computing_for_SemCom'22}. Meanwhile, simulation results demonstrate that the auction considerably improves revenue compared with the baseline while gathering almost zero individual rationality and IC penalties \cite{Luong_Edge_Computing_for_SemCom'22}.

We now proceed to discuss miscellaneous state-of-the-art algorithmic developments in SemCom. 

\subsection{Miscellaneous Algorithmic Developments in SemCom}
\label{subsec: Miscellaneous_Devs_SemCom}
The authors of \cite{Chen_Neuromorphic_Cognition'22} attempt to unify neuromorphic sensing, processing, and communications by introducing an architecture for wireless cognition named \textit{NeuroComm}. NeuroComm's end-to-end design \cite{Chen_Neuromorphic_Cognition'22} is based on supervised learning via surrogate gradient descent (GD) methods. On the other hand, the authors of \cite{Shao_PAPR_Reduction'22} consider SemCom with discrete-time
analog transmission and validate that DeepJSCC-based SemCom's notable image reconstruction performance can be maintained while the transmitted peak-to-average power ratio is suppressed to an acceptable level, which is important for the practical implementation of DeepJSCC-based SemCom systems \cite{Shao_PAPR_Reduction'22}. In light of DeepJSCC and the DL model’s overfitting property, the authors of \cite{Dai_Adaptive_SemCom'22} demonstrate the feasibility of overfitting neural-enhancement on SemCom systems and the effectiveness of overfitting when combined with online learning. Consequently, the authors put forward an adaptive SemCom framework that employs online learning to \textit{overfit} the instant source data sample and CSI. Furthermore, the authors of \cite{SemCom_with_Conceptual_Spaces'22} employ the conceptual space theory of semantics \cite{Conceptual_Spaces_Book'00,The_Geometry_of_Meaning_Book'14} to propose a model for SemCom with conceptual spaces \cite[Fig. 2]{SemCom_with_Conceptual_Spaces'22} wherein functional compression is proposed to obtain optimal encoding schemes. The authors simulate image transmission using their SemCom system and confirm its potential to faithfully convey meaning with a gigantic reduction in communication rate. 

By approximating the semantic similarity metric by a generalized logistic function, the authors of \cite{Mu_SemCom_and_Bitcom'22} propose a framework for heterogeneous SemCom and bit communication (BitCom) in which an access point simultaneously transmits the semantic stream to a semantics-interested user and the bit stream to a bit-interested user. Following the work in \cite{Mu_SemCom_and_Bitcom'22}, the authors of \cite{Mu_SemCom_for_NOMA'22} propose a semantics-empowered two-user uplink non-orthogonal multiple access (NOMA) framework in which a primary near user (N-user) and a secondary far user (F-user) communicate with the access point using BitCom and SemCom, respectively. The authors investigate this semantic-empowered NOMA framework over fading channels and put forward an opportunistic SemCom and BitCom scheme to allow the secondary F-user to exploit the advantages of both technologies whenever it is admitted into the NOMA framework (at each fading state). This opportunistic scheme is demonstrated to outperform other baseline schemes \cite{Mu_SemCom_for_NOMA'22}. Following the work in \cite{Mu_SemCom_for_NOMA'22} and in \cite{Mu_SemCom_and_Bitcom'22}, the authors of \cite{Mu_SemCom_in_MU_Wireless_Networks'22} propose a heterogeneous semantic and bit multi-user framework. Concerning this framework, they uncover that the interplay between NOMA and SemCom shows promise for supporting NOMA-enabled SemCom and SemCom-enhanced NOMA. When it comes to NOMA-enabled SemCom, the authors of \cite{Mu_SemCom_in_MU_Wireless_Networks'22} proffer a semi-NOMA-enabled heterogeneous SemCom and BitCom scheme that combines conventional OMA and NOMA as its special cases and offers flexible transmission options.

The authors of \cite{Hu_Robust_SemCom_with_Masked_VQ-VAE'22} proffer a framework for robust end-to-end SemCom systems that combats semantic noise and develop an adversarial training with weight perturbation by incorporating samples with semantic noise in the training dataset. They propose to mask a portion of the input wherein the semantic noise appears frequently when employing the training dataset and design a masked vector quantized-variational AE. The results of simulations conducted with this robust image SemCom technique demonstrate it improves the system's robustness against semantic noise \cite{Hu_Robust_SemCom_with_Masked_VQ-VAE'22}.

The design of conventional technical communication (TechnicCom) is largely based on stochastic modeling and manipulation in startling contrast with SemCom, which uses semantic elements that can be logically connected \cite{Choi_Unified_View_SemInfo'22,Choi_Unified_Approach'22}. To address this knowledge gap, the authors of \cite{Choi_Unified_View_SemInfo'22,Choi_Unified_Approach'22} propose a unified approach to semantic information and communication by leveraging probabilistic logic \cite{NILSSON_ProbLogic'86} via the interplay of SemCom and TechnicCom. More specifically, the authors of \cite{Choi_Unified_View_SemInfo'22,Choi_Unified_Approach'22} combine the existing TechnicCom layer with a SemCom layer that uses communicating parties' KBs to exchange (text) semantic information. The authors of \cite{Cognitive_SemCom_Systems'22} develop a KG-driven \textquotedblleft cognitive'' text SemCom framework for a corresponding text SemCom system. More specifically, they propose a simple, general, and interpretable solution for their text SemCom scheme to detect semantic information. 

The authors of \cite{SemCom_for_In-Cabin_Scenarios'22} leverage advancements in 6G research and SemCom to proffer a 6G SemCom technique that is based on intelligent fabrics for in-cabin transportation scenarios. The authors then propose a DL-based end-to-end SemCom scheme for time-series data that exploits DL for semantic sensing and information extraction \cite{SemCom_for_In-Cabin_Scenarios'22}. The authors of \cite{Xiao_Semantic-driven'22} put forward a DeepSC-based network service framework that combines a SemCom system and intelligent fabrics for a smart healthcare system in intelligent fabric. In light of this framework, the authors establish a combined service offloading and bandwidth allocation optimization model for resource-efficient semantic-aware networks with enhanced QoS. The authors of \cite{LLL_Reasoning_Based_SemCom'22} introduce a reasoning-based SemCom architecture wherein semantic information is represented by KG. To convert the KG-based representation, which is high-dimensional, to a low-dimensional representation, the authors develop an embedding-based semantic interpretation framework. They then propose function-based inference to infer hidden information that cannot be directly observed from the received message \cite{LLL_Reasoning_Based_SemCom'22}. The authors of \cite{Xiao_Reasoning_on_the_Air'22} incorporate reasoning into SemCom and develop a framework for representing, modeling, and interpreting implicit semantic meaning. The authors then develop an implicit SemCom architecture in which a reasoning procedure can be trained at the receiving user with the help of the transmitting user \cite{Xiao_Reasoning_on_the_Air'22}.

The authors of \cite{Guo_Signal_Shaping_for_SemCom'22} introduce a signal-shaping method to minimize semantic loss as quantified by a pretrained BERT (bidirectional encoder representations from transformers \cite{Peters_BERT_Paper'18}) model in SemCom systems with a few message candidates. The authors provide a solution that is based on an efficient projected GD method. The authors of \cite{Dong_Innovative_SemCom'22} put forward an image SemCom system that transmits not only semantic information but also a semantic decoder. The authors of \cite{zhang_Context-Based_SemCom'22} propose a context-aware text SemCom framework wherein the transmitter and the receiver use their respective KB for coding and decoding by developing a part-of-speech-based encoding strategy and a context-based decoding strategy. To investigate the impact of adaptive bit lengths on semantic coding under various SNRs, the authors of \cite{Zhou_Adaptive_Bit_Rate_Control_SemCom'22} put forward progressive semantic HARQ schemes that use incremental knowledge by designing a semantic encoding solution with multibit length selection. The authors of \cite{Jiang_Reliable_SemCom'22} develop a KG-based text SemCom system that adaptively adjusts the transmitted content per the channel quality and allocates more resources to important triplets to enhance the reliability of communication. In light of this text SemCom technique whose performance needs to be enhanced for low-SNR conditions, the authors of \cite{Enhancing_SemCom_Reliablity'22} investigate reasoning and decoding at the semantic level instead of the grammar level. Employing reasoning and decoding at the semantic level to improve communication reliability, the authors of \cite{Enhancing_SemCom_Reliablity'22} propose a text SemCom scheme that incorporates a language model, prior information, and parts of speech and wherein the language model and prior information are deployed to enhance the receiver's semantic reasoning.

Despite their promises, SemCom techniques are difficult to realize when source signals are used for a variety of tasks (e.g., as wireless sensing data for localization and activity detection) due to increased processing complexity \cite{Du_RIS-aided_Encoding'22}. To address this challenge, the authors of \cite{Du_RIS-aided_Encoding'22} devise a new SemCom paradigm named \textit{inverse SemCom} wherein task-related source messages are encoded into a hyper-source message for data transmission or storage rather than extracting semantic information from messages. Concerning this SemCom paradigm, the authors of \cite{Du_RIS-aided_Encoding'22} develop an inverse semantic-aware wireless sensing framework by proposing three algorithms for data sampling, RIS-aided encoding, and self-supervised decoding. At the same time, the authors of \cite{Du_RIS-aided_Encoding'22} design RIS hardware for encoding multiple signal spectrums into one \textit{MetaSpectrum}. Regarding MetaSpectrum, the authors propose a semantic hash sampling method for selecting task-related signal spectrums and a self-supervised learning method for decoding MetaSpectrums. Their framework reduced the data volume by $95\%$ as reported by the authors in comparison with the volume before encoding without impacting the accomplishment of sensing tasks \cite{Du_RIS-aided_Encoding'22}. 

In many DL-based SemCom systems, DNNs have substituted various building blocks of conventional communication systems so the entire system can be termed an analog communication system. However, a digital communication system with digital modulation has many advantages over an analog one despite DNN-based digital modulation being a huge challenge \cite{Learning_Based_Digital_SemCom'22}. The challenge stems from the fact that DNN-based digital modulation is based on mapping the continuous output of a DNN-based encoder into discrete constellation symbols, which requires a non-differentiable mapping function that cannot be trained using GD algorithms \cite{Learning_Based_Digital_SemCom'22}. To overcome this challenge, the authors of \cite{Learning_Based_Digital_SemCom'22} devise a joint coding-modulation scheme for a digital SemCom with binary phase shift keying modulation. This digital SemCom scheme is shown to outperform existing digital modulation methods in SemCom over a wide range of SNRs and DNN-based analog modulation in low SNR regimes \cite{Learning_Based_Digital_SemCom'22}. As another DL-based digital SemCom technique, the authors of \cite{Fu_VQ_SemCom'22} put forward a DL-enabled vector quantized digital image SemCom system for image transmission that is dubbed \textit{VQ-DeepSC}. VQ-DeepSC deploys a CNN-based transceiver to extract an image's multi-scale semantic features and adversarial training to improve the quality of received images. Meanwhile, simulation results corroborate that VQ-DeepSC outperforms traditional image transmission methods, especially in low SNR regimes \cite{Fu_VQ_SemCom'22}.    

In light of the rapidly emerging SemCom trends and use cases that already substantiate SemCom's potential in building next-generation 6G networks, not all users can be served by SemCom systems, for such systems are mainly specialized to handle specific applications \cite{Lee_SemCom_Impact_Analysis'22}. This fact calls for a network design that rigorously addresses the inevitable coexistence of SemCom systems and BitCom systems. Toward this end, the authors of \cite{Lee_SemCom_Impact_Analysis'22} examine from a network vantage point how introducing emerging SemCom systems impacts the performance of existing BitCom systems. They do so by formulating a max-min fairness problem concerning the coexistence of SemCom and BitCom systems. Regarding this coexistence problem, extensive numerical results corroborate that SemCom systems are indeed a promising next-generation communication alternative \cite{Lee_SemCom_Impact_Analysis'22}. Similar to the work in \cite{Lee_SemCom_Impact_Analysis'22}, the authors of \cite{LEE_Perf_analysis_2022} study the impact of BitCom and SemCom system coexistence on network performance by analyzing sum-rate maximization with a minimum required SNR constraint for SemCom users. For this problem, the authors provide a power control algorithm whose numerical results substantiate \cite{LEE_Perf_analysis_2022} that introducing a SemCom system enhances the sum-rate of BitCom users whilst offering the same or a higher SNR to SemCom users with less transmit power.

Many of the existing SemCom techniques are one-to-one or point-to-point SemCom techniques that do not consider the one-to-many broadcasting scenario. For this scenario, the authors of \cite{Hu_One-to-Many_SemCom'22} develop a DNN-enabled one-to-many SemCom system dubbed \textit{MR$\_$DeepSC}. The MR$\_$DeepSC system comprises a transmitter made up of a semantic encoder and a channel decoder that is coupled with multiple receivers that each consists of a channel decoder, a semantic decoder, and a semantic recognizer \cite[Fig. 1]{Hu_One-to-Many_SemCom'22}. The semantic recognizer is designed to distinguish different users based on a pre-trained model by leveraging their semantic features \cite{Hu_One-to-Many_SemCom'22}. In this \textit{one-to-many SemCom system}, transfer learning is adopted to accelerate the training of new receiver networks, and simulation results demonstrate that MR DeepSC performs significantly better -- especially in low SNR regimes -- than other benchmarks under a variety of channel conditions \cite{Hu_One-to-Many_SemCom'22}. The authors of \cite{Yu_Optical_SemCom'22} introduce and experimentally demonstrate an optical SemCom system wherein DL is employed to extract semantic information that is then directly transmitted through an optical fiber. This \textit{optical SemCom system} achieves greater information compression and more stable performance than a bit-based optical communication system, especially in low received optical power regimes \cite{Yu_Optical_SemCom'22}. The optical SemCom system in \cite{Yu_Optical_SemCom'22} also improves robustness against optical link impairments \cite{Yu_Optical_SemCom'22}. Furthermore, the authors of \cite{Zhang_Semantic_Sensing_and_Commun'22} propose a joint semantic sensing, rendering, and communication framework for wireless ultimate XR that is inspired by SemCom and its
advancements. This framework involves three components: $1)$ semantic sensing is employed to enhance sensing efficiency by exploiting the semantic information's spatial-temporal distributions; $2)$ semantic rendering is intended to reduce the cost of semantically-redundant pixels; and $3)$ SemCom is adopted for high-efficiency data transmission in wireless ultimate XR \cite{Zhang_Semantic_Sensing_and_Commun'22}. This joint SemCom scheme is demonstrated to be effective via two case studies \cite{Zhang_Semantic_Sensing_and_Commun'22}.

We now continue to the major state-of-the-art trends and use cases of SemCom.

\section{Major State-of-the-art Trends and Use Cases of SemCom}
\label{sec: SemCom_Major_Trends_Use-Cases}
Multiple state-of-the-art trends and use cases of SemCom are emerging, and many research communities are putting intensive efforts into research on 6G. Thus, we now discuss the major trends and major use cases of SemCom, starting with the major trends.

\subsection{Major Trends of SemCom}
\label{subsec: major_trends_SemCom}
In this section, we detail the following major trends of SemCom: JSCC schemes \cite{Eirina_JSCC'19,Kurka_Deep_JSCC-f'20,Bandwidth_Agile_Image_Transmission'21,Tung_DeepWiVe'21}; DeepSC and its variants \cite{Xie_DL-based_SemCom'21,Xie_Lite_distributed_SemCom'21,Weng_SemCom_Sys_Speech_Trans'21}; joint semantics-noise coding (JSNC) \cite{Rethinking_modern_com_Lu_2022}; SemCom in the S-PB layer \cite{Zhang_Wisdom_Evolutionary_6G'21}; an understand-first-and-then-transmit SemCom framework \cite{Shi_to_Semantic_Fidelity'21}; context-based SemCom \cite{Engineering_SemCom'22}; semantic coded transmission \cite{Commun_Beyond_Transmitting_Bits'22}; neuromorphic wireless cognition \cite{Chen_Neuromorphic_Cognition'22}; a cognitive SemCom system that is driven by KG \cite{Cognitive_SemCom_Systems'22}; implicit SemCom \cite{Xiao_Reasoning_on_the_Air'22}; innovative SemCom \cite{Dong_Innovative_SemCom'22}; a reliable SemCom system that is enabled by KG \cite{Jiang_Reliable_SemCom'22}; an AE-based SemCom system \cite{Luo_SemCom_with_relay'21}; a semantic-aware speech-to-text SemCom system with redundancy removal \cite{Han_Semantic-aware_Speech2Text_Transmission'22}; cross-modal SemCom \cite{Li_Cross-Modal_SemCom'22}; and encrypted SemCom \cite{Luo_Encrypted_SemCom'22}. We begin with JSCC schemes.

\begin{figure}[t!]
	\centering
	\hspace{-6mm}\includegraphics[scale=0.28]{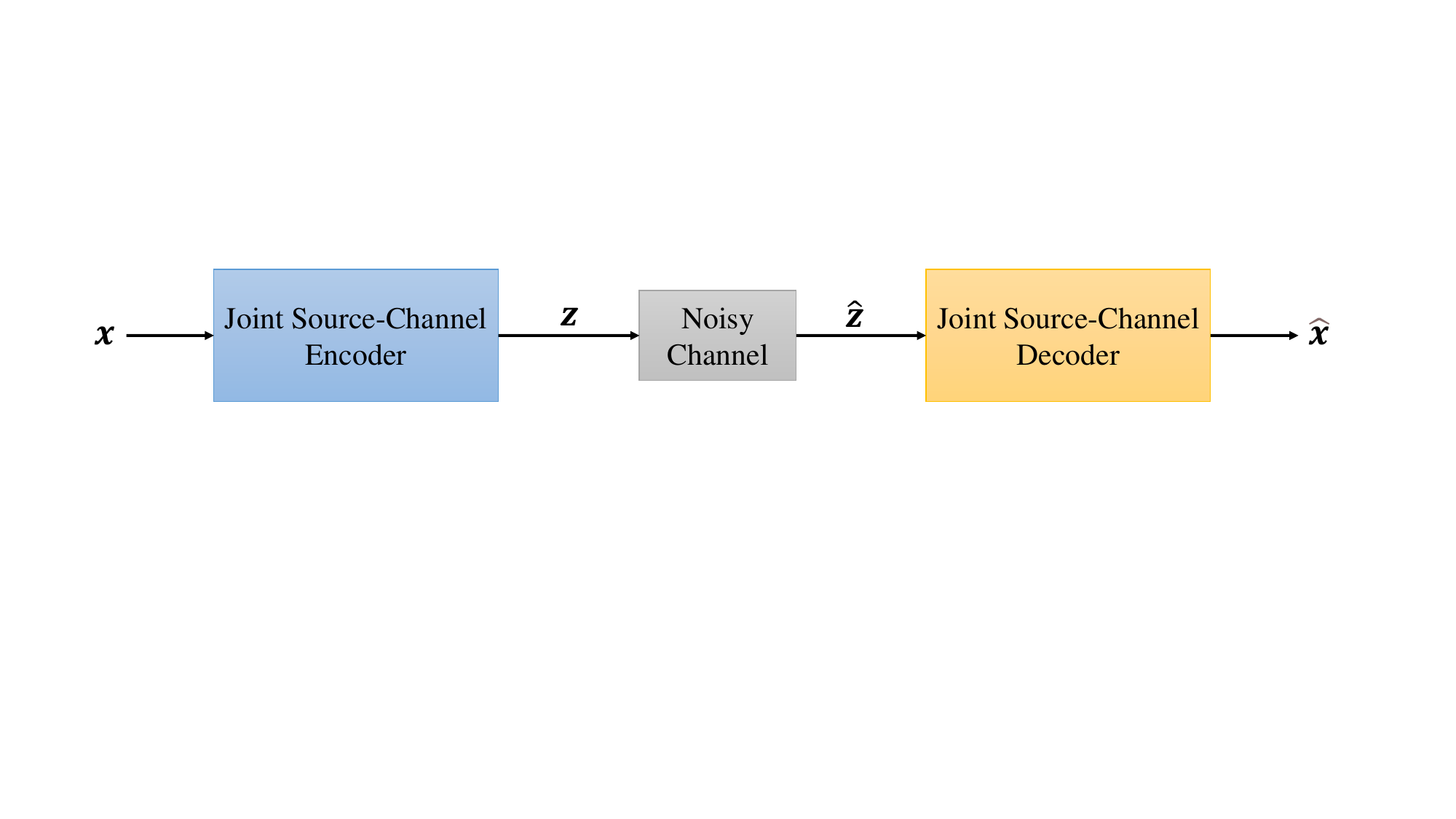}  \vspace{-2.0cm} 
	\caption{Components of deep JSCC \cite[Fig. 1(b)]{Eirina_JSCC'19}.}
	\label{fig: JSCC_Block_Diagram_20221115}
\end{figure}

\subsubsection{JSCC Schemes}
unlike separation-based source and channel coding schemes which are known to suffer from the cliff effect, the DL-based JSCC technique depicted in Fig. \ref{fig: JSCC_Block_Diagram_20221115} doesn't suffer from the cliff effect, which makes it a major trend in video SemCom and image SemCom. When it comes to image SemCom, deep JSCC \cite{Eirina_JSCC'19}, DeepJSCC-f \cite{Kurka_Deep_JSCC-f'20}, and DeepJSCC-l \cite{Bandwidth_Agile_Image_Transmission'21} all use a DL-based end-to-end-trained joint source-channel encoder and joint source-channel decoder, and provide a considerable performance gain over conventional separation-based schemes. When it comes to video SemCom, on the other hand, DeepWiVe \cite{Tung_DeepWiVe'21} is a DL-based end-to-end JSCC video transmission scheme that avoids the cliff effect while achieving a graceful degradation in channel quality and producing superior video quality in comparison with state-of-the-art video transmission techniques \cite{Tung_DeepWiVe'21}.

Apart from JSCC, DeepSC \cite{Xie_DL-based_SemCom'21} and its variants are another major trend in SemCom.

\begin{figure*}[t!]
	\centering
	\includegraphics[scale=0.50]{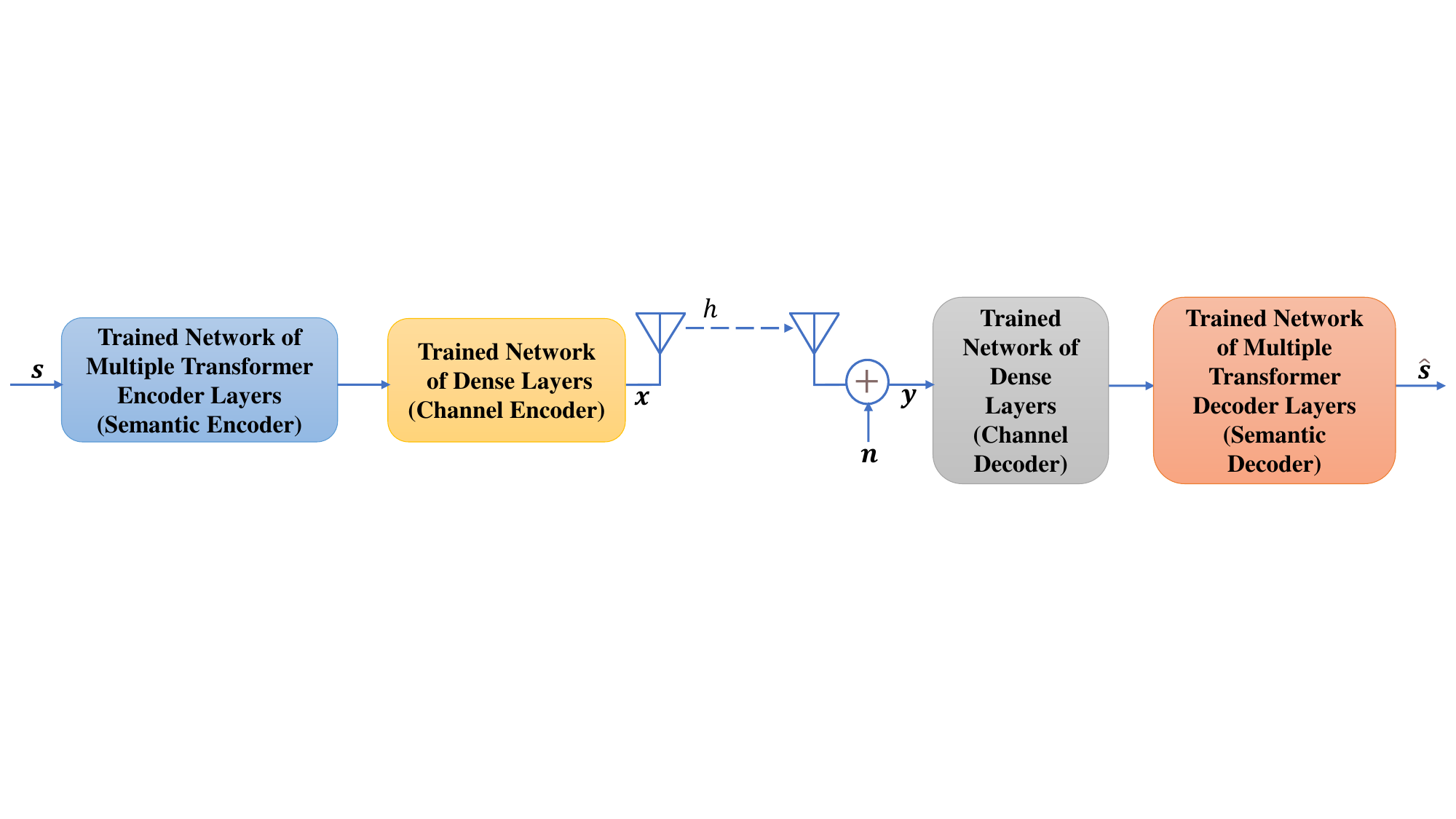}  \vspace{-2.5cm} 
	\caption{An end-to-end trained DeepSC \cite{Xie_DL-based_SemCom'21}.}
	\label{fig: DeepSC_Original_Model_20221103}
\end{figure*}

\subsubsection{DeepSC and its Variants}
a major trend in text SemCom, a DL-based and end-to-end-trained DeepSC architecture that is a well-known text SemCom technique is shown in Fig. \ref{fig: DeepSC_Original_Model_20221103}. As seen on the left side of Fig. \ref{fig: DeepSC_Original_Model_20221103}, the DeepSC transmitter comprises a semantic encoder that extracts semantic information from the source's text using several Transformer encoder layers \cite{Peters_BERT_Paper'18} that feed the extracted semantic information to a channel encoder made of dense layers with different units that produce semantic symbols to be transmitted to the DeepSC receiver \cite[Sec. IV]{Xie_DL-based_SemCom'21}. The DeepSC receiver -- depicted on the right side of Fig. \ref{fig: DeepSC_Original_Model_20221103} -- is composed of a channel decoder that is made of dense layers with different units whose output is inputted in a semantic decoder built from multiple Transformer decoder layers \cite[Sec. IV]{Xie_DL-based_SemCom'21}. The Transformer decoder layers (of the semantic decoder) and the dense layers (of the channel decoder) are used for text recovery and (semantic) symbol detection, respectively. For text recovery applications, end-to-end-trained DeepSC outperforms various contemporary conventional communication system benchmarks -- especially in low SNR regimes -- for both the additive white Gaussian noise channel and Rayleigh fading channel \cite{Xie_DL-based_SemCom'21}. In light of DeepSC's significant performance gain for low SNR regimes, several variants of DeepSC are proposed in the literature for both text SemCom and audio SemCom: L-DeepSC \cite{Xie_Lite_distributed_SemCom'21} as a text SemCom technique for IoT networks (considering the limited power and computing capabilities of the IoT devices); R-DeepSC \cite{Peng_Robust_DL-Based_SemCom'22} as a text SemCom technique that improves system robustness in a variety of wireless environments; DeepSC-S \cite{Weng_SemCom_Sys_Speech_Trans'21,SemCom_for_speech_signals'20} as an audio SemCom technique to improve transmission efficiency by transmitting only the semantic information; DeepSC-SR as an audio SemCom scheme for speech recognition; and DeepSC-ST \cite{Weng_DL-enabled_SemCom'22} for speech recognition and synthesis.

\begin{figure*}[htb!]
	\centering
	\includegraphics[scale=0.50]{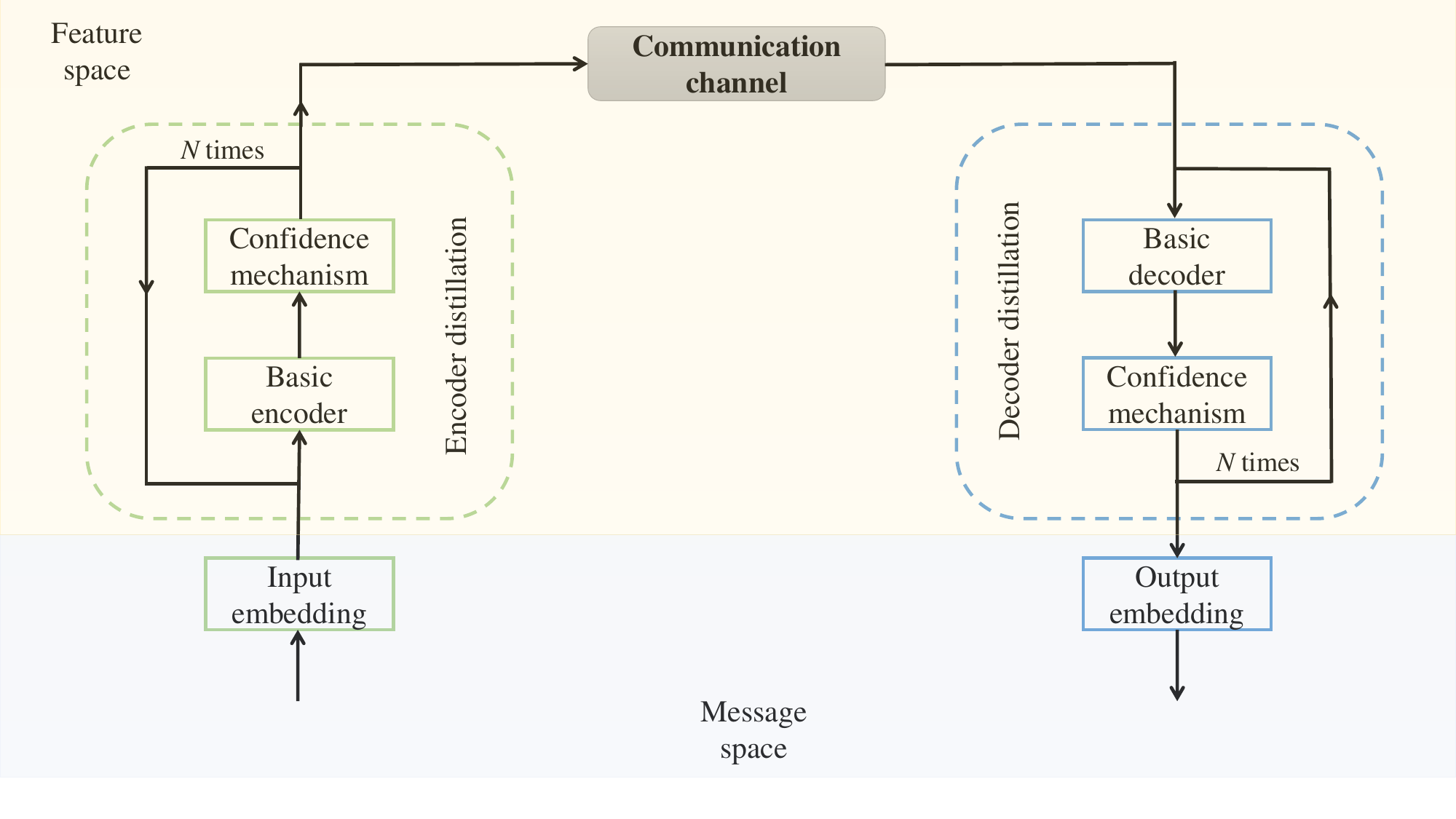}  \vspace{-0.05cm} 
	\caption{A joint semantic-noise coding (JSNC) mechanism \cite[Fig. 2]{Rethinking_modern_com_Lu_2022}.}
	\label{fig: JSNC_architecture_20221117}
\end{figure*}

Following DeepSC and its variants, we proceed with our brief discussion on a text SemCom technique named JSNC \cite{Rethinking_modern_com_Lu_2022}.  

\begin{figure*}[t!]
	\centering
	\includegraphics[scale=0.50]{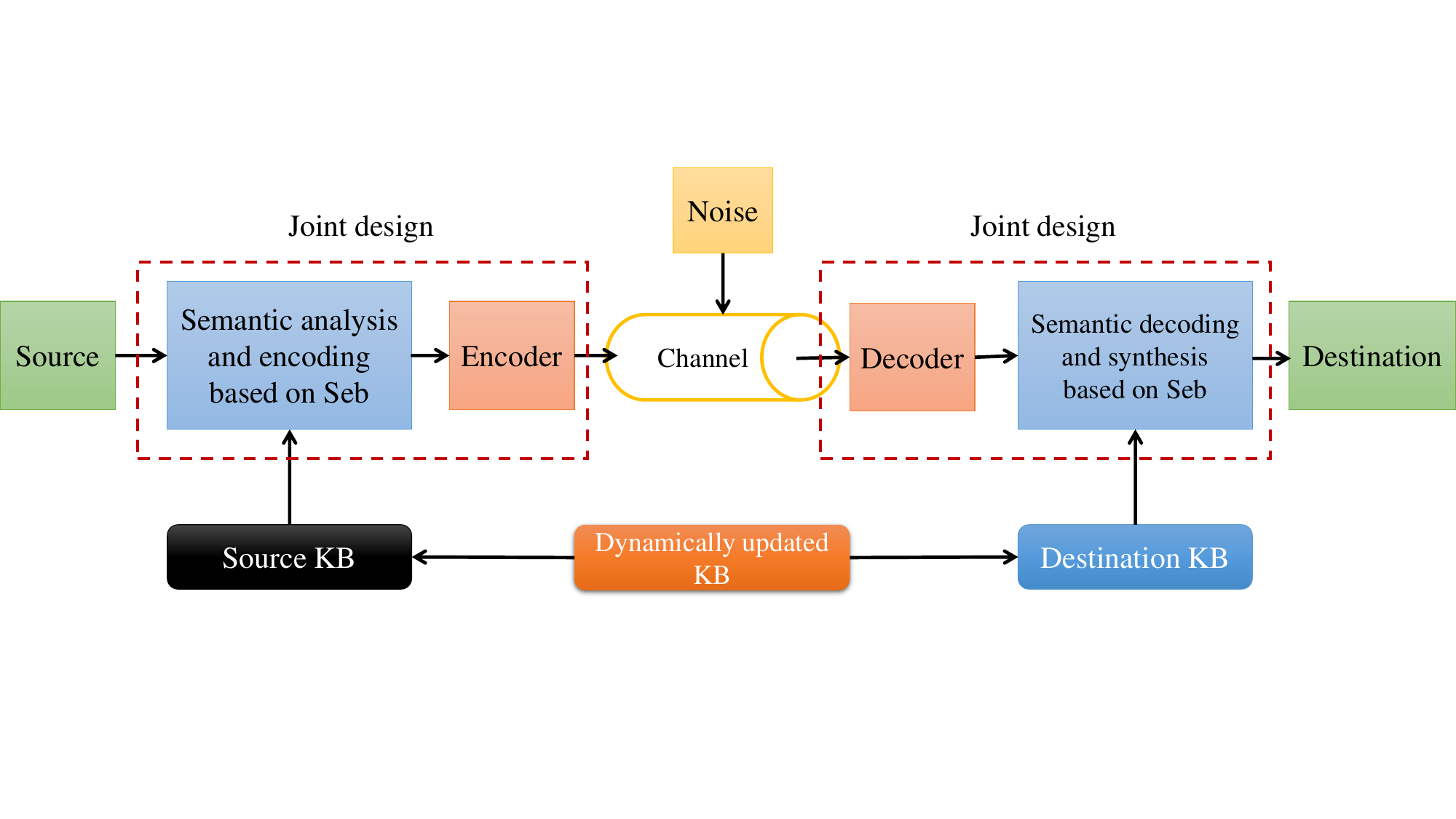}  \vspace{-2.0cm} 
	\caption{A Text SemCom framework in the S-PB layer \cite[Fig. 4]{Zhang_Wisdom_Evolutionary_6G'21}: S-PB layer -- semantic-empowered physical-bearing layer; Seb -- semantic base.}
	\label{fig: SemCom_in_S-PB_layer_20221115}
\end{figure*}

\subsubsection{JSNC}
JSNC -- as it is schematized in Fig. \ref{fig: JSNC_architecture_20221117} -- can be used as a text SemCom technique and is put forward by the authors of \cite{Rethinking_modern_com_Lu_2022} to address a varying communication channel and incorporate a mechanism for the possible interpretation of semantic meaning. For the former goal, the authors propose encoder distillation and decoder distillation mechanisms -- both DNN-based -- that aim to refine the \textit{embeddings} in the encoder and decoder, as seen in Fig. \ref{fig: JSNC_architecture_20221117}. As for the second goal, of semantic meaning interpretation, the authors incorporate DNN-based confidence-based mechanisms (, which are also shown in Fig. \ref{fig: JSNC_architecture_20221117}, at both the transmitter and the receiver) that assess the quality of semantic representation while guiding encoder distillation and decoder distillation. Once a given message is projected into the feature space as depicted in Fig. \ref{fig: JSNC_architecture_20221117}, distillation at the transmitter and the receiver is triggered by the semantic confidence module provided that semantic confidence doesn't reach a pre-defined threshold \cite{Rethinking_modern_com_Lu_2022}. Otherwise, the JSNC mechanism discharges the processed semantic information for processing further downstream \cite{Rethinking_modern_com_Lu_2022}. Moreover, as shown in Fig. \ref{fig: JSNC_architecture_20221117}, the maximum distillation round is $N$, which amounts to a JSNC system being done with the SE and ready for further processing \cite{Rethinking_modern_com_Lu_2022}.

We now move on to our brief discussion on the SemCom framework in the S-PB layer that is proposed by the authors of \cite{Zhang_Wisdom_Evolutionary_6G'21}.

\subsubsection{SemCom in the S-PB layer}
the authors of \cite{Zhang_Wisdom_Evolutionary_6G'21} propose a text SemCom framework that comprises semantic analysis and encoding at the transmitter and semantic decoding and synthesis at the receiver, all of which are based on semantic base (\textit{Seb}) -- as seen in Fig. \ref{fig: SemCom_in_S-PB_layer_20221115} -- which is the basic representation framework for semantic information that the authors introduce. The transmitter's and receiver's Seb-based processing is dictated by the source KB and destination KB, respectively, which are generally different. To overcome this challenge, the authors consider a dynamically updated KB that is shared between the source KB and destination KB in their design depicted in Fig. \ref{fig: SemCom_in_S-PB_layer_20221115}. The authors also suggest  joint semantic encoding and channel encoding at the transmitter and joint channel decoding and semantic decoding at the receiver. 

We now proceed to our discussion on the understand-first-and-then-transmit SemCom framework \cite[Fig. 2]{Shi_to_Semantic_Fidelity'21}.

\begin{figure*}[t!]
	\centering
	\includegraphics[scale=0.50]{ 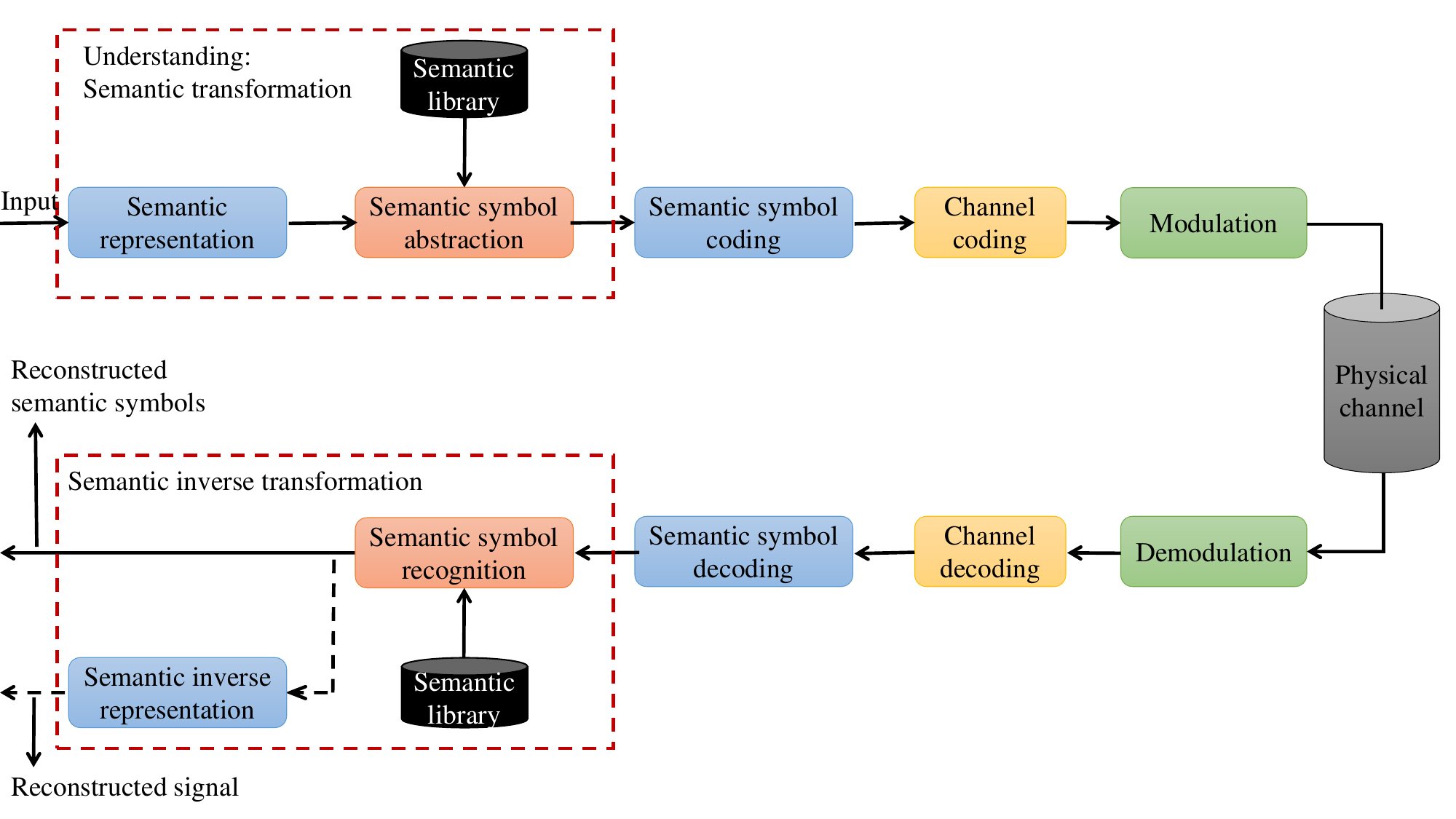}  \vspace{2mm} 
	\caption{A Text SemCom framework named communication toward semantic fidelity (CTSF) \cite[Fig. 2]{Shi_to_Semantic_Fidelity'21}.}
	\label{fig:  CTSF_SemCom_20221116}
\end{figure*}

\begin{figure}[htb!]
	\centering
	\includegraphics[scale=0.26]{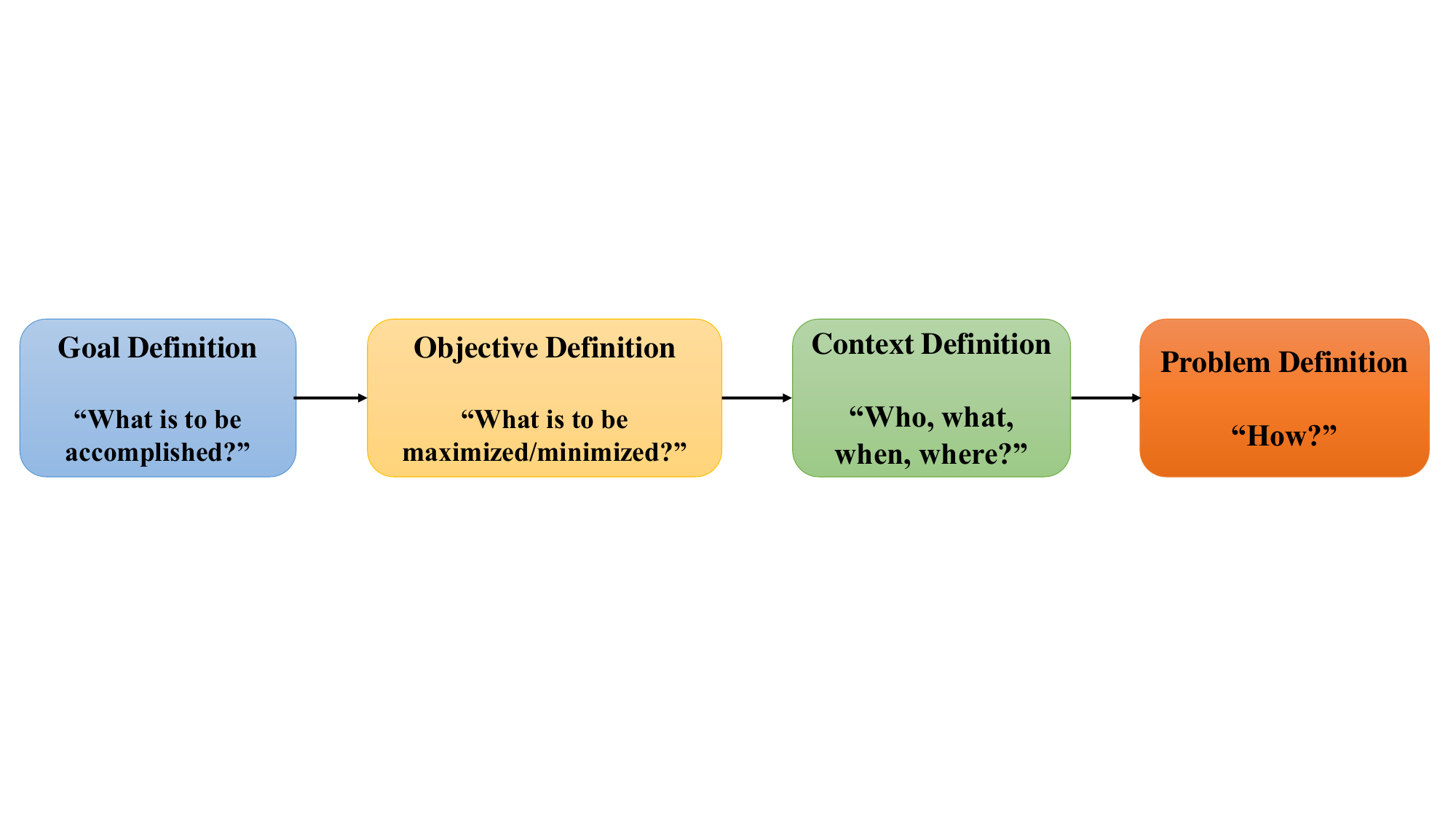}  \vspace{-1.5cm} 
	\caption{Design flow for context-based SemCom systems \cite[Fig. 9]{Engineering_SemCom'22}.}
	\label{fig: Context-Based_SemCom_20221103}
\end{figure}

\subsubsection{Understand First and then Transmit SemCom Framework}
the authors of \cite{Shi_to_Semantic_Fidelity'21} put forward an understand-first-and-then-transmit text SemCom framework -- as seen in Fig. \ref{fig:  CTSF_SemCom_20221116} -- named \textit{communication toward semantic fidelity} (CTSF). Per CTSF, the source converts the input signal into semantic symbols through semantic transformation, which is based on an understanding of a source's semantic library (also a determinant of transmission efficiency \cite{Shi_to_Semantic_Fidelity'21}). Semantic transformation is followed by semantic symbol abstraction. The abstracted semantic symbols are then transformed by semantic symbol encoding, channel encoding, and then modulation prior to their transmission through the communication channel. The receiver receives the channel's output, as seen in Fig. \ref{fig:  CTSF_SemCom_20221116}, and undoes the transmitter's signal processing through demodulation followed by channel decoding and then semantic symbol decoding. The receiver's semantic inverse transformation process -- as shown in Fig. \ref{fig:  CTSF_SemCom_20221116} -- uses the decoder's output to deliver the reconstructed signal and the reconstructed semantic symbols through the destination's semantic library-dictated semantic symbol recognition process and the semantic inverse representation process, respectively.
 
We now continue with the design flow for context-based SemCom systems \cite[Fig. 9]{Engineering_SemCom'22}.

\subsubsection{Context-Based SemCom}
the authors of \cite{Engineering_SemCom'22} proffer a design flow for context-based SemCom systems, which is depicted in Fig. \ref{fig: Context-Based_SemCom_20221103}. The authors' proposed design flow comprises goal definition, objective definition, context definition, and problem definition \cite[Fig. 9]{Engineering_SemCom'22}. Goal definition and objective definition determine the \textit{why} aspect of the context and what exactly is to be optimized \cite{Engineering_SemCom'22}, respectively. Once the optimization objective has been determined, context definition defines the remaining aspects of the context (i.e., who, what, where, and when), and problem definition determines the optimization problem w.r.t. the objective function chosen and the constraints (derived from goal definition, objective definition, and context definition) \cite{Engineering_SemCom'22}.

We now move on to the semantic coded transmission (SCT) SemCom technique \cite[Fig. 1]{Commun_Beyond_Transmitting_Bits'22}.

 \begin{figure*}[t!]
 	\centering
 	\includegraphics[scale=0.50]{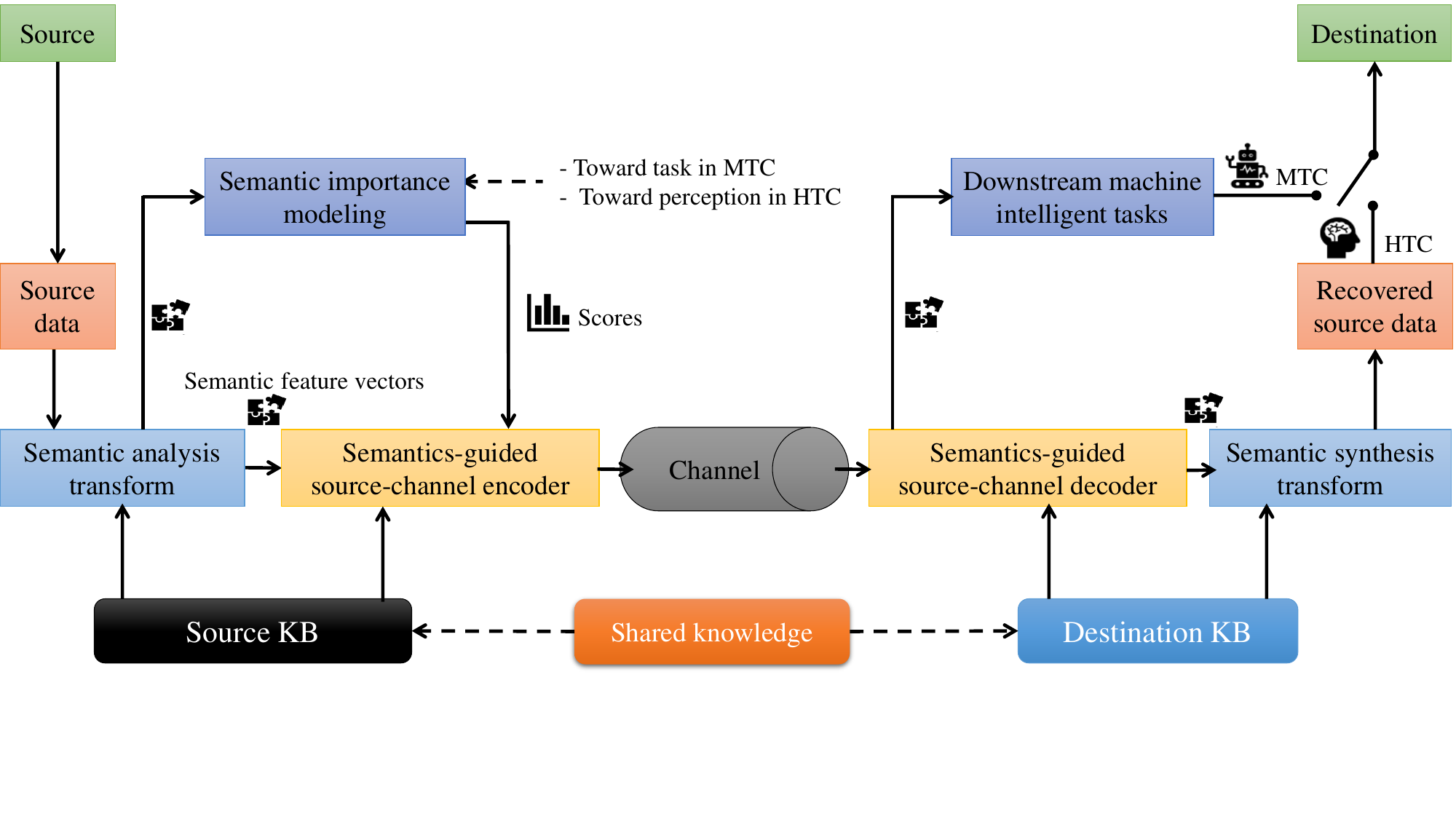}  \vspace{-1.0cm} 
 	\caption{Semantic coded transmission (SCT) \cite[Fig. 1]{Commun_Beyond_Transmitting_Bits'22}: MTC -- machine-type communication; HTC -- human-type communication.}
 	\label{fig: SCT_system_20221117}
 \end{figure*}
 
\subsubsection{Semantic Coded Transmission}
the SCT SemCom technique is introduced by the authors of \cite{Commun_Beyond_Transmitting_Bits'22} and schematized in Fig. \ref{fig: SCT_system_20221117}. As shown in Fig. \ref{fig: SCT_system_20221117}, the transmitter consists of the following modules: a semantic analysis transform, a semantic importance modeling module, and a semantics-guided source-channel encoder. The encoder's output is then transmitted to the channel whose output is processed by the receiver that comprises the semantics-guided source-channel decoder and semantic synthesis transform modules. The functions of these receiver modules and the above-mentioned transmitter modules are itemized below:
\begin{itemize}
	\item The semantic analysis transform module extracts the source data's semantic features and produces semantically annotated messages that are segmented as multiple semantic channels \cite{Commun_Beyond_Transmitting_Bits'22}, each of which comprises a \textit{semantic feature vector} (SFV) whose elements relate to the same semantic object \cite{Commun_Beyond_Transmitting_Bits'22}.

	\item The semantic importance modeling module evaluates each SFV's semantic value, which is determined based on the communication purpose of a scenario, such as human-type communication (HTC) or machine-type communication (MTC) \cite{Commun_Beyond_Transmitting_Bits'22}.

	\item The semantics-guided source-channel encoder is guided by the semantic importance scores and acts on each SFV to ensure the reliable transmission of SemCom signals over the wireless communication channel \cite{Commun_Beyond_Transmitting_Bits'22}.

	\item At the receiver, the semantics-guided source-channel decoder reconstructs the SFVs by performing the inverse operation of the semantics-guided source-channel encoder \cite{Commun_Beyond_Transmitting_Bits'22}.

	\item The semantic synthesis transform module takes the output of the semantics-guided source-channel decoder and performs the inverse operation of the transmitter's semantic analysis transform \cite{Commun_Beyond_Transmitting_Bits'22}. Thereafter, semantic feature fusion is used to either recover the source data or drive downstream machine intelligence tasks \cite{Commun_Beyond_Transmitting_Bits'22}.	 
\end{itemize}
In light of the work in \cite{Commun_Beyond_Transmitting_Bits'22}, the authors of \cite{Yao_SCT_over_MIMO'22} devise a novel versatile SCT system over MIMO fading channels that is dubbed \textit{VST-MIMO}. VST-MIMO supports parallel versatile rate transmission and multiple-stream transmission \cite{Yao_SCT_over_MIMO'22}. To this end, the authors of \cite{Yao_SCT_over_MIMO'22} design an adaptive spatial multiplexing module that guides rate allocation and stream mapping, and effectively couples the source semantics and channel states \cite{Yao_SCT_over_MIMO'22}. 
 
We now move on to our brief discussion on neuromorphic wireless cognition (\textit{NeuroComm}) \cite[Fig. 2(a)]{Chen_Neuromorphic_Cognition'22}.
 \begin{figure*}[t!]
 	\centering
 	\includegraphics[scale=0.50]{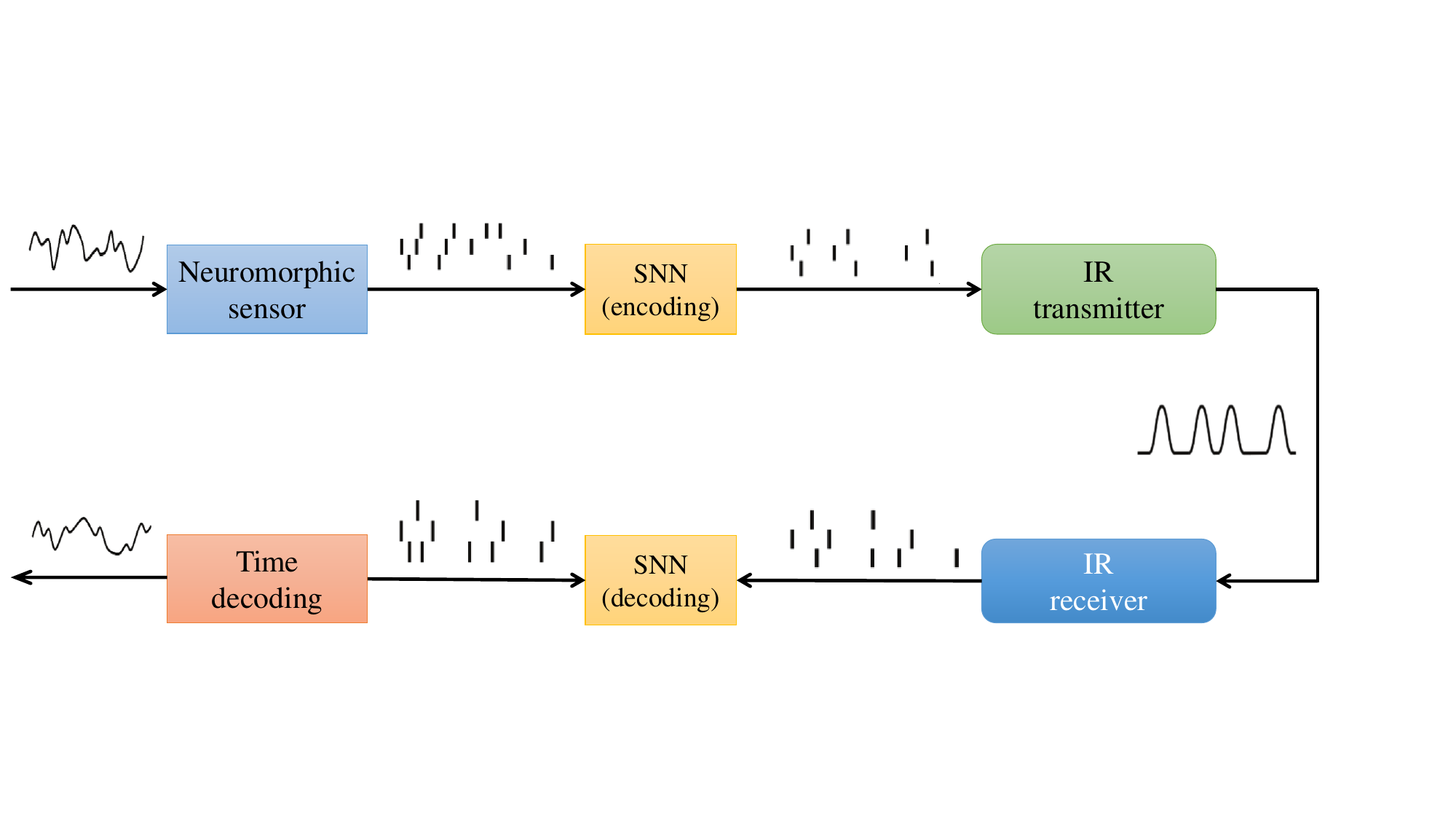}  \vspace{-1.5cm} 
 	\caption{Neuromorphic communication system (NeuroComm) \cite[Fig. 2(a)]{Chen_Neuromorphic_Cognition'22}: SNN -- spiking neural network; IR -- impulse radio.}
 	\label{fig: NeuroComm_20221115_ed}
 \end{figure*}

\subsubsection{Neuromorphic Wireless Cognition}
NeuroComm is proposed by the authors of \cite{Chen_Neuromorphic_Cognition'22} and aims to combine neuromorphic sensing, processing, and communications by introducing a wireless cognition architecture. As schematized in Fig. \ref{fig: NeuroComm_20221115_ed}, NeuroComm's transmitter consists of a neuromorphic sensor whose output is fed to an end-to-end-trained encoding spiking neural network (SNN) followed by an impulse radio (IR) transmitter (see \cite{Moe_Win_IR_TCom'00}). The IR transmitter's output is then fed to the IR receiver followed by an end-to-end-trained decoding SNN whose output is, in turn, undergoes time decoding. In view of these transceiver operations, NeuroComm's main innovations are \textit{semantic-aware energy consumption} and \textit{enhanced time-to-efficiency} \cite{Chen_Neuromorphic_Cognition'22} \cite{Chen_Neuromorphic_Cognition'22}. These goals are fulfilled by leveraging neuromorphic sensing and computing -- which are predominantly event-driven by nature -- coupled with the synergy between spike processing and pulse-based transmission through IR \cite{Moe_Win_IR_TCom'00}. Accordingly, NeuroComm reflects patterns of activity in the monitored scene because neuromorphic sensors, SNNs, and IR consume energy only when spikes are produced \cite{Chen_Neuromorphic_Cognition'22}.

We now move on to highlight a cognitive SemCom system that is driven by KG \cite[Fig. 1]{Cognitive_SemCom_Systems'22}.

\begin{figure*}[t!]
	\centering
	\includegraphics[scale=0.5]{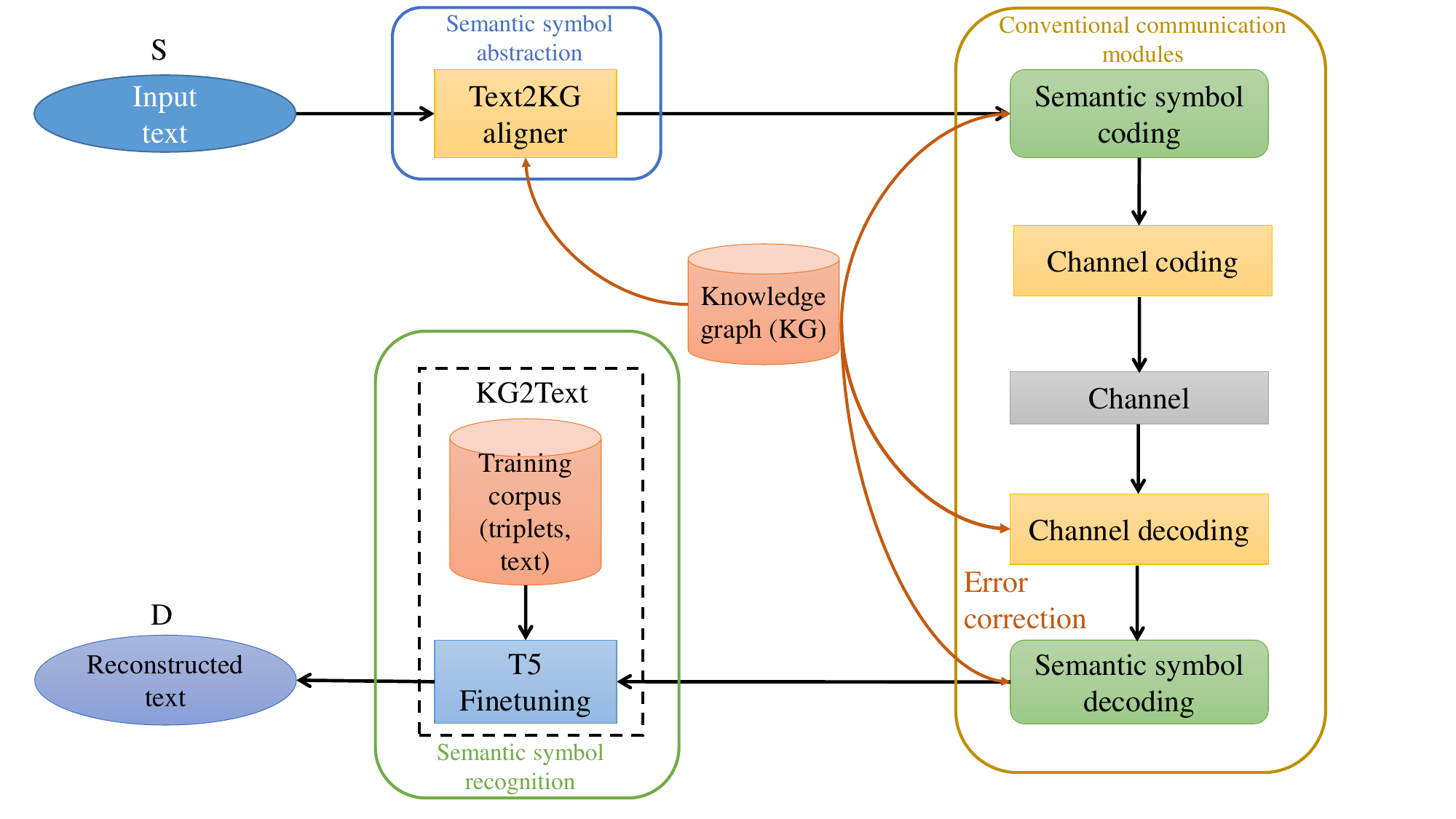}  \vspace{5mm} 
	\caption{A cognitive SemCom framework \cite[Fig. 1]{Cognitive_SemCom_Systems'22}: S -- sender; D -- destination.}
	\label{fig: Cognitive_SemCom_20221115}
\end{figure*}

\subsubsection{A Cognitive SemCom System that is Driven by KG}
the authors of \cite{Cognitive_SemCom_Systems'22} put forward a KG-based cognitive SemCom framework, which is shown in Fig. \ref{fig: Cognitive_SemCom_20221115}. As seen in Fig. \ref{fig: Cognitive_SemCom_20221115}, the proposed framework encompasses a SemCom transceiver made up of a semantic symbol abstraction module, followed by conventional communication system modules, and then a semantic symbol recognition module. For semantic symbol recognition in view of reconstructing a segment of text, the sender's text is first abstracted into semantic symbols in accordance with KG using the Text2KG aligner. The aligner's output is then fed to a semantic symbol coding module followed by a channel coding module before being transmitted over the wireless communication channel. The channel's output is received by a receiving antenna and then processed by the channel decoder, which also exploits KG to correct errors \cite{Cognitive_SemCom_Systems'22}. Once any errors have been corrected, the symbols are fed to the semantic symbol decoder, which produces an estimated semantic symbol (of the transmitted semantic symbol), which can naturally suffer from inherent semantic ambiguity. To alleviate semantic ambiguity and implement the triple-to-text conversion, the authors of \cite{Cognitive_SemCom_Systems'22} fine-tuned a pre-trained model named \textit{text-to-text transfer Transformer} (T5) on the training corpus of \cite{Agarwal_KG-Based_Corpus'20}. The fine-tuned T5 model is then deployed by the receiver to reconstruct the transmitted text. 

We now move on to our brief discussion on an implicit SemCom architecture \cite{Xiao_Reasoning_on_the_Air'22}.
 
\begin{figure*}[t!]
	\centering
	\includegraphics[scale=0.50]{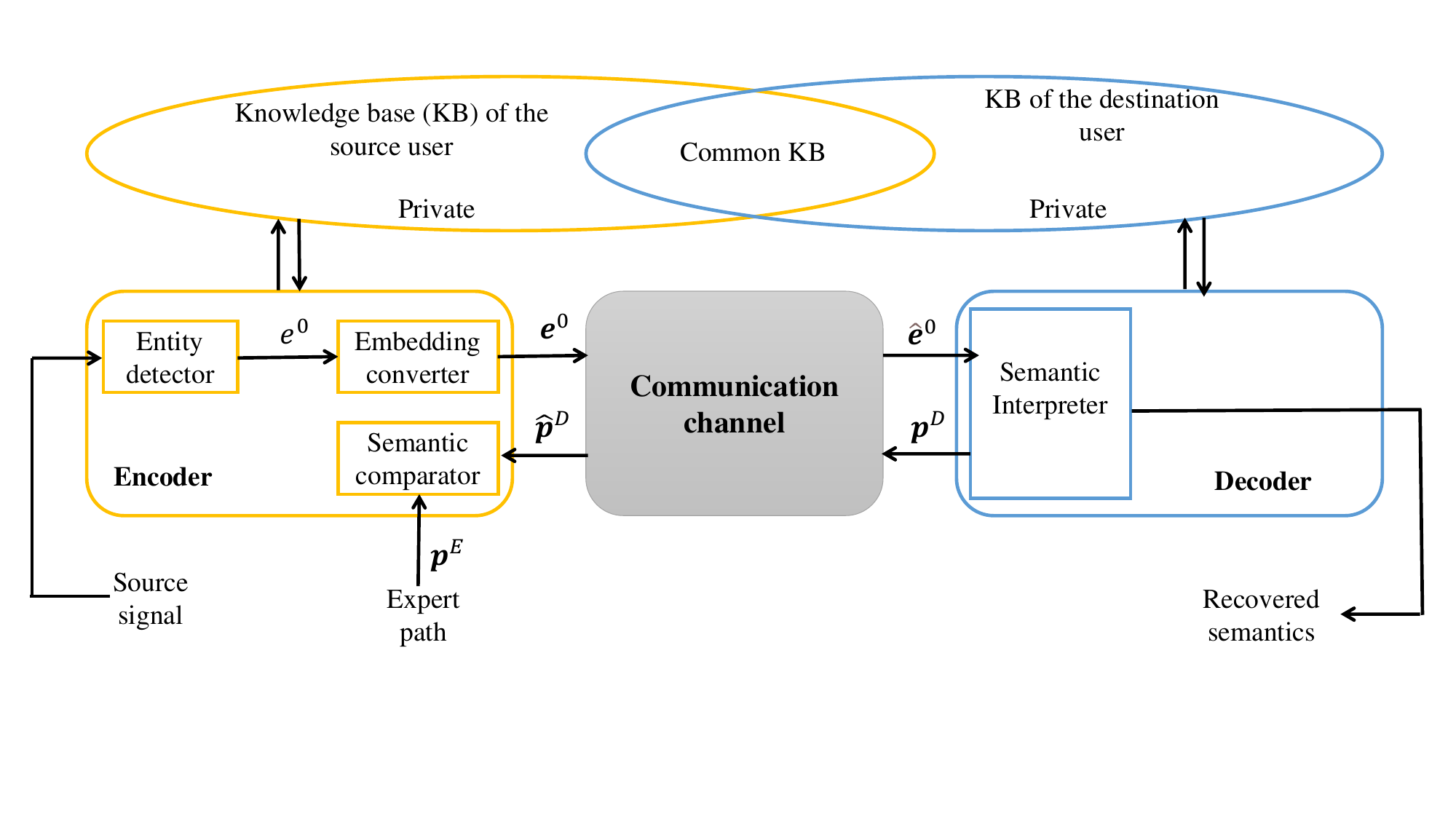}  \vspace{-1cm} 
	\caption{Implicit SemCom architecture \cite[Fig. 1]{Xiao_Reasoning_on_the_Air'22}.}
	\label{fig: iSemCom_architecture_20221117}
\end{figure*}

\subsubsection{An Implicit Semantic Communication Architecture}
the authors of \cite{Xiao_Reasoning_on_the_Air'22} introduce an implicit SemCom architecture for representing, communicating, and interpreting implicit semantic meaning, which is depicted in Fig. \ref{fig: iSemCom_architecture_20221117}. The implicit SemCom architecture is made up of a private source KB, a private destination KB, and a common KB that guides the implicit SemCom architecture's encoder and decoder. The implicit SemCom encoder is made up of an entity detector, an embedding converter, and a semantic comparator as schematized in Fig. \ref{fig: iSemCom_architecture_20221117}. As shown in Fig. \ref{fig: iSemCom_architecture_20221117}, the implicit SemCom decoder encompasses a semantic interpreter that delivers the recovered semantics of the message sent by the implicit SemCom encoder. The implicit SemCom encoder must first identify one or more key entities from the source signal using its entity detector \cite{Xiao_Reasoning_on_the_Air'22}. The entities are then processed by the embedding converter -- as shown in Fig. \ref{fig: iSemCom_architecture_20221117} -- and transmitted over the communication channel. The channel output received by the implicit SemCom decoder is then fed to the semantic interpreter. The semantic interpreter is designed to recover a reasoning path $\eta^D$ that represents its interpretation of the implicit meaning associated with the key entities of the transmitted message \cite{Xiao_Reasoning_on_the_Air'22}. If $\bm{p}^E$ and $\bm{p}^D$ are the embeddings corresponding to expert paths and those corresponding to paths generated by the decoder, respectively, as shown in Fig. \ref{fig: iSemCom_architecture_20221117}, the semantic comparator is trained to properly distinguish the semantic meaning of expert paths and that of paths generated by the decoder \cite{Xiao_Reasoning_on_the_Air'22}. Furthermore, the implicit SemCom decoder is designed to generate a reasoning path $\eta^D$ that has the shortest semantic distance from the original meaning $\eta^E$ of the source signal, and thus the underneath optimization problem \cite{Xiao_Reasoning_on_the_Air'22}:
\begin{equation}
\label{opt_problems_iSemCom_interpreter}
\min_{\theta} \Gamma_{\theta} \big( \eta^E, \eta^D \big),
\end{equation}
where $\theta$ denotes the latent parameters of the implicit SemCom decoder's semantic interpreter \cite{Xiao_Reasoning_on_the_Air'22}. In solving (\ref{opt_problems_iSemCom_interpreter}), the authors of \cite{Xiao_Reasoning_on_the_Air'22} provide a generative adversarial imitation learning-based reasoning mechanism learning (GAML) algorithm \cite[Algorithm 1]{Xiao_Reasoning_on_the_Air'22}.
 
We now discuss a SemCom technique named innovative SemCom \cite{Dong_Innovative_SemCom'22}. 

 \begin{figure*}[t!]
  	\centering
  	\includegraphics[scale=0.50]{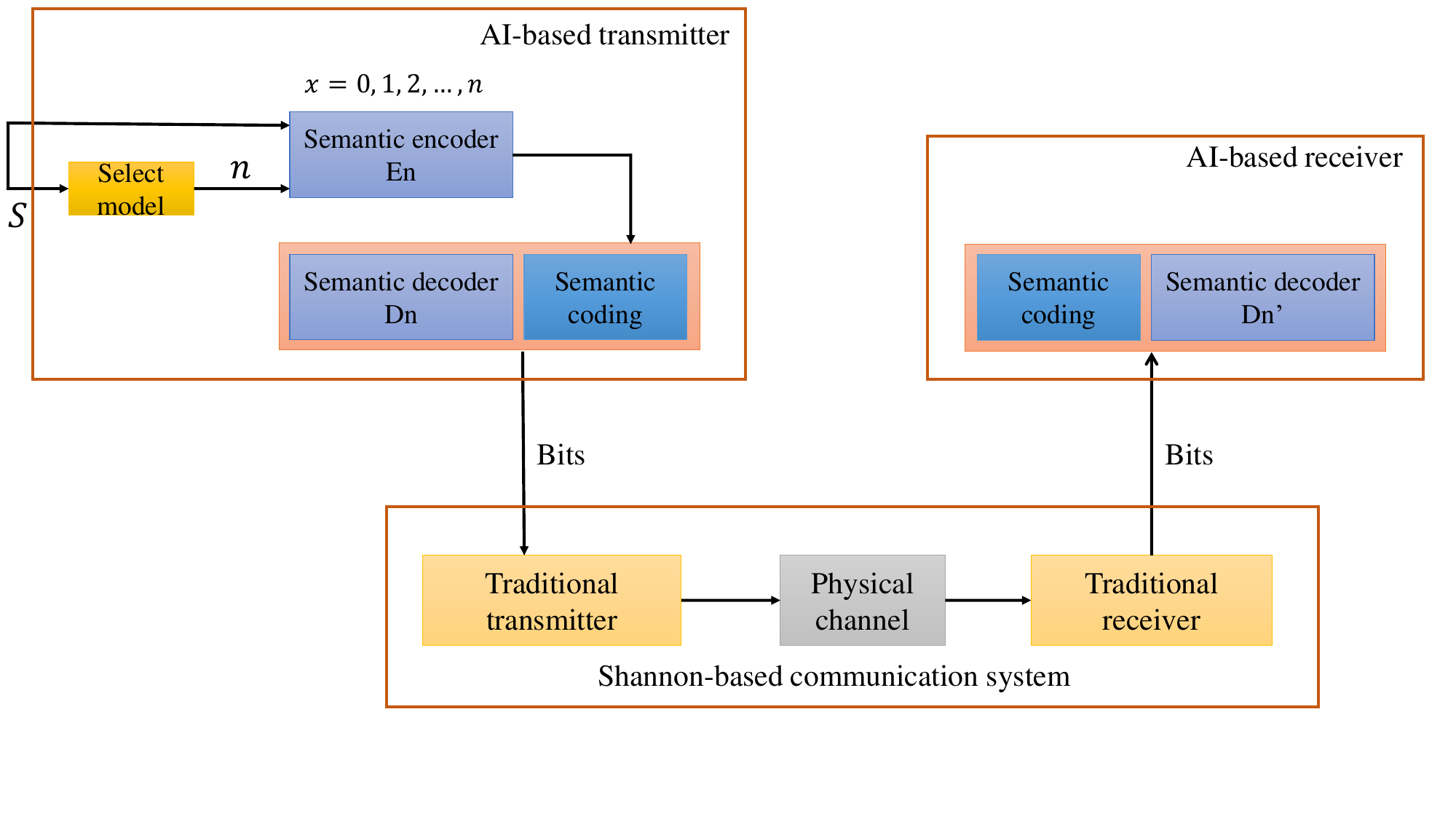}  \vspace{-0.5cm} 
  	\caption{Innovative SemCom system \cite[Fig. 1]{Dong_Innovative_SemCom'22}.}
  	\label{fig: Innovative_SemCom_20221116}
 \end{figure*}

\begin{figure*}[t!]
	\centering
	\includegraphics[scale=0.50]{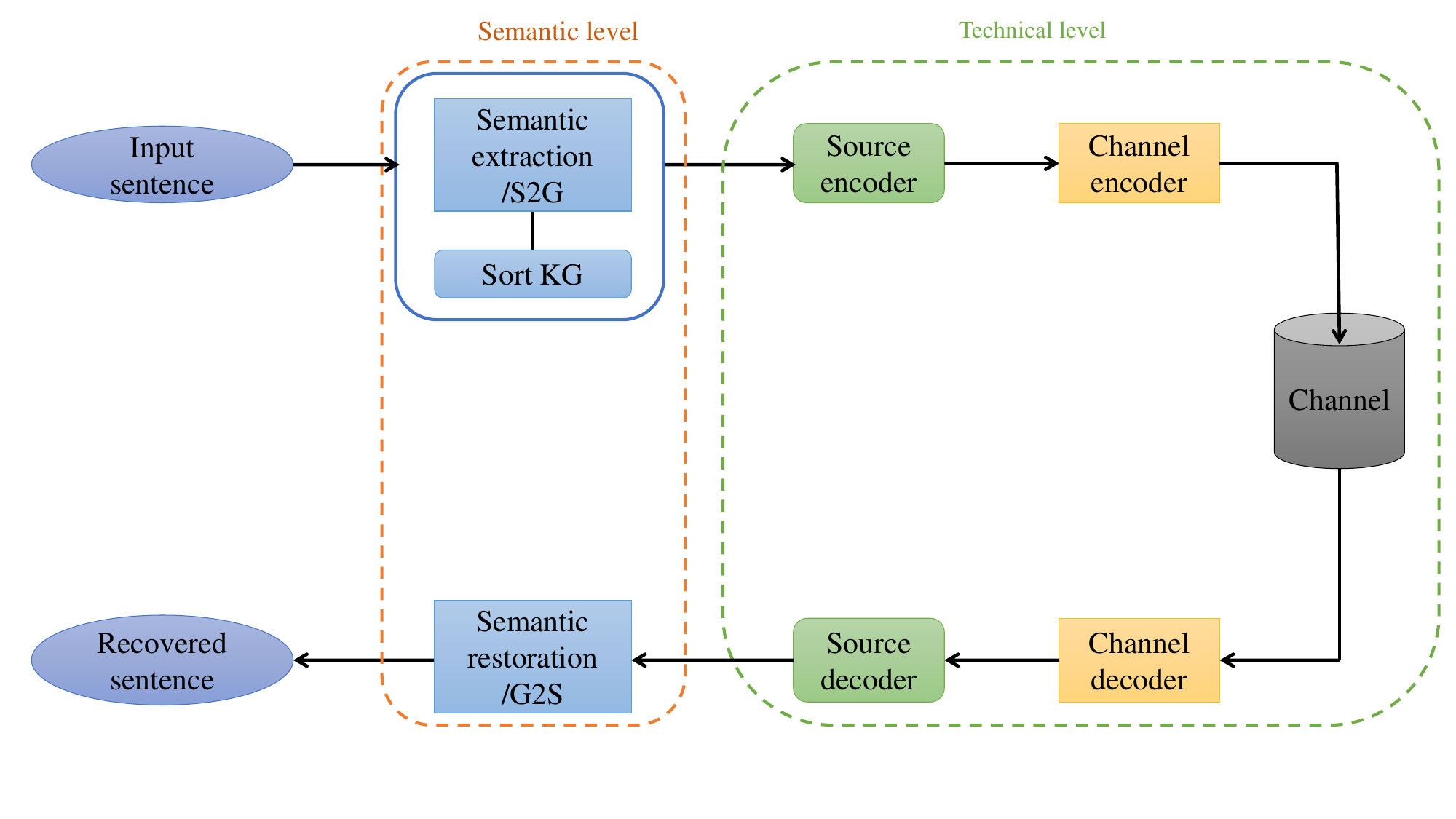}  \vspace{-0.05cm} 
	\caption{A reliable text SemCom system \cite[Figure 1]{Jiang_Reliable_SemCom'22}: S2G -- a mapping from a sentence to knowledge graph (KG); G2S -- a mapping from KG to a sentence.}
	\label{fig: Reliable_SemCom_20221116}
\end{figure*}

\subsubsection{Innovative SemCom}
the authors of \cite{Dong_Innovative_SemCom'22} employ a Shannon theory-based traditional communication system to embed an AI model into the transmitter and receiver of PHY for effective SemCom. This SemCom system is named an innovative SemCom system and is depicted in Fig. \ref{fig: Innovative_SemCom_20221116}. As seen on the left side of Fig. \ref{fig: Innovative_SemCom_20221116}, the authors pre-train a semantic encoder and a semantic decoder at the AI-based transmitter instead of deploying AI models in advance at all nodes. Per the AI-based transmitter's upper-level requirement, the authors suggest transmitting the selected decoder and its respective semantic coding using a Shannon-based communication system. This system's traditional receiver then recovers the received semantic coding using the AI-based receiver's decoder \cite{Dong_Innovative_SemCom'22}.

At the AI-based transmitter shown in Fig. \ref{fig: Innovative_SemCom_20221116}, the category of the information source needs to be identified and an AI model needs to be selected in accordance with the requirements of a classification or clustering algorithm \cite{Dong_Innovative_SemCom'22}. The AI model then extracts and compresses the source semantics and packages the semantic coding with the AI model to generate a stream of bits \cite{Dong_Innovative_SemCom'22} for the Shannon-based communication system. As shown in Fig. \ref{fig: Innovative_SemCom_20221116}, the information received by the traditional receiver includes semantic information, the AI model, and environmental information \cite{Dong_Innovative_SemCom'22}. When it comes to environmental information, the spectrum environment and the electromagnetic environment can be interpreted from a Shannon-based traditional PHY \cite{Dong_Innovative_SemCom'22}. 
  
We now proceed with our discussion on a reliable text SemCom system that is enabled by KG \cite{Jiang_Reliable_SemCom'22}.

\subsubsection{Reliable SemCom System that is Enabled by KG}
the authors of \cite{Jiang_Reliable_SemCom'22} propose a reliable text SemCom system that comprises an SE module, a traditional communication architecture, and a semantic restoration module, and can be broken down as shown in Fig. \ref{fig: Reliable_SemCom_20221116} into a semantic level and a technical level \cite{Jiang_Reliable_SemCom'22}. The semantic level is introduced by the authors of \cite{Jiang_Reliable_SemCom'22}, and the technical level is essentially identical to that of the Shannon theory-based traditional communication system \cite{Jiang_Reliable_SemCom'22}. As is shown in Fig. \ref{fig: Reliable_SemCom_20221116}, the transmitter's SE module feeds this technical level and extracts the KG of the input sentence to be transmitted and sorts it in order of semantic importance to represent the input sentence's semantics \cite{Jiang_Reliable_SemCom'22}. The semantics are then processed first by the traditional source encoder and then by the traditional channel encoder before they are transmitted over a communication channel. The channel's output is then received by the receiver, whose output is processed by the traditional channel decoder followed by the traditional source decoder. The source decoder's output is then fed to the semantic restoration module, which recovers the transmitted sentence per the received KG \cite{Jiang_Reliable_SemCom'22}.

We now move on to briefly discuss an AE-based SemCom system with relay channels \cite{Luo_SemCom_with_relay'21}.

\begin{figure*}[t!]
	\centering
	\includegraphics[scale=0.50]{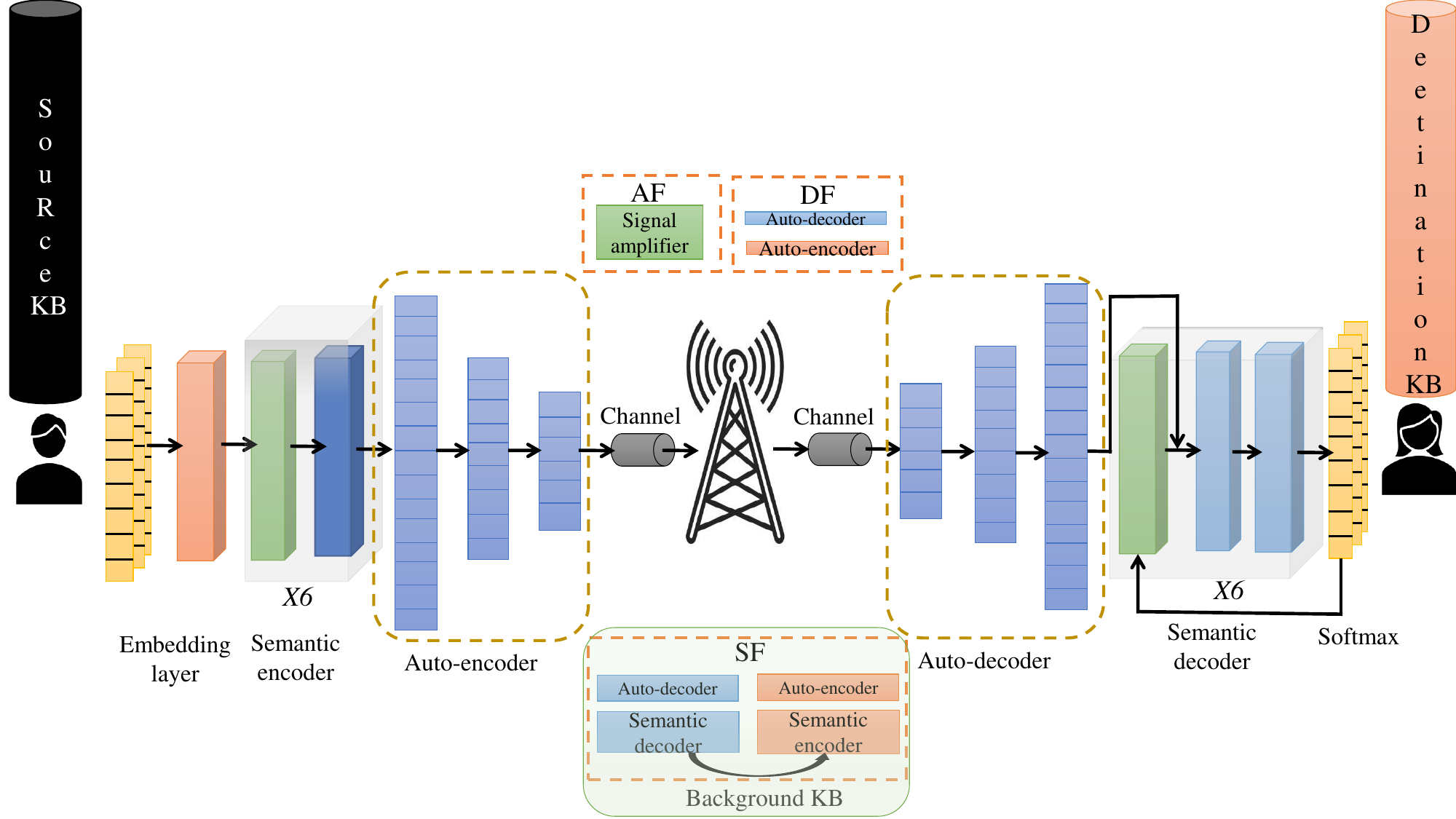}  
	\caption{An autoencoder-based SemCom system \cite[Figure 1]{Luo_SemCom_with_relay'21}: AF -- amplify-and-forward; DF -- decode-and-forward; and SF -- semantic forward.}
	\label{fig: AE-Based_SemCom_20221117}
\end{figure*}

\subsubsection{Autoencoder-Based SemCom with Relay Channels}
the authors of \cite{Luo_SemCom_with_relay'21} introduce an AE-based SemCom system with one-way relay channels, which is shown in Fig. \ref{fig: AE-Based_SemCom_20221117}, to enable SemCom between a source node and a destination node that have no common KB. As can be seen in Fig. \ref{fig: AE-Based_SemCom_20221117}, the source node transmits its information to the destination node via a relay node using SemCom that incorporates both a transmission level and a semantic level \cite{Luo_SemCom_with_relay'21}. The semantic level contains a semantic encoder (a Transformer encoder) for extracting semantic information and a semantic decoder (which is made up of three sublayers) for analyzing semantic information \cite{Luo_SemCom_with_relay'21}. The transmission level, on the other hand, aims to ensure the semantic information is accurately transmitted over the wireless channel \cite{Luo_SemCom_with_relay'21}.

The channel's output is received by a one-way relay which transmits the source's information over another wireless communication channel to the destination. When the destination's background KB and the source's background KB are different, traditional relay protocols fail to convey the source's semantic information to the destination. To overcome this design challenge, the authors of \cite{Luo_SemCom_with_relay'21} introduce a relay forward protocol named SF. The SF protocol -- which is shown in Fig. \ref{fig: AE-Based_SemCom_20221117} -- consists of two consecutive steps: $i)$ the relay node executes semantic decoding based on a background KB that is shared between the source node and itself to recover the source's information from the signal it receives, and $ii)$ the relay node semantically encodes the recovered information based on another background KB that is shared between the source node and itself in a way that the destination node can decode and understand \cite{Luo_SemCom_with_relay'21}. 

We now highlight an audio SemCom technique entitled semantic-aware speech-to-text transmission with redundancy removal \cite{Han_Semantic-aware_Speech2Text_Transmission'22}.

\begin{figure*}[t!]
	\centering
	\includegraphics[scale=0.50]{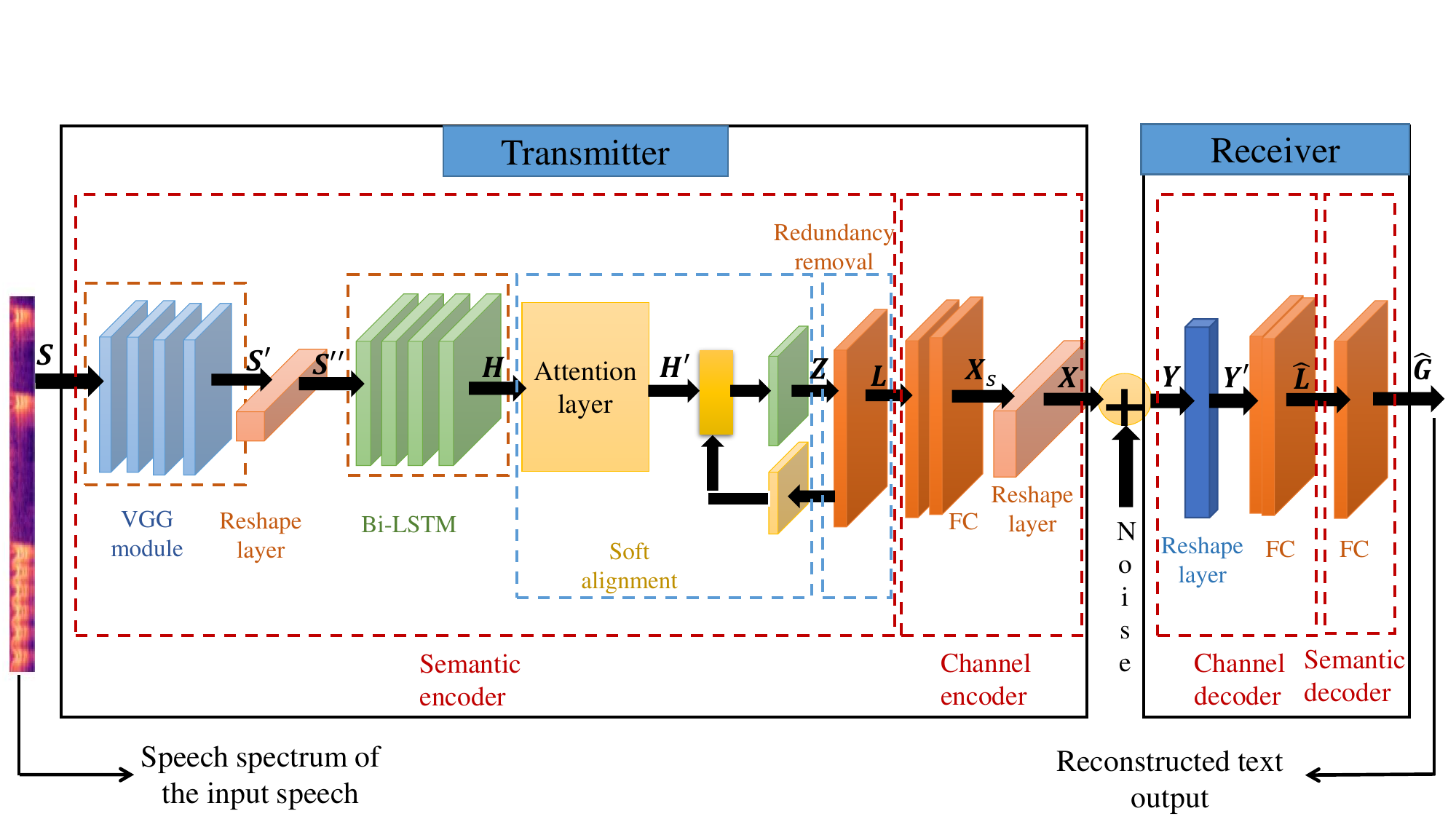}  \\ [2mm]   
	\caption{An architecture of a speech to text SemCom system with redundancy removal \cite[Fig. 1]{Han_Semantic-aware_Speech2Text_Transmission'22}: VGG -- Visual Geometry Group; Bi-LSTM -- bidirectional long short term memory; and FC -- fully connected.}
	\label{fig: Speech2Text_SemCom_20221117}
\end{figure*}

\subsubsection{Semantic-Aware Speech-to-Text Transmission with Redundancy Removal}
the speech-to-text SemCom system architecture with redundancy removal that is depicted in Fig. \ref{fig: Speech2Text_SemCom_20221117} was developed by the authors of \cite{Han_Semantic-aware_Speech2Text_Transmission'22}. As can be seen in Fig. \ref{fig: Speech2Text_SemCom_20221117}, the proposed audio SemCom transceiver consists of a receiver made up of a semantic decoder and a channel decoder as well as a transmitter composed of a channel encoder and a semantic encoder. The semantic encoder is first fed the input speech spectrum $\bm{S}$, which is obtained by applying \cite{Han_Semantic-aware_Speech2Text_Transmission'22} a 25 ms Hamming window and a 10 ms shift to the input speech signal followed by fast Fourier transforms (FFTs) to get the coefficients as well as the first- and second-order derivatives of 40 filter banks (see \cite{Furui_Cepstral_analysis'81}). This speech spectrum is then processed by the semantic decoder, with its sequence of four components: the VGG module, the Bi-LSTM module, the soft alignment module, and the redundancy removal module, as shown in Fig. \ref{fig: Speech2Text_SemCom_20221117}. The semantic decoder delivers the latent semantic representations $\bm{L}$ in which the semantic redundancy has been reduced by the redundancy removal module \cite{Han_Semantic-aware_Speech2Text_Transmission'22}. Using $\bm{L}$ as the input, the channel encoder produces a sequence of symbols $\bm{X}$ that is transmitted over the physical channel \cite{Han_Semantic-aware_Speech2Text_Transmission'22}. The channel's output is then received by the receiver, whose received signal $\bm{Y}$ is fed to the channel decoder to acquire the estimated latent semantic representation sequence $\hat{\bm{L}}$ \cite{Han_Semantic-aware_Speech2Text_Transmission'22}. $\hat{\bm{L}}$ is then inputted into the semantic decoder, as seen in Fig. \ref{fig: Speech2Text_SemCom_20221117}, which eventually decodes it and produces the predicted transcription $\hat{\bm{G}}$ \cite{Han_Semantic-aware_Speech2Text_Transmission'22}. Moreover, it is worth underscoring that the end-to-end-trained speech-to-text SemCom architecture shown in Fig. \ref{fig: Speech2Text_SemCom_20221117} outperforms DeepSC-SR (see \cite{Weng_SemCom_Speech_Recognition'21}), especially in low SNR regimes \cite[Sec. IV]{Han_Semantic-aware_Speech2Text_Transmission'22}, because it removes redundant content.

We now move on to discuss cross-modal SemCom \cite{Li_Cross-Modal_SemCom'22}.

\begin{figure*}[htb!]
	\centering
	\includegraphics[scale=0.50]{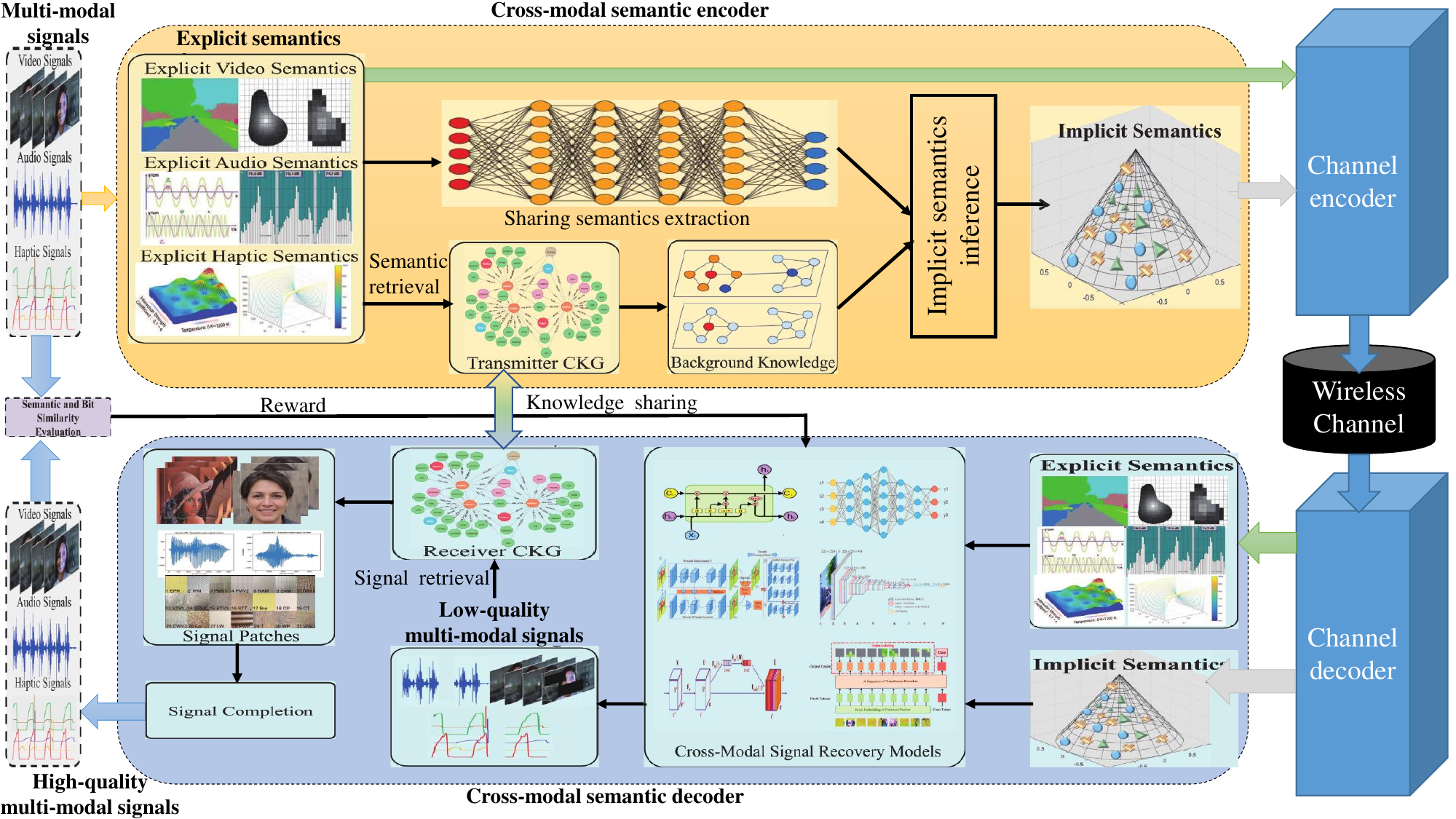}  \\ [4mm]   
	\caption{A cross-modal SemCom paradigm -- modified from \cite[Fig. 2]{Li_Cross-Modal_SemCom'22}: CKG -- cross-modal KG.}
	\label{fig: Multi-modal_SemCom_20221223.pdf}
\end{figure*}

\subsubsection{Cross-Modal SemCom}
the authors of \cite{Li_Cross-Modal_SemCom'22} propose a cross-modal SemCom system -- which is schematized in Fig. \ref{fig: Multi-modal_SemCom_20221223.pdf} -- whose purpose is to make the receiver understand what the transmitter is trying to convey and recover multi-modal source signals as precisely as possible. This SemCom paradigm focuses on both semantic- and bit-level message delivery, and is thus more universal than many state-of-the-art SemCom  systems \cite{Li_Cross-Modal_SemCom'22}. The cross-modal SemCom shown in Fig. \ref{fig: Multi-modal_SemCom_20221223.pdf} is made up of a cross-modal semantic encoder/decoder and a CKG, which are analogous to a semantic encoder/decoder and a background KB (building blocks of conventional SemCom), respectively \cite{Li_Cross-Modal_SemCom'22}. This multi-modal SemCom paradigm uses video, audio, and haptic signals as multi-modal input signals, which is also shown in Fig. \ref{fig: Multi-modal_SemCom_20221223.pdf}.

The video, audio, and haptic signals are fed to the cross-modal semantic encoder, which performs explicit semantic extraction and implicit semantic inference \cite{Li_Cross-Modal_SemCom'22}. Explicit semantic extraction leads to explicit semantics (clearly expressed / readily observable semantics) and implicit semantic inference leads to implicit semantics\footnote{Implicit semantics can reflect multi-modal signals' \textquotedblleft true meaning'' \cite{Li_Cross-Modal_SemCom'22}, which may be vital to reduce polysemy. However, a hacker can then steal implicit semantics and cause a considerable privacy problem. This possibility affirms that privacy protection can be a much more severe problem in (cross-modal) SemCom than in conventional communication and therefore needs to be given immense attention \cite{Li_Cross-Modal_SemCom'22}.} (indirectly expressed / intention-related semantics), as can be seen in Fig. \ref{fig: Multi-modal_SemCom_20221223.pdf}. The generation of implicit semantics is facilitated by the transmitter CKG that the authors of \cite{Li_Cross-Modal_SemCom'22} propose to construct using a sequence of \textit{multi-modal knowledge extraction},  \textit{cross-modal knowledge fusion}, and \textit{information storage and application} \cite[Fig. 3]{Li_Cross-Modal_SemCom'22}. Meanwhile, the explicit and implicit semantics that are generated are fed to the channel decoder, whose output is transmitted over a wireless channel to a cross-modal SemCom receiver, as shown in Fig. \ref{fig: Multi-modal_SemCom_20221223.pdf}.

At the cross-modal SemCom receiver, the received signal is fed to the channel decoder, whose output of recovered explicit and implicit semantics are inputted to the cross-modal semantic decoder. The decoder's cross-modal signal recovery module transforms the received explicit and implicit semantics into multi-modal signals using GAN-based signal recovery models \cite{Li_Cross-Modal_SemCom'22}. These models' output may be incomplete because of several distortions. As seen in Fig. \ref{fig: Multi-modal_SemCom_20221223.pdf}, the authors of \cite{Li_Cross-Modal_SemCom'22} propose the signal completion module to complete the missing parts by retrieving similar signal patches from the receiver CKG. Doing so considerably improves the quality and completeness of the recovered multi-modal signals \cite{Li_Cross-Modal_SemCom'22}. Nevertheless, the recovered signals can suffer from semantic ambiguity. The cross-modal semantic decoder must therefore be designed to minimize semantic ambiguity by ensuring the recovered multi-modal signals are accurate at the semantic level as well as the bit level \cite{Li_Cross-Modal_SemCom'22}. To this end, the authors of \cite{Li_Cross-Modal_SemCom'22} propose to optimize the cross-modal signal recovery models in an RL manner while using both semantic similarity and bit similarity between the input and recovered multi-modal signals as rewards \cite{Li_Cross-Modal_SemCom'22}. Finally, the authors implement the DQN algorithm to optimize the cross-modal signal recovery model whose results lead to the following conclusion in \cite{Li_Cross-Modal_SemCom'22}: the combination of semantic similarity and bit similarity is more effective in SemCom applications that require precise signal recovery \cite[Sec. II]{Li_Cross-Modal_SemCom'22}. 

We now proceed with our brief discussion of an emerging SemCom system named encrypted SemCom.

\begin{figure*}[htb!]
	\centering
	\includegraphics[scale=0.50]{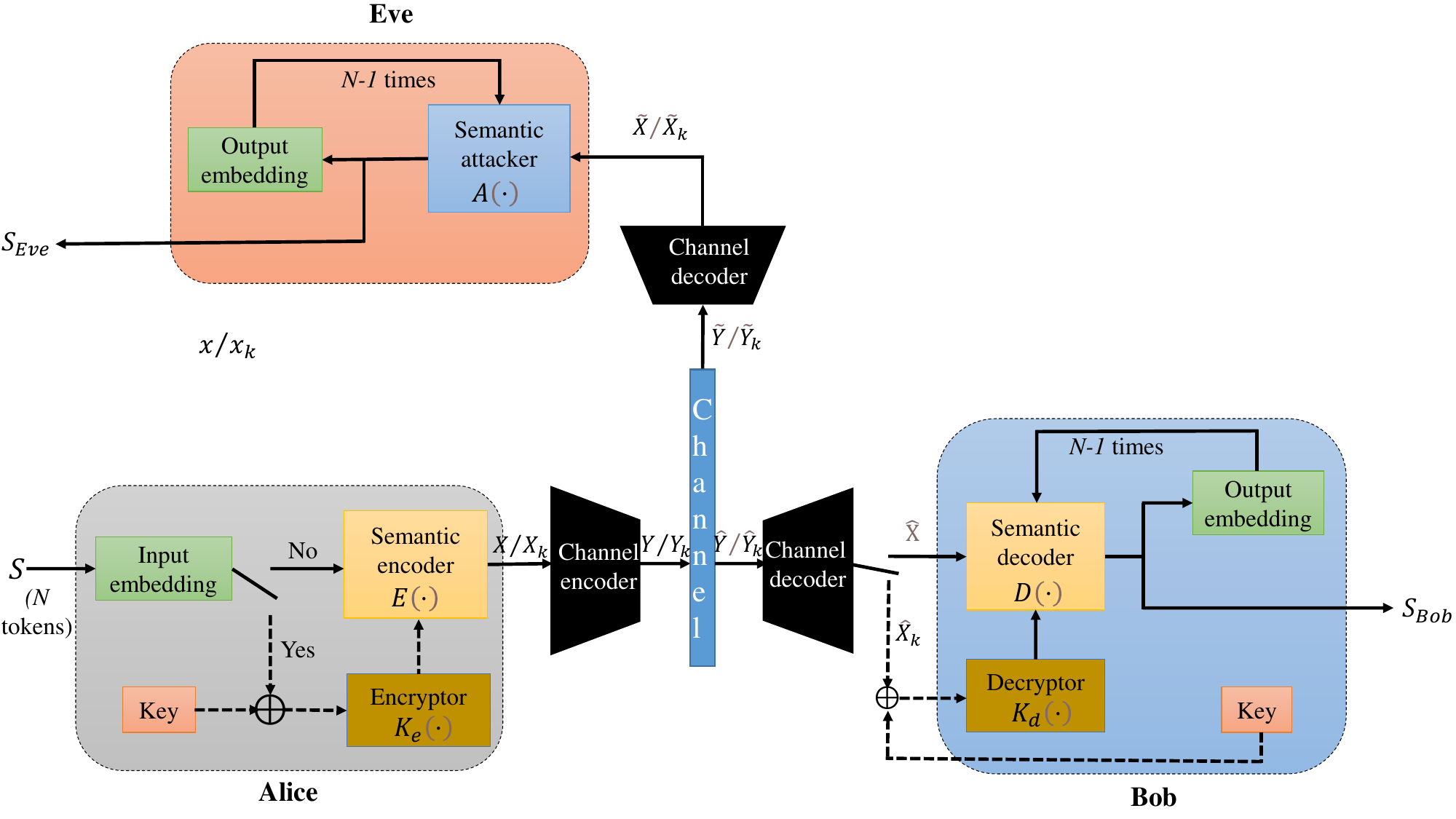}  \\ [4mm]   
	\caption{Encrypted SemCom system \cite[Fig. 1]{Luo_Encrypted_SemCom'22}.}
	\label{fig: Encrypted_SemCom_20221223.pdf}
\end{figure*}

\subsubsection{Encrypted SemCom}
many conventional SemCom systems require that background KBs be shared between the transmitter and the receiver. Many existing SemCom systems therefore assume a private communication model between two communication agents to jointly train a private semantic encoder and decoder \cite{Luo_Encrypted_SemCom'22}. In this vein, most state-of-the-art SemCom works advocate for centralized SemCom systems and unified multi-user SemCom systems that are trained based on one or more common background KBs \cite{Luo_Encrypted_SemCom'22}. This design philosophy inevitably leads to an important \textit{privacy leakage problem} \cite{Luo_Encrypted_SemCom'22}. Accordingly, balancing the generality and confidentiality of SemCom is a major challenge of SemCom design \cite{Luo_Encrypted_SemCom'22}. To alleviate this challenge, the author of \cite{Luo_Encrypted_SemCom'22} put forward an encrypted SemCom system that is schematized in Fig. \ref{fig: Encrypted_SemCom_20221223.pdf}.

The proposed SemCom system provides \textit{encrypted and unencrypted} modes of semantic transmission without needing to change the semantic encoder or semantic decoder \cite{Luo_Encrypted_SemCom'22}. When no privacy protection is required, the text SemCom system shown in Fig. \ref{fig: Encrypted_SemCom_20221223.pdf} transmits the embedded input sentence in an unencrypted manner without the need for a secret key, encryptor, or decryptor. When encryption is needed for privacy protection, the input sentence is first tokenized as a one-hot vector whose length is the size of the word dictionary in the background KB \cite{Luo_Encrypted_SemCom'22}. Each token is then mapped via the word embedding layer to a fixed-dimensional vector of floats, and the output is denoted by $\tilde{S}$ \cite{Luo_Encrypted_SemCom'22}. $\tilde{S}$ is then inputted into the encryptor $K_e(\cdot)$ along with the secret key for encryption, as schematized in Fig. \ref{fig: Encrypted_SemCom_20221223.pdf}, and the encrypted message is then fed to the semantic encoder $E(\cdot)$ for semantic encoding \cite{Luo_Encrypted_SemCom'22}. The semantic encoder encodes the semantic message (whether it is encrypted or not) and produces the semantic vector $X/X_k$ (with $X_k$ being an encrypted and semantically encoded output)\footnote{To be consistent with the notation used in Fig. \ref{fig: Encrypted_SemCom_20221223.pdf} and by the authors of \cite{Luo_Encrypted_SemCom'22}, we abuse our notation rules and represent vectors with uppercase italic letters.}  as seen in Fig. \ref{fig: Encrypted_SemCom_20221223.pdf}. Vector $X/X_k$ is then fed to the channel encoder, whose output is transmitted over the wireless channel to Bob's receiver (the legitimate receiver with the secret key).

Bob can then decrypt the received encrypted message by first running it through the decryptor $K_d(\cdot)$, as seen in Fig. \ref{fig: Encrypted_SemCom_20221223.pdf}, and then decoding it using the semantic decoder $D(\cdot)$. It should be noted that an attacker like Eve (see Fig. \ref{fig: Encrypted_SemCom_20221223.pdf}) cannot recover the transmitted encrypted and semantically encoded message even if it has the same semantic decoder as Bob provided that Eve does not have the secret key or the decryptor $K_d(\cdot)$. Meanwhile, Bob's semantic decoder decodes word by word so that the first $N-1$ outputs of the output embedding layer can be used as another input for Bob's semantic decoder \cite{Luo_Encrypted_SemCom'22} -- as seen in Fig. \ref{fig: Encrypted_SemCom_20221223.pdf}. As for the encrypted SemCom system in Fig. \ref{fig: Encrypted_SemCom_20221223.pdf}, the authors of \cite{Luo_Encrypted_SemCom'22} design the structure of the secret key, encryptor, and decryptor, and use simulations to confirm that their proposed encrypted SemCom system considerably enhances a SemCom system's privacy protection ability.  

We end our detailed discussion of the major trends in SemCom with the above-discussed encrypted SemCom system. We now move on to discuss major use cases of SemCom.

\subsection{Major Use Cases of SemCom}
\label{subsec: major_use_cases_SemCom}	
{We highlight below the following major use cases of SemCom: H2H SemCom, H2M SemCom, M2M SemCom, and KG-based SemCom -- along with their respective applications. We begin with H2H SemCom.  

\subsubsection{H2H SemCom}
consistent with the semantic level of Weaver’s framework (see Fig. \ref{fig: Three_lavels_of_communication}), H2H SemCom aims to deliver accurate meanings over a channel for message exchange between two human beings \cite{Qiao_What_is_SemCom'21}.

We now move on to H2M SemCom. 
 
\subsubsection{H2M SemCom}
H2M SemCom concerns communication between a human and a machine through the interface of human and machine intelligence by involving the second and third levels of Weaver’s framework -- the semantic level and the effectiveness level \cite{Qiao_What_is_SemCom'21}. At both levels, H2M SemCom's success depends on two design elements: $i)$ a message sent by a human must be correctly interpreted by a machine to trigger the desired action (the effectiveness problem), and $ii)$ a message sent by a machine should be meaningful to the receiving human (the semantic problem) \cite{Qiao_What_is_SemCom'21}. In light of these design goals, H2M SemCom has numerous applications as diverse as: $i)$ \textit{human-machine symbiosis systems} (e.g., AI-assisted systems, interactive ML, worker-AI collaboration); $ii)$ \textit{recommendation systems} (e.g., social network applications such as emotional health monitoring, travel recommendations for mobile tourists, remote healthcare, TV channel recommendations, video and music recommendations, UAV-assisted recommendations for location-based social networks, and distributed recommendations for privacy preservation); $iii)$ \textit{human sensing and care} (e.g., elderly monitoring, a super soldier system, general human activity recognition systems, remote healthcare systems, and smart-home monitoring systems); $iv)$ \textit{virtual reality (VR) / augmented reality (AR) techniques and applications}; $v)$ \textit{latent semantic analysis}; $vi)$ \textit{computation offloading for edge computing}; and $vii)$ \textit{decentralization for privacy preservation} \cite{Qiao_What_is_SemCom'21}. For more details about these applications, the reader is referred to \cite[Section 3]{Qiao_What_is_SemCom'21}. 

We now continue with major use cases of M2M SemCom.

\subsubsection{M2M SemCom}
M2M SemCom deals with the connection and coordination of multiple machines without human involvement to carry out a computing task \cite{Qiao_What_is_SemCom'21}. Carrying out a computing task, consequently, is more of an effectiveness problem (level three communication) than a semantic problem (level two communication) \cite{Qiao_What_is_SemCom'21}. In view of the effectiveness problem, M2M SemCom has myriad applications as varied as: $i)$ \textit{distributed learning} (e.g., effectiveness encoding, local gradient computation, over-the-air computing, over-the-air FL, importance-aware radio resource management, differential privacy); $ii)$ \textit{split inference} (e.g., feature extraction, importance-aware quantization and radio resource management, and effectiveness encoding and transmission for SplitNet); $iii)$ \textit{distributed consensus} (e.g., vehicle platooning, blockchain, local-state estimation and prediction, semantic difference transactions, and practical Byzantine fault tolerance consensus); $iv)$ \textit{machine-vision cameras} (e.g., camera-side feature extraction, effectiveness encoding based on regions of interest, surveillance, production-line inspection, and aerial and space sensing) \cite{Qiao_What_is_SemCom'21}. For more details about these applications, the reader is referred to \cite[Section 4]{Qiao_What_is_SemCom'21}. In the meantime, some use cases of M2M SemCom are described below.
\begin{itemize}
	
	\item \textit{IoT networks}: since SemCom consumes few radio resources and is relatively robust to channel noise, it is promising for accurate and instant wireless transmission in IoT networks \cite{Luo_SemCom_Overview'22,Xie_Lite_distributed_SemCom'21}. Nonetheless, the main challenge of deploying SemCom for IoT networks stems from the limited computation and storage capabilities of IoT devices, which makes the on-board use of complex DNNs unfeasible \cite{Luo_SemCom_Overview'22,Xie_Lite_distributed_SemCom'21}. As a result, determining how to optimally train and fine-tune an IoT device's semantic encoder (or decoder) and channel encoder (or decoder) in an IoT device is a major challenge.
	 
	\item \textit{Connected autonomous vehicles (connected AVs)}: in networks of connected AVs, which have multiple on-board sensors, tens or even thousands of gigabytes of data are generated per day of videos and images containing traffic information \cite{Luo_SemCom_Overview'22}. Most of the data are processed at the AVs, and the remainder is uploaded to roadside units (RSUs) and cloud/edge servers, which leads to considerable uploading latency \cite{Luo_SemCom_Overview'22}. SemCom is very promising since it transmits only semantically relevant information by design and is relatively robust against channel noise and interference \cite{Luo_SemCom_Overview'22}. Such robustness, consequently, can make SemCom promising for the design and realization of interference-resistant 6G wireless communication \cite{arXiv_Getu_DeepSC_Performance_Limits'23}.
	
	\item \textit{Device-to-device (D2D) vehicular communication}: vehicles employing D2D-based vehicular communication share radio resources with cellular users in an underlay fashion, which can lead to possibly severe co-channel interference \cite{Luo_SemCom_Overview'22}. SemCom can be used to minimize this interference by exploiting the diversity of KBs to understand the transmitted messages' meaning \cite{Luo_SemCom_Overview'22}.
	
	\item \textit{Smart factories}: in futuristic smart factories with real-time control and monitoring, the semantic features of the factories' monitored information -- such as machines' status, the temperature, and the humidity -- can be extracted and uploaded to a central controller or a cloud/edge server in order to analyze the status of materials and the quality of products \cite{Luo_SemCom_Overview'22}.
	
	\item \textit{Video communication}: the latest video coding standards, such as H.266/VVC and AV1 \cite{Devs_in_Video_Coding'21}, reportedly improve coding efficiency by 30\%-50\% \cite{Shi_to_Semantic_Fidelity'21}. However, achieving this level of improvement for high-fidelity video communication over a wireless channel with ultra-low bandwidth will be next to impossible \cite{Shi_to_Semantic_Fidelity'21}. SemCom helps to overcome this challenge by shedding light on achieving high-quality video communication over a wireless channel with low bandwidth via semantic representation and a powerful semantic library \cite{Shi_to_Semantic_Fidelity'21}.
	
	\item \textit{Holographic stereo video communication}: holographic stereoscopic video represents information with 5D data encompassing all the human senses (visual, auditory, tactile, smell, and	taste) and has the potential to deliver a truly immersive remote interaction experience \cite{Shi_to_Semantic_Fidelity'21}. However, holographic communication using multiple-view cameras requires data rates in the terabits/second \cite{ECSSCD_6G_19}. Edge intelligence \cite{Dustdar_Edge_AI'20,Dustdar_Edge_AI_Convergence'19,Letaief_Edge_AI_Vision'22,Zhou_Edge_AI'19} can be employed to alleviate this ultra-high data rate requirement by transmitting/recovering only the parts of the scene that users are interested in \cite{Shi_to_Semantic_Fidelity'21}. Nevertheless, this requires an accurate prediction of users’ behavior \cite{Shi_to_Semantic_Fidelity'21}, which is often difficult to obtain in real-time, and the viable solution is therefore to transmit semantic information using a powerful semantic library \cite{Shi_to_Semantic_Fidelity'21}. Accordingly, SemCom is a potential enabler of holographic stereoscopic video communication by reducing the volume of data to be transmitted so that the user experience is greatly enhanced. 
\end{itemize}

We now move on to some use cases of KG-based SemCom.   

\subsubsection{KG-based SemCom}
KG can be employed to realize faithful M2M SemCom, H2M SemCom, and H2H SemCom. For H2H SemCom, a KG symbolizing knowledge about the domains of the conversing parties can be injected into a semantic encoder to boost SemCom efficiency and robustness \cite{Qiao_What_is_SemCom'21}. For H2M SemCom, a KG helps a machine to understand the semantic information and its context embedded in the messages it receives from humans and to respond intelligently \cite{Qiao_What_is_SemCom'21}. For M2M SemCom, KGs can provide a platform for developing large-scale IoT networks such as logistics networks, smart cities, and vehicular networks \cite{Qiao_What_is_SemCom'21}. KGs can also act as a SemCom management tool to facilitate service selection, resource allocation, and work flow recommendation \cite{Qiao_What_is_SemCom'21}. Accordingly, KG-based SemCom  has the potential to have many applications.

KG-based SemCom is generally useful for enhancing AI applications such as frequently asked questions, virtual assistants, dialogue, and recommendation systems \cite{Qiao_What_is_SemCom'21}. KG-based M2M SemCom specifically is applicable for KG construction and updating KG-based network management, and interpretation for cross-domain applications \cite{Qiao_What_is_SemCom'21}. For much more details about these applications, the reader is referred to \cite[Section 5]{Qiao_What_is_SemCom'21}. 

We now proceed with state-of-the-art theories of SemCom.   

\section{Theories of SemCom}
\label{sec: theories_of_SemCom}
Several theories using different approaches have been developed to incorporate semantics into Shannon’s communication theory \cite{Shannon_Math_Theory_of_Communication'1948,Shannon_Weaver_Math_Theory_Commun'49}. Some of these approaches include probabilistic logic, complexity theory, semantic coding and communication games, and rate distortion theory \cite{Liu_Indirect_Rate_Distortion_Characterization'22,Liu_ISIT_Rate-Distortion_Framework'21,Xiao_RD_Theory_Strategic_SemCom'22,Stavrou_Fidelity_goal_priented_SemCom'22,Stavrou_Rate_distortion_approach_SemCom'22}. We discuss an information-theoretic approach to SemCom. More specifically, we discuss and put into context the latest crucial developments in SemCom theory by deploying the information-theoretic concepts of entropy, relative entropy, and mutual information that are detailed in Appendix \ref{Info_theory_basics}. We start with our discussion of recent SemCom theory \cite{Basu_Perserving_QoI'12,Bao_Towards_Theory_SemCom'11,Bao_Towards_Theory_of_SemCom'11} using the logical probability of messages. 

\begin{figure*}[t!]
	\centering
	\hspace{-0.20cm}\includegraphics[scale=0.48]{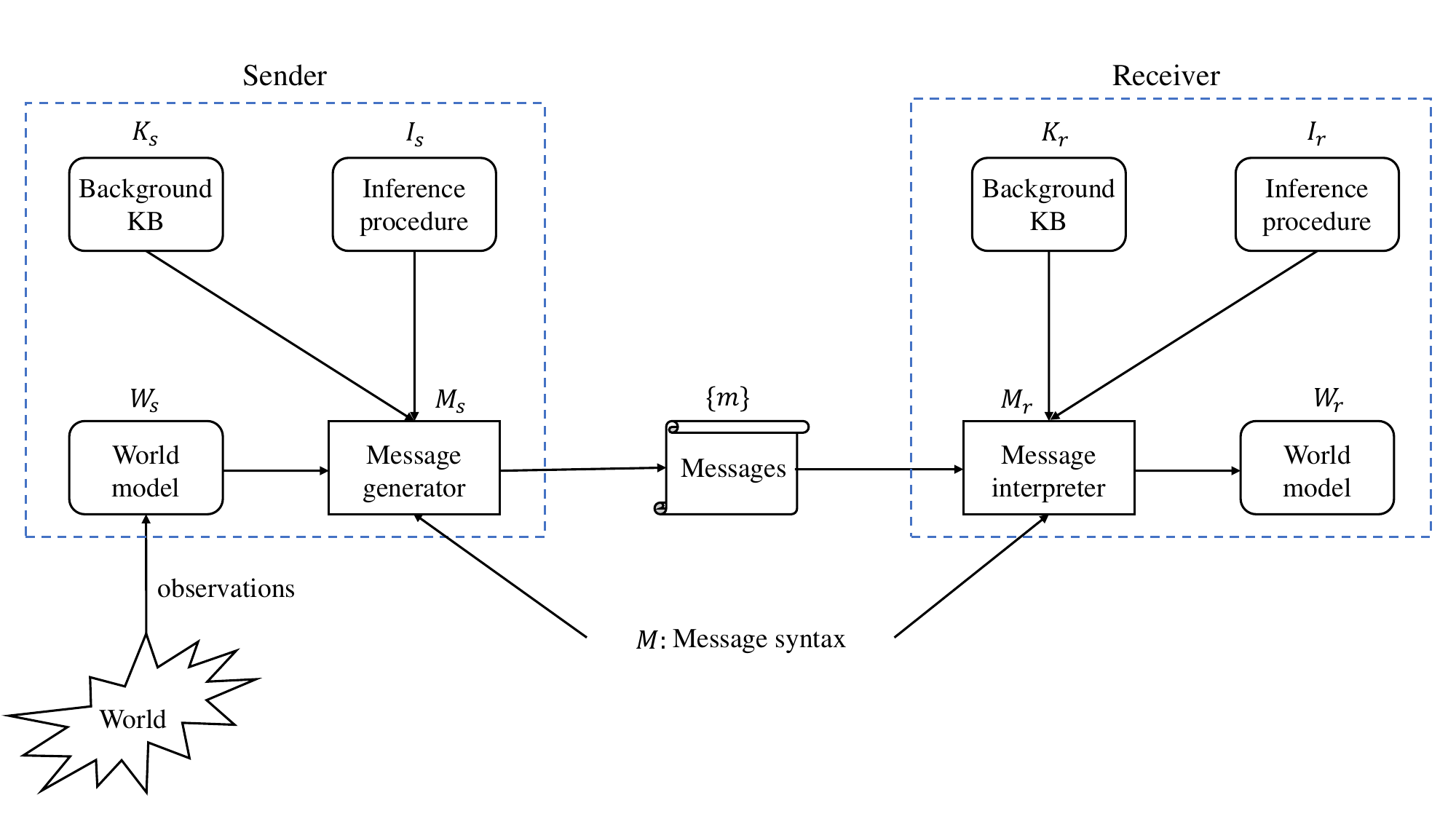} \vspace{-0.05cm}  
	\caption{A SemCom (system) model that comprises semantic information source and destination -- modified from \cite[Fig. 2]{Bao_Towards_Theory_SemCom'11}: KB -- knowledge base.}
	\label{fig: SemCom_with_world_model.pdf}
\end{figure*}

\subsection{SemCom Theory using the Logical Probability of Messages}
\label{subsec: SemCom_with_ProbLog}
By exploiting the logical probability of messages, which is a classical SIT developed by Carnap and Bar-Hillel \cite{Tech_Report_Theory_of_Sem_Info'52,Bar_Carnap_Theory_SemInfo'1954}, the authors of \cite{Basu_Perserving_QoI'12,Bao_Towards_Theory_SemCom'11,Bao_Towards_Theory_of_SemCom'11} recently developed a theory of SemCom. This theory relies on the system model of SemCom that is schematized in Fig. \ref{fig: SemCom_with_world_model.pdf}. Fig. \ref{fig: SemCom_with_world_model.pdf} shows a SemCom system that comprises a semantic information source (or semantic sender/transmitter) and a semantic information destination (receiver). A semantic information source is a tuple $(W_s,K_s, I_s, M_s)$, where $W_s$ is the model of the worlds that are possibly observable by the source, $K_s$ is the source's background KB, $I_s$ is the inference procedure employed by the source, and $M_s$ is the source message generator that is used to encode the source's message \cite{Bao_Towards_Theory_SemCom'11}.

Similar to the source, a semantic information destination (or semantic receiver) -- as seen in Fig. \ref{fig: SemCom_with_world_model.pdf} -- is a tuple $(W_r,K_r, I_r,M_r)$, where \cite{Bao_Towards_Theory_SemCom'11} $W_r$ is the world model of the destination, $K_r$ is the destination's background KB, $I_r$ is the inference procedure deployed by the destination, and $M_r$ is the destination's message interpreter (semantic decoder). In view of the tuples $(W_r,K_r, I_r,M_r)$ and $(W_s,K_s, I_s, M_s)$, a SemCom error happens if the message to be conveyed is \textquotedblleft true'' at the sender (w.r.t. $W_s$, $K_s$, and $I_s$), but the received message is \textquotedblleft false'' at the receiver (w.r.t. $W_r$, $K_r$, and $I_r$) \cite{Bao_Towards_Theory_of_SemCom'11}. Errors generally occur because of source coding losses, channel noise, semantic noise, decoding losses, or a combination thereof \cite{Bao_Towards_Theory_of_SemCom'11}. This highlights the challenge associated with designing a SemCom system -- versus a conventional communication system -- whose realization is guided by a rigorous SemCom theory. To inspire the development of such a theory and much more discussion, we hereinafter present results on semantic source coding \cite{Bao_Towards_Theory_of_SemCom'11}, semantic channel capacity \cite{Bao_Towards_Theory_of_SemCom'11}, and semantic compression \cite{Basu_Perserving_QoI'12}. We start with semantic source coding.

\subsubsection{Semantic Source Coding}
\label{subsubsec: Semantic_Source_Coding}
the design of semantic source coding deals with the design of the sender's message generator, as viewed in Fig. \ref{fig: SemCom_with_world_model.pdf}. We thus drop the subscript \textquotedblleft s'' (when there is no confusion) and start by defining the Shannon entropy of $W$, i.e., $H(W) \eqdef H(W_s)$. For a probability measure $\mu(\cdot)$, the Shannon entropy of $W$ is defined as \cite{Bao_Towards_Theory_of_SemCom'11}
\begin{equation}
\label{Shannon_entropy_of_W}
H(W) \eqdef - \sum_{ w \in \mathcal{W}}  \mu(w) \log_2 \mu(w),
\end{equation}
where $\mathcal{W}$ is the alphabet of $W$ and $\mu(\cdot)$ is a probability measure such that $\sum_{ w \in \mathcal{W}}  \mu(w)=1$. $H(W)$ is the entropy of the source provided that the source is classical with $\mathcal{W}$ as the symbol set \cite{Bao_Towards_Theory_SemCom'11}. In this case, $H(W)$ is called the \textit{model entropy} of the semantic source \cite{Bao_Towards_Theory_SemCom'11}.

In the design of a semantic encoder for a given interface language, a semantic coding strategy needs to achieve two potentially conflicting goals: $i)$ maximize the expected faithfulness in symbolizing the observed worlds, and $ii)$ minimize the expected coding length (the amount of data to be transmitted \cite{Bao_Towards_Theory_SemCom'11}. Accordingly, a semantic coding strategy is a conditional probability distribution $p(X|W)$ given that $X$ is a finite set of allowed messages (messages allowed by the message generator) \cite{Bao_Towards_Theory_SemCom'11}. Meanwhile, deterministic coding is a type coding in which each $w \in \mathcal{W}$ has at most one possible coded message \cite{Bao_Towards_Theory_SemCom'11}. For a $\mu(\mathcal{W})$ and a $p(X|W)$ that are known \textit{a priori}, the distribution of the generated messages can be determined as \cite{Bao_Towards_Theory_SemCom'11}
\begin{equation}
\label{message_distribution_eqn}
p(x)= \sum_{ w \in \mathcal{W}} \mu(w) p(x|w).
\end{equation}
Using $p(x)$ as defined in (\ref{message_distribution_eqn}), the Shannon entropy of the messages in $X$ is defined as \cite{Bao_Towards_Theory_SemCom'11} 
\begin{equation}
\label{shannon_entropy_of_X_eqn}
H(X) \eqdef - \sum_{ x \in \mathcal{X}} p(x) \log_2 p(x), 
\end{equation}
where $\mathcal{X}$ is the alphabet of $X$.

Interestingly, the message entropy as defined in (\ref{shannon_entropy_of_X_eqn}) and the model entropy as defined in (\ref{Shannon_entropy_of_W}) can be related to one another. Consequently, the following theorem links the model (semantic) entropy and the message (syntactic) entropy of a source \cite{Bao_Towards_Theory_SemCom'11}.
\begin{theorem}[{\textbf{Relationship between the model entropy and the message entropy \cite[Theorem 1]{Bao_Towards_Theory_SemCom'11}}}]
\label{entropy_relationship_thm}
The message entropy $H(X )$ and the model entropy $H(W)$ that are defined in (\ref{shannon_entropy_of_X_eqn}) and (\ref{Shannon_entropy_of_W}), respectively, are related as follows: 
\begin{equation}
\label{entropy_relationship_eqn}
H(X) = H(W) + H(X|W) - H(W|X).
\end{equation}
\proof The proof is provided in Appendix \ref{proof_entropy_relationship_thm}.
\end{theorem}

Intuitively, $H(X|W)$ and $H(W|X)$ quantify the semantic redundancy of the coding and the semantic ambiguity of the coding, respectively \cite{Bao_Towards_Theory_SemCom'11}. This intuition and Theorem \ref{entropy_relationship_thm} leads to the following remark.
\begin{remark}
\label{remark_entropy_relation}
Theorem \ref{entropy_relationship_thm} affirms that message entropy can be larger or smaller than model entropy depending on whether semantic redundancy or semantic ambiguity is larger.
\end{remark}

With this remark in mind, we now move on to our brief discussion on semantic channel capacity. 

\subsubsection{Semantic Channel Capacity}
\label{subsubsec: Semantic_Channel_Capacity}
to formally present the semantic channel capacity theorem derived by the authors of \cite{Bao_Towards_Theory_SemCom'11,Bao_Towards_Theory_of_SemCom'11}, we first define the following parameters:
\begin{itemize}
	\item For a semantically coded message $X$ and its corresponding semantically decoded message $Y$, $I(X; Y) \eqdef H(X) - H(X|Y )$ is the mutual information between $X$ and $Y$. $I(X; Y )$ represents syntactical channel ambiguity due to non-literal semantic transmission, technical noise, or semantic noise \cite{Bao_Towards_Theory_SemCom'11}.
	
	\item Given the transmitter's local knowledge $K_s$ and inference procedure $I_s$, $H_{K_s,I_s}(W|X)$ is the uncertainty of the semantic encoder \cite{Bao_Towards_Theory_SemCom'11}. A larger $H_{K_s,I_s}(W|X)$ implies larger semantic ambiguity (in semantic coding) \cite{Bao_Towards_Theory_SemCom'11}.
		
	\item Given the receiver’s local knowledge $K_r$ and inference procedure $I_r$, $\overline{H_{s;K_r,I_r}(Y) } = -\sum_{y \in \mathcal{Y} } p(y) H_s(y)$\footnote{For $m(y)$ being the logical probability of a message (sentence) $y$ as defined by Carnap and Bar-Hillel \cite{Tech_Report_Theory_of_Sem_Info'52,Bar_Carnap_Theory_SemInfo'1954}, $H_s(y)$ is the semantic entropy of $y$ and $H_s(y) \eqdef -\log_2 (m(y))$.} (for $\mathcal{Y}$ being the alphabet of $Y$) is the average logical information of the received messages. The bigger $\overline{H_{s;K_r,I_r}(Y) }$ is, the better the receiver is able to interpret the received messages \cite{Bao_Towards_Theory_SemCom'11}.
\end{itemize}

Under the assumptions $K_s = K_r$ and $I_s = I_r$ (which is less realistic), the following theorem (with no subscripts) holds.
\begin{theorem}[{\textbf{Semantic channel coding theorem \cite[Theorem 3]{Bao_Towards_Theory_SemCom'11}}}]
\label{Semantic_channel_coding_theorem}
For a discrete memoryless channel, the semantic channel capacity $C_s$ given by
\begin{equation}
\label{semantic_C_s_def}
C_s = \sup _{p(X|W)}  \big\{ I(X; Y) - H(W|X)  + \overline{H_s(Y)}  \big\}
\end{equation}
has the following property: there exists a block coding strategy for any $\epsilon > 0$ and $R < C_s$ such that the maximum probability of semantic error is less than $\epsilon$.
\proof The proof is in \cite[Appendix]{Bao_Towards_Theory_of_SemCom'11}.
\end{theorem}
In (\ref{semantic_C_s_def}), the argument $\sup$ is the semantic coding strategy and $\sup I(X; Y )$ is the engineering channel capacity \cite{Bao_Towards_Theory_SemCom'11}. The engineering channel capacity can be larger or smaller than the semantic channel capacity -- as asserted by Theorem \ref{Semantic_channel_coding_theorem} -- depending on whether $H(W|X)$  or $\overline{H_s(Y)}$ is larger.

Despite the less realistic underlying assumption, Theorem \ref{Semantic_channel_coding_theorem} is informative and may signify the best-case semantic capacity scenario, as both the background KB and inference procedure of the sender and receiver are assumed to be the same. This leads us to our discussion on the important result of semantic compression.

\subsubsection{Semantic Compression}
\label{subsubsec: Semantic_Compression}
semantic compression is carried out by the semantic encoder, which attempts to \textit{efficiently encode} only the semantically relevant information of the source's message. To this end, a fundamental question of semantic compression is: \textquotedblleft To what extent is semantic compression possible?'' \cite{Bao_Towards_Theory_SemCom'11}. Answering this question in part, the authors of \cite{Basu_Perserving_QoI'12} provide an informative theorem \cite[Theorem 1]{Basu_Perserving_QoI'12}. Before presenting this theorem, the following definitions from \cite{Basu_Perserving_QoI'12} are in order.
\begin{definition}[\textbf{\cite[Definition 1]{Basu_Perserving_QoI'12}}]
\label{Shannon_info_source_defn}
A (statistical/syntactic) source is a process that stochastically generates symbols from some alphabet $\Delta_{\tilde{X}}$. It is represented by an RV $\mathcal{\tilde{X}} $ that is drawn from $\mathcal{\tilde{X}} \eqdef \{\tilde{x}_1, \tilde{x}_2, \ldots, \tilde{x}_n\}$ with a probability mass function (PMF) $p(\cdot)$. 
\end{definition}
\begin{definition}[\textbf{\cite[Definition 2]{Basu_Perserving_QoI'12}}]
\label{Shannon_entropy_defn}
The Shannon entropy of an RV $\tilde{X}$ is defined as \cite[eq. (1)]{Basu_Perserving_QoI'12}
\begin{equation}
\label{Shannon_entropy_eqn}
H(\tilde{X})  \eqdef - \sum_{\tilde{x} \in \mathcal{\tilde{X}}} p(\tilde{x}) \log_2 p(\tilde{x}).
\end{equation}
\end{definition}
\begin{definition}[\textbf{\cite[Definition 5]{Basu_Perserving_QoI'12}}]
\label{semantic_information_source}
Stochastically generating messages with related meanings, a semantic information source $S$ is symbolized by a tuple $(\tilde{M}, \tilde{X}, P, L)$, where $L$ is the formal language, $\tilde{X}$ is an RV drawn from $\mathcal{\tilde{X}}$ and each instance of $\tilde{X}$ is an expression in $L$, $\tilde{M}$ is a RV that takes values from $\mathcal{\tilde{M}}$\footnote{Both $\mathcal{\tilde{X}}$ and $\mathcal{\tilde{M}}$ may be countably infinite \cite{Basu_Perserving_QoI'12}.} and each instance of $\tilde{M}$ is an interpretation of $L$, and $P$ is the joint distribution of $(\tilde{M}, \tilde{X})$.   
\end{definition}
\begin{definition}[\textbf{\cite[Definition 8]{Basu_Perserving_QoI'12}}]
\label{Logical_probability_of_messages_defn}
For $S \eqdef (\tilde{M}, \tilde{X}, P, L)$ being the semantic information source whose message is denoted by $\tilde{x} \in \tilde{\mathcal{X}}$, the logical probability of $\tilde{x}$ is denoted by $P_{\tilde{M}}(\tilde{x})$ and defined as \cite[eq. (2)]{Basu_Perserving_QoI'12}
\begin{equation}
\label{logical_probability_eqn}
P_{\tilde{M}}(\tilde{x}) \eqdef \sum_{ \tilde{m} \models \tilde{x}, \tilde{m} \in \tilde{\mathcal{M}} } P_{\tilde{M}}(\tilde{m}).
\end{equation}
\end{definition}
\begin{definition}[\textbf{\cite[Definition 9]{Basu_Perserving_QoI'12}}]
\label{(Semantic_information_of_messages_defn}
For $S \eqdef (\tilde{M}, \tilde{X}, P, L)$ being the semantic information source whose message is denoted by $\tilde{x} \in \tilde{\mathcal{X}}$, the semantic information of $\tilde{x}$ is denoted by $H_s(\tilde{x})$ and defined as \cite[eq. (3)]{Basu_Perserving_QoI'12}
\begin{equation}
\label{Semantic_information_of_x}
H_s(\tilde{x}) \eqdef - \log_2 P_{\tilde{M}}(\tilde{x}), 
\end{equation}
where $H_s(\tilde{x})$ quantifies the element of surprise in finding $\tilde{x}$ to be true \cite{Basu_Perserving_QoI'12}.
\end{definition}
\begin{definition}[\textbf{\cite[Definition 10]{Basu_Perserving_QoI'12}}]
\label{(Semantic_entropy_of_a_source_defn}
For $S \eqdef (\tilde{M}, \tilde{X}, P, L)$ being the semantic information source, its semantic entropy is denoted by $H_s(\tilde{X})$ and defined as \cite[eq. (4)]{Basu_Perserving_QoI'12} 
\begin{equation}
\label{Semantic_entropy_of_a_source_defn_eqn}
H_s(\tilde{X}) \eqdef \sum_{ \tilde{x} \in \mathcal{\tilde{X}} }  p(\tilde{x})H_s(\tilde{x}).
\end{equation}
\end{definition}
\begin{definition}[\textbf{\cite[Definition 11]{Basu_Perserving_QoI'12}}]
\label{Model_entropy_defn}
For $S \eqdef (\tilde{M}, \tilde{X}, P, L)$ being the semantic information source, its model entropy is denoted by $H(\tilde{M})$ and defined as \cite[eq. (5)]{Basu_Perserving_QoI'12}
\begin{equation}
\label{Model_entropy_defn_eqn}
H(\tilde{M}) \eqdef - \sum_{ \tilde{m} \in \mathcal{\tilde{M}} }  P_{\tilde{M}}(\tilde{m}) \log_2 P_{\tilde{M}}(\tilde{m}). 
\end{equation}
\end{definition}

In light of Definitions \ref{Shannon_info_source_defn}-\ref{Model_entropy_defn}, semantic compression can be thought of -- hypothetically -- as a \textit{transformation by a semantic channel} inside a source whose inputs are models $\tilde{M}$ and outputs are messages $\tilde{X}$ (see \cite[Fig. 1]{Basu_Perserving_QoI'12}). This transformation is characterized by the following theorem.
\begin{theorem}[\textbf{\cite[Theorem 1]{Basu_Perserving_QoI'12}}]
\label{semantic_entropy_bound_thm}
The semantic entropy of a source $S \eqdef (\tilde{M}, \tilde{X}, P, L)$ is bounded from above by the mutual information between its models $\tilde{M}$ and messages $\tilde{X}$ \cite[eq. (8)]{Basu_Perserving_QoI'12}:
\begin{equation}
\label{semantic_entropy_upper_bound_eqn}
 H_s( \tilde{X} )  \leq I(\tilde{M}; \tilde{X}), 
\end{equation}
where $I(\tilde{M}; \tilde{X}) \eqdef H(\tilde{M})-H(\tilde{M}|\tilde{X})=H(\tilde{X})-H(\tilde{X}|\tilde{M})$.
\proof The proof is in \cite[p. 195-196]{Basu_Perserving_QoI'12}.
\end{theorem}

We now continue with our discussion of indirect rate distortion characterization for semantic sources \cite{Liu_RD-Characterization_TCOM'22,Liu_Indirect_Rate_Distortion_Characterization'22}.

\begin{figure}[t!]
	%\centering
	\hspace{-0.10cm}\includegraphics[scale=0.28]{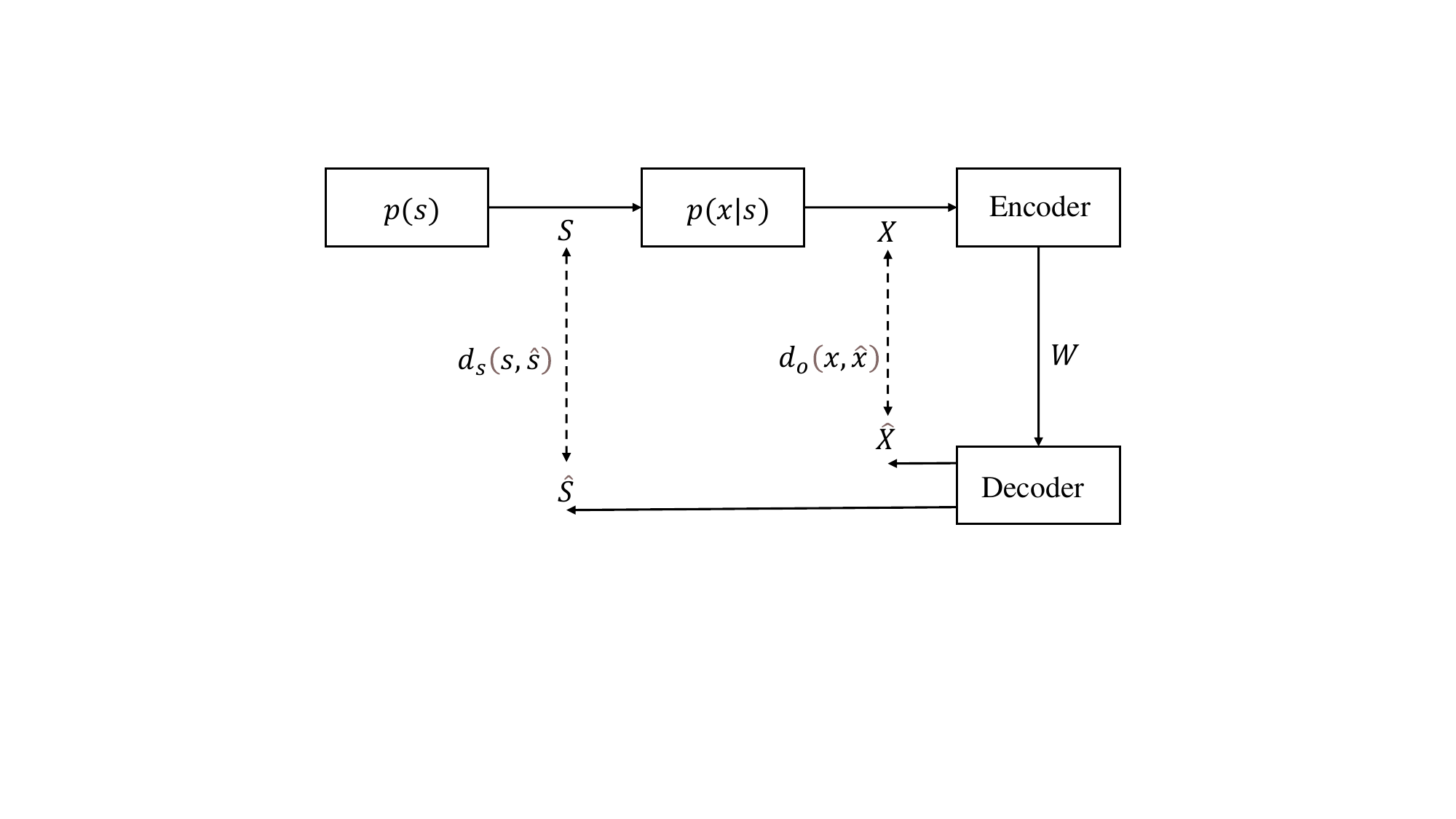} \vspace{-2.0cm}  
	\caption{An illustrating schematic of a semantic source and its loss compression \cite[Fig. 1]{Liu_RD-Characterization_TCOM'22}.}
	\label{fig: Semantic_RDT_pic.pdf}
\end{figure}

\subsection{Rate Distortion Characterization for Semantic Sources}
\label{subsec: semantic_rate_distortion}
The authors of \cite{Liu_RD-Characterization_TCOM'22} took inspiration from the classical rate distortion theory \cite[Ch. 10]{Cover_Elements_of_Info_Theory'06} and propose an indirect rate distortion theory for a source model that comprises an intrinsic state part and an extrinsic observation part. This source model, which is in Fig. \ref{fig: Semantic_RDT_pic.pdf}, is quite relevant in the context of SemCom because the intrinsic state corresponds to the semantic feature $S$ of the source, which is generally unobservable and can only be inferred from the extrinsic observation $X$ \cite{Liu_RD-Characterization_TCOM'22}. The pair of RVs $(S,X)$ that are correlated with the joint probability distribution $p(s, x)$ are employed to model this memoryless semantic source \cite{Liu_RD-Characterization_TCOM'22}, as schematized in Fig. \ref{fig: Semantic_RDT_pic.pdf}. As is also shown in Fig. \ref{fig: Semantic_RDT_pic.pdf}, the encoder has access to only a length-$n$ block of the extrinsic observation sequence $X^n$ and its output is fed to the decoder. The decoder has two major tasks, as seen in Fig. \ref{fig: Semantic_RDT_pic.pdf}: $i)$ replicate the intrinsic state block as $\hat{S}^n$ under a state distortion measure $d_s$, and $ii)$ replicate the extrinsic observation block as $\hat{X}^n$ under an observation distortion measure $d_o$ \cite{Liu_RD-Characterization_TCOM'22}. Moreover, the encoder and decoder are linked via a bit pipe in which the codeword $W$ of $nR$ bits -- with $R$ hence being the code rate -- is transferred from the encoder to the decoder \cite{Liu_RD-Characterization_TCOM'22}.

For the system setup mentioned and Fig. \ref{fig: Semantic_RDT_pic.pdf}, let $d_s : \mathcal{S} \times \hat{\mathcal{S}} \rightarrow \mathbb{R}_+$ and $d_o : \mathcal{X} \times \hat{\mathcal{X}} \rightarrow \mathbb{R}_+$ be two distortion measures that are defined over the source product alphabet $\mathcal{S} \times \mathcal{X}$ and the reproduction product alphabet $\hat{\mathcal{S}} \times \hat{\mathcal{X}}$, respectively \cite{Liu_RD-Characterization_TCOM'22}. The extended block-wise distortion measures are given by \cite[eqs. (1) and (2)]{Liu_RD-Characterization_TCOM'22}  % 
\begin{subequations}
\begin{align}
\label{d_s_measure_eqn}
d_s (s^n, \hat{s}^n)  &  =  \frac{1}{n} \sum_{i=1}^n d_s (s_i, \hat{s}_i)     \\
\label{d_o_measure_eqn}
d_0 (x^n, \hat{x}^n)   & =  \frac{1}{n} \sum_{i=1}^n  d_o (x_i, \hat{x}_i)  .
\end{align}
\end{subequations}
The authors of \cite{Liu_RD-Characterization_TCOM'22} claim that a tuple $(R, D_s, D_o)$ is achievable -- for any $\epsilon > 0$ and all sufficiently large $n$ -- if the encoding function, the state decoding function, and the observation decoding function defined in \cite[p. 5948]{Liu_RD-Characterization_TCOM'22} exist. This overall setup's goal is to characterize the region of all achievable $(R, D_s, D_o)$ tuples, and its semantic rate distortion function is defined as \cite[eq. (5)]{Liu_RD-Characterization_TCOM'22}
\begin{equation}
\label{semantic_rate_distortion_function_defn}
R(D_s, D_o) \eqdef \inf \{ R: (R, D_s, D_o) \hspace{2mm} \textnormal{is achievable}   \}.
\end{equation}
Regarding the characterization of the problem in (\ref{semantic_rate_distortion_function_defn}), the authors of \cite{Liu_RD-Characterization_TCOM'22} derive the following theorem. 
\begin{theorem}[\textbf{\cite[Theorem 1]{Liu_RD-Characterization_TCOM'22}}]
\label{achievable_semantic_rate_region_thm}
For a semantic source modeled by a pair of RVs $(S,X) \sim p(s, x)$ over the source alphabet $\mathcal{S} \times \mathcal{X}$, the reproduction alphabet $\hat{\mathcal{S}} \times \hat{\mathcal{X}}$, and distortion measures $d_s$ and $d_o$, the semantic rate distortion function $R(D_s,D_o)$ is equated as \cite[eqs. (9)-(11)]{Liu_RD-Characterization_TCOM'22}
\begin{subequations}
\begin{align}
R(D_s,D_o) \quad &= \min_{p(\hat{s}, \hat{x} | x)}  I (X; \hat{S}, \hat{X}) \\
\textnormal{s.t.} \quad & \mathbb{E} \{ d_0 (X, \hat{X}) \}  \leq D_o \\
&\mathbb{E} \{ \hat{d}_s (X, \hat{S}) \}  \leq D_s ,
\end{align}
\end{subequations}
where $S, X, \hat{S}, \hat{X}$ constitute a Markov chain $S \leftrightarrow X \leftrightarrow (\hat{S}, \hat{X})$ and \cite[eq. (12)]{Liu_RD-Characterization_TCOM'22}
\begin{equation}
\label{distortion_defn_eqn}
\hat{d}_s (x, \hat{s}) = \mathbb{E} \{ d_s(S, \hat{s}) | x\} = \sum_{s \in \mathcal{S}} p(s|x) d_s (s, \hat{s}).
\end{equation}
\proof The proof is in \cite[Appendix I]{Liu_RD-Characterization_TCOM'22}.
\end{theorem}

We now discuss the newest SemCom formulation dubbed \textit{semantic language utilization and design} \cite{Shao_Theory_of_SemCom'22}. 

\subsection{Semantic Language Utilization and Design}
\label{subsec: language_utilization_and design}
The authors of \cite{Shao_Theory_of_SemCom'22} formulate two fundamental problems related to SemCom: \textit{language design} and \textit{language utilization}.\footnote{Throughout the work in \cite{Shao_Theory_of_SemCom'22}, \textquotedblleft language'' refers to \textit{semantic language}, which is often formed commonly through interactions and is much richer in interpretation and expression: \textquotedblleft a meaning can be expressed by multiple messages and a message can be interpreted as multiple meanings'' \cite{Shao_Theory_of_SemCom'22}.} The language design problem deals with the design of common languages or codebooks between the transmitter and receiver to efficiently convey meaning and can be resolved by applying a JSCC theory. Below is the formal definition of this problem.

\begin{problem}[\textbf{Language design \cite[Problem 2]{Shao_Theory_of_SemCom'22}}]
\label{Language_design_problem}
Having presumed that both the transmitter and receiver are allowed to negotiate prior to transmission, how can the semantic language and technical language be designed to efficiently communicate the meaning of a semantic source?
\end{problem}

Unlike Problem \ref{Language_design_problem}, Weaver's SemCom vision \cite[Ch. 1]{Shannon_Weaver_Math_Theory_Commun'49} is concerned with the interpretation of meaning by the receiver as compared with the sender's intended meaning \cite{Shao_Theory_of_SemCom'22}. To this end, the only things that the transmitter and receiver agree on are the semantic and technical languages \cite{Shao_Theory_of_SemCom'22} that should be leveraged by the SemCom system designer to reduce misinterpretation by the receiver. This underscores the following language utilization problem.

\begin{problem}[\textbf{Language utilization \cite[Problem 1]{Shao_Theory_of_SemCom'22}}]
\label{Language_utilization_problem}
Once the transmitter and receiver have agreed on semantic and technical languages, the following problems arise while communicating an intended meaning: $1)$ how to minimize the receiver's misinterpretations from the transmitter’s perspective (semantic encoding)? $2)$ how to minimize the receiver's misinterpretations from the receiver’s perspective (semantic decoding)?
\begin{itemize}
	\item \textbf{Semantic encoding problem:} while minimizing the communication cost, how can the transmitter generate a message such that its intended meaning can be recovered at the receiver as accurately as possible? \cite{Shao_Theory_of_SemCom'22}
	
	\item \textbf{Semantic decoding problem:} given a received message and no prior information about the predetermined meaning of the transmitter, how can the receiver decode the intended meaning of the transmitter? \cite{Shao_Theory_of_SemCom'22}
\end{itemize}
Combing the two mentioned problems, the following \textit{combined semantic encoding and decoding} (CSED) problem ensues.
\begin{itemize}
	\item \textbf{The CSED problem:} while acting separately in their own ways, what if the transmitter and receiver simultaneously perform semantic encoding and semantic decoding, respectively?  
\end{itemize}
\end{problem}
	
The above language utilization problems are discussed below. We begin with some clarifying definitions from \cite{Shao_Theory_of_SemCom'22}.

\begin{definition}[\textbf{Words and syntax \cite[Definition 3.1]{Shao_Theory_of_SemCom'22}}]
\label{Words_and_syntax_defn}
The smallest elements of a message in a given language are words. Words can be employed on their own or together (with other words) to form a message. Syntax is a set of rules that establish the grouping of words in a message.
\end{definition}   

\begin{definition}[\textbf{The set of messages \cite[Definition 3.2]{Shao_Theory_of_SemCom'22}}]
\label{Set_of_messages_defn}
Let $\mathcal{S}$ denote the set of all possible messages that are determined by the word and syntax of a language. Assuming that $\mathcal{S}$ is finite or countably infinite, we let $\mathcal{S} \eqdef \big\{ s_m: m= 1, 2, \ldots, M \big\}$, where $s_m$ and $M$ are the $m$-th message and the number of all possible messages, respectively. 
\end{definition} 

\begin{definition}[\textbf{The set of meanings \cite[Definition 3.3]{Shao_Theory_of_SemCom'22}}]
\label{Set_of_meanings_defn}
Let the messages in $\mathcal{S}$ convey a finite or countably infinite number of meanings. The set of all possible meanings is denoted by $\mathcal{W}$ and defined as $\mathcal{W} \eqdef \big\{ w_n: n= 1, 2, \ldots, N \big\}$, where $w_n$ and $N$ represent one meaning and the number of meanings, respectively. Meanwhile, the probability of the transmitter's intended meaning being $w_n$ is denoted by $p(w_n)$.
\end{definition} 

\begin{definition}[\textbf{Expression \cite[Definition 3.4]{Shao_Theory_of_SemCom'22}}]
\label{Expression_defn}
Defining a mapping from the set of meanings to the set of messages, the expression of a language is defined as \cite[eq. (2)]{Shao_Theory_of_SemCom'22}:
\begin{equation}
\label{language_expression_eqn}
\big\{ p(s|w) \in [0,1]: w\in \mathcal{W}, s\in \mathcal{S},  \sum_{s} p(s|w)=1 \big\}.
\end{equation}
These mappings form a matrix $\bm{P} \in \mathbb{R}^{N \times M}$ such that $(\bm{P})_{n, m} \eqdef p(s_m | w_n)$ and $\sum_{m=1}^M (\bm{P})_{n, m}=1$ for all $n\in[N]$.
\end{definition} 

\begin{definition}[\textbf{Interpretation \cite[Definition 3.5]{Shao_Theory_of_SemCom'22}}]
\label{Interpretation_defn}
Defining a mapping from the set of messages to the set of meanings, the interpretation of a language is defined as \cite[eq. (3)]{Shao_Theory_of_SemCom'22}:
\begin{equation}
\label{language_interpretation_eqn}
\big\{ q(w|s) \in [0,1]: w\in \mathcal{W}, s\in \mathcal{S},  \sum_{w} q(w|s)=1 \big\}.
\end{equation}
It is also defined as a matrix $\bm{Q} \in \mathbb{R}^{M \times N}$ such that $\sum_{n=1}^N (\bm{Q})_{m, n}=1$ for all $m\in[M]$.
\end{definition}

Per Definitions \ref{Words_and_syntax_defn}-\ref{Interpretation_defn}, a semantic language is denoted as a 4-tuple $(\mathcal{W}, \mathcal{S}, \bm{P}, \bm{Q})$ \cite{Shao_Theory_of_SemCom'22}. We now proceed to state the definition of \textit{semantic channel}.

\begin{definition}[\textbf{Semantic channel \cite[Definition 3.7]{Shao_Theory_of_SemCom'22}}]
\label{Semantic_channel_defn}
Considering communication between a transmitter and a receiver, let the transmitted and the received message be $s \in \mathcal{S}$ and $\hat{s} \in \mathcal{S}$, respectively. The semantic channel is characterized by the transition probabilities from $s$ to $\hat{s}$ and defined as \cite[eq. (5)]{Shao_Theory_of_SemCom'22}
\begin{equation}
\label{Semantic_channel_eqn}
\big\{  c(\hat{s}|s) \in [0,1]: s, \hat{s} \in \mathcal{S}, \sum_{\hat{s}} c(\hat{s}|s) =1  \big\}.
\end{equation}
A semantic channel is also expressed as a matrix $\bm{C} \in \mathbb{R}^{M \times M}$ and said to be error free if and only if $ c(\hat{s}|s)=1$ for $s=\hat{s}$ and $ c(\hat{s}|s)=0$ for $s \neq \hat{s}$.
\end{definition}

With Definitions \ref{Words_and_syntax_defn}-\ref{Semantic_channel_defn} in hand, we now move on to discuss the language utilization problem of semantic encoding.

\subsubsection{Language Utilization -- Semantic Encoding}
\label{subsubsec: language_utilization_SE}
in light of the transmitter’s perspective that the semantic decoder is dictated by the \textit{interpretation of the agreed language}, the semantic channel, and the semantic decoder, the semantic encoding problem revolves around encoding the intended meaning to minimize misinterpretation by the receiver \cite{Shao_Theory_of_SemCom'22}. Accordingly, the following definitions of \textit{semantic encoding scheme} and \textit{semantic distortion of semantic encoding} ensue.
\begin{definition}[\textbf{Semantic encoding schemes \cite[Definition 3.8]{Shao_Theory_of_SemCom'22}}]
\label{Semantic_encoding_scheme_defn}
A semantic encoding scheme is a mapping from the meaning set $\mathcal{W}$ to the message set $\mathcal{S}$ and is defined as \cite[eq. (7)]{Shao_Theory_of_SemCom'22}
\begin{equation}
\label{semantic_encoding_eqn}
\big\{ u(s|w) \in [0,1]: w\in \mathcal{W}, s\in \mathcal{S},  \sum_{s} u(s|w)=1 \big\}, 
\end{equation}
where (\ref{semantic_encoding_eqn}) constitutes the matrix $\bm{U} \in \mathbb{R}^{N \times M}$.
\end{definition}

\begin{definition}[\textbf{Semantic distortion of semantic encoding \cite[Definition 3.9]{Shao_Theory_of_SemCom'22}}]
\label{semantic_distrotion_defn}
Suppose $w \in\mathcal{W}$ and $\hat{w} \in \mathcal{W}$ be the transmitted and reconstructed meanings at the transmitter and receiver, respectively. The average semantic distortion $D_{\bm{U}, \bm{Q}}$ attained by a semantic encoding scheme $\bm{U}$ can be expressed as \cite[eq. (8)]{Shao_Theory_of_SemCom'22}
\begin{equation}
\label{Average_semantic_distortion_eqn}
D_{\bm{U}, \bm{Q}} \eqdef \sum_{w, s, \hat{s},\hat{w}} p(w) u(s|w) c(\hat{s}|s) q(\hat{w}|\hat{s})d(w, \hat{w}) , 
\end{equation}
where $d(w, \hat{w}): \mathcal{W} \times  \mathcal{W} \to \mathbb{R}_+$ is a semantic distortion measure \cite{Shao_Theory_of_SemCom'22}.
\end{definition}

W.r.t. the average semantic distortion as expressed in (\ref{Average_semantic_distortion_eqn}) and the semantic cost defined in \cite[eq. (9)]{Shao_Theory_of_SemCom'22}, the authors of \cite{Shao_Theory_of_SemCom'22} devise insightful characterizations of the distortion-cost function of semantic encoding \cite[Sec. IV-A]{Shao_Theory_of_SemCom'22} and the distortion-cost region of semantic encoding \cite[Theorem 4.2]{Shao_Theory_of_SemCom'22}.

We now move on to the language utilization problem of semantic decoding.

\subsubsection{Language Utilization -- Semantic Decoding}
\label{subsubsec: language_utilization_SD}
in view of the receiver’s perspective that the semantic encoder is dictated by the \textit{expression of the agreed language}, the semantic channel, and the semantic encoder, the semantic decoding problem concerns how to decode a received message in order to minimize semantic distortion \cite{Shao_Theory_of_SemCom'22}. Consequently, the following definitions of \textit{semantic decoding scheme} and \textit{semantic distortion of semantic decoding} ensue.

\begin{definition}[\textbf{Semantic decoding schemes \cite[Definition 3.11]{Shao_Theory_of_SemCom'22}}]
\label{Semantic_decoding_scheme_defn}
A semantic decoding scheme is a mapping from the message set $\mathcal{S}$ to the meaning set $\mathcal{W}$ and defined as \cite[eq. (10)]{Shao_Theory_of_SemCom'22}
\begin{equation}
\label{semantic_decoding_eqn}
\big\{ v(w|s) \in [0,1]: w\in \mathcal{W}, s\in \mathcal{S},  \sum_{w} v(w|s)=1 \big\}, 
\end{equation}
where (\ref{semantic_decoding_eqn}) constitutes the matrix $\bm{V} \in \mathbb{R}^{M \times N}$.
\end{definition}

\begin{definition}[\textbf{Semantic distortion of semantic decoding \cite[Definition 3.12]{Shao_Theory_of_SemCom'22}}]
\label{semantic_distrotion_semantic_decoding_defn}
Suppose $w \in\mathcal{W}$ and $\hat{w} \in \mathcal{W}$ be the transmitted and recovered meanings at the transmitter and receiver, respectively. The average semantic distortion $D_{\bm{P}, \bm{V}}$ attained by a semantic decoding scheme $\bm{V}$ can be expressed as \cite[eq. (11)]{Shao_Theory_of_SemCom'22}
\begin{equation}
\label{Average_semantic_distortion_decodig_eqn}
D_{\bm{P}, \bm{V}} \eqdef \sum_{w, s, \hat{s},\hat{w}} p(w) p(s|w) c(\hat{s}|s) v(\hat{w}|\hat{s})d(w, \hat{w}).
\end{equation}
\end{definition}

W.r.t. the average distortion as given in (\ref{Average_semantic_distortion_decodig_eqn}) and the achievable cost per \cite[eq. (23)]{Shao_Theory_of_SemCom'22}, the authors of \cite{Shao_Theory_of_SemCom'22} derive the distortion-cost region of semantic decoding \cite[Theorem 5.1]{Shao_Theory_of_SemCom'22}, semantic decoding with an inaccurate prior \cite[Proposition 5.2]{Shao_Theory_of_SemCom'22}, and semantic decoding with the Hamming distortion \cite[Proposition 5.3]{Shao_Theory_of_SemCom'22}, among other results. 

We now proceed with our brief discussion of the CSED language utilization problem. 

\subsubsection{Language Utilization -- CSED}
\label{subsubsec: language_utilization_CSED}
in CSED problem, both the transmitter and the receiver act individually per their own perspectives: the transmitter and receiver simultaneously carry out semantic encoding and semantic decoding, respectively. This problem is formally defined in \cite[Definition 5.4]{Shao_Theory_of_SemCom'22}. In light of this problem, the semantic distortion of CSED and the semantic cost of CSED are given by \cite[eqs. (46) and (47)]{Shao_Theory_of_SemCom'22}
\begin{subequations}
\begin{align}
\label{Average_semantic_distortion_CSED_eqn}
D_{\bm{U}, \bm{V}_q^{*}} & \eqdef \sum_{w, s, \hat{s},\hat{w}} p(w) u(s|w) c(\hat{s}|s) v(\hat{w}|\hat{s})d(w, \hat{w})    \\
\label{Average_semantic_cost_CSED_eqn}
L_{\bm{U}} & \eqdef  \sum_{w, s} p(w) u(s|w) \ell(s),
\end{align}
\end{subequations}
where $\bm{V}_q^{*}$ is the optimal semantic decoding strategy and $\ell(s): \mathcal{S} \to \mathbb{R}^{+}$ is a cost function \cite{Shao_Theory_of_SemCom'22}. In light of (\ref{Average_semantic_distortion_CSED_eqn}) and (\ref{Average_semantic_cost_CSED_eqn}), the authors of \cite{Shao_Theory_of_SemCom'22} derive the \textit{distortion-cost region of CSED} \cite[Theorem 5.5]{Shao_Theory_of_SemCom'22} and \textit{CSED with an error-free semantic channel} \cite[Theorem 5.6]{Shao_Theory_of_SemCom'22}. The result in \cite[Theorem 5.6]{Shao_Theory_of_SemCom'22} is particularly insightful in that the CSED scheme would comprise the semantic encoding scheme $\bm{U}$, which is optimized based on the interpretation $\bm{Q}$, and the decoding scheme $\bm{V}$, which is optimized based on the expression $\bm{P}$ of the language \cite{Shao_Theory_of_SemCom'22}.

Apart from the advancements in SemCom theory that are discussed in Sections \ref{subsec: SemCom_with_ProbLog} through \ref{subsec: language_utilization_and design}, there have been several other recent developments and perspectives, such as the equivalence of SemCom and online learning \cite{SemCom_Equivalent'11}; the rate distortion theory for strategic SemCom \cite{Xiao_RD_Theory_Strategic_SemCom'22}; universal SemCom \cite{Juba_Universal_SemCom_I'08,Juba_Universal_SemCom'08}; an IB viewpoint of SemCom \cite{Beck_SemCom_Info_Bottleneck_View'22}; a probabilistic logic approach to SemCom \cite{Choi_Unified_Approach'22,Choi_Unified_View_SemInfo'22}; and compatibility among various vantage points of SemCom \cite{Juba_Compatibility_among_diversity'13}. These theoretical advancements and the above-discussed SemCom theories have their respective limitations. Hence, existing SemCom theories are not the most rigorous and complete of theories (though they sure are interesting!) due to the many fundamental and major challenges of SemCom, which are detailed below.

\section{Fundamental and Major Challenges of SemCom}
\label{sec: fundamental_and_major_challenges_of_SemCom}
When it comes to realizing high-fidelity SemCom for 6G and beyond, the research field of SemCom is fraught with fundamental and major challenges in the theoretical, algorithmic, and realization/implementation-related research frontiers. These challenges are discussed in detail below, beginning with the challenges in the development of fundamental SemCom theories.

\subsection{Challenges in the Development of Fundamental SemCom Theories}
\label{subsec: challenges_of_SemCom_fundamental_theories}
In what follows, we present (in no particular order) the fundamental and major challenges related to -- but not limited to -- the development of fundamental SemCom theories.  

\subsubsection{Lack of any Commonly Accepted Definition of Semantics / Semantic Information}
although several notions of semantics have been put forward to define what SemCom could be, none have been satisfactory to date \cite[p. 125]{Tong_Zhu__6G'21}. This has hugely restricted further progress in SemCom theories \cite[p. 125]{Tong_Zhu__6G'21}. Amid the lack of adequate mathematical quantification of semantic information, DL-based SE has emerged as a popular enabler of SemCom. However, deploying a black box as the foundation of system design and optimization is not fundamentally convincing \cite{SemCom_for_6G_Future_Internet'22}. This definitely hinders the advancement of SemCom theory (as well as algorithm and realization).

\subsubsection{Quantifying Semantic Noise, Semantic Interference, and the Effect of Semantic Noise and Semantic Interference}
semantic noise and semantic interference can occur in (and hence impact) text SemCom, audio SemCom, image SemCom, video SemCom, multi-modal SemCom, and cross-modal SemCom. Accordingly, quantifying semantic noise, semantic interference, and the effect of semantic noise and semantic interference can pave the way for a rigorous SemCom theory. However, they have the following the respective fundamental challenges: $i)$ \textit{how to quantify semantic noise?} $ii)$ \textit{how to quantify semantic interference?} and $iii)$ \textit{how to quantify the effect of semantic noise and semantic interference?}

\subsubsection{Fundamental Performance Analysis of SemCom}
the fundamental non-asymptotic performance analysis of SemCom is fundamentally challenging for the following reasons \cite{arXiv_Getu_DeepSC_Performance_Limits'23}: $i)$ the lack of a commonly agreed-upon definition of semantics / semantic information \cite[Ch. 10, p. 125]{Tong_Zhu__6G'21}; $ii)$ the fundamental lack of interpretability/explainability of \textit{optimization}, \textit{generalization}, and \textit{approximation} in DL models \cite{Poggio_Theo_Issues_Dnets_2020}; and $iii)$ the lack of a comprehensive mathematical foundation for SemCom \cite[Sec. IV]{hTong_Li_Nine_Challenges'21}.
	
\subsubsection{Fundamental Performance Analysis of a SemCom System under (Semantic) Interference}
in a SemCom system, interference or semantic interference can cause considerable semantic noise, to the extent that the faithfulness of SemCom is destroyed \cite{arXiv_Getu_DeepSC_Performance_Limits'23}. Owing to a lack of understanding of how to quantify semantic noise and the aforementioned three fundamental challenges that hinder the fundamental non-asymptotic performance analysis of SemCom, analyzing the fundamental performance of a SemCom system that is subjected to (semantic) interference is indeed fundamentally challenging. The authors of \cite{arXiv_Getu_DeepSC_Performance_Limits'23} have made progress toward analyzing the asymptotic performance of a DL-based text SemCom system with interference. However, there is a long way to go for the non-asymptotic performance analysis of a SemCom system with interference. 

\subsubsection{Performance Analysis of DL-based SemCom Systems}
DL-based SemCom systems such as DeepSC \cite{Xie_DL-based_SemCom'21} benefit from a joint DL-based source and channel coding technique. Nevertheless, the rigorous non-asymptotic performance analysis of DL-based SemCom systems is hampered by the \textit{fundamental lack of interpretability/explainability} \cite{Toward_Science_of_Interpretable_ML'17,Guo_XAI_6G'20} that is inherent in (trained) DL models.

\subsubsection{Devising a Reasonable Semantic Channel Model in View of SemCom's Attributes}
it is widely recognized that background KB mismatch between a semantic encoder and a semantic decoder can definitely cause semantic ambiguity that leads to information distortion. To this end, a major challenge is devising a reasonable semantic channel model based on different degrees of background KB matching from the perspective of SIT \cite{Xia_Wireless_SemCom'22}. 

\subsubsection{ Semantic-Enabled Intelligence Evolution}
to orchestrate and achieve overarching semantic-enabled networked intelligence regardless of the situation, SemCom systems and semantic networks need to be able to autonomously evolve to an enhanced level of intelligence \cite{Elements_Causal_Inference'17} that may require greater \textit{conciseness} \cite{Morin_Self_Awareness_06,Understanding_consciousness'21,Dehaene_What_is_consciousness'17,Tononi_IIT_Nat_Reviews'16,Consciousness_and_the_Universe'17}. This calls for \textit{semantic-enabled lifelong autonomy} and \textit{lifelong autonomously networked evolving intelligence}, which are both fraught with innumerable fundamental challenges.

\subsubsection{Fundamental Limits of SemCom}
in SemCom PHY design, the overarching goal is to optimize semantic information transmission over various channels with relevant background KBs \cite{Zhang_Wisdom_Evolutionary_6G'21}. When it comes to the background KBs, the fundamental limits of SemCom are contingent on not only their respective PHY constraints but also on the contextual constraints \cite{Zhang_Wisdom_Evolutionary_6G'21}. As for contextual constraints, the degree of mutual understanding between any pair of communication parties can influence the interaction, the signaling strategies, and the volume of SemCom \cite{Zhang_Wisdom_Evolutionary_6G'21}. Accordingly, a suitable measure of \textit{intent-achieving efficiency} -- which is generally abstract and complex -- must be established to address this question: \textit{what is the most efficient SemCom strategy to attain a given intent?} \cite{Zhang_Wisdom_Evolutionary_6G'21} Strategies worth considering include: semantic-aware joint source–channel coding in PHY and semantic-linked processing in higher layers \cite{Zhang_Wisdom_Evolutionary_6G'21}. o these ends, $i)$ some theories and coding schemes should first be established to materialize the new measure framework and $ii)$ achievable bounds should be derived for the fundamental limits of SemCom \cite{Zhang_Wisdom_Evolutionary_6G'21}.

\subsubsection{Fundamental Limits of SemCom-Enabled Distributed Model Training}
distributed training such as FL is key for enabling distributed AI and edge AI, especially in memory- and computing capacity-limited IoT devices. However, the performance of the resulting edge AI will depend on not only rigorous model training / model tweaking but also the quality/informativeness of the devices' semantic information, provided that SemCom-enabled distributed model training/retraining is the aim. Determining the fundamental limits of this setup is a significant and relevant fundamental challenge.

\subsubsection{Deriving the Capacity of a Semantic-Aware Network}
since semantic-aware networks are more complex, their capacity can be closely related to knowledge sharing among users \cite{Shi_From_SemCom_to_Sematic-aware_Networking'20}. Accordingly, developing comprehensive mathematical foundations for the performance limits of a SemCom network is a crucial fundamental challenge. 

\subsubsection{Unified Fundamental Theory of Semantic Information}
SIT is a bedrock of SemCom that can provide insight into semantic information bounds and serve as a crucial framework for evaluating semantic abstraction \cite{Rethinking_modern_com_Lu_2022}. On the other hand, unlike conventional communication systems that mainly rely on a fixed set of transmitted symbols, SemCom has to grapple with the extensibility and openness of semantics, which lead to symbols changing dynamically \cite{Shi_to_Semantic_Fidelity'21}. How to model a dynamic set and its impact on channel capacity remains unknown \cite{Shi_to_Semantic_Fidelity'21}. Moreover, despite recent and emerging progress on the theory front (as discussed in Section \ref{sec: theories_of_SemCom}), a unified fundamental theory of semantic information remains elusive.

\subsubsection{Unified SemCom Theory}
employing logical probability \cite{Bar_Carnap_Theory_SemInfo'1954}, existing and emerging SemCom theories attempt to develop a theory for semantic entropy, semantic channel capacity, semantic-level rate distortion theory, and the relationship between inference accuracy and transmission rate \cite{Qin_Sem_Com_Principles_Apps'22}. Nevertheless, it is not clear whether this path could lead to a unified SemCom theory. On the other hand, in sharp contrast with conventional communication systems that often deploy a fixed set of transmitted symbols, SemCom hinges on information semantics whose flexibility and complexity causes the semantic symbol set to change dynamically while possibly exhibiting polysemy \cite{Sem_Empowered_Commun'22}. As a result, how to process and model dynamic semantic symbol sets is a fundamental challenge \cite{Sem_Empowered_Commun'22} that stands in the way of a unified SemCom theory. The authors of \cite{Rodoplu_Challenges_Directions_SemCom'07} took inspiration from the fundamental synaptic plasticity \cite{Kandel_Principles_of_Neural_Science'21,Neuronal_Dynamics_Book'14,Bear_MF_Neuroscience'16} of our brains and assert that the fundamental limit of effective semantic information communication hinges on the plasticity of our brains or the substrates on which the semantic domains are constructed. This calls for an \textit{extraordinarily fundamental understanding} of brain computation/operation -- at not only the system level, but also the molecular, cellular, and network levels -- prior to developing a unified SemCom theory. Furthermore, per the information ecosystem model (see \cite[Fig. 1]{hong_Theory_Semantic_Info'17} and \cite[Fig. 1]{Zhong_Theory_Sem_Information_Book_Chapter}) proposed by the author of \cite{hong_Theory_Semantic_Info'17}, a theory of semantic information must take into account the emerging \textit{rate-distortion-perception tradeoff} \cite{Blau_Rethinking_Loss_Compression'19} -- a triple tradeoff rather than Shannon's rate-distortion tradeoff -- whose significance is affirmed by the prevalence of DL.

We now proceed to fundamental and major challenges in the development of fundamental SemCom algorithms.

\subsection{Challenges in the Development of Fundamental SemCom Algorithms}
\label{subsec: challenges_of_SemCom_fundamental_algorithms}
In what follows, we point out (in no particular order) the fundamental and major challenges related to -- but not limited to -- the development of fundamental SemCom algorithms. 

\subsubsection{Inevitability of Semantic Mismatch}
although both the source KB and the destination KB can learn from the perceived environment and continuously expand as well as update their entries through training and sharing via communications, the KBs at the source and destination can be quite different as a result of observing different environments (hence worlds) with unequal abilities to understand things \cite{Luo_SemCom_Overview'22}. Accordingly, \textit{semantic mismatch} is inevitable to the extent that it can fundamentally constrain the performance of SemCom-based wireless systems.

\subsubsection{The Need for Additional SemCom Performance Assessment Metrics}
even though a variety of metrics have been employed in early algorithmic developments of SemCom, additional performance assessment metrics, such as the ones to evaluate the amount of semantic information that has been preserved or missing are required \cite{Qin_Sem_Com_Principles_Apps'22}. For semantic-enabled networked intelligence, on the other hand, a comprehensive evaluation framework is needed that takes into account objective and subjective metrics that can capture the efficiency and potential of systems/networks in achieving intent \cite{Zhang_a_New_Paradigm'22}.

\subsubsection{Lack of Unified Semantic Performance Assessment Metrics}
unified SemCom performance assessment metrics\footnote{A proper semantic similarity metric is necessary to define a loss function and pre-train the parameters of DNNs for various tasks \cite{Sem_Empowered_Commun'22}.} are needed to fairly compare and contrast existing and prospective SemCom techniques \cite{Getu_Metrics_of_SemCom_and_GO_SemCom_2023}. When it comes to unified metrics, the major challenge is to establish concrete metrics that can capture source and network dynamics as well as any potentially non-trivial interdependencies among information attributes \cite{Kountouris_Semantics_EmpoweredCF'21}. Meanwhile, it is worth underscoring that the lack of unified SemCom performance assessment metrics can hinder the advancement of SemCom research, standardization, and deployment in 6G. 

\subsubsection{Semantic Transformation}
semantic transformation is a core SemCom process in the understand-first-and-then-transmit SemCom framework \cite[Fig. 2]{Shi_to_Semantic_Fidelity'21} at both a transmitter and a receiver. At the receiver, the semantic symbol recognition module can be fundamentally limited by semantic ambiguity, which is an open challenge, especially when there is no context \cite{Shi_to_Semantic_Fidelity'21}, and a hard problem in the field of NLP \cite{Daniel_NLP'00}. One way to overcome this fundamental challenge is to transmit more symbols to achieve such disambiguation that the sender needs to eliminate the inherent ambiguity in as few symbols as possible \cite{Shi_to_Semantic_Fidelity'21}. For the achievement of this goal, devising the optimal symbols is an open problem \cite{Shi_to_Semantic_Fidelity'21}.

\subsubsection{Lack of Interpretability in DL-Based SE}
differentiable loss functions that are widely deployed by DL-based SE techniques, such as CE and MSE, give equal importance to the semantic contributions of all bits, which is inconsistent with human perception \cite{Yang_SemCom_meets_Edge_Intelligence'22}. This corroborates the fundamental lack of interpretability in DL-based SE. 

\subsubsection{Lack of Interpretability in DL-Based SemCom}
there is a fundamental lack of interpretability in DL-based SemCom techniques due to the fundamental lack of interpretability/explainability that is inherent in DL models \cite{Toward_Science_of_Interpretable_ML'17,Guo_XAI_6G'20}.

\subsubsection{Addressing Time- and Frequency-Selective Channels in DL-Based SemCom Systems} 
a number of existing works on DL-based SemCom \cite{Xie_DL-based_SemCom'21,Farsad_DL_JSCC'18} demonstrate the visible gain that can be achieved by deploying DL using mainly fixed channel conditions (or slow fading channels). However, wireless channels are usually time- and frequency-selective channels whose inherent doubly selective fading will challenge any DL-based design \cite{Getu_DSFC_Estimation'22}. This challenge will be significant because DL is used to train and test datasets that are drawn from the same distribution, which is in sharp contrast to a realistic wireless communication scenario, wherein the testing distribution can be very different from the training distribution.

\subsubsection{Semantic Communications and Networking over Wireless Fading Channels}
semantically-encoded information is often intended to be transmitted over wireless channels, which are naturally fading channels \cite{Simon_Alouni_DC_over_Fading_Channels'05}. In a realistic SemCom scenario, the semantic decoding of a semantically-encoded message received over wireless fading channels can result in a complete loss of meaning at the receiver. This leads to the following fundamental question worth attacking: \textit{how to design an optimal semantic transceiver (optimal SemCom system) so that the loss of meaning at the receiver is minimized?} 

\subsubsection{Bandwidth Allocation in SemCom}
semantic information is unevenly distributed in SemCom and semantic-aware networks. This uneven distribution needs to be taken into consideration, and more bandwidth should be allocated to agents that wish to transmit more semantic information \cite{SemCom_for_6G_Future_Internet'22}. Nevertheless, quantifying semantic information is fundamentally challenging. 

\subsubsection{Different Device Capacities}
individual communication devices have different computational power, communication resources, and storage capacity, all of which are limited. In light of this natural limitation, the design of SemCom networks must not assume that all devices have sufficient capacity \cite{SemCom_for_6G_Future_Internet'22}. To this end, devising effective methods for balancing heterogeneous devices' performance and cost requirements is a key challenge for the design of robust SemCom networks \cite{SemCom_for_6G_Future_Internet'22}.

\subsubsection{Semantic-Aware Multiple Access}
designing an optimal semantic-aware multiple access scheme for plenty of devices that transmit signals in a time- or event-triggered process-aware manner to convey multi-attribute information to a remote destination is challenging \cite{Kountouris_Semantics_EmpoweredCF'21}. The challenge arises from the following requirement regarding the optimal utilization of the shared medium: the devices have to adapt their access patterns depending on not only the arrival of exogenous traffic (and other nodes’ status), but also source/process variability, information semantics, and the needs of applications \cite{Kountouris_Semantics_EmpoweredCF'21}.

\subsubsection{Semantic-Aware Multiple and Random Access}
while the celebrated multiple access scheme named NOMA \cite{Makki_NOMA_Survey'20,Dai_NOMA_Survey'18} improves spectral efficiency, this improvement hardly translates into better performance w.r.t. freshness or other semantic attributes \cite{SemCom_Net_Systems'21}. Not only is there scheduled multiple access, but there is also random access (CSMA, CSA, E-SSA, frameless ALOHA) for next-generation massive MTC \cite{mahmood2020white} and IoT applications, for which determining semantic principles remains a fundamental problem \cite{SemCom_Net_Systems'21}. This calls for novel protocols that incorporate sampling and data generation \cite{SemCom_Net_Systems'21}. Thus, rethinking random or scheduled access in relation to broader semantic metrics remains an open problem \cite{SemCom_Net_Systems'21}.

\subsubsection{Designing Semantics Networks}
SemCom can be a pivotal component of distributed intelligent networks in 6G and beyond due to its minimal bandwidth consumption, significantly decreased data transmission, and ability to exploit more knowledge \cite{Sem_Empowered_Commun'22}. Nevertheless, semantic network design is affected by the following significant challenges: $i)$ lack of a specific/comprehensive blueprint definition of semantic network; and $ii)$ the complexity of transmission/computation to disseminate KBs to distributed devices and adapt to the network's ultra-heterogeneity \cite{Sem_Empowered_Commun'22}.

\subsubsection{Balancing the Generality and Confidentiality of SemCom} 
most existing SemCom works promote centralized SemCom systems and unified multi-user SemCom systems that are trained based on one or more shared background KBs \cite{Luo_Encrypted_SemCom'22}. Nonetheless, this design philosophy certainly causes \textit{privacy leakage} \cite{Luo_Encrypted_SemCom'22}. Thus, balancing the generality and confidentiality of SemCom is a major challenge of SemCom \cite{Luo_Encrypted_SemCom'22} from an algorithmic standpoint, but also possibly from a realization vantage point.

We now move on to detailing fundamental and major challenges in the realization of SemCom.

\subsection{Challenges in the Realization of SemCom}
\label{subsec: challenges_of_SemCom_realization}
We describe (in no particular order) the fundamental and major challenges related to -- but not limited to -- the realization of SemCom.

\subsubsection{Huge Time Consumption and Complication in Semantic Index Assignment}
semantic index assignment is an intuitive approach to preserving semantic similarity that assigns a binary codeword to each word \cite{Luo_SemCom_Overview'22} and wherein semantically similar words and semantically dissimilar (independent) words are coded with the shortest Hamming distance and the longest Hamming distance, respectively \cite{SemCom_Game'18,Guler_Semantic_index_assignment'14}. Nevertheless, because the length of a codeword is exponentially proportional to the number of words, the semantic index assignment process is extremely time-consuming and complicated \cite{Luo_SemCom_Overview'22}.

\subsubsection{Real-Time Requirement}
with its promises of minimum power usage, bandwidth consumption, and transmission delay in addition to some level of inherent security, SemCom is indeed an enabler of 6G and hence attractive for many 6G IoT applications \cite{Xie_Lite_distributed_SemCom'21}. Nonetheless, the vital incorporation of semantic reasoning for correcting transmission errors incurs more delay in the overall SemCom transceivers, which are relatively more complex \cite{Sem_Empowered_Commun'22}. Accordingly, satisfying the ultra-low end-to-end latency requirements -- i.e., real-time requirements -- of 6G (and beyond) is a major challenge\footnote{To overcome the major challenge of satisfying the ultra-low end-to-end latency requirement of 6G, it is useful to develop \textit{lightweight} SemCom algorithms and improve SemCom transceiver/hardware design \cite{Sem_Empowered_Commun'22}.} for the realization of SemCom.

\subsubsection{Scalability}
even though some multi-modal SemCom (e.g., \cite{Zhang_Unified_Multi-Task_SemCom'22,WANG_Multimodal_SemCom'23}) / cross-modal SemCom (e.g., \cite{Li_Cross-Modal_SemCom'22}) systems that work for the transmission of multi-modal / cross-modal data show promise, it is very challenging to grapple with more complex data types in legacy OSI models \cite{Sem_Empowered_Commun'22}. One scalability challenge is that a general semantic-level framework for distinct types of sources has not yet been developed \cite{Sem_Empowered_Commun'22}. Another scalability challenge is the fact that sharing, updating, and maintaining KBs at the source and destination would necessitate additional storage costs and algorithm design \cite{Sem_Empowered_Commun'22}. Therefore, SemCom involves considerable computational as well as storage costs. Consequently, guaranteeing the scalability of SemCom remains a challenge \cite{Sem_Empowered_Commun'22}.  

\subsubsection{Knowledge Evolution Tracking}
it is well-known that humans' knowledge evolves continuously throughout their lives. Regarding such temporal variations, modeling and keeping track of each piece of knowledge (e.g., aggregating new knowledge entities and relationships while discarding obsolete ones in the context of KGs \cite{Shi_From_SemCom_to_Sematic-aware_Networking'20}) is fundamentally important for improving SemCom efficiency and reducing the probability of error in semantic information delivery \cite{Shi_From_SemCom_to_Sematic-aware_Networking'20}. Nevertheless, the basic neuroscientific understanding of knowledge, knowledge evolution, and knowledge tracking are very difficult fundamental problems. 

\subsubsection{Networked KB Updating and Upgrading}
in semantic-enabled networked intelligence, a networked KB plays a predominant role in and is an enabler of intelligence \cite{Zhang_a_New_Paradigm'22}. Nonetheless, networked KB updating and upgrading is a huge challenge for massive communication objects -- in 6G and beyond -- with highly heterogeneous computation and storage capabilities \cite{Zhang_a_New_Paradigm'22}. Consequently, how to build a networked KB and how to efficiently update and upgrade it are major challenges for semantic-enabled networked intelligence in particular \cite{Zhang_a_New_Paradigm'22} and SemCom in general. 

\subsubsection{Compatibility with Existing Communication Infrastructure}
any SemCom realization effort must ensure that SemCom is compatible with the existing communication infrastructure \cite{Lee_EQ2SEQ-SC'22}. To this end, extensive link-level simulations must be performed to verify the realistic end-to-end performance of SemCom \cite{Lee_EQ2SEQ-SC'22}.

\subsubsection{Generalizability of the Semantic Network}
the semantic network has to be generalized to work with any dataset rather than only specific datasets \cite{Lee_EQ2SEQ-SC'22}. This calls for DL-based semantic transceivers that can generalize widely over numerous datasets. Deep networks face fundamental limitations in this regard, as they are able to learn only stationary data distributions \cite{Synthe_lect_DLLR_16,GIP_LLDL_17,Parisi_CLLL_19,DeLange_CL_Survey'22}.

\subsubsection{Coexistence of SemCom and Technical/Conventional Communication}
future networks in 6G and beyond should support highly heterogeneous data transmission and multiple RANs. Accordingly, networks in 6G and beyond must be able to support not only users of SemCom, but also users of traditional BitCom, as the latter cannot be fully replaced by the former \cite{Lee_EQ2SEQ-SC'22}. This calls for a robust coexistence design for networks in 6G and beyond network that must be able to deliver information both \textquotedblleft as is'' and modified with high semantic similarity (depending on the communication scenario) \cite{Lee_EQ2SEQ-SC'22}.  

In light of the above-detailed realization challenges of SemCom, we refer the reader to the work in \cite{Yoo_Demo'22} that implements a real-time image SemCom system whose feasibility in actual wireless environments is demonstrated. Meanwhile, because \textit{challenges are always opportunities}, some of the above-detailed fundamental and major challenges of SemCom are also big opportunities for novel future directions of SemCom, as discussed below.

\section{Future Directions of SemCom}
\label{sec: future_directions_of_SemCom}
In light of the fundamental and major challenges of SemCom that are detailed in Section \ref{sec: fundamental_and_major_challenges_of_SemCom}, the developments in SemCom theory that are presented in Section \ref{sec: theories_of_SemCom}, and the many proposals of state-of-the-art SemCom algorithms that are surveyed in Section \ref{sec: SemCom_research_landscape}, we offer some novel future directions or SemCom theory, algorithm, and realization. We begin with some novel future directions for SemCom theory.

\subsection{Future Directions for SemCom Theory}
\label{subsec: Future_directions_in_SemCom_theories}
We point out (in no distinct order) some novel future directions related to -- but not limited to -- SemCom theory.

\subsubsection{(General) Semantic Information Theory}
the following would be needed for a fundamental SIT: investigations of SIT with interference channels and a specific definition of semantic channel and its capacity among other things \cite{Luo_SemCom_Overview'22}. For a general fundamental SIT, on the other hand, the following research directions are crucial research topics: $i)$ the theory of semantic security and robustness; $ii)$ the theory of the tradeoff between semantic efficiency and generalization; $iii)$ the theory of \textit{semantic computability} \cite{Sem_Empowered_Commun'22}.

\subsubsection{A General Framework for DL-Based SemCom}
for faithful and interpretable DL-based SemCom, a generic framework that encompasses a suitable DNN architecture, proper performance assessment metrics, and the likes should be explored \cite{Luo_SemCom_Overview'22}.

\subsubsection{The Tradeoff Between Semantic Extraction Accuracy and Communication Overhead} 
in DL-enabled SemCom, the training of accurate SE models hinges on complete KBs -- at both senders and receivers -- which requires sufficient storage. In case of adequate storage, each user’s local KB has to be constantly updated as the communication context evolves \cite{SemCom_for_6G_Future_Internet'22}. To this end, ensuring that updates to the local KB of each communicating party can be shared in real-time is quite challenging, to the extent that it will cause significant communication overhead \cite{SemCom_for_6G_Future_Internet'22}. The communication overhead and the real-time update/sharing challenge will be particularly significant in scenarios in which there is a massive number of participating users that are geographically distant. Accordingly, devising insightful tradeoffs between SE accuracy and communication overhead is a crucial future research direction for SemCom.

\subsubsection{The Tradeoff Between SemCom Performance and Security} 
because SemCom transmits only semantically-encoded data -- therefore, less data -- and the decoding of semantic information depends of the intended receiver's KB, SemCom itself has also been regarded as a potential method of secure wireless communication \cite{Basu_Perserving_QoI'12,Choi_Unified_View_SemInfo'22,Choi_Unified_Approach'22}. When it comes to secure wireless communication, any potential eavesdropper can be made uncertain whether SemCom is being used by introducing AI into the PHY \cite{SemCom_for_6G_Future_Internet'22}. A PHY with interference, meanwhile, will harm the transmission quality of semantic information, hence the tradeoff between \textit{signal covertness and signal quality} \cite{SemCom_for_6G_Future_Internet'22}. This calls for future research into an optimal tradeoff between SemCom performance and security. When it comes to ensuring security via SemCom, however, SemCom's secrecy performance under adversarial attacks remains largely unknown \cite{Mu_SemCom_in_MU_Wireless_Networks'22} and novel research is therefore called for.

\subsubsection{The Impact of Semantic KBs}
semantic KBs at the source and destination directly impact the faithfulness of SemCom. To this end, the following are research questions worth addressing: $i)$ \textit{to what extent do shared KBs affect the SemCom process?} $ii)$ \textit{how can the semantic flow in KBs that are partially shared be modeled quantitatively?} \cite{Sem_Empowered_Commun'22}

We now continue to some novel future directions for SemCom algorithms.
 
\subsection{Future Directions for SemCom Algorithms}
\label{subsec: Future_directions_in_SemCom_algorithms}
We call attention (in no particular order) to promising future directions related to -- but not limited to -- SemCom algorithms.

\subsubsection{Multi-User SemCom and Multi-User SemCom Signal Detection}
multi-user SemCom signals exploit the diversity of the KBs that comprise different SemCom systems and can be transmitted using the same radio resources -- such as frequency or time slot -- so that considerable bandwidth can be saved in the interest of spectral-efficient 6G \cite{Luo_SemCom_Overview'22}. To this end, optimal multi-user SemCom signal detection and the complexity of message interpretation at the receiver are critical research themes \cite{Luo_SemCom_Overview'22}.
 
\subsubsection{Multi-User Interpretation Algorithm Design} 
exploiting the diversity in their KBs might be able to utilize the same frequency or time slot to transmit semantic information over a wireless channel \cite{Luo_SemCom_Overview'22}. Nevertheless, this gain introduces the following two major challenges: $i)$ since semantic information interpretation at a receiver in a multi-user environment should consider joint multi-user detection, channel decoding, and semantic decoding, the complexity will be very high; $ii)$ a receiver's KB should include various types of data in order to rigorously distinguish different users’ messages \cite{Luo_SemCom_Overview'22}. To overcome these challenges, more effective yet efficient interpretation algorithms must be devised for the joint semantic-channel decoding of an intended user \cite{Luo_SemCom_Overview'22}. To this end, the design of a low-complexity multi-user interpretation algorithm is an essential future direction for SemCom \cite{Luo_SemCom_Overview'22}.
 
\subsubsection{Joint Design of the Technical Level and Semantic Level} 
to understand the impact the data transmission rate has on SemCom and how much semantic information can be transmitted over a wireless channel, the technical level and semantic level of communication -- per Weaver's three levels of communication, which is schematized in Fig. \ref{fig: Three_lavels_of_communication} --  must be jointly designed \cite{Luo_SemCom_Overview'22}. In view of a joint design, the interference-resistant and robust SemCom (\textit{IR$^2$ SemCom}) advocated for by the authors of \cite{arXiv_Getu_DeepSC_Performance_Limits'23} also requires that the technical and semantic levels of communication be jointly designed.
 
\subsubsection{Addressing the SNR Uncertainty Affecting DL-Based SemCom Systems}
SNR uncertainty can arise from the uncertainty in the noise power, the inevitability of interference, and the variation in transmission power \cite{SemCom_for_6G_Future_Internet'22}. Accordingly, unlike in the fixed SNR approaches that are typically adopted to train current SemCom models, SNR uncertainty's effect on SemCom system performance has to be considered. In this vein, ensuring that the trained semantic model dapts to a variable SNR and understanding/quantifying its generalization ability require further investigation \cite{SemCom_for_6G_Future_Internet'22}.

\subsubsection{The Design and Realization of Secure SemCom}
secure SemCom can be realized using semantic noise \cite{SemCom_for_6G_Future_Internet'22} in view of deploying artificial noise to ensure secure wireless transmission \cite{Goeal_guaranteeing_secrecy'08,Zhou_secure_transmission_using_artif_noise'10}. To this end, using a background KB to enhance semantic noise is a promising approach for designing and realizing secure SemCom.
 
\subsubsection{Blending SemCom and Semantic Caching}
in contrast to traditional data caching, which mainly focuses on the hit rate of the data content, semantic caching focuses on whether the semantic information in the cache can be accurately inferred by the requester \cite{SemCom_for_6G_Future_Internet'22}. Determining which semantic information to cache requires prior knowledge, such as the popularity of particular semantic information, since a variety of semantic information could exist for the same data content \cite{SemCom_for_6G_Future_Internet'22}. Meanwhile, since the context of SemCom changes continually, the lifetime of semantic information is difficult to determine and new estimate refreshing algorithms for semantic caching are required \cite{SemCom_for_6G_Future_Internet'22}.
 
\subsubsection{Reasoning in Implicit SemCom} 
although most existing SemCom works focus on the transmission of explicit semantic information, communication between users involves not only explicit information, but also rich implicit information that is difficult to express, recognize, or recover \cite{Xiao_Reasoning_on_the_Air'22} (see also \cite{LLL_Reasoning_Based_SemCom'22}). Since explicit semantic information is generally dominant and communication resources should be allocated proportionally between explicit and implicit semantic information, joint optimization algorithms need to be designed \cite{SemCom_for_6G_Future_Internet'22}. This calls for the design and realization of joint implicit and explicit SemCom with rigorous reasoning.
	
We now move on to some novel future directions for SemCom realization.

\subsection{Future Directions for SemCom Realization}
\label{subsec: Future_directions_in_SemCom_realization}
We highlight (in no definite order) some useful future directions related to -- but not limited to -- SemCom realization.

\subsubsection{SemCom Implementation}  
state-of-the-art \textit{system-on-chip} technologies cannot meet the ultra-low latency requirements of wireless communication in 6G networks \cite{Luo_SemCom_Overview'22}. To overcome this major challenge, more advanced microelectronic and chip technologies are needed \cite{Luo_SemCom_Overview'22}. 

\subsubsection{The Impact of Inconsistent KBs at the Semantic Source and Destination} 
although most state-of-the-art SemCom works presume real-time knowledge sharing to make the KBs at the source and destination consistent, the KBs are naturally inconsistent \cite{Luo_SemCom_Overview'22}. Therefore, how semantic information can be communicated, shared, and inferred when the KBs are inconsistent are open issues for SemCom design as well as realization \cite{Luo_SemCom_Overview'22}.

We now move on to our extensive discussion on the state-of-the-art research landscape of goal-oriented SemCom.

\begin{figure*}[htb!]
	\centering
	\includegraphics[scale=0.50]{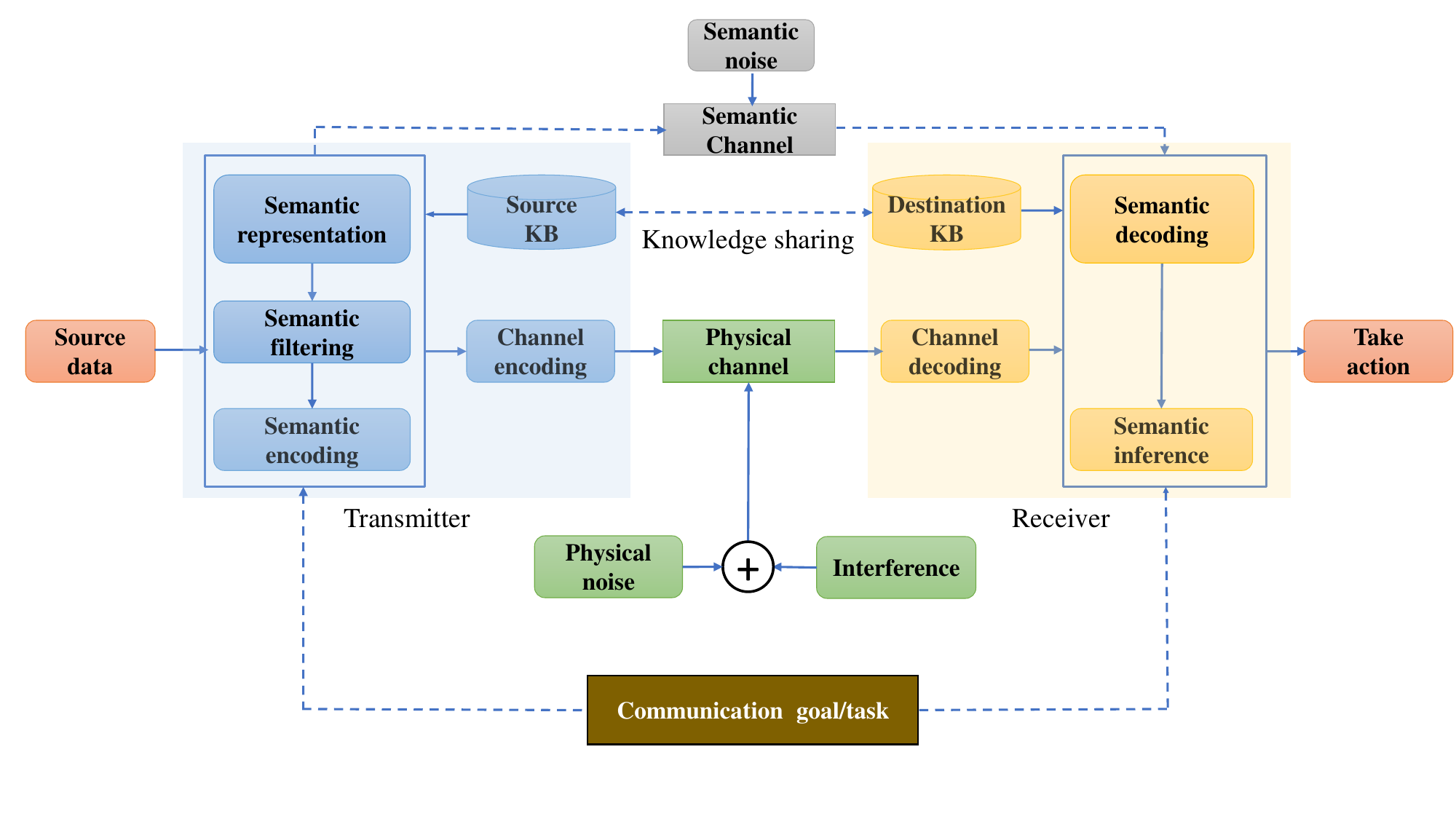} \vspace{-0.4cm}  
	\caption{System model for goal-oriented SemCom -- adopted from \cite[Fig. 6(c)]{SemCom_for_6G_Future_Internet'22}.}
	\label{fig: Goal_oriented_SemCom_system_model}
\end{figure*}

\section{State-of-the-Art Research Landscape of Goal-Oriented SemCom}
\label{sec: goal_oriented_SemCom_research_landscape}
Goal-oriented SemCom aims to enable interested communicating parties to achieve a joint communication goal/task \cite{SemCom_for_6G_Future_Internet'22,Juba_Universal_SemCom'08}. To complete a joint communication goal/task, Fig. \ref{fig: Goal_oriented_SemCom_system_model} illustrates a system model for goal-oriented SemCom. The effectiveness-level SemCom's transmitter transforms the source data into semantically encoded information via semantic representation, semantic filtering, and semantic encoding in a sequential process. This process is carried out using the source KB w.r.t. a given communication goal/task. W.r.t. a communication goal/task and a destination KB that largely share common knowledge with a source KB, the receiver aims to take a desired action by acting on the output of the channel decoder via semantic decoding followed by semantic inference. The inference module's output -- for instance, in self-driving autonomous cars -- includes action execution instructions for accelerating and braking; changing the angle for the steering wheel and flashing the headlights; and responding to pedestrians, roadblocks, and traffic signal changes, among other actions \cite{SemCom_for_6G_Future_Internet'22}. At the receiver, each of these goals requires (possibly application/goal-tailored) SE followed by semantic filtering and semantic post-processing prior to source signal transmission \cite{Kalfa_Toward_GO_Semantic_Signal_Processing'21}, as depicted in Fig. \ref{fig: Goal_oriented_SSP_framework}. This figure shows the  task/goal-oriented semantic signal processing framework put forward by the authors of \cite{Kalfa_Toward_GO_Semantic_Signal_Processing'21}.

\begin{figure}[htb!]
	\centering
	\includegraphics[scale=0.26]{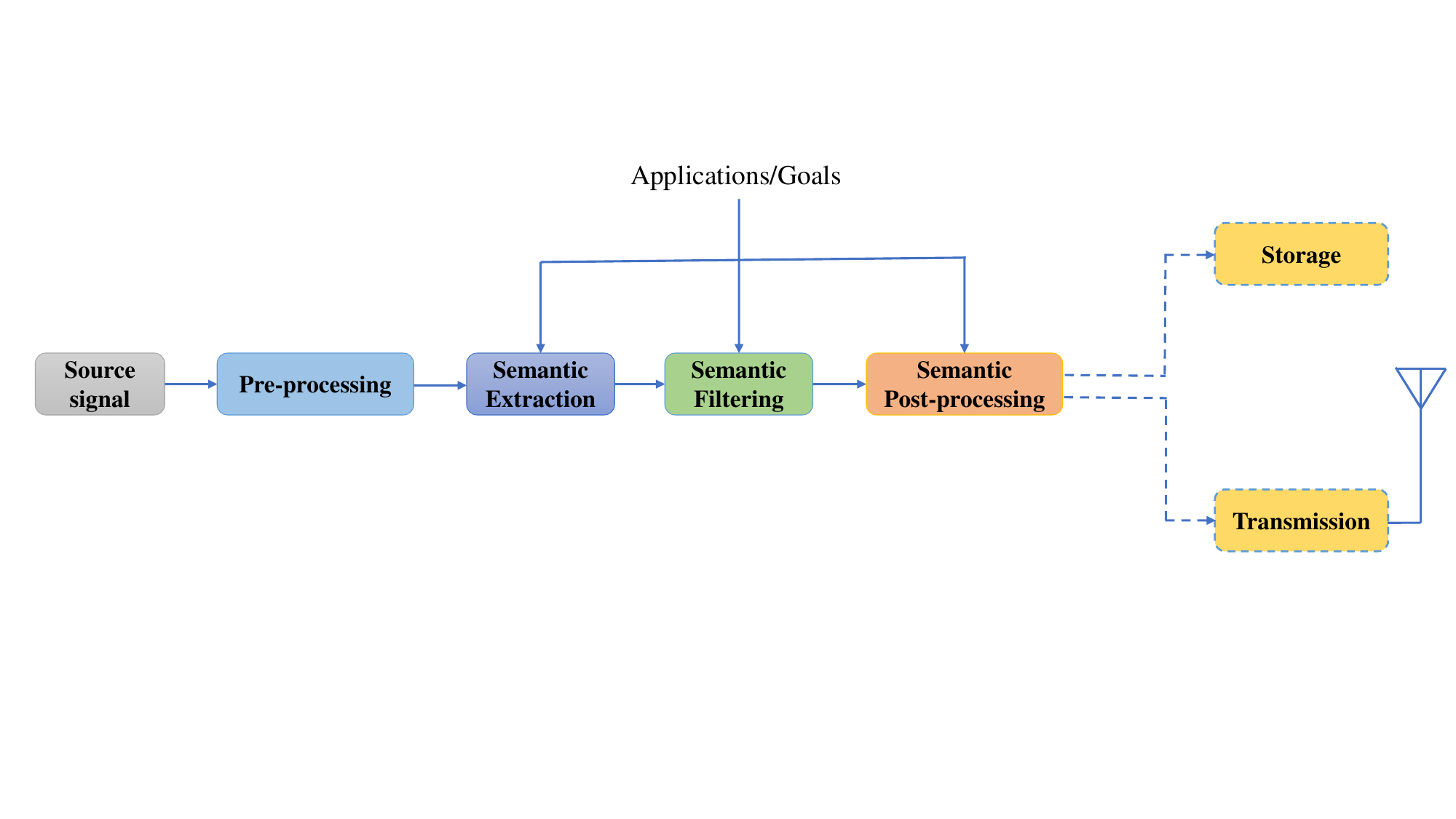} \vspace{-1cm}  
	\caption{Goal-oriented semantic signal processing framework \cite[Figure 12]{Kalfa_Toward_GO_Semantic_Signal_Processing'21}, \cite[Fig. 6]{Qin_Sem_Com_Principles_Apps'22}.}
	\label{fig: Goal_oriented_SSP_framework}
\end{figure}

The authors propose a framework that comprises pre-processing, SE,\footnote{In goal-oriented SemCom, SE ust necessarily capture pragmatic information \cite{SemCom_for_6G_Future_Internet'22}, whereas in SemCom, it revolves around semantic information.} semantic filtering, semantic post-processing, and storage/transmission in a sequence. When it comes to storage/transmission scheduling, the pre-processing block first transforms the input signal -- following a possible pre-filtering to reduce noise and/or interference -- into an appropriate domain for efficient component detection/classification \cite{Kalfa_Toward_GO_Semantic_Signal_Processing'21}. The SE block employs this transformed input under a time-varying application/goal and generates the corresponding multi-graph description and attribute sets \cite{Kalfa_Toward_GO_Semantic_Signal_Processing'21}. Thereafter, the semantic filtering block carries out semantic filtering per the local and time-varying goals to produce semantic data \cite{Kalfa_Toward_GO_Semantic_Signal_Processing'21}. The goal-filtered semantic data are then fed to the semantic post-processing block (see Fig. \ref{fig: Goal_oriented_SSP_framework}). The semantic post-processing block finally schedules -- while incorporating the (time-varying) local goals -- either transmission or storage per the receiver's communication goals/tasks \cite{Kalfa_Toward_GO_Semantic_Signal_Processing'21,Kalfa_Reliable_Extraction'23}. In the context of this goal-oriented semantic signal processing framework, the principal signal processing problems encountered in IoT networks, such as data compression, data clustering, data estimation, and ML, are related to the paradigm of goal-oriented SemCom \cite{Zhang_Goal-Oriented_Commun'22}.  

\begin{figure*}[htb!]
	\centering
	\includegraphics[scale=0.51]{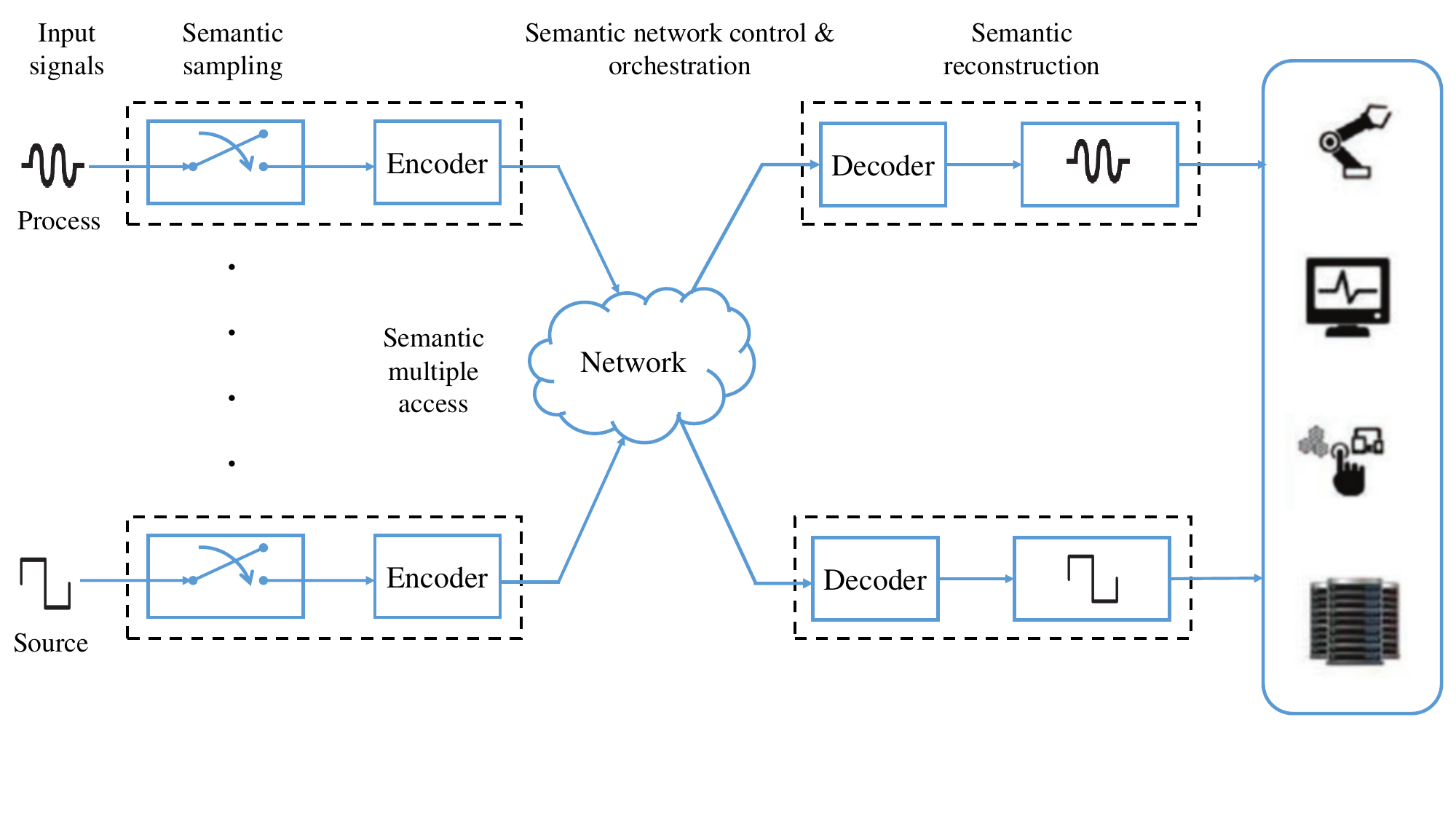} \vspace{-1.0cm}  
	\caption{End-to-end goal-oriented SemCom model \cite[FIGURE 1]{Kountouris_Semantics_EmpoweredCF'21}.}
	\label{fig: End_to_end_goal-oriented_SemCom_20221103}
\end{figure*}

When looking beyond conventional wireless connectivity, it is worth underscoring that communication is not an end in itself, but a means to achieving definite goals \cite{Kountouris_Semantics_EmpoweredCF'21}. The end-to-end goal-oriented SemCom model that is proposed by the authors of \cite{Kountouris_Semantics_EmpoweredCF'21} and depicted in Fig. \ref{fig: End_to_end_goal-oriented_SemCom_20221103} is therefore crucial. This figure comprises the following four building blocks.
\begin{itemize}
	\item \textit{Multiple continuous or discrete signals (stochastic processes);} various (possibly correlated) signals illustrating time-varying real-world physical phenomena in space are observed by spatially distributed smart devices \cite{Kountouris_Semantics_EmpoweredCF'21}. These devices are empowered by heterogeneous sensing, computational, and learning/inference capabilities \cite{Kountouris_Semantics_EmpoweredCF'21}. 
	
	\item \textit{A shared communication medium:} a shared medium is used jointly by smart devices to send data samples -- e.g., their observations, measurements, and updates -- to one or more destinations, such as a fusion center or a control unit \cite{Kountouris_Semantics_EmpoweredCF'21}. Their respective samples are generated using process-aware (non-uniform active) sampling n accordance with the communication characteristics, the semantic-aware applications’ requirements, and source variability (in terms of changes, innovation rate, autocorrelation, and self-similarity) \cite{Kountouris_Semantics_EmpoweredCF'21}.

	\item \textit{Preprocessing of source samples:} prior to being encoded and scheduled for transmission over noisy and delay-prone (error-prone) communication channels, source samples could be preprocessed \cite{Kountouris_Semantics_EmpoweredCF'21}. This preprocessing may incorporates quantization, compression, and feature extraction, among other processes \cite{Kountouris_Semantics_EmpoweredCF'21}. For goal-oriented SemCom, meanwhile, scheduling is performed per the semantic information's value and priority, which are extracted from the input data \cite{Kountouris_Semantics_EmpoweredCF'21}.
	
	\item \textit{Signal reconstruction:} the input signals are eventually reconstructed from causally or non-causally received samples at their respective destinations to serve the purpose of an application such as collision avoidance, remote state estimation, control and actuation, situation awareness, and learning model training \cite{Kountouris_Semantics_EmpoweredCF'21}.   
\end{itemize}

Apart from the aforementioned early works on goal-oriented SemCom, the authors of \cite{Phopovski_SE_Filtering'19} proffer a goal-oriented SemCom architecture (see Fig. \ref{fig: Architecture_with_SE_20220311.pdf}) with a \textit{semantic-effectiveness plane} \cite[Fig. 3]{Zhang_Wisdom_Evolutionary_6G'21} whose functionalities address both the semantic and effectiveness problems. When it comes to these problems, and as schematized in Fig. \ref{fig: Architecture_with_SE_20220311.pdf}, the architecture proposed in \cite[Fig. 1(b)]{Phopovski_SE_Filtering'19} supports not only information extraction, but also direct control.

\begin{figure}[htb!]
	%\centering
	\hspace{-1.35cm}\includegraphics[scale=0.35]{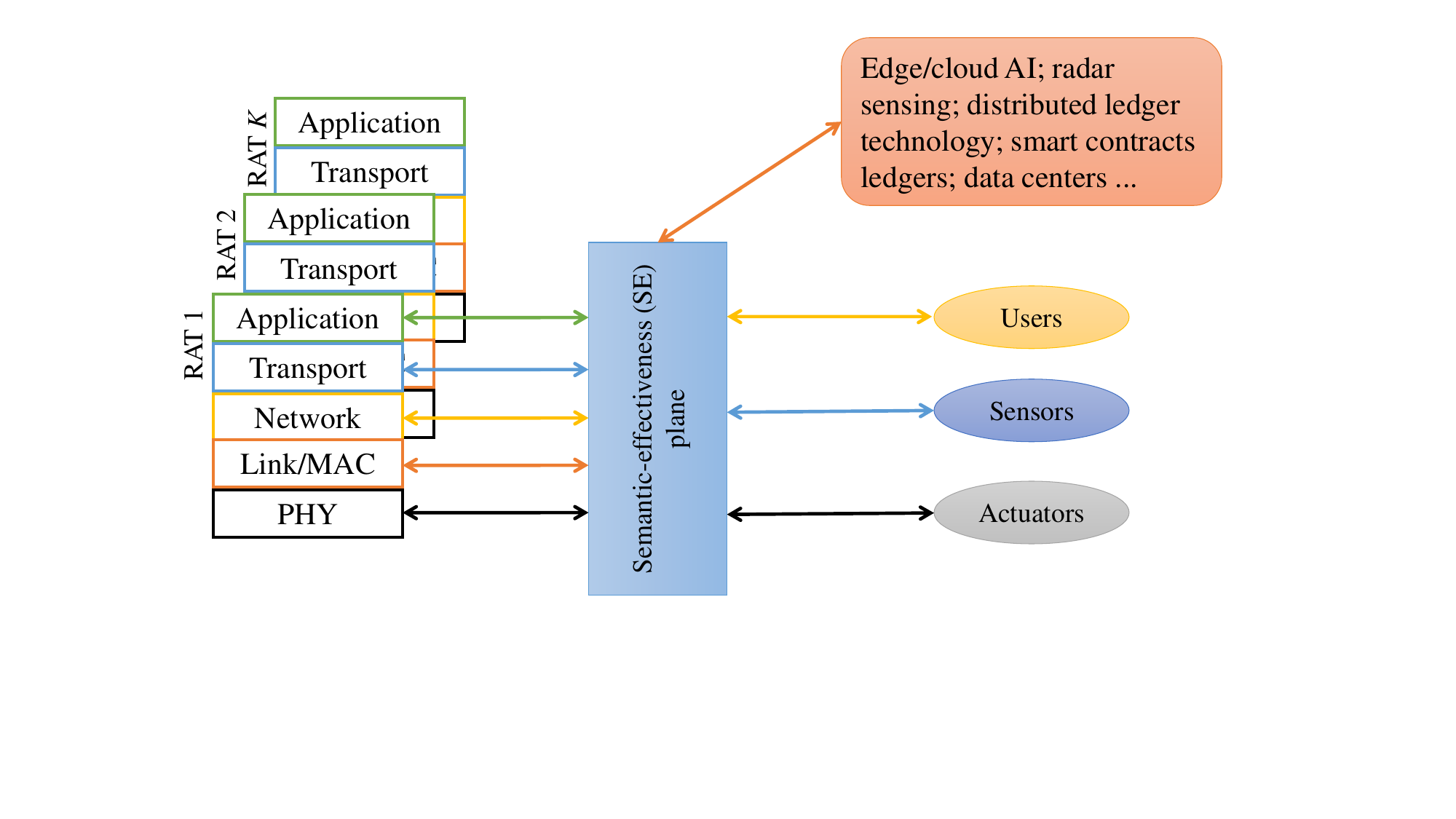}  \vspace{-1.5cm} 
	\caption{A goal-oriented SemCom architecture with semantic-effectiveness filtering -- \cite[Fig. 1]{Phopovski_SE_Filtering'19}: RAT: radio access technology.}
	\label{fig: Architecture_with_SE_20220311.pdf}
\end{figure}

We now proceed to state-of-the-art vision and tutorial works on goal-oriented SemCom.
	
\subsection{Vision and Tutorial Works on Goal-Oriented SemCom}
\label{subsec: vision_turotial_works_goal_oriented_SemCom}
We highlight below vision and tutorial works on goal-oriented SemCom, beginning with vision works.

\subsubsection{Vision Works on Goal-Oriented SemCom}
\label{subsubsec: vision_works_goal_oriented_SemCom}
the authors of \cite{Kountouris_Semantics_EmpoweredCF'21} envision a communication paradigm shift that requires the goal-oriented unification of information generation, information transmission, and information reconstruction while taking into account multiple factors such as process dynamics, data correlation, signal sparsity, and semantic information attributes. The authors of \cite{SemCom_Net_Systems'21} present a vision of a new paradigm shift that targets joint optimal information gathering, information dissemination, and decision-making policies in NCSs that incorporate the semantics of information based on the significance of the messages -- but not necessarily the meaning of the messages, and possibly with a real-time constraint -- w.r.t. the purpose of the data exchange. The authors of \cite{Strinati_Beyond_Shannon'20} present their vision of 6G wireless networks, wherein SemCom and goal-oriented SemCom are promising technologies that derive a crucial paradigm shift away from Shannon's information-theoretic framework. This paradigm shift underscores the fact that the success of task execution at a given destination (the effectiveness problem) is more of the essence than achieving error-free communication at the symbol level (the technical problem) \cite{Strinati_Beyond_Shannon'20}.

To ensure the concrete representation and efficient processing of the semantic information, the authors of \cite{Kalfa_Toward_GO_Semantic_Signal_Processing'21} introduce a formal graph-based semantic language and a goal filtering method for goal-oriented signal processing. Expanding upon this framework, the authors of \cite{Kalfa_Reliable_Extraction'23} introduce a semantic information extraction framework wherein the extracted graph-based imperfect semantic signals can be improved for better fidelity and filtered for reduced semantic source noise. The authors of \cite{Sana_Learning_Semantics'21} put forward an architecture that makes it possible to learn the representation of semantic symbols for goal-oriented SemCom (effectiveness-level SemCom) and design objective functions, which would help train effective semantic encoders/decoders. The authors of \cite{Effective_Commun_for_6G'22} present the challenges and opportunities related to goal-oriented SemCom networks while advocating goal-oriented SemCom as an enabler of 6G use cases.

We now proceed to highlight the existing tutorial works on goal-oriented SemCom.
 
\subsubsection{Tutorial Works on Goal-Oriented SemCom}
\label{subsubsec: turotial_works_goal_oriented_SemCom}
the authors of \cite{SemCom_for_6G_Future_Internet'22} provide a partial review of the fundamentals, applications, and challenges of goal-oriented SemCom. The authors of \cite{Gunduz_Beyond_Transmitting_Bits'22} offer a tutorial -- for communication theorists and practitioners -- that provides an introduction to state-of-the-art tools and advancements in goal-oriented SemCom. The authors of \cite{Zhang_Goal-Oriented_Commun'22} offer a partial overview of recent research developments in goal-oriented SemCom while focusing on goal-oriented data compression for IoT applications. The authors of \cite{Goal-oriented_SemCom_Thesis} review goal-oriented SemCom and semantic transformations.

Apart from the aforementioned vision and tutorial works on goal-oriented SemCom, the rapidly evolving state-of-the-art works on goal-oriented SemCom investigate numerous goal-oriented SemCom techniques, trends, and use cases such as task-oriented communication with digital modulation \cite{Xie_Robust_IB'22}; goal-oriented SemCom with AI tasks \cite{Yang_SemCom_with_AI_Tasks'21}; intent-based goal-oriented SemCom \cite{Thomas_NeuroSymb_AI_SemCom'22,Thomas_Neuro-Symbolic_Causal_Reasoning'22}; multi-user goal-oriented SemCom \cite{Xie_Task-Oriented_MU-SemCom'22}; and cooperative SemCom \cite{Xu_SemCom_for_IoV'22}.

We now move on to state-of-the-art algorithmic developments in goal-oriented SemCom. 

\subsection{Algorithmic Developments in Goal-Oriented SemCom}
\label{subsec: algorithmic_devs_goal_oriented_SemCom}
In this section, we detail state-of-the-art algorithms for single-user/single-task goal-oriented SemCom and multi-user/multi-task goal-oriented SemCom, starting with the former.

\subsubsection{Algorithms for Single-User/Single-Task Goal-Oriented SemCom}
\label{subsubsec: single-user_or_single-task_goal_oriented_SemCom}
the authors of \cite{Sem_aware_commun_SSP'21} aim to devise a joint sampling and communication scheme over a wireless multiple access channel to compute the empirical probability measure of a quantity of interest at the destination and put forward a goal-oriented SemCom strategy that encompasses both $(i)$ semantic-aware active sampling for goal-oriented signal reconstruction (at a fusion center) and $(ii)$ a transmission scheme to access the shared communication medium. The authors of \cite{Guo_Learning_Precoding'22}, on the other hand, propose a semantic information-aware policy for a MIMO-OFDM (orthogonal frequency division multiplexing) system -- whose goal is to classify images -- that is employed to transmit images to multiple users. In this goal-oriented SemCom system that is made up of a CNN-based transmitter and a CNN-based receiver, a graph neural network that is fed modulated symbols is deployed to learn a precoding policy \cite{Guo_Learning_Precoding'22}. The policy is demonstrated to outperform regularized zero-forcing precoding and zero-forcing precoding when it comes to minimizing the bandwidth consumed by required data to attain an expected level of classification accuracy \cite{Guo_Learning_Precoding'22}.

The authors of \cite{Yang_SemCom_with_AI_Tasks'21} underscore the premise that SemCom must take AI tasks into account and put forward a goal-oriented SemCom paradigm dubbed \textit{SemCom paradigm with AI tasks} (SC-AIT), which is schematized in Fig. \ref{fig: SemCom_with_AI_tasks_20221126}. Inspired by this goal-oriented SemCom systems (among others), the authors of \cite{Kang_TO_Image_Transmission'22} investigate a goal-oriented SemCom scheme for image classification task offloading in aerial systems in addition to proffering a joint SE-compression model. Their system is demonstrated to deliver an optimal SE under various channel states while taking into consideration the system's optimization objective that comprises the uplink transmission latency and the classification accuracy of the back-end target model \cite{Kang_TO_Image_Transmission'22}. Moreover, the authors of \cite{Farshbafan_Goal_Oriented_SemCom'21} proffer a curriculum learning-based SemCom framework for goal-oriented task execution. Building on this work, the authors of \cite{Farshbafan_Cur_Learning_Goal_SemCom'22} introduce a goal-oriented SemCom model that incorporates a speaker and a listener who wish to jointly execute a set of tasks for task execution in a dynamic environment with the objective of jointly optimizing task execution time, transmission cost, inference cost, and resource efficiency. To solve this optimization problem, the authors of \cite{Farshbafan_Cur_Learning_Goal_SemCom'22} provide an RL-based bottom-up curriculum learning framework that is shown to outperform traditional RL in terms of convergence time, task execution cost and time, reliability, and belief efficiency \cite{Farshbafan_Cur_Learning_Goal_SemCom'22}.

In view of emerging 6G applications such as  AR/VR online role-playing game, the authors of \cite{WANG_Multimodal_SemCom'23} proffer a MEC structure for goal-oriented multimodal SemCom, wherein the proposed structure deploys a bidirectional caching task model (a realistic model for emerging AI-enabled applications). More specifically, the authors of \cite{WANG_Multimodal_SemCom'23} put forward an offloading scheme with cache enhancement to minimize a system's computation cost by formulating the cache-computational resource coordination problem as a mixed integer non-linear programming problem. As a result, they develop the content popularity-based DQN caching algorithm (CP-DQN) to make quasi-optimal caching decisions and the cache-computing coordination algorithm (CCCA) to achieve a tradeoff between using computing resources and caching \cite{WANG_Multimodal_SemCom'23}. The CP-DQN and CCCA algorithms are shown to perform optimally w.r.t. cache hit rate, cache reward, and system cost reduction \cite{WANG_Multimodal_SemCom'23}. On the other hand, many current works are geared toward designing advanced algorithms for high-performance goal-oriented SemCom \cite{Kang_Personalized_Saliency'23}. Nonetheless, energy-hungry and efficiency-limited image retrieval and semantic encoding without considering user personality are major challenges for UAV image-sensing-driven goal-oriented SemCom scenarios \cite{Kang_Personalized_Saliency'23}. To overcome these challenges, the authors of \cite{Kang_Personalized_Saliency'23} devise an energy-efficient goal-oriented SemCom framework that uses a triple-based scene graph for image information. Meanwhile, the authors of \cite{Kang_Personalized_Saliency'23} develop a personalized attention-based mechanism to achieve the differential weight encoding of triplets for crucial information following user preferences and ensure personalized SemCom. This scheme's ability to achieve personalized SemCom is corroborated by numerical results \cite{Kang_Personalized_Saliency'23}.

The authors of \cite{Shao_IB_for_Edge_Inference'22} leverage the IB framework (see Appendix \ref{sec: on_IB}) to formalize a rate-distortion tradeoff between the encoded feature's informativeness and the inference performance and design a goal-oriented SemCom system. They incorporate variational approximation -- named variational IB (VIB) -- in their system to build a tractable upper bound w.r.t. IB optimization, which is computationally prohibitive for high-dimensional data. Meanwhile, their system is shown to achieve a better rate-distortion tradeoff than baseline methods while considerably reducing feature transmission latency in dynamic channel conditions \cite{Shao_IB_for_Edge_Inference'22}. Building on the work in \cite{Shao_IB_for_Edge_Inference'22}, the authors of \cite{TWC_Shao_Task_Oriented_Commun'21} put forward a goal-oriented SemCom strategy for multi-device cooperative edge inference wherein a group of edge devices transmit task-relevant features to an edge server for aggregation and processing by leveraging the IB principle \cite{IB_method_Tishby'00} and the distributed IB (DIB) framework \cite{Aguerri_Distributed_VRL'18}. The IB principle and the DIB framework are exploited in \cite{TWC_Shao_Task_Oriented_Commun'21} for feature extraction and distributed feature encoding, respectively. This IB- and DB-based goal-oriented SemCom technique is shown to significantly reduce communication overhead in comparison with conventional data-oriented communication and, in turn, enable low-latency cooperative edge inference \cite{TWC_Shao_Task_Oriented_Commun'21}. Building on the work in \cite{TWC_Shao_Task_Oriented_Commun'21} and in \cite{Shao_IB_for_Edge_Inference'22}, the authors of \cite{Shao_TO_Commun'22} study goal-oriented SemCom for edge video analytics by exploiting the deterministic IB principle \cite{Strouse_DIB'16} for feature extraction and the temporal entropy model for encoding. This goal-oriented SemCom scheme outperforms conventional data-oriented communication strategies in terms of its rate-performance tradeoff \cite{Shao_TO_Commun'22}.

Corresponding to the effectiveness level of Weaver's three levels of communication (see Fig. \ref{fig: Three_lavels_of_communication}), the authors of \cite{Tung_Effective_Commun'21} investigate a multi-agent partially observable Markov decision process (MA-POMDP), wherein agents not only interact with the environment but also communicate with each other over a noisy communication channel. In light of this multi-agent RL (MARL) framework, the authors of \cite{Tung_Effective_Commun'21} demonstrate that the joint policy that is learned by all the agents is far better than the one that is obtained by treating  the communication and principal MARL problems separately.

To minimize the amount of semantic information needing to be transmitted for a given task, many works on goal-oriented SemCom aim to transmit only task-relevant information without introducing any redundancy. Nevertheless, doing so with a JSCC-based design causes robustness issues in learning due to channel variation and JSCC, while mapping the source data directly to continuous channel input symbols poses compatibility issues with existing digital communication systems \cite{Xie_Robust_IB'22}. To address these challenges while examining the inherent tradeoff between the informativeness of the encoded representations and the robustness of the received representations to information distortion, the authors of \cite{Xie_Robust_IB'22} devise a goal-oriented SemCom system with digital modulation that is dubbed \textit{discrete task-oriented JSCC} (DT-JSCC). In DT-JSCC, the transmitter encodes the extracted input features into a discrete representation and transmits it to the receiver using digital modulation \cite{Xie_Robust_IB'22}. As for the DT-JSCC scheme's improved robustness to channel variation, the authors of \cite{Xie_Robust_IB'22} develop an IB-based encoding framework named \textit{robust IB} (RIB) and derive a tractable variational upper bound for the RIB objective function using variational approximation \cite{Xie_Robust_IB'22}. Consequently, DT-JSCC is shown to be robust against channel variation with better inference performance than low-latency baseline methods \cite{Xie_Robust_IB'22}.

The authors of \cite{Pappas_Goal-oriented_Commun'21} leverage the significance and effectiveness of messages to devise new goal-oriented sampling and communication policies as a means of generating and transmitting only the \textquotedblleft most informative samples'' for real-time tracking in autonomous systems. For these systems and the use cases mentioned, the results reported by the authors of \cite{Pappas_Goal-oriented_Commun'21} demonstrate that semantics-empowered policies considerably reduce real-time reconstruction error, the cost of actuation error, and the amount of ineffective updates \cite[Sec. 5]{Pappas_Goal-oriented_Commun'21}.

We now continue to state-of-the-art algorithms for multi-user/multi-task goal-oriented SemCom.

\subsubsection{Algorithms for Multi-User/Multi-Task Goal-Oriented SemCom}
\label{subsubsec: multi-user_or_multi-task_goal_oriented_SemCom}
in single-user goal/task-oriented SemCom, either the trained model has to be updated once the task is altered or several trained models need to be stored to serve different tasks \cite{Zhang_Unified_Multi-Task_SemCom'22}. To overcome this limitation, the authors of \cite{Zhang_Unified_Multi-Task_SemCom'22} develop a unified DL-enabled SemCom system named \textit{U-DeepSC}. U-DeepSC is a unified end-to-end framework that is designed to serve various tasks with multiple modalities \cite{Zhang_Unified_Multi-Task_SemCom'22,Zhang_Multi-Task_SemCom_with_Domain_Adaptation'22}. Moreover, the authors of \cite{Zhang_Unified_Multi-Task_SemCom'22} devise a multi-exit architecture in U-DeepSC to provide early-exit results for relatively simple tasks and design a unified codebook for feature representation to serve different tasks with reduced transmission overhead. 

Aiming to exploit multimodal data from multiple users, the authors of \cite{Xie_TO_MU_SemCom'21} propose a multi-user task-oriented SemCom system for visual question answering (VQA), named \textit{MU-DeepSC}, to exploit multimodal data from multiple users. MU-DeepSC is a DL-enabled goal-oriented SemCom system whose transceiver is designed and optimized to jointly capture features from the correlated multimodal data of multiple users \cite{Xie_TO_MU_SemCom'21}. Consequently, MU-DeepSC is demonstrated to be more robust to channel variation than traditional communication systems, especially in low SNR regimes \cite{Xie_TO_MU_SemCom'21}. Building on the work in \cite{Xie_TO_MU_SemCom'21}, the authors of \cite{Xie_Task-Oriented_MU-SemCom'22} design and implement multi-user task-oriented SemCom systems for the transmission of both data with one modality and data with multiple modalities. The authors consider image retrieval / machine translation for their single-modal task and VQA for their multimodal task \cite{Xie_Task-Oriented_MU-SemCom'22}. The authors of \cite{Xie_Task-Oriented_MU-SemCom'22} develop three Transformer-based transceivers for their systems, which are dubbed \textit{DeepSC-IR}, \textit{DeepSC-MT}, and \textit{DeepSC-VQA}, that share the same transmitter structure but have different receiver structures \cite{Xie_Task-Oriented_MU-SemCom'22}. When these transceivers are trained jointly in an end-to-end manner using the training algorithms in \cite{Xie_Task-Oriented_MU-SemCom'22}, they are corroborated to outperform traditional transceivers, especially in low SNR regimes \cite{Xie_Task-Oriented_MU-SemCom'22}.
		
Apart from the above-detailed algorithmic developments in goal-oriented SemCom, useful algorithmic developments have also been made in goal-oriented SemCom resource allocation, which we discuss below.
 
\subsection{Algorithmic Developments in Goal-Oriented SemCom Resource Allocation}
\label{subsec: Go_SemCom_resource_allocation}
The authors of \cite{Binucci_Dynamic_Resource_Allocation'22} consider a multi-user goal-oriented SemCom system at the wireless edge and exploit the Lyapunov optimization framework to devise a joint computation and transmission management strategy for their overall system. More specifically, the authors of \cite{Binucci_Dynamic_Resource_Allocation'22} develop a multi-user minimum energy resource allocation strategy that ensures energy-efficient optimal resource allocation for edge devices and edge server. This resource allocation strategy's simulation results demonstrate there is an edge-ML trade-off between energy, latency, and accuracy \cite{Binucci_Dynamic_Resource_Allocation'22}. Extending the work in \cite{Binucci_Dynamic_Resource_Allocation'22}, the authors of \cite{Binucci_adaptive_resource_optimization'22} investigate the trade-offs between energy, latency, and accuracy in a goal-oriented SemCom-enabled edge learning system. More specifically, they develop two resource optimization strategies (that also exploit the Lyapunov stochastic optimization framework) to jointly optimize the communication parameters and the computation resources while aiming for an optimal trade-off between energy, latency, and accuracy for the edge learning task \cite{Binucci_adaptive_resource_optimization'22}. These proposed strategies are corroborated -- by extensive simulations -- to provide adaptation capabilities and to be effective for edge learning with goal-oriented SemCom \cite{Binucci_adaptive_resource_optimization'22}.

When it comes to personalized saliency-based goal-oriented SemCom in UAV image sensing scenarios, the authors of \cite{Kang_Personalized_Saliency'23} investigate SemCom personalization and its corresponding optimal resource allocation. For the former, the authors theoretically analyze the effects of wireless fading channels on SemCom, and for the latter, they} put forward a game-based model for multi-user resource allocation (to efficiently utilize UAV resources). The resource allocation framework is confirmed to improve UAV resource utilization \cite{Kang_Personalized_Saliency'23}.

The authors of \cite{Liu_Adaptable_Semantic_Compression'22} propose a multi-user goal-oriented SemCom framework that aims o enable users to effectively extract, compress, and transmit the semantics of input data to the edge server. The edge server then executes the intelligence task based on the received semantics and delivers results to users \cite{Liu_Adaptable_Semantic_Compression'22}. Meanwhile, the authors of \cite{Liu_Adaptable_Semantic_Compression'22} also propose a new approach dubbed \textit{adaptable semantic compression} (ASC) to compress the extracted semantics based on semantic importance, which helps to reduce the communication burden. However, ASC faces the following problem in a multi-user setting: a higher compression ratio requires fewer channel resources but causes considerable semantic distortion, while a lower compression ratio calls on more channel resources and hence results in transmission failure due to the delay constraint (especially in delay-intolerant systems) \cite{Liu_Adaptable_Semantic_Compression'22}. In light of this problem, the authors of \cite{Liu_Adaptable_Semantic_Compression'22} formulate a resource allocation and compression ratio optimization problem that aims to maximize the \textit{success probability of tasks}\footnote{The success probability of tasks is defined to quantify the performance of goal-oriented SemCom systems and is given by \cite[eq. (11)]{Liu_Bandwidth_and_Power_Allocation'22}.} under bandwidth and power constraints. In addressing this non-convex problem, the authors of \cite{Liu_Adaptable_Semantic_Compression'22} develop two algorithms that achieve greater task performance gains than the baseline algorithms do while significantly reducing the volume of data transmitted \cite[Sec. VI]{Liu_Adaptable_Semantic_Compression'22}.

\begin{figure*}[t!]
	\centering
	\includegraphics[scale=0.5]{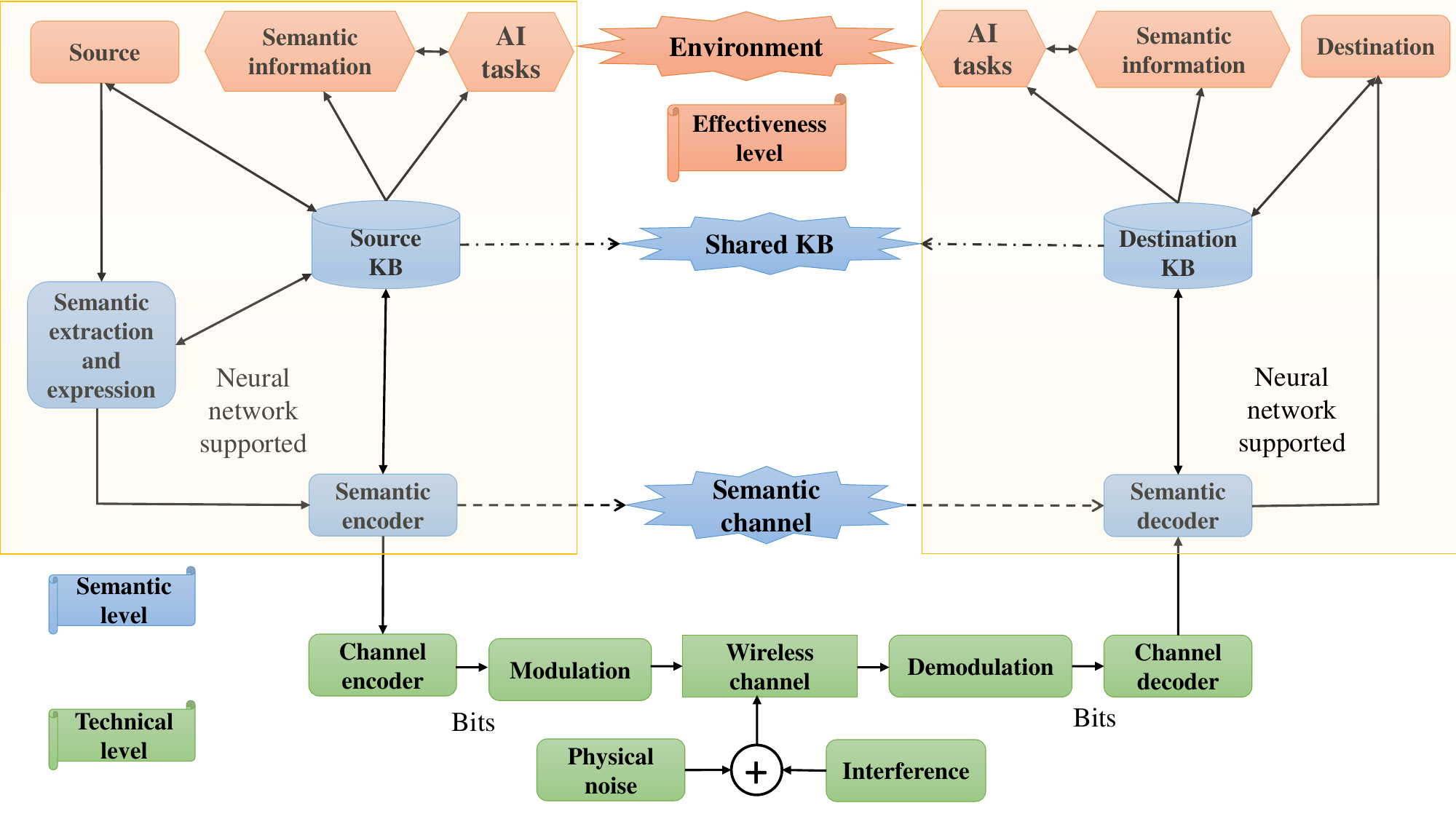}  \vspace{3mm} 
	\caption{Goal-oriented SemCom with AI tasks -- modified from \cite[Fig. 2]{Yang_SemCom_with_AI_Tasks'21}.}
	\label{fig: SemCom_with_AI_tasks_20221126}
\end{figure*}

We now continue to major state-of-the-art trends and use cases of goal-oriented SemCom.

\section{Major State-of-the-art Trends and Use Cases of Goal-Oriented SemCom}
\label{sec: Goal_oriented_SemCom_Major_Trends_Use-Cases}
In this section, we present the major state-of-the-art trends and use cases related to goal-oriented SemCom, beginning with the major trends.

\subsection{Major Trends of Goal-Oriented SemCom}
\label{subsec: major_trends_goal_oriented_SemCom}
We discuss the following major trends of goal-oriented SemCom: \textit{goal-oriented SemCom with AI tasks} \cite{Yang_SemCom_with_AI_Tasks'21}, \textit{neuro-symbolic AI for intent-based goal-oriented SemCom} \cite{Thomas_NeuroSymb_AI_SemCom'22}, \textit{multi-user goal-oriented SemCom} \cite{Xie_Task-Oriented_MU-SemCom'22}, and \textit{cooperative goal-oriented SemCom} \cite{Xu_SemCom_for_IoV'22}. We start our discussion with goal-oriented SemCom with AI tasks.   

\subsubsection{Goal-Oriented SemCom with AI Tasks}
the authors of \cite{Yang_SemCom_with_AI_Tasks'21} were the first to assert that semantic information is closely related to the target AI task. This assertion is indeed reasonable when one considers the detection of a dog from a transmitted image that comprises both a dog and a cat (see \cite[Fig. 1]{Yang_SemCom_with_AI_Tasks'21}), since the information related to the cat is no longer relevant. For this goal-oriented SemCom scenario, the authors of \cite{Yang_SemCom_with_AI_Tasks'21} put forward a goal-oriented SemCom system dubbed goal-oriented SemCom with AI tasks, which is shown in Fig. \ref{fig: SemCom_with_AI_tasks_20221126}. Fig. \ref{fig: SemCom_with_AI_tasks_20221126} shows the technical and semantic levels -- per Weaver's vision (shown in Fig. \ref{fig: Three_lavels_of_communication}) -- and a newly proposed effectiveness level. In their effectiveness level design, the authors propose to minimize the redundancy in the semantic information based on the contribution of raw information to the successful execution of AI tasks by discarding the information that is irrelevant to the success of AI tasks. This process can be conducted per the knowledge stored in the source KB that can be designed to account for the relationships between the AI tasks and the semantic information \cite{Yang_SemCom_with_AI_Tasks'21}.

Once encoded using a semantic encoder, the semantic information is then is then channel-encoded and modulated prior to its transmission over a wireless channel. The received semantic information, which may be contaminated by physical noise and interference is then demodulated and channel-decoded, as seen in Fig. \ref{fig: SemCom_with_AI_tasks_20221126}. This information is fed to a semantic-level receiver (as shown in Fig. \ref{fig: SemCom_with_AI_tasks_20221126}), whose semantic decoder is employed to recover the transmitted semantic information in accordance with the destination KB. The destination KB can \textit{synchronize} its knowledge elements with those of the source KB through a shared KB that can be either stored in an authoritative third party or a virtual KB \cite{Yang_SemCom_with_AI_Tasks'21}.

We now proceed to neuro-symbolic AI for intent-based goal-oriented SemCom \cite{Thomas_NeuroSymb_AI_SemCom'22}.

\begin{figure*}[t!]
	\centering
	\includegraphics[scale=0.5]{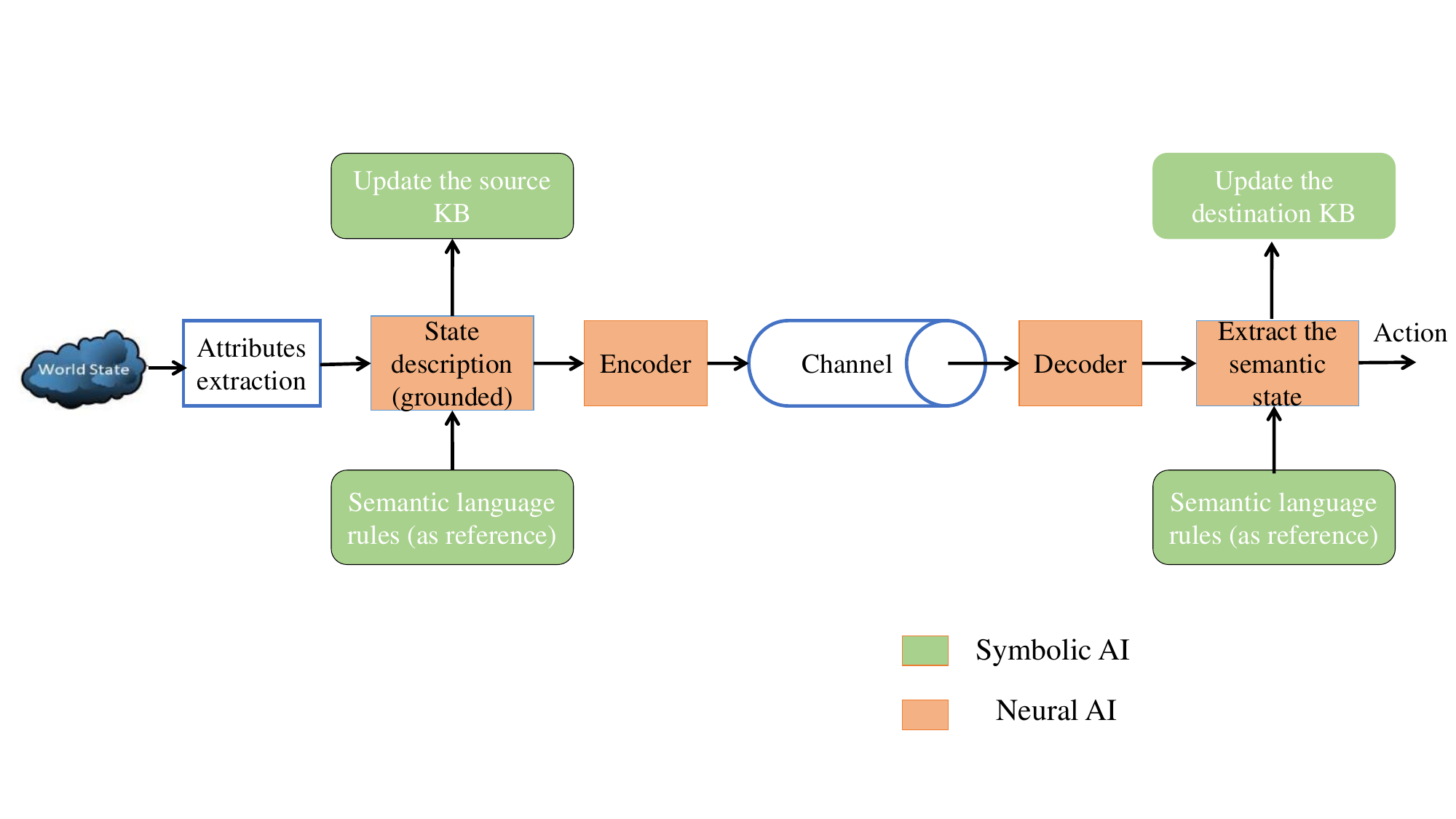}  \vspace{-0.25cm} 
	\caption{Intent-based goal-oriented SemCom (NeSy AI) \cite[Fig. 1]{Thomas_NeuroSymb_AI_SemCom'22}.}
	\label{fig: Intent-based_SemCom_20221116}
\end{figure*}

\subsubsection{Neuro-Symbolic AI for Intent-Based Goal-Oriented SemCom} in contrast with the state-of-the-art works on goal-oriented SemCom that characteristically lack data explainability, the work in \cite{Thomas_NeuroSymb_AI_SemCom'22} leverages \textit{neuro-symbolic AI} (NeSy AI) \cite{besold2017neuralsymbolic,Garcez_Neurosymbolic_AI'20} and generative flow networks (GFlowNets) \cite{Bengio_FlowNB_Gen_Models'21} to introduce a goal-oriented SemCom model named \textit{NeSy AI} that aims to bring intelligence to the end nodes and is depicted in Fig. \ref{fig: Intent-based_SemCom_20221116}. As is shown in Fig. \ref{fig: Intent-based_SemCom_20221116}, NeSy AI's transmitter comprises an attribute extraction module, a state description module, and an encoder. When it comes to the latter two components, the state description module is learnable using neural AI and grounded in real logic according to the semantic language rules that are embedded in symbolic AI, and the encoder is realizable using neural AI and translates the states to (optimal) physical messages \cite{Thomas_NeuroSymb_AI_SemCom'22}. The receiver, on the other hand, is made up of a decoder and a semantic state extraction module, as shown in Fig. \ref{fig: Intent-based_SemCom_20221116}. As can be seen in Fig. \ref{fig: Intent-based_SemCom_20221116}, the decoder (which is designed using neural AI) transforms the received message into an estimated state description that is fed to the SE module (which is also designed using neural AI) that effectively recovers the transmitted semantic states in accordance with the reference semantic language rules (which are realizable using symbolic AI) \cite{Thomas_NeuroSymb_AI_SemCom'22}.

In NeSy AI, the symbolic part is elaborated by the KB, and the reasoning -- from learning the probabilistic structure that generates the data -- is enabled by the GFlowNet \cite{Bengio_FlowNB_Gen_Models'21}. The authors of \cite{Thomas_NeuroSymb_AI_SemCom'22}, thus, formulate an optimization problem for causal structure learning -- from the data and the optimal encoder/decoder functions -- whose simulation results indicate it needs to transmit considerably fewer bits than a conventional communication system to reliably convey the same meaning \cite{Thomas_NeuroSymb_AI_SemCom'22}. Building on the work in \cite{Thomas_NeuroSymb_AI_SemCom'22} and NeSy AI, the authors of \cite{Thomas_Neuro-Symbolic_Causal_Reasoning'22} introduce a goal-oriented SemCom framework named \textit{emergent semantic communication} (ESC). ESC is made up of a signaling game for emergent language design and a NeSy AI approach for causal reasoning \cite{Thomas_Neuro-Symbolic_Causal_Reasoning'22}. To design an emergent language that is compositional and semantic-aware, the authors of \cite{Thomas_Neuro-Symbolic_Causal_Reasoning'22} solve the signaling game -- using alternating maximization between the transmit and receive nodes' utilities -- and characterize the generalized Nash equilibrium. The authors also deploy GFlowNet \cite{Bengio_FlowNB_Gen_Models'21} to induce causal reasoning at the nodes and prove analytically that ESC systems can encode data with minimal bits in comparison with a classical system that does not employ causal reasoning \cite{Thomas_Neuro-Symbolic_Causal_Reasoning'22}.

We now move on to our discussion of multi-user goal-oriented SemCom systems \cite{Xie_Task-Oriented_MU-SemCom'22}.
\begin{figure*}[t!]
	\centering
	\includegraphics[scale=0.5]{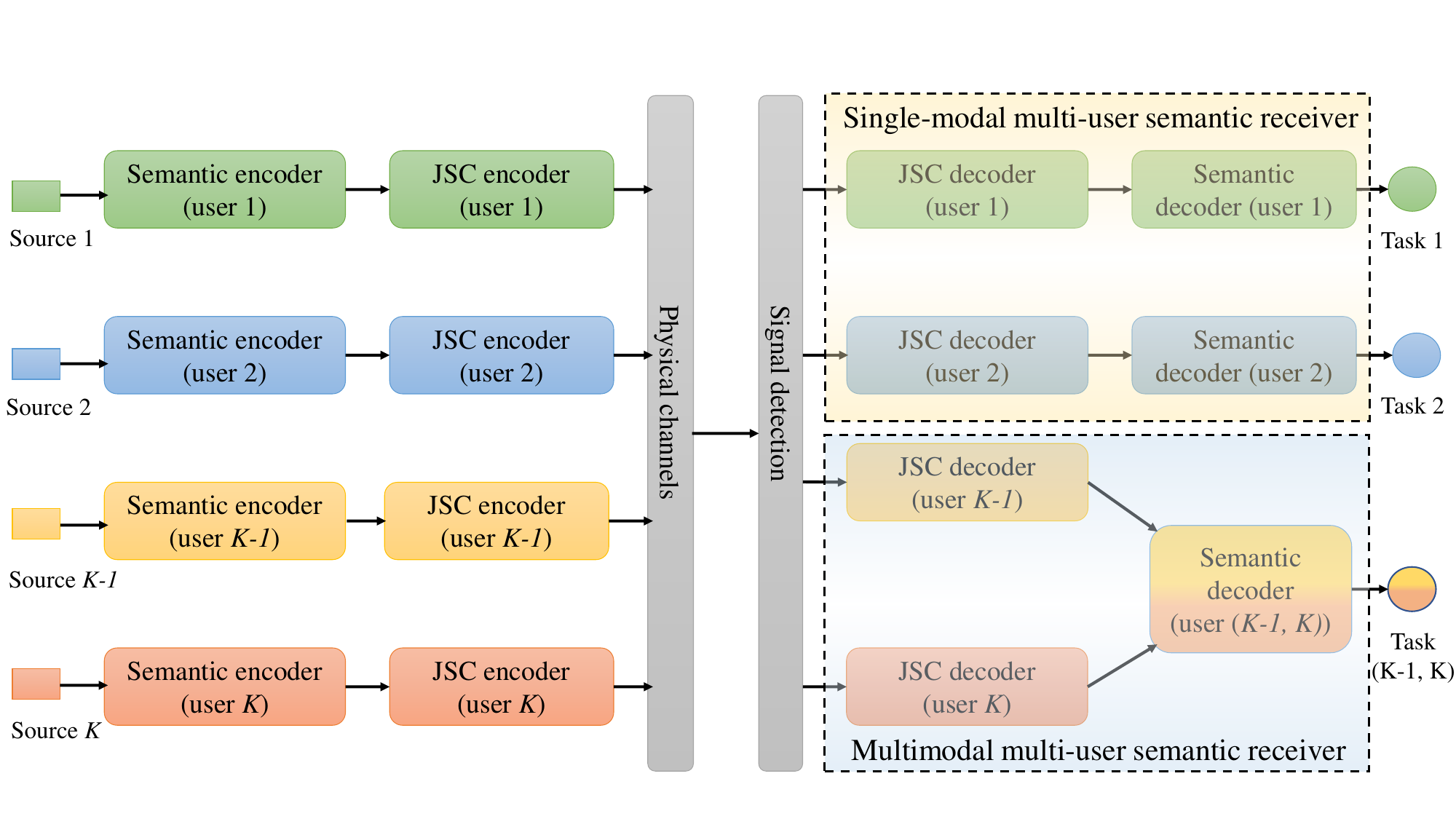}  \vspace{3mm} 
	\caption{Multi-user goal-oriented SemCom systems \cite[Fig. 1]{Xie_Task-Oriented_MU-SemCom'22}: JSC -- joint source-channel.}
	\label{fig: MU-SemCom_Systems_20221126}
\end{figure*}

\subsubsection{Multi-User Goal-Oriented SemCom}
the authors of \cite{Xie_Task-Oriented_MU-SemCom'22} devise a multi-user goal-oriented SemCom system, which is depicted in Fig. \ref{fig: MU-SemCom_Systems_20221126}, to extend the benefits of single-user, single-modal goal-oriented SemCom to multiple users. Their proposed system is a multi-user MIMO system that is made up of a receiver equipped with $M$ antennas and $K$ single-antenna transmitters \cite{Xie_Task-Oriented_MU-SemCom'22}. Each of the transmitters consists of a DL-based semantic encoder and a JSC encoder (both of which are learnable in an end-to-end fashion), and accepts image, text, video, or speech signals as input \cite{Xie_Task-Oriented_MU-SemCom'22}. The receiver, on the other hand, can either be a \textit{single-modal multi-user semantic receiver} and enable single-modal multi-user data transmission or be a \textit{multimodal multi-user semantic receiver} and enable multimodal multi-user data transmission, as can be seen in Fig. \ref{fig: MU-SemCom_Systems_20221126}  \cite{Xie_Task-Oriented_MU-SemCom'22}.

Single-modal multi-user transmission means that each user independently transmits its extracted semantic information to carry out its task \cite{Xie_Task-Oriented_MU-SemCom'22}. Multimodal multi-user transmission, on the other hand, means that the data from different users are semantically complementary \cite{Xie_Task-Oriented_MU-SemCom'22}. Each of these goal-oriented SemCom systems relies on the linear minimum MSE (L-MMSE) detector to recover signals with estimated CSI \cite{Xie_Task-Oriented_MU-SemCom'22}. Each user's JSC decoder is designed/trained to decompress the received semantic information following L-MMSE detection while mitigating the effects of channel distortion and inter-user interference \cite{Xie_Task-Oriented_MU-SemCom'22}. When the JSC decoder is used in sequence with the semantic decoder to form a single-modal multi-user semantic receiver, as schematized in Fig. \ref{fig: MU-SemCom_Systems_20221126}, each user's semantic information is exploited to perform different tasks independently \cite{Xie_Task-Oriented_MU-SemCom'22}. This single-modal multi-user goal-oriented SemCom system can be used for the joint performance of an image retrieval task and a machine translation task, as in \cite[Fig. 2]{Xie_Task-Oriented_MU-SemCom'22}. Moreover, as is shown in Fig. \ref{fig: MU-SemCom_Systems_20221126}, the final task that corresponds to a multimodal semantic receiver is completed by merging the different users' semantic information \cite{Xie_Task-Oriented_MU-SemCom'22}. This multi-user goal-oriented SemCom system is useful for realizing a multimodal multi-user goal-oriented SemCom system with a \textit{DeepSC-VQA} transceiver, as shown in \cite[Fig. 3]{Xie_Task-Oriented_MU-SemCom'22}.

The authors of \cite{Xu_SemCom_for_IoV'22} build on the multimodal multi-user goal-oriented SemCom system that is depicted in Fig. \ref{fig: MU-SemCom_Systems_20221126} and put forward a goal-oriented SemCom system named cooperative goal-oriented SemCom, which we discuss below.

\begin{figure*}[t!]
	\centering
	\includegraphics[scale=0.5]{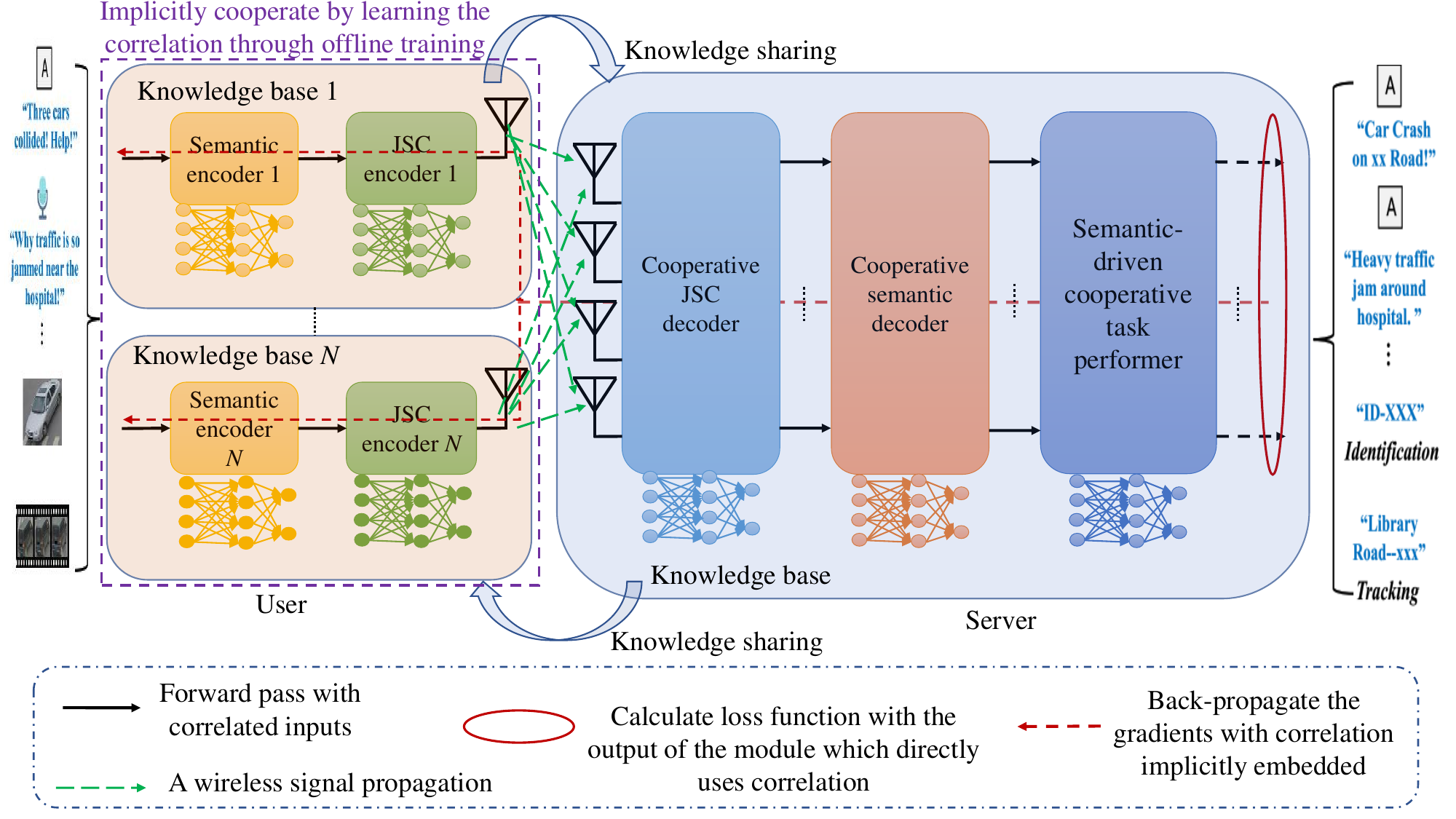}  \vspace{3mm} 
	\caption{An architecture for a general cooperative goal-oriented SemCom -- modified from \cite[Fig. 2]{Xu_SemCom_for_IoV'22}: JSC -- joint source and channel.}
	\label{fig: Coperative_SemCom_20230302}
\end{figure*}
\subsubsection{Cooperative Goal-Oriented SemCom}
it is proposed for IoV applications such as pedestrian detection, traffic analysis, and vehicle tracking \cite{Xu_SemCom_for_IoV'22}. A general cooperative goal-oriented SemCom architecture is shown in Fig. \ref{fig: Coperative_SemCom_20230302}. As can be seen in Fig. \ref{fig: Coperative_SemCom_20230302}, cooperative goal-oriented SemCom comprises a semantic encoder and a cooperative semantic decoder, a JSC encoder and a cooperative JSC decoder, and a semantic-driven cooperative task performer. Interestingly, the correlation among users is \textit{pre-learned} and \textit{embedded} in the cooperative goal-oriented SemCom architecture, including the encoders at the transmitters and the cooperative modules at the receiver \cite{Xu_SemCom_for_IoV'22}. The different modules' functions are itemized below.
\begin{itemize}
	\item \textit{Semantic encoder:} it is designed to extract semantic information from the source data with a focus on meaning and goal-relevance \cite{Xu_SemCom_for_IoV'22}. 
	
	\item \textit{Cooperative semantic decoder:} it recovers the source data according to the specific goals set while leveraging semantic-level correlation among users \cite{Xu_SemCom_for_IoV'22}.
	
	\item \textit{JSC encoder:} it is applied to encode the extracted semantic information (the output of the semantic encoder) as channel input symbols \cite{Xu_SemCom_for_IoV'22}.
	
	\item \textit{Cooperative JSC decoder:} it is realized/trained to jointly recover the transmitted semantic information of multiple users, as depicted in Fig. \ref{fig: Coperative_SemCom_20230302}.
	
	\item \textit{Semantic-driven cooperative task performer:} it is used to achieve specific tasks/actions while adapting its structure to a specific task based on the semantic information recovered from multiple users (its input) \cite{Xu_SemCom_for_IoV'22}. It also leverages semantic-level correlation and distinctive user attributes while cooperatively performing a task by combining information provided by different users \cite{Xu_SemCom_for_IoV'22}.
\end{itemize}

The cooperative goal-oriented SemCom scheme shown in Fig. \ref{fig: Coperative_SemCom_20230302} requires that knowledge be shared between the transmitters and a receiver and that each user have a background KB \cite{Xu_SemCom_for_IoV'22}. The background KBs are presumed to be shared between users and the server by jointly training the whole DNN offline with a common dataset \cite{Xu_SemCom_for_IoV'22}.

We now move on to major use cases of goal-oriented SemCom. 

\subsection{Major Use Cases of Goal-Oriented SemCom}
\label{subsec: major_use_cases_goal_oriented_SemCom}
Similar to H2H SemCom, H2M SemCom, and M2M SemCom \cite{Qiao_What_is_SemCom'21}, the major use cases of goal-oriented SemCom can be classified as \textit{H2H goal-oriented SemCom}, \textit{H2M goal-oriented SemCom}, and \textit{M2M goal-oriented SemCom}.\footnote{Although the state-of-the-art goal-oriented SemCom literature comprises little to none information on the use cases of \textit{H2H goal-oriented SemCom} and \textit{H2M goal-oriented SemCom}, they are applicable for designing 6G H2H and H2M communication systems, respectively.} Major use cases of M2M goal-oriented SemCom include autonomous transportation, consumer robotics, environmental monitoring, telehealth, smart factories, and NCSs, which are highlighted below. We begin our discussion with autonomous transportation, consumer robotics, environmental monitoring, and telehealth. 

\subsubsection{Autonomous Transportation, Consumer Robotics, Environmental Monitoring, and Telehealth}
supporting the scalability of future massive networked intelligent systems, semantic-empowered communication will support the scalability of future massive networked intelligent systems and enhance network resource usage, energy consumption, and computational efficiency significantly, and thus pave the way for the design of next-generation real-time data networking \cite{Kountouris_Semantics_EmpoweredCF'21}. This type of semantic networking will make it possible to transmit only informative data samples and convey only the information that is relevant, useful, and valuable for achieving its defined goals \cite{Kountouris_Semantics_EmpoweredCF'21}. Accordingly, goal-oriented SemCom will provide the foundational technology for dozens of socially beneficial services, including autonomous transportation, consumer robotics, environmental monitoring, and telehealth \cite{Kountouris_Semantics_EmpoweredCF'21}. 

We now proceed to our brief discussion of smart factories.   

\subsubsection{Smart Factories}  
in smart factories of the future, it will be crucial to limit the operation of machines to performing specific actions \cite{Luo_SemCom_Overview'22}. In this vein, goal-oriented SemCom can be designed and employed to convey only the semantic information of the control signals \cite{Luo_SemCom_Overview'22}. Thus, smart factories can reduce their communication cost and improve their operational efficiency by deploying goal-oriented SemCom \cite{Luo_SemCom_Overview'22}.

We now continue with our brief discussion of NCSs.

\subsubsection{NCSs}
emerging and futuristic NCSs are a major use case of goal-oriented SemCom and require the joint optimization of the communication and control objectives \cite{SemCom_Net_Systems'21}. State-of-the-art communication technologies, on the other hand, are agnostic to control objectives and pursue communication network and control system optimization separately, which is likely to yield suboptimal solutions by narrowing the solution space of the problem in both areas \cite{SemCom_Net_Systems'21}. Accordingly, massive-scale NCSs can be enabled by unifying control and communication techniques under the umbrella of semantics of information. To this end, fundamentally re-designing the techniques for information generation, transmission, transport, and reconstruction to optimize the performance of applications that would utilize this information \cite{SemCom_Net_Systems'21} is of paramount importance.

To summarize, goal-oriented SemCom also has many other applications and use cases, including fault detection \cite{Stamatakis_Semantic-Aware_Active_Fault_Detection'22}. We now move on to our discussion of state-of-the-art theories of goal-oriented SemCom.   

\section{Theories of Goal-Oriented SemCom}
\label{subsec: theories_of_Goal_Oriented_SemCom}
 In this section, we discuss major developments in goal-oriented SemCom theory. We detail below the rate-distortion approach to goal-oriented SemCom \cite{Stavrou_Fidelity_goal_priented_SemCom'22,Stavrou_Rate_distortion_approach_SemCom'22}; the extended rate-distortion approach to goal-oriented SemCom \cite{Liu_Task-oriented_SemCom'22}; and goal-oriented quantization (GOQ) \cite{Zou_JSAC_GO-Quantization'23,Zou_GO-Quantization_WiOpt'22}. We begin with the rate-distortion approach to goal-oriented SemCom and, in particular the role of fidelity in goal-oriented SemCom \cite{Stavrou_Fidelity_goal_priented_SemCom'22}. 

\subsection{Rate-Distortion Approach to Goal-Oriented SemCom}
\label{subsec: Rate_distortion_approach_GO_SemCom}
The authors of \cite{Stavrou_Fidelity_goal_priented_SemCom'22} develop a theory that asserts that choosing the type of individual distortion measures (or context-dependent fidelity criteria) per the application/task requirements can considerably affect the semantic source's remote reconstruction. The authors develop their theory by adopting the problem setup proposed by the authors of \cite{Liu_RD-Characterization_TCOM'22}, which is schematized in Fig. \ref{fig: Semantic_RDT_pic.pdf}. The authors of \cite{Stavrou_Fidelity_goal_priented_SemCom'22} consider a memoryless source that is represented by the tuple $(\bm{x}, \bm{z})$ and has a joint probability density function (PDF) $p(x, z)$ in the product alphabet space $\mathcal{X} \times \mathcal{Z}$. Here, $\bm{x}$ is the source's semantic or intrinsic information (directly unobservable) and $\bm{z}$ is the noisy observation of the source at the encoder side \cite{Stavrou_Fidelity_goal_priented_SemCom'22}.    

The system model that the authors of \cite{Stavrou_Fidelity_goal_priented_SemCom'22} adopted in the above-mentioned setup is shown in \cite[Fig. 1]{Stavrou_Fidelity_goal_priented_SemCom'22}. Per \cite[Fig. 1]{Stavrou_Fidelity_goal_priented_SemCom'22} and its accompanying assumption, an information source is a sequence of $n$ independent and identically distributed (i.i.d.) RVs $(\bm{x}^n, \bm{z}^n)$, and the PDFs $p(x)$ and $p(z|x)$ are assumed to be known \cite{Stavrou_Fidelity_goal_priented_SemCom'22}. Meanwhile, the encoder ($E$) and the decoder ($D$) are defined through the following mappings \cite[eq. (1)]{Stavrou_Fidelity_goal_priented_SemCom'22}:
\begin{subequations}
\begin{align}
\label{f_E_def}
f^E: & \mathcal{Z}^n \to \mathcal{W}     \\
\label{g_o_D_def}
g_o^D: &  \mathcal{W} \to \hat{\mathcal{Z}}^n  \\
\label{g_s_D_def}
g_s^D: &  \mathcal{W} \to \hat{\mathcal{X}}^n,  
\end{align}
\end{subequations}
where $\mathcal{W} \in [M]$, $g_o^D$ is the observation decoder, and $g_s^D$ is the semantic information decoder. If one now considers two per-letter distortion measures that are defined by $d_s: \mathcal{X} \times \hat{\mathcal{X}} \rightarrow [0, \infty)$ and $d_o: \mathcal{Z} \times \hat{\mathcal{Z}} \rightarrow [0, \infty)$, the corresponding average per-symbol distortions are then given by \cite[eqs. (2) and (3)]{Stavrou_Fidelity_goal_priented_SemCom'22}:
\begin{subequations}
\begin{align}
\label{d_s_distortion_defn}
d_s^n(x^n, \hat{x}^n ) \eqdef  &  \frac{1}{n} \sum_{i=1}^n d_s (x_i, \hat{x}_i)    \\
\label{d_o_distortion_defn}
d_o^n(z^n, \hat{z}^n ) \eqdef  &  \frac{1}{n} \sum_{i=1}^n d_o (z_i, \hat{z}_i).  
\end{align}
\end{subequations} 
Using (\ref{d_s_distortion_defn}) and (\ref{d_o_distortion_defn}), the fidelity criteria of the semantic information and the observable information are defined as \cite{Stavrou_Fidelity_goal_priented_SemCom'22}
\begin{equation}
\label{fidelity_criterion}
\Delta_s \eqdef  \mathbb{E}\{d_s^n(x^n, \hat{x}^n )\}   \hspace{2mm}  \textnormal{and}   \hspace{2mm}    \Delta_o \eqdef \mathbb{E}\{d_o^n(z^n, \hat{z}^n )\},
\end{equation}
respectively. Using (\ref{d_s_distortion_defn})-(\ref{fidelity_criterion}), we state the following definition of the achievable rates and the infimum of all achievable rates.
\begin{definition}[{\textbf{\cite[Definition 1]{Stavrou_Fidelity_goal_priented_SemCom'22}}}]
\label{achievable_rates_defn}
For two distortion levels $D_o, D_s \geq 0$, $R$ is said to be $(D_o, D_s)$--achievable w.r.t. an arbitrary $\epsilon >0$, there exists -- for a very large $n$ -- a semantic-aware lossy source code $(n, M, \Delta_o, \Delta_s)$ with $M \leq 2^{n(R+\epsilon)}$ given that $\Delta_o \leq D_0+\epsilon$ and $\Delta_s \leq D_s+\epsilon$. Furthermore, considering that sequences of distortion functions $\{(d_o^n, d_s^n): n=1, 2, \ldots \}$ are given, then \cite[eq. (5)]{Stavrou_Fidelity_goal_priented_SemCom'22}
\begin{equation}
\label{R_D_o_D_s_def}
R(D_o, D_s) \eqdef \inf \{R: (R, D_o, D_s) \hspace{1mm} \textnormal{is achievable}\}.
\end{equation}
\end{definition} 
Per Definition \ref{achievable_rates_defn}, the information-theoretic characterization of (\ref{R_D_o_D_s_def}) is captured by the following lemma.
\begin{lemma}[{\textbf{\cite[Lemma 1]{Stavrou_Fidelity_goal_priented_SemCom'22}}}]
\label{Lemma_info-theoretic_characterization}
For a given $p(x)$ and $p(z|x)$, the semantic rate distortion function of the system model in \cite[Fig. 1]{Stavrou_Fidelity_goal_priented_SemCom'22} can be expressed as \cite[eq. (6)]{Stavrou_Fidelity_goal_priented_SemCom'22}
\begin{subequations}
\begin{align}
\label{SRDF_eqn}
R(D_s,D_o) \quad &= \inf_{q(\hat{z}, \hat{x} | z)}  I (\bm{z}; \hat{\bm{z}}, \hat{\bm{x}}) \\
\label{SRDF_eqn_const_1}
\textnormal{s.t.} \quad & \mathbb{E} \{ \hat{d}_s (\bm{z}, \hat{\bm{x}}) \}  \leq D_s \\
\label{SRDF_eqn_const_2}
&\mathbb{E} \{ d_o (\bm{z}, \hat{\bm{z}}) \}  \leq D_0 ,
\end{align}
\end{subequations}
where $\hat{d}_s (z, \hat{x}) \eqdef \sum_{x \in \mathcal{X}} p(x|z) d_s(x, \hat{x})$, $D_s \in [0, \infty]$, $D_o \in [0, \infty]$, and \cite[eq. (7)]{Stavrou_Fidelity_goal_priented_SemCom'22}
\begin{equation}
\label{I_z_hat_z_hat_z_defn}
I (\bm{z}; \hat{\bm{z}}, \hat{\bm{x}}) \eqdef \mathbb{E}\bigg\{ \log \bigg( \frac{q(\hat{\bm{z}}, \hat{\bm{x}} | \bm{z})}{ \nu(\hat{\bm{z}}, \hat{\bm{x}})} \bigg)\bigg\}.
\end{equation}
\end{lemma}

The constrained optimization problem in (\ref{SRDF_eqn})-(\ref{SRDF_eqn_const_2}) can be written as an unconstrained optimization problem through the \textit{Lagrange duality theorem} as follows \cite[eq. (15)]{Stavrou_Fidelity_goal_priented_SemCom'22}:
\begin{multline}
\label{unconst_optimization_problem}
\hspace{-2mm}R(D_s,D_o) = \quad \max_{s_1, s_2 \leq 0} \hspace{1.5mm} \min_{q(\hat{z}, \hat{x} | z) \geq 0,  \sum_{\hat{z}, z} q(\hat{z}, \hat{x} | z) =1}  \big\{   I (\bm{z}; \hat{\bm{z}}, \hat{\bm{x}}) \\ - s_1 \big( \mathbb{E} \{ \hat{d}_s (\bm{z}, \hat{\bm{x}}) \}- D_s\big) - s_2 \big( \mathbb{E} \{ d_o (\bm{z}, \hat{\bm{z}}) \} - D_o\big)  \big\},
\end{multline}
where $s_1, s_2 \leq 0$ are the Lagrange multipliers. The authors of \cite{Stavrou_Fidelity_goal_priented_SemCom'22} solve (\ref{unconst_optimization_problem}) and state the following main result.
\begin{theorem}[{\textbf{\cite[Theorem 1]{Stavrou_Fidelity_goal_priented_SemCom'22}}}]
\label{Thm_main_result}
Given that $p(x)$ and $p(z|x)$ are known, the underneath parametric solutions follow for the optimization problem in (\ref{SRDF_eqn})-(\ref{SRDF_eqn_const_2}):
\begin{itemize}
	\item If $s_1, s_2 <0$, the implicit optimal form of the minimizer that attains the minimum is given by \cite[eq. (16)]{Stavrou_Fidelity_goal_priented_SemCom'22}. In addition, the optimal parametric solution when $R(D_s^{*}, D_o^{*}) >0$ is expressed by \cite[eq. (17)]{Stavrou_Fidelity_goal_priented_SemCom'22}.
	
	\item If $s_1 <0, s_2=0$, and $R(D_s^{*}, D_o^{*}) >0$, $R(D_s^{*}, D_o^{*})$ is given by \cite[eq. (20)]{Stavrou_Fidelity_goal_priented_SemCom'22}.
	
	\item If $s_1=0, s_2<0$, and $R(D_s^{*}, D_o^{*}) >0$, $R(D_s^{*}, D_o^{*})$ is characterized by \cite[eq. (21)]{Stavrou_Fidelity_goal_priented_SemCom'22}.
	
	\item If $s_1=s_2=0$, $R(D_s^{*}, D_o^{*})=0$.
\end{itemize}
\proof The proof is given in \cite[Appendix A]{Stavrou_Fidelity_goal_priented_SemCom'22}.
\end{theorem}
Theorem \ref{Thm_main_result} is useful for deriving analytical expressions of the constrained optimization problem in (\ref{SRDF_eqn})-(\ref{SRDF_eqn_const_2}) and constructing generalizations of the \textit{Blahut--Arimoto algorithm} (BA algorithm) \cite{Cover_Elements_of_Info_Theory'06}.

We now move on to discuss an extended rate-distortion approach to goal-oriented SemCom \cite{Liu_Task-oriented_SemCom'22}.

\subsection{Extended Rate-Distortion Approach to Goal-Oriented SemCom}
\label{subsec: Extended_rate_distortion_approach_GO_SemCom}
The authors of \cite{Liu_Task-oriented_SemCom'22} put forward a JSCC-based goal-oriented SemCom system that incorporates a semantic reconstruction scheme while focusing on predicting the precision and generalizability of multiple goals/tasks. This goal-oriented SemCom system is composed of a JSCC encoder, a quantizer, a wireless channel, a JSCC decoder, and a network of AI tasks at the receiver \cite[Fig. 2]{Liu_Task-oriented_SemCom'22}. When the system is fed input $X$, which denotes an RV pertaining to the source image space, let an RV $Y$ be the desired output of an AI task. As can be seen in \cite[Fig. 2]{Liu_Task-oriented_SemCom'22}, the JSCC encoder maps the input to semantic representations that are subsequently quantized by the quantizer to minimize the transmission cost. The quantized symbols $Z$ are then transmitted over a wireless channel to the receiver \cite{Liu_Task-oriented_SemCom'22}. At the receiver, the JSCC decoder maps the noisy received symbols to the reconstructed image $\hat{X}$. Eventually, the AI task network uses $\hat{X}$ as an input and produces its prediction $\hat{Y}$. This overall goal-oriented SemCom scheme is formulated as an extended rate-distortion problem \cite{Liu_Task-oriented_SemCom'22}, and its analytical characterization is presented below.

To ensure that the reconstructed images can perform the given AI task properly, IB distortion \cite{IB_method_Tishby'00} must be minimized. To this end, the IB distortion between $x$ and $\hat{X}$ amounts to the Kullback--Leibler (KL) divergence $D_{KL}(p(y|x) || p(y|\hat{x}))$, which is given by \cite[eq. (1)]{Liu_Task-oriented_SemCom'22}   
\begin{equation}
\label{d_IB_defn}
d_{IB}(x, \hat{x}) \eqdef \sum_{y \in \mathcal{Y}} p(y|x) \log \frac{p(y|x)}{p(y|\hat{x})}, 
\end{equation}
where $\mathcal{Y}$ is the alphabet of $Y$. For (\ref{d_IB_defn}), $D_{IB}(X, \hat{X}) \eqdef\mathbb{E}\big\{ d_{IB}(x, \hat{x}) \big\}$ is the conditional mutual information $I(X; Y|\hat{X})$ \cite{DL_and_IB_Tishby'15} and defined as \cite[eq. (2)]{Liu_Task-oriented_SemCom'22}
\begin{equation}
\label{D_IB_defn}
D_{IB}(X, \hat{X}) = \sum_{x \in \mathcal{X}} \sum_{\hat{x} \in \mathcal{\hat{X}}} \sum_{y \in \mathcal{Y}} p(x, \hat{x}) p(y|x) \log \frac{p(y|x)}{p(y|\hat{x})}, 
\end{equation}
where $\mathcal{X}$, $\mathcal{\hat{X}}$, and $\mathcal{Y}$ are the alphabet of $X$, $\hat{X}$, and $Y$, respectively. The definition in (\ref{D_IB_defn}) then leads to the following theorem.
\begin{theorem}[{\textbf{\cite[Theorem 1]{Liu_Task-oriented_SemCom'22}}}]
$D_{IB}(X, \hat{X})$ as defined in (\ref{D_IB_defn}) can also be expressed as \cite[eq. (3)]{Liu_Task-oriented_SemCom'22}
\begin{equation}
\label{D_IB_equivalent_defn_1}
D_{IB}(X, \hat{X}) = I(X;Y)- I(\hat{X};Y).
\end{equation}
\proof The proof is given in \cite[Appendix A]{Liu_Task-oriented_SemCom'22}.
\end{theorem}
The expression in (\ref{D_IB_equivalent_defn_1}) intuitively illustrates the reduction of useful information \cite{Liu_Task-oriented_SemCom'22}. To improve the generalizability among different AI tasks, one must also minimize the reconstruction distortion $D_{RD}(X, \hat{X})$ that is equated as \cite[eq. (4)]{Liu_Task-oriented_SemCom'22}
\begin{equation}
\label{D_RD_defn}
D_{RD}(X, \hat{X}) \eqdef  \sum_{x \in \mathcal{X}} \sum_{\hat{x} \in \mathcal{\hat{X}}} p(x, \hat{x})  d_{RD}(x, \hat{x}),
\end{equation}
where $d_{RD}(x, \hat{x})=(x - \hat{x})^2$ \cite[eq. (5)]{Liu_Task-oriented_SemCom'22}. Meanwhile, the authors of \cite{Liu_Task-oriented_SemCom'22} take into account the natural tradeoff between $D_{IB}(X, \hat{X})$ and $D_{RD}(X, \hat{X})$, and define the semantic distortion measurement $D_{S}(X, \hat{X})$ as \cite[eq. (6)]{Liu_Task-oriented_SemCom'22}
\begin{equation}
\label{D_S_defn}
D_{S}(X, \hat{X}) \eqdef D_{RD}(X, \hat{X}) + \beta D_{IB}(X, \hat{X}), 
\end{equation}
where $\beta$ controls the tradeoff between the AI task’s prediction accuracy and the goal-oriented SemCom system’s generalizability \cite{Liu_Task-oriented_SemCom'22}. Using (\ref{D_S_defn}), the goal-oriented SemCom system proposed in \cite{Liu_Task-oriented_SemCom'22} can be formulated as an extended rate-distortion optimization problem that is given by \cite[eq. (10)]{Liu_Task-oriented_SemCom'22} 
\begin{subequations}
\begin{align}
\label{opt_prob_obj}
 \min_{p(\hat{x} | x)} &   D_{RD}(X, \hat{X}) + \beta D_{IB}(X, \hat{X})   \\
 \label{opt_prob_cons_1}
\textnormal{s.t.} \quad &  I (X; \hat{X})  \leq I_C \\
\label{opt_prob_cons_2}
&\sum_{\hat{x}} p(\hat{x} | x) =1 ,
\end{align}
\end{subequations}
where the constraints in (\ref{opt_prob_cons_1}) and (\ref{opt_prob_cons_2}) correspond to the maximum channel capacity $I_C$ and the normalization constraint of the conditional PMF $p(\hat{x} | x)$, respectively \cite{Liu_Task-oriented_SemCom'22}. Substituting (\ref{D_IB_equivalent_defn_1}) into (\ref{opt_prob_obj}) and discarding $I(X;Y)$ -- since it is constant for a given dataset -- leads to the following optimization problem:
\begin{subequations}
	\begin{align}
		\label{opt_prob_obj_2}
		\min_{p(\hat{x} | x)} &   D_{RD}(X, \hat{X}) - \beta I(\hat{X};Y)    \\
		\label{opt_prob_cons_1_1}
		\textnormal{s.t.} \quad &  I (X; \hat{X})  \leq I_C \\
		\label{opt_prob_cons_2_1}
		&\sum_{\hat{x}} p(\hat{x} | x) =1.
	\end{align}
\end{subequations}
The fact that the authors of \cite{Liu_Task-oriented_SemCom'22} solve this optimization problem using the Lagrange multiplier technique leads to the following theorem.   
\begin{theorem}[{\textbf{\cite[Theorem 2]{Liu_Task-oriented_SemCom'22}}}]
\label{optimal_mapping_thm}
The optimal mapping from the source images $X$ to the semantically-reconstructed images $\hat{X}$ must satisfy \cite[eqs. (12)-(15)]{Liu_Task-oriented_SemCom'22}:
\begin{subequations}
\begin{align}
p(\hat{x} | x) &= \frac{ p(\hat{x})e^{-\lambda^{-1}d_S(x, \hat{x})}  }{\mu(x)}   \\
p(\hat{x}) &= \sum_{x \in \mathcal{X} } p(x) p(\hat{x} | x)   \\
p(y | \hat{x}) &=\sum_{x \in \mathcal{X} } p(y|x) p(x|\hat{x}),
\end{align}
\end{subequations}
where 
\begin{subequations}
\begin{align}
\mu(x) &= \sum_{\hat{x} \in \hat{\mathcal{X}}} p(\hat{x})  e^{-\lambda^{-1}d_S(x, \hat{x})}  \\
d_S(x, \hat{x}) &= d_{RD}(x, \hat{x}) + \beta  d_{IB}(x, \hat{x}). 
\end{align}
\end{subequations}

\proof The proof is provided in \cite[Appendix B]{Liu_Task-oriented_SemCom'22}. 
\end{theorem}
The optimal distributions $p(\hat{x} | x)$, $p(\hat{x})$, and $p(y | \hat{x})$ can be obtained \cite{Liu_Task-oriented_SemCom'22} using the BA algorithm \cite{Cover_Elements_of_Info_Theory'06}.

We now continue with our discussion of GOQ \cite{Zou_JSAC_GO-Quantization'23,Zou_GO-Quantization_WiOpt'22}. 

\subsection{Goal-Oriented Quantization}
\label{subsec: GO_Quantization}
GOQ is quite useful for many applications, including controlled networks that are built on a communication network, wireless resource allocation, and 6G systems \cite{Zou_JSAC_GO-Quantization'23,Zou_GO-Quantization_WiOpt'22}. In this vein, a general GOQ framework wherein the goal/task of a receiver is modeled by a generic optimization problem that comprises both decision variables and parameters is illustrated in \cite[Fig. 1]{Zou_JSAC_GO-Quantization'23}. More specifically, the goal is modeled as a minimization problem of a general goal function $f(\bm{x};\bm{g})$ for $\bm{x}$ (with dimension $d$) being the decision that has to be made from a quantized version of the parameters $\bm{g}$ (with dimension $p$) \cite{Zou_JSAC_GO-Quantization'23}. In view of this problem, we state the following two definitions.

\begin{definition}[{\textbf{\cite[Definition II.1]{Zou_JSAC_GO-Quantization'23}}}]
\label{Definition_II.1}
Suppose $M, d \in \mathbb{N}_{\geq 1}$ and $\mathcal{G} \in \mathbb{R}^d$. An $M$--quantizer $\mathcal{Q}_M$ is completely decided by a piecewise constant function $\mathcal{Q}_M: \mathcal{G} \to \mathcal{G}$. This mapping is defined as $\mathcal{Q}_M(g)=z_m$ for all $z_m \in \mathcal{G}_m$ given that $m\in[M]$; $\mathcal{G}_1, \ldots, \mathcal{G}_M$ are the quantization regions that define a partition of $\mathcal{G}$; and $z_1, \ldots, z_M$ are the region representatives.
\end{definition}

\begin{definition}[{\textbf{\cite[Definition II.2]{Zou_JSAC_GO-Quantization'23}}}]
\label{Definition_II.2}
Suppose $p\in \mathbb{N}_{\geq 1}$ and $g$ is a fixed parameter. Let $\chi(g)$ be a decision function that provides the minimum points for the goal function $f(\bm{x};g)$, whose decision variable is $\bm{x} \in \mathbb{R}^P$ \cite[eq. (1)]{Zou_JSAC_GO-Quantization'23}: 
\begin{equation}
\label{decision_function_defn}
\chi(g) \in \argmin_{\bm{x} } f(\bm{x};g).
\end{equation}
The optimality loss induced by quantization is equated as \cite[eq. (2)]{Zou_JSAC_GO-Quantization'23}:
\begin{equation}
\label{L_Q_f_defn}
L(Q; f) \eqdef \alpha_f \int_{g \in \mathcal{G}} \big[ f(\chi(\mathcal{Q}(g)); g) - f(\chi(g); g) \big] \phi(g) dg,
\end{equation}
where $\phi(\cdot)$ is the PDF of $g$ and $\alpha_f >0$ denotes a scaling factor that is independent of Q.
\end{definition}
From Definition \ref{Definition_II.2}, the following remarks follow.

\begin{remark}
The conventional quantization approach can be derived from the GOQ approach by observing that the second term of $L(Q; f)$ -- as defined in (\ref{L_Q_f_defn}) -- is independent of $Q$ and specifying $f$ as $f(\bm{x};\bm{g}) = \| \bm{x}-\bm{g} \|^2$ \cite{Zou_JSAC_GO-Quantization'23}. 
\end{remark}
\begin{remark}
Unlike the conventional quantization approach that aims to provide a version of $g$ that resembles $g$, what matters in the GOQ approach is the quality of the end decision made \cite{Zou_JSAC_GO-Quantization'23}.
\end{remark}
\begin{remark}
\label{rem: GOQ_quantizer_deign}
The design of a GOQ quantizer constitutes a major difference w.r.t. the conventional quantization approach and thus hinges on the mathematical properties of $f$ and the underlying decision function $\chi(\cdot)$ \cite{Zou_JSAC_GO-Quantization'23}.
\end{remark}
When it comes to Remark \ref{rem: GOQ_quantizer_deign}, quantifying the relationship between the nature of $f$ and the quantization performance is a challenging problem \cite{Zou_JSAC_GO-Quantization'23}. Meanwhile, for a scalar GOQ such that $d=p=1$ and $\rho(\cdot)$ being a density function, the number of quantization intervals over $[a, b]$ can be approximated by $M \displaystyle\int_{a}^b \rho(g) dg$ \cite{Zou_JSAC_GO-Quantization'23}. Accordingly, the problem of finding a GOQ in the high-resolution regime boils down to finding the density function that minimizes the optimality loss that is denoted by $L(\rho; f)$ \cite{Zou_JSAC_GO-Quantization'23}. This leads to the following proposition.
\begin{proposition}[{\textbf{\cite[Proposition III.1]{Zou_JSAC_GO-Quantization'23}}}]
\label{Proposition_III.1}
Suppose $f$ is a fixed goal function that is assumed to be $\kappa$ times differentiable and $\chi$ differentiable with \cite[eq. (4)]{Zou_JSAC_GO-Quantization'23}
\begin{equation}
\label{kappa_defn}
\kappa = \min \Big\{ i \in \mathbb{N}: \forall g, \frac{\partial^i f(x;g)}{\partial x^i}  \bigg|_{x=\chi(g)}   \neq   \textnormal{a.s.} \Big\}.
\end{equation}
In the high resolution regime, the optimality loss $L(\rho; f)$ is minimized by employing the underneath quantization interval density function \cite[eq. (5)]{Zou_JSAC_GO-Quantization'23}:
\begin{equation}
\label{opt_PDF_S_GQO}
\rho^{\star}(g) = C \bigg[ \bigg( \frac{d\chi(g)}{dg}\bigg)^{\kappa} \frac{\partial^{\kappa} f(\chi(g);g)}{\partial x^{\kappa}}  \phi(g) \bigg]^{\frac{1}{\kappa+1}},
\end{equation}
where $\frac{1}{C} = \displaystyle \int_{\mathcal{G}}  \bigg[ \bigg( \frac{d\chi(t)}{dt}\bigg)^{\kappa} \frac{\partial^{\kappa} f(\chi(t);t)}{\partial x^{\kappa}}  \phi(t) \bigg]^{\frac{1}{\kappa+1}} dt$.

\proof The proof is provided in \cite[Appendix A]{Zou_JSAC_GO-Quantization'23}.
\end{proposition}

On the other hand, when $d, p \in \mathbb{N}_{\geq 1}$, the goal-oriented quantization problem becomes a vector GOQ problem \cite{Zou_JSAC_GO-Quantization'23}, which the following proposition is derived for. 
\begin{proposition}[{\textbf{\cite[Proposition IV.1]{Zou_JSAC_GO-Quantization'23}}}]
\label{Proposition_IV.1}
Assume $d, p \in \mathbb{N}_{\geq 1}$; $\kappa=1$; and $f$ and $\chi$ are twice differentiable. Let $\bm{H}_f(\bm{x};\bm{g})$ and $\bm{J}_{\chi}(\bm{g}$ be the Hessian matrix of $f$ and the Jacobian matrix of $f$ evaluated for an optimal decision $\chi(\bm{g})$, respectively. In the regime of large $M$, the optimality loss function $L(Q;f)$ -- defined in (\ref{L_Q_f_defn}) -- can be approximated as \cite[eq. (9)]{Zou_JSAC_GO-Quantization'23}:
\begin{multline}
\label{L_Q_f_approximation}
L(Q;f) = \underbrace{\alpha_f \sum_{m=1}^M \int_{\mathcal{G}_m} (\bm{g}-\bm{z}_m)^T \bm{A}_{f, \chi}(\bm{g})(\bm{g}-\bm{z}_m)\phi(\bm{g})d\bm{g}  }_{=\hat{L}_M(Q;f)}   \\ +\mathcal{O}(M^{-2/p}), 
\end{multline}
where $\bm{A}_{f, \chi}(\bm{g})=\bm{J}_{\chi}^T(\bm{g}) \bm{H}_f(\chi(\bm{g});\bm{g}) \bm{J}_{\chi}(\bm{g})$. Moreover, $\hat{L}_M(Q;f)$ -- as expressed in  (\ref{L_Q_f_approximation}) -- can be bounded as $ L_M^{\textnormal{min}}(Q;f) \leq \hat{L}_M(Q;f) \leq L_M^{\textnormal{max}}(Q;f)$, where $L_M^{\textnormal{min}}(Q;f)$ and $L_M^{\textnormal{max}}(Q;f)$ are given in \cite[eq. (10)]{Zou_JSAC_GO-Quantization'23} and \cite[eq. (11)]{Zou_JSAC_GO-Quantization'23}, respectively.

\proof The proof is provided in \cite[Appendix B]{Zou_JSAC_GO-Quantization'23}.
\end{proposition}

Apart from the above-discussed goal-oriented SemCom theories, there have also been other theoretical developments such as the \textit{theory of goal-oriented communication} \cite{Goldreich_Goal_Oriented_Com_Theory'12} and \textit{universal SemCom {II}} \cite{Juba_Universal_SemCom'08}. This leads us to our in-depth discussion of the fundamental and major challenges of goal-oriented SemCom. It is worth noting, however, that the above-discussed goal-oriented SemCom theories have their corresponding limitations and are hence not the most rigorous and complete of theories (though they are interesting!). This is attributed to the numerous fundamental and major challenges of goal-oriented SemCom, which are detailed below.

\section{Fundamental and Major Challenges of Goal-Oriented SemCom}
\label{sec: fundamental_and_major_challenges_of_goal_oriented_SemCom}
When it comes to realizing high-fidelity goal-oriented SemCom for 6G and beyond, the research field of goal-oriented SemCom is fraught with fundamental and major challenges in the theoretical, algorithmic, and realization/implementation-related research frontiers. These challenges are discussed in detail below, beginning with the challenges in the development of fundamental goal-oriented SemCom theories.

\subsection{Challenges in the Development of Fundamental Goal-Oriented SemCom Theories}
\label{subsec: challenges_of_GO-SemCom_fundamental_theories}
We detail below (in no specific order) the fundamental and major challenges related to -- but not limited to -- the development of fundamental goal-oriented SemCom theories.

\subsubsection{Lack of a Commonly Accepted Definition of Semantics / Semantic Information}
despite the many definitions of semantics / semantic information that exist, there is no commonly agreed upon definition. This is a fundamental challenge that hinders the advancement of goal-oriented SemCom theory (as well as algorithm and realization).

\subsubsection{Fundamental Performance Analysis of Goal-Oriented SemCom}
as is the case for SemCom, analyzing the fundamental non-asymptotic performance of goal-oriented SemCom is fundamentally challenging for the following reasons \cite{arXiv_Getu_DeepSC_Performance_Limits'23}: $i)$ the lack of a commonly agreed-upon definition of semantics / semantic information \cite[Ch. 10, p. 125]{Tong_Zhu__6G'21}; $ii)$ the fundamental lack of interpretability/explainability of optimization, generalization, and approximation in DL models \cite{Poggio_Theo_Issues_Dnets_2020}; and $iii)$ the lack of a comprehensive mathematical foundation for goal-oriented SemCom \cite[Sec. IV]{hTong_Li_Nine_Challenges'21}. Moreover, since a system's goal may not be explicitly represented by a utility function, it can be fundamentally challenging to rigorously analyze a goal-oriented SemCom system's performance.

\subsubsection{Performance Analysis of DL-based Goal-Oriented SemCom Systems}
DL-based goal-oriented SemCom systems such as cooperative goal-oriented SemCom \cite{Xu_SemCom_for_IoV'22} rely on a joint DL-based source and channel coding technique. The rigorous non-asymptotic performance analysis of DL-based goal-oriented SemCom systems is therefore hindered by the fundamental lack of interpretability/explainability \cite{Toward_Science_of_Interpretable_ML'17,Guo_XAI_6G'20} that is inherent in DL models.

\subsubsection{Fundamental Limits of Goal-Oriented SemCom Systems}
the fundamental limits of goal-oriented SemCom depend on not only the type of DL-based semantic encoder and semantic decoder used, but also the type of goal, and hence the goal function. The goal function can hardly be detailed enough to capture all aspects of a goal, and DL-based goal-oriented SemCom techniques suffer from a fundamental lack of interpretability (the same as DL-based SemCom schemes).

\subsubsection{Semantic Compressed Sensing and Optimal Sampling Theory}
in stark contrast to the state-of-the-art techniques that pursue a \textquotedblleft sample-then-compress'' structure, \textit{semantic compressed sensing} is a computationally lighter scheme that gathers only the minimum volume of data needed to reconstruct the signal of interest at the desired resolution, as determined by the application requesting the data \cite{SemCom_Net_Systems'21}. It carries out certain signal processing operations directly in the \textquotedblleft compressed domain'' without complete signal reconstruction \cite{SemCom_Net_Systems'21}. This calls for tackling the formidable challenge of developing an \textit{optimal sampling theory} that unifies signal sparsity and aging/semantics for real-time prediction/reconstruction under communication and delay constraints \cite{SemCom_Net_Systems'21}.
	
We now carry on with fundamental and major challenges in the development of fundamental goal-oriented SemCom algorithms.

\subsection{Challenges in the Development of Fundamental Goal-Oriented SemCom Algorithms}
\label{subsec: challenges_of_GO-SemCom_fundamental_algorithms}
We detail below (in no specific order) the fundamental and major challenges related to –- but not limited to –- the development of fundamental goal-oriented SemCom algorithms.

\subsubsection{Inevitability of Semantic Mismatch}
source KB and the destination KB can be quite different because they observe different worlds with unequal abilities to understand things \cite{Luo_SemCom_Overview'22}. Consequently, semantic mismatch is unavoidable to the extent that it can fundamentally constrain the performance of wireless systems that are based on goal-oriented SemCom.

\subsubsection{Lack of Unified Semantic Performance Assessment Metrics}
despite the numerous metrics that have been proposed for goal-oriented SemCom, there is a lack of unified/universal performance assessment metrics for goal-oriented SemCom \cite{Getu_Metrics_of_SemCom_and_GO_SemCom_2023}. When it comes to unified metrics, the major challenge is to establish concrete metrics that can capture source and network dynamics, as well as any potentially non-trivial interdependencies among information attributes \cite{Kountouris_Semantics_EmpoweredCF'21}.

\subsubsection{Lack of Interpretability in DL-Based Goal-Oriented SemCom}
there is a fundamental lack of interpretability in DL-based goal-oriented SemCom algorithms due to the fundamental lack of interpretability/explainability that is inherent in trained DL models \cite{Toward_Science_of_Interpretable_ML'17,Guo_XAI_6G'20}. 

\subsubsection{Optimal Semantic-Aware Joint Sampling, Transmission, and Reconstruction of Multidimensional Signals}
in a number of conventional communication systems, transmission is optimized on the basis of QoS metrics -- e.g., delay, rate, timeliness -- while ignoring source variations, the fact that samples may be received on time but contain no useful information; or the fact that samples can even be misleading about the system’s true state \cite{Kountouris_Semantics_EmpoweredCF'21}. This scenario highlights the implicit structural links that exist between sampling and communication, which are generally inseparable in SemCom and goal-oriented SemCom \cite{Kountouris_Semantics_EmpoweredCF'21}. For reliable goal-oriented SemCom that enables timely decision-making and satisfies the stringent requirements of real-time NCSs, the foundational challenge is therefore to develop a theory for optimal semantic-aware joint active sampling, transmission, and reconstruction of multidimensional signals, especially under stringent timing constraints \cite{Kountouris_Semantics_EmpoweredCF'21}.

\subsubsection{Resource Allocation for Goal-Oriented SemCom}
from the vantage point of optimal resource allocation, goal-oriented SemCom systems face many fundamental challenges, some of which have led to the following major research problems: \textit{how can a generic resource allocation problem be optimized for different goal-oriented SemCom systems?} \textit{How can a resource allocation policy be optimized while maximizing goal-oriented SemCom's efficiency?}  \cite{Qin_Sem_Com_Principles_Apps'22}.

\subsubsection{Goal-Oriented Resource Orchestration}
in emerging cyber-physical and autonomous networked systems, semantic-aware real-time data networking requires effective scheduling and resource allocation policies for gathering (often correlated) multi-source multi-modal information \cite{Kountouris_Semantics_EmpoweredCF'21}. The objectives in the networked applications could be achieved by using an alternative set of multi-quality data \cite{Kountouris_Semantics_EmpoweredCF'21}. These goal-oriented resource orchestration problems fall into the realm of real-time scheduling with multiple choices \cite{Kountouris_Semantics_EmpoweredCF'21}. It is therefore challenging to devise online algorithms that can select which piece of information -- from where and when -- to gather and transmit under communication and processing constraints \cite{Kountouris_Semantics_EmpoweredCF'21}.

\subsubsection{Multi-Objective Stochastic Optimization}
when it comes to goal-oriented end-user-perceived utilities that estimate the relative degree of priority of different information attributes, semantic-aware data gathering and prioritization require multi-criteria optimization \cite{Kountouris_Semantics_EmpoweredCF'21}. In view of multi-criteria optimization and overcoming its challenges, multi-objective stochastic optimization based on the cumulative prospect theory -- which incorporates semantic information via risk-sensitive measures and multi-attribute entropy-based utility functions -- holds promise \cite{Kountouris_Semantics_EmpoweredCF'21}.

We now proceed to discuss the fundamental and major challenges in the realization of goal-oriented SemCom.

\subsection{Challenges in the Realization of Goal-Oriented SemCom}
\label{subsec: challenges_of_GO-SemCom_realization}
In what follows, we discuss (in no specific order) the fundamental and major challenges related to -- but not limited to -- the realization of goal-oriented SemCom.

\subsubsection{Real-Time Requirement}
several major use cases of goal-oriented SemCom, such as autonomous transportation, telehealth, smart factories, and NCSs, have real-time requirements for goal-oriented communication/control. However, incorporating semantic reasoning into the goal-oriented SemCom use cases mentioned incurs extra delay in goal-oriented SemCom's overall transceivers \cite{Sem_Empowered_Commun'22}. Satisfying the ultra-low end-to-end latency requirements (i.e., real-time requirements) of 6G (and beyond) is therefore a major realization challenge for goal-oriented SemCom.

\subsubsection{Scalability}
as is the case for SemCom, the realization of goal-oriented SemCom is hampered by several scalability challenges, such as: $i)$ the lack of a general semantic-level framework for distinct types of sources; $ii)$ sharing, updating, and maintaining KBs at the source and destination definitely necessitate additional storage costs and algorithm design \cite{Sem_Empowered_Commun'22}; and $iii)$ realizing goal-oriented SemCom involves significant computational as well as storage costs.

\subsubsection{Knowledge Evolution Tracking}
many existing goal-oriented SemCom techniques rely on the dynamic sharing of knowledge sharing between the source KB and the destination KB. To this end, modeling and keeping track of each piece of knowledge is fundamentally important for improving the efficiency and reliability of goal-oriented SemCom. Nonetheless, the basic neuroscientific understanding of knowledge, knowledge evolution, and knowledge tracking are very difficult fundamental problems.

\subsubsection{Compatibility with Existing Communication Infrastructure}
since BitCom systems and services will still be in use when goal-oriented SemCom systems and services are rolled out in 6G networks and systems, any implementation of goal-oriented SemCom should ensure that futuristic goal-oriented SemCom systems are compatible with the existing communication infrastructure. To this end, extensive link-level simulations must be performed to verify the realistic end-to-end performance of goal-oriented SemCom.

\subsubsection{Efficient Knowledge Sharing in Multi-User MIMO Goal-Oriented SemCom Systems}
a multi-user MIMO goal-oriented SemCom system such as cooperative goal-oriented SemCom \cite{Xu_SemCom_for_IoV'22} -- which is schematized in Fig. \ref{fig: Coperative_SemCom_20230302} -- needs knowledge to be shared between the receiver with multiple antennas and a number of goal-oriented SemCom users that are equipped with either a single antenna or multiple antennas. However, achieving efficient global knowledge sharing in multi-user MIMO goal-oriented SemCom systems is challenging.    

Because challenges are always opportunities, some of the above-detailed fundamental and major challenges of goal-oriented SemCom are also big opportunities for novel future directions for goal-oriented SemCom, as discussed below.

\section{Future Directions of Goal-Oriented SemCom}
\label{sec: future_directions_of_goal-oriented_SemCom}
In light of the fundamental and major challenges of goal-oriented SemCom that are detailed in Section \ref{sec: fundamental_and_major_challenges_of_goal_oriented_SemCom}, the developments in goal-oriented SemCom that are discussed in Section \ref{subsec: theories_of_Goal_Oriented_SemCom}, and the many proposals of state-of-the-art goal-oriented SemCom algorithms that are surveyed in Section \ref{sec: goal_oriented_SemCom_research_landscape}, we offer some novel future directions for goal-oriented SemCom theory, algorithm, and realization. We begin with some novel future directions for goal-oriented SemCom theory.

\subsection{Future Directions for Goal-Oriented SemCom Theory}
\label{subsec: Future_directions_in_goal-oriented_SemCom_theories}
We highlight (in no particular order) some novel future directions related to -- but not limited to -- goal-oriented SemCom theory.

\subsubsection{A Fundamental Theory and the Fundamental Limits of Actionable Intelligence}
actionable intelligence is on-time and accurate intelligence that would help decision-makers make an optimal/well-informed decision \cite{Keith_Actionable_Intelligence'14}. Representing decision in the context of goal-oriented SemCom, the reconstructed signals of a communicating smart device can alter the recipients’ states and initiate specific actions at the receivers \cite{Kountouris_Semantics_EmpoweredCF'21}. The limits of actionable intelligence must be well-understood before deploying any goal-oriented SemCom system. To this end, a fundamental theory and the fundamental limits of actionable intelligence -- in the context of DL, big data, or a combination thereof -- are critical future research directions for goal-oriented SemCom.

\subsubsection{A Fundamental Theory of Optimal Semantic-Aware Joint Active Sampling, Transmission, and Reconstruction of Multidimensional Signals}
a theory of optimal semantic-aware joint active sampling, transmission, and reconstruction of multidimensional signals -- especially, under stringent timing constraints -- is needed to enable timely decision-making and efficiently meet the requirements of real-time networked applications \cite{Kountouris_Semantics_EmpoweredCF'21}.

We now proceed to highlight some novel future directions for goal-oriented SemCom algorithms.

\subsection{Future Directions for Goal-Oriented SemCom Algorithms}
\label{subsec: Future_directions_in_goal-oriented_SemCom_algorithms}
We point out (in no specific order) the promising future directions related to -- but not limited to -- goal-oriented SemCom algorithms.  

\subsubsection{Semantic-Aware Networking}
in semantic-aware goal-oriented networks, the major operations include local goal-oriented information acquisition, representation, and semantic value inference; data prioritization; in-network processing (e.g., fusion, compression); semantic reception; and semantic reconstruction \cite{Kountouris_Semantics_EmpoweredCF'21}. These operations will require optimal or nearly-optimal algorithms for semantic filtering, semantic preprocessing, semantic reception, and semantic control \cite{Kountouris_Semantics_EmpoweredCF'21}.    

\subsubsection{Goal-Oriented SemCom with Time-Evolving Goals}
although most state-of-the-art goal-oriented SemCom works consider fixed goals, it is often the case that one task is followed by one or more other tasks in different systems that include smart devices \cite{Zhang_Goal-Oriented_Commun'22}. This enforces the design constraint that a new task needs to be executed seamlessly once the previous task has ended \cite{Zhang_Goal-Oriented_Commun'22}. Nevertheless, retraining from scratch for every goal not only takes time but also wastes resources \cite{Zhang_Goal-Oriented_Commun'22}. Consequently, a unified goal-oriented SemCom framework that takes into account multiple -- often causally related -- goals while maximizing the expected goal accomplishment \cite{Zhang_Goal-Oriented_Commun'22} is a research direction worth pursuing.

\subsubsection{Goal-Oriented Coding and Control}
source coding or JSCC models could be implemented to characterize goal-oriented compression and its performance limits \cite{Zhang_Goal-Oriented_Commun'22}. Whenever a goal relies on not only the state, but also the decision made, in the current time slot as well as previous time slots, formulating dynamic system models can lead to a promising solution \cite{Zhang_Goal-Oriented_Commun'22}. For this scenario, there are two possible ways to design optimal goal-oriented coding and control:
\begin{itemize}
	\item Resorting to differential equations to explore the evolution of the transmitted messages and the goal \cite{Zhang_Goal-Oriented_Commun'22}.
	
	\item Revisiting the sampling process by tailoring the sampling problem to a general utility function \cite{Zhang_Goal-Oriented_Commun'22}. 
\end{itemize}

We now move on to some crucial future directions for goal-oriented SemCom realization.

\subsection{Future Directions for Goal-Oriented SemCom Realization}
\label{subsec: Future_directions_in_goal-oriented_SemCom_realization}
In what follows, we point out (in no particular order) some useful future directions related to -- but not limited to -- goal-oriented SemCom realization.

\subsubsection{The Coexistence of BitCom and Goal-Oriented SemCom Users}
since BitCom service and infrastructure will still be in use when goal-oriented SemCom is implemented in 6G and beyond, the coexistence of BitCom users and goal-oriented SemCom users must be investigated through the lens of not only measurements, but also theory. Regarding theory, the coexistence of BitCom users and goal-oriented SemCom users should be studied in detail from the vantage points of optimal resource allocation and interference mitigation.

\subsubsection{The Impact of Inconsistent KBs at the Source and Destination} 
although most state-of-the-art goal-oriented SemCom proposals resort to the assumption that knowledge is shared in real time to consider consistent KBs at the source and destination, the source KB and the destination KB are fundamentally inconsistent \cite{Luo_SemCom_Overview'22}. Therefore, how to design and realize novel (multi-user) goal-oriented SemCom systems with inconsistent KBs are an open issue in goal-oriented SemCom design and realization.

At last, we continue with this tutorial-cum-survey paper's concluding summary and research outlook.

\section{Concluding Summary and Research Outlook}
\label{sec: conc_summary_and_research_outlook}
Driven by the many highly heterogeneous 6G applications and use cases that exist, numerous researchers in academia, industry, and national laboratories have disseminated several 6G proposals and roadmaps. Despite the abundance of proposals and roadmaps, realizing 6G -- as it is presently envisaged -- is fraught with many fundamental IMT challenges that are interwoven with several technological challenges and uncertainties. To alleviate some of these technological challenges and uncertainties, SemCom and goal-oriented SemCom (effectiveness-level SemCom) have emerged as promising technological enablers of 6G. SemCom and goal-oriented SemCom enable 6G because they are designed to transmit only semantically-relevant information. This semantic-centric design helps to minimize power usage, bandwidth consumption, and transmission delay in 6G, which attests to the criticality of SemCom and goal-oriented SemCom for 6G. 6G is also critical for the realization of major SemCom use cases (e.g., H2H SemCom, H2M SemCom, M2M SemCom, and KG-based SemCom) and major goal-oriented SemCom use cases (e.g., autonomous transportation, consumer robotics, environmental monitoring, telehealth, smart factories, and NCSs). The paradigms of \textit{6G for SemCom and goal-oriented SemCom} and \textit{SemCom and goal-oriented SemCom for 6G} call for the tighter integration and marriage of 6G, SemCom, and goal-oriented SemCom.

While underscoring an overarching paradigm shift that can change the status quo that wireless connectivity is an opaque data pipe carrying messages whose context-dependent meaning and effectiveness have been ignored, this holistically comprehensive tutorial-cum-survey paper aims to facilitate and inspire a tighter integration and marriage of 6G, SemCom, and goal-oriented SemCom. For this purpose, this article first explained the fundamentals of semantics and semantic information, semantic representation, theories of semantic information, and definitions of semantic entropy. It then built on this understanding by detailing the state-of-the-art research landscape of SemCom, presenting the major state-of-the-art trends and use cases of SemCom, discussing state-of-the-art SemCom theories, uncovering fundamental and major challenges (of SemCom theory, algorithm, and realization), and offering novel future research directions (for SemCom theory, algorithm, and realization). This article also documented the state-of-the-art research landscape of goal-oriented SemCom, provided major state-of-the-art trends and use cases of goal-oriented SemCom, discussed state-of-the-art goal-oriented SemCom theories, exposed the fundamental and major challenges (for goal-oriented SemCom theory, algorithm, and realization), and provided novel future research directions for goal-oriented SemCom theory, algorithm, and realization. 

By proffering fundamental and major challenges as well as novel future research directions for SemCom and goal-oriented SemCom, this comprehensive tutorial-cum-survey article fittingly inspires astronomical lines of research on SemCom and goal-oriented SemCom theory, algorithm, and implementation for 6G and beyond. For 6G and beyond, at last, this article calls for novel System 2-type SemCom design and realization in sharp contrast to many existing and discussed SemCom works that are System 1-type by design.

\appendices

\section{On Entropy, Relative Entropy, and Mutual Information} 
\label{Info_theory_basics}
To lay the groundwork for our discussion of existing SemCom and goal-oriented SemCom theories, we offer a brief discussion on the basics of entropy, relative entropy, and mutual information. We begin by defining the entropy of a discrete RV.
\begin{definition}
	\label{entropy_defn}
	For a discrete RV $X$, its entropy $H(X)$ is defined by \cite[eq. (2.1)]{Cover_Elements_of_Info_Theory'06}
	\begin{equation}
		\label{entropy_defn_discrete_RV}
		H(X) \eqdef - \sum_{x \in \mathcal{X} } p(x) \log_2 p(x),
	\end{equation}
	where $\mathcal{X}$ is the alphabet, $p(x) \eqdef p_X(x)=\mathbb{P}(\{X=x\})$ is the PMF of $X$, and the entropy $H(X)$ is expressed in bits \cite{Cover_Elements_of_Info_Theory'06}.
\end{definition}

The entropy defined in (\ref{entropy_defn_discrete_RV}) is often referred to as \textit{Shannon entropy}, and $H(X) \geq 0$ \cite[Lemma 2.1.1]{Cover_Elements_of_Info_Theory'06}. Meanwhile, if $X \sim p(x)$, the expected value of the RV $g(X)$ is equated as \cite[eq. (2.2)]{Cover_Elements_of_Info_Theory'06}
\begin{equation}
	\label{expectation_g_X_defn}
	\mathbb{E}\{ g(X) \} \eqdef \sum_{ x \in \mathcal{X}} g(x) p(x).
\end{equation}
Thus, it follows from (\ref{expectation_g_X_defn}) and (\ref{entropy_defn_discrete_RV}) that 
\begin{equation}
	\label{entropy_defn_discrete_RV_2}
	H(X) = - \mathbb{E}\{ \log_2 p(X) \} =  \mathbb{E}\{ 1/ \log_2 p(X) \}.
\end{equation}	
As a generalization of the entropy definitions in (\ref{entropy_defn_discrete_RV_2}) and (\ref{entropy_defn_discrete_RV}), we provide below the definition of \textit{joint entropy}.
\begin{definition}
	\label{joint_entropy_defn_1}
	For a pair of discrete RVs $(X,Y)$ with a joint PMF $p(x,y)$, their joint entropy $H(X,Y)$ is defined as \cite[eq. (2.8)]{Cover_Elements_of_Info_Theory'06}  
	\begin{equation}
		\label{joint_entropy_defn_eqn_1}
		H(X,Y) \eqdef - \sum_{x \in \mathcal{X} }\sum_{y \in \mathcal{Y} } p(x,y) \log_2 p(x,y),
	\end{equation}
	where $\mathcal{X}$ and $\mathcal{Y}$ are the alphabets of $X$ and $Y$, respectively, and $p(x,y)\eqdef P_{X,Y}(x, y) = \mathbb{P}(X = x, Y = y)$.
\end{definition}
To express the right-hand side (RHS) of (\ref{joint_entropy_defn_eqn_1}) using expectation, we provide the following definition of the expectation of a function of multi-variate RVs: if $X \sim p(x)$ and $Y \sim p(y)$, the expected value of the RV $g(X,Y)$ takes the form \cite{DPJN08} 
\begin{equation}
	\label{expectation_g_X__Y_defn}
	\mathbb{E}\{ g(X,Y) \} \eqdef \sum_{ x \in \mathcal{X}} \sum_{ y \in \mathcal{Y}} g(x,y) p(x,y).
\end{equation}
Thus, using (\ref{expectation_g_X__Y_defn}), the joint entropy -- as it is defined in (\ref{joint_entropy_defn_eqn_1}) -- can also be expressed as \cite[eq. (2.9)]{Cover_Elements_of_Info_Theory'06}
\begin{equation}
	\label{joint_entropy_defn_eqn_2}
	H(X,Y) = -\mathbb{E} \{ \log_2 p(X,Y) \}.
\end{equation}
We now move on to define the \textit{conditional entropy} of an RV given another RV. 
\begin{definition}
	\label{conditional_entropy_defn}
	For a pair of discrete RVs $(X,Y) \sim p(x,y)$, the conditional entropy $H(Y|X)$ is defined as \cite[eq. (2.10)]{Cover_Elements_of_Info_Theory'06}
	\begin{equation}
		\label{conditional_entropy_defn_eqn_1}
		H(Y|X) \eqdef \sum_{x \in \mathcal{X}} p(x) H(Y|X=x).
	\end{equation}
\end{definition}
The RHS of (\ref{conditional_entropy_defn_eqn_1}) can then be simplified as
\begin{subequations}
	\begin{align}
		\label{conditional_entropy_defn_eqn_2}
		H(Y|X) \stackrel{(a)}{=} & -\sum_{x \in \mathcal{X}} p(x) \sum_{y \in \mathcal{Y}}  p(y|x) \log_2 p(y|x)   \\
		\label{conditional_entropy_defn_eqn_3}
		\stackrel{(b)}{=} & -\sum_{x \in \mathcal{X}} \sum_{y \in \mathcal{Y}} p(x)   p(y|x) \log_2 p(y|x)   \\
		\label{conditional_entropy_defn_eqn_4}
		\stackrel{(c)}{=} & -\sum_{x \in \mathcal{X}} \sum_{y \in \mathcal{Y}} p(x,y)   \log_2 p(y|x)   \\
		\label{conditional_entropy_defn_eqn_5}
		\stackrel{(d)}{=} & -\mathbb{E} \{ \log_2 p(Y|X) \},
	\end{align}
\end{subequations}
where $(a)$ is due to the entropy definition in (\ref{entropy_defn_discrete_RV}), $(b)$ follows from rearranging the RHS of (\ref{conditional_entropy_defn_eqn_2}), $(c)$ is for the definition of the conditional PMF $p(y|x)$ with $ p(y|x) \eqdef  p_{Y|X}(y|x) =\mathbb{P} (Y=y | X=x) =p(x,y)/p(x) $ \cite{DPJN08}, and $(d)$ is because of the definition in (\ref{expectation_g_X__Y_defn}). It is intuitive from (\ref{conditional_entropy_defn_eqn_5}) that $H(Y|X) \neq H(X|Y)$ \cite{Cover_Elements_of_Info_Theory'06}.

If we now simply add (\ref{conditional_entropy_defn_eqn_5}) and (\ref{joint_entropy_defn_eqn_2}), it follows that 
\begin{multline}
	\label{add_enrtopies_rev_1}
	\hspace{-3mm} H(Y|X)+H(X,Y)= -\big[\mathbb{E} \{ \log_2 p(Y|X) \} + \mathbb{E} \{ \log_2 p(X,Y)  \} \big] \\ \stackrel{(a)}{=} -\sum_{x \in \mathcal{X}} \sum_{y \in \mathcal{Y}} p(x,y)  [\log_2 p(y|x) +    \log_2 p(x,y)]      \\
	\stackrel{(b)}{=} -\sum_{x \in \mathcal{X}} \sum_{y \in \mathcal{Y}} p(x,y)  [\log_2 p(y|x) p(x,y)]     \\ 
	\stackrel{(c)}{=} -\sum_{x \in \mathcal{X}} \sum_{y \in \mathcal{Y}} p(x,y)  [\log_2  [p(x,y)]^2/p(x)]   \\
	\stackrel{(d)}{=} -2\sum_{x \in \mathcal{X}} \sum_{y \in \mathcal{Y}} p(x,y) \log_2  p(x,y) + \sum_{x \in \mathcal{X}} \sum_{y \in \mathcal{Y}} p(x,y)  \log_2 p(x)  \\
	\stackrel{(e)}{=} -2\sum_{x \in \mathcal{X}} \sum_{y \in \mathcal{Y}} p(x,y) \log_2  p(x,y) + \sum_{x \in \mathcal{X}}  p(x)  \log_2 p(x)  \\
	\stackrel{(f)}{=}  2 H(X,Y) - H(X),
\end{multline}
where $(a)$ is due to (\ref{conditional_entropy_defn_eqn_4}) and (\ref{joint_entropy_defn_eqn_1}), $(b)$ is due to the property of the logarithm, $(c)$ is for $ p(y|x) \eqdef p(x,y)/p(x) $ \cite{DPJN08}, $(d)$ is also because of the property of the logarithm, $(e)$ is due to the property of the joint PMF $p(x,y)$ with $p(x)=\sum_{y \in \mathcal{Y}} p(x,y)$ \cite{DPJN08}, and $(f)$ follows from (\ref{joint_entropy_defn_eqn_1}) and (\ref{entropy_defn_discrete_RV}). Rearranging (\ref{add_enrtopies_rev_1}) gives the result
\begin{equation}
	\label{add_enrtopies_7}
	H(Y|X)+ H(X)=  H(X,Y).  
\end{equation}
This is an important result that is widely known as the \textit{chain rule} \cite{Cover_Elements_of_Info_Theory'06} and formalized below.
\begin{theorem}[{\textbf{Chain rule \cite[Theorem 2.2.1]{Cover_Elements_of_Info_Theory'06}}}]
	For a pair of discrete RVs $(X,Y) \sim p(x,y)$,
	\begin{equation}   
		\label{chain_rule}
		H(X,Y) = H(X) + H(Y|X).
	\end{equation}
\end{theorem}
From (\ref{chain_rule}), the following corollary \cite[eq. (2.21)]{Cover_Elements_of_Info_Theory'06} follows.
\begin{corollary}
	\begin{equation}
		\label{chain_rule_corollary}
		H(X,Y|Z) = H(X|Z) + H(Y|X, Z).
	\end{equation}
\end{corollary}

We now proceed to define the \textit{relative entropy} \cite{Cover_Elements_of_Info_Theory'06}.
\begin{definition}
	\label{relative_entropy_defn}
	The relative entropy or KL distance between two PMFs $p(x)$ and $q(x)$ is given\footnote{Throughout this paper, we follow definitions w.r.t. the logarithm to the base two. However, the logarithm to the base ten are generally used, as it is also the case with some of the literature \cite{Cover_Elements_of_Info_Theory'06}.} by \cite[eqs. (2.26) and (2.27)]{Cover_Elements_of_Info_Theory'06}
	\begin{equation}
		\label{relative_entropy_eqn_1}
		D(p || q) \eqdef \sum_{x \in \mathcal{X}} p(x) \log_2 \frac{p(x)}{q(x)} = \mathbb{E}_{p(x)} \Big\{ \log_2 \frac{p(X)}{q(X)} \Big\},
	\end{equation}
	where the conventions $0\log_2 \frac{0}{0}=0$, $0\log_2 \frac{0}{q}=0$, and $p\log_2 \frac{p}{0}=\infty$ are used \cite{Cover_Elements_of_Info_Theory'06}.
\end{definition}
W.r.t. the relative entropy defined in Definition \ref{relative_entropy_defn}, the \textit{mutual information} between two RVs is defined below.
\begin{definition}
	\label{mutual_information_defn}
	For two discrete RVs $X$ and $Y$ with a joint PMF $p(x,y)$ and marginal PMFs $p(x)$ and $p(y)$, respectively, their mutual information $I(X;Y)$ is the relative entropy between $p(x,y)$ and the product distribution $p(x) p(y)$ \cite[eqs. (2.28)-(2.30)]{Cover_Elements_of_Info_Theory'06}:
	\begin{subequations}
		\begin{align}
			\label{mutual_information_eqn_1}
			I(X;Y) &= \sum_{ x \in \mathcal{X}} \sum_{ y \in \mathcal{Y}} p(x,y) \log_2 \frac{p(x,y)}{p(x) p(y)}   \\
			\label{mutual_information_eqn_2}
			&= D(p(x,y) || p(x) p(y))   \\
			\label{mutual_information_eqn_3}
			&= \mathbb{E}_{p(x,y)}  \Big\{ \log_2 \frac{p(X,Y)}{p(X) p(Y)} \Big\}.
		\end{align}
	\end{subequations}
\end{definition}
From (\ref{mutual_information_eqn_3}), it directly follows that
\begin{equation}
	\label{mutual_information_eqn_3_2}
	I(Y;X) =\mathbb{E}_{p(y,x)}  \Big\{ \log_2 \frac{p(Y,X)}{p(Y) p(X)} \Big\}=I(X;Y).
\end{equation}	
The equality in (\ref{mutual_information_eqn_3_2}) states the \textit{symmetrical} nature of mutual information: i.e., $X$ says as much about $Y$ as $Y$ says about $X$ \cite{Cover_Elements_of_Info_Theory'06}. Meanwhile, simplifying the RHS of (\ref{mutual_information_eqn_1}), the mutual information $I(X;Y)$ can also be expressed as \cite[eq. (2.39)]{Cover_Elements_of_Info_Theory'06}
\begin{equation}
	\label{mutual_information_eqn_4}
	I(X;Y)=H(X)-H(X|Y).
\end{equation}
Thus, it follows from (\ref{mutual_information_eqn_4}) and (\ref{mutual_information_eqn_3_2}) that
\begin{equation}
	\label{mutual_information_eqn_5}
	I(X;Y)=\overbrace{ H(Y)-H(Y|X) }^{=I(Y;X)}.
\end{equation}
From the chain rule as expressed in (\ref{chain_rule}), $H(Y|X)=H(X,Y)- H(X)$. Substituting this inequality into the RHS of (\ref{mutual_information_eqn_5}) leads to the relationship \cite[eq. (2.41)]{Cover_Elements_of_Info_Theory'06}
\begin{equation}
	\label{mutual_information_eqn_6}
	I(X;Y)= H(X) + H(Y) - H(X,Y).
\end{equation}
At last, we note that \cite[eq. (2.42)]{Cover_Elements_of_Info_Theory'06}
\begin{equation}
	\label{mutual_information_eqn_7}
	I(X;X)=H(X)-H(X|X) \stackrel{(a)}{=} H(X)-0=H(X), 
\end{equation}
where $(a)$ follows through Definition \ref{conditional_entropy_defn} and (\ref{conditional_entropy_defn_eqn_4}) due to the probabilistic fact\footnote{Intuitively, $H(X|X)=0$ is the reflection of the fact that there is no any uncertainty about $x \in \mathcal{X}$ provided that $x$ is already known/given.} that $p(x|x)=1$ for all $x \in \mathcal{X}$. In summary, the information-theoretic results in (\ref{mutual_information_eqn_3_2})-(\ref{mutual_information_eqn_7}) are formalized in the following theorem. 
\begin{theorem}[{\textbf{Mutual information and entropy \cite[Theorem 2.4.1]{Cover_Elements_of_Info_Theory'06}}}]
	\label{Multual_information_results_thm}
	The underneath results \cite[eqs. (2.43)-(2.47)]{Cover_Elements_of_Info_Theory'06} are valid concerning the relationship between mutual information and entropy: 
	\begin{subequations}
		\begin{align}
			\label{mutual_information_thm_eqn_1}
			I(X;Y) &= H(X)-H(X|Y)   \\
			\label{mutual_information_thm_eqn_2}
			I(X;Y) &= H(Y)-H(Y|X)     \\
			\label{mutual_information_thm_eqn_3}
			I(X;Y)&= H(X) + H(Y) - H(X,Y)  \\
			\label{mutual_information_thm_eqn_4}
			I(X;Y) &= I(Y;X)     \\
			\label{mutual_information_thm_eqn_5}
			I(X;X) &= H(X).
		\end{align}
	\end{subequations}
\end{theorem}

We build on the chain rule as stated in (\ref{chain_rule}) and continue with the chain rules for entropy and mutual information. Beginning with the former, we state the following chain rule for the entropy of a collection of RVs.
\begin{theorem}[{\textbf{Chain rule for the entropy of a collection of RVs \cite[Theorem 2.5.1]{Cover_Elements_of_Info_Theory'06}}}]
	\label{chain_rule_for_entropy_thm}
	For discrete RVs $X_1, X_2, \ldots, X_n$ drawn according to $p(x_1, x_2, \ldots, x_n)$, their joint entropy $H(X_1, X_2, \ldots, X_n)$ can be expressed as \cite[eq. (2.48)]{Cover_Elements_of_Info_Theory'06}
	\begin{equation}
		\label{joint_entropy_multiple_RVs_relation}
		H(X_1, X_2, \ldots, X_n) = \sum_{i=1}^n H( X_i| X_{i-1}, \ldots, X_1 ).
	\end{equation}
	\proof The proof is provided in \cite[p. 22-23]{Cover_Elements_of_Info_Theory'06}. 
\end{theorem}
We define \textit{conditional mutual information} below \cite{Cover_Elements_of_Info_Theory'06}.
\begin{definition}
	\label{Cond_MI_defn}
	For discrete RVs $X, Y, Z \sim p(x,y,z)$, the conditional mutual information of $X$ and $Y$ given $Z$ is defined by \cite[eqs. (2.60) and (2.61)]{Cover_Elements_of_Info_Theory'06}
	\begin{subequations}
		\begin{align}
			\label{Cond_MI_eqn_1}
			I(X; Y | Z) & \eqdef  H(X|Z) - H(X | Y, Z)     \\
			\label{Cond_MI_eqn_2}
			&= \mathbb{E}_{p(x, y, z)}  \Big\{ \log_2 \frac{p(X, Y|Z)}{p(X|Z)p(Y|Z)} \Big\}. 
		\end{align}
	\end{subequations}
\end{definition}
Definition \ref{Cond_MI_defn} and Theorem \ref{chain_rule_for_entropy_thm} then lead to the following theorem on the chain rule for mutual information.

\begin{theorem}[{\textbf{Chain rule for mutual information \cite[Theorem 2.5.2]{Cover_Elements_of_Info_Theory'06}}}]
	\label{chain_rule_for_MI_thm}
	The following result is valid for the mutual information of multiple RVs \cite[eq. (2.62)]{Cover_Elements_of_Info_Theory'06}:
	\begin{equation}
		\label{chain_rule_for_MI_eqn_1}
		I(X_1, X_2, \ldots, X_n; Y) = \sum_{i=1}^n I (X_i; Y | X_{i-1}, X_{i-2}, \ldots, X_1).
	\end{equation}
\end{theorem}
\proof Theorem \ref{chain_rule_for_MI_thm} follows from Theorem \ref{Multual_information_results_thm} and (\ref{mutual_information_thm_eqn_1}) that 
\begin{multline}
	\label{proof_eqn_1}
	\hspace{-3mm} I(X_1, X_2, \ldots, X_n; Y)=H(X_1,\ldots, X_n)-  H(X_1,  \ldots, X_n | Y) \\ \stackrel{(a)}{=} \sum_{i=1}^n  \big[ H( X_i| X_{i-1}, \ldots, X_1 ) -  H( X_i| X_{i-1}, \ldots, X_1, Y ) \big] \\ \stackrel{(b)}{=} \sum_{i=1}^n I(X_i; Y | X_{i-1}, X_{i-2}, \ldots, X_1), 
\end{multline} 
where $(a)$ is due to (\ref{joint_entropy_multiple_RVs_relation}) and $(b)$ follows from (\ref{Cond_MI_eqn_1}). The last equation on the RHS of (\ref{proof_eqn_1}) is the RHS of (\ref{chain_rule_for_MI_eqn_1}). This completes the proof of Theorem \ref{chain_rule_for_MI_thm}.   \QEDclosed

\section{Proof of Theorem \ref{entropy_relationship_thm}}
\label{proof_entropy_relationship_thm}
Without providing a proof, the authors of \cite{Bao_Towards_Theory_SemCom'11} wrote that Theorem \ref{entropy_relationship_thm} follows from the definitions of entropy and conditional entropy. For the sake of completeness and insight, we provide below our own proof of Theorem \ref{entropy_relationship_thm}.

It follows from Definition \ref{conditional_entropy_defn} and (\ref{conditional_entropy_defn_eqn_1}) that the conditional entropies $H(X|W)$ and $H(W|X)$ can be determined as
\begin{equation}
\label{conditional_entropy_X|W_eqn_1}
H(X|W) \eqdef \sum_{w \in \mathcal{W}} p(w) H(X|W=w).
\end{equation}
\begin{equation}
\label{conditional_entropy_W|X_eqn_1}
H(W|X) = \sum_{x \in \mathcal{X}} p(x) H(W|X=x).
\end{equation}
To proceed, we simplify the RHS of (\ref{conditional_entropy_W|X_eqn_1}). To this end, we first note that $H(W)$ -- as it is defined in (\ref{Shannon_entropy_of_W}) -- is independent of all $x \in \mathcal{X}$. Consequently,
\begin{equation}
\label{conditional_entropy_W|X_indepent_cases}
H(W|X=x) = H(W).
\end{equation}
Substituting the equality in (\ref{conditional_entropy_W|X_indepent_cases}) into the RHS of (\ref{conditional_entropy_W|X_eqn_1}), we obtain
\begin{subequations}
\begin{align}
\label{conditional_entropy_W|X_eqn_2}
H(W|X)  &= \sum_{x \in \mathcal{X}} p(x) H(W) = H(W) \sum_{x \in \mathcal{X}} p(x)     \\
\label{conditional_entropy_W|X_eqn_3}
& \stackrel{(a)}{=} H(W) \sum_{x \in \mathcal{X}} \sum_{ w \in \mathcal{W}} \mu(w) p(x|w)  ,
\end{align}
\end{subequations}
where $(a)$ is due to (\ref{message_distribution_eqn}). Meanwhile, applying \textit{Bayes' rule (Bayes' theorem)} \cite{DPJN08} to $p(x|w)$, it follows that
\begin{equation}
\label{Bayes_rule_p_x|w}
p(x|w)= \frac{p(w|x) p(x)}{p(w)}.
\end{equation}
Plugging (\ref{Bayes_rule_p_x|w}) into the RHS of (\ref{conditional_entropy_W|X_eqn_3}) and rearranging give
\begin{subequations}
\begin{align}
\label{conditional_entropy_W|X_eqn_4}
H(W|X)&=H(W) \sum_{ w \in \mathcal{W}} \frac{\mu(w)}{p(w)}\sum_{x \in \mathcal{X}} p(w|x) p(x)  \\
\label{conditional_entropy_W|X_eqn_5}
&\stackrel{(a)}{=} H(W) \sum_{ w \in \mathcal{W}} \mu(w)  , 
\end{align}
\end{subequations}
where $(a)$ is due to the conditional PMF property that $\sum_{x \in \mathcal{X}} p(w|x) p(x)=p(w)$ \cite{DPJN08}. Since $\mu(\cdot)$ is a probability measure, it is evident that $\sum_{ w \in \mathcal{W}}  \mu(w)=1$. Replacing this value in the RHS of (\ref{conditional_entropy_W|X_eqn_5}) then leads to the relationship
\begin{equation}
\label{conditional_entropy_W|X_eqn_6}
H(W|X) = H(W).
\end{equation}

To move further forward, we now simplify the RHS of (\ref{conditional_entropy_X|W_eqn_1}). In doing so, $H(X|W=w)$ simplifies through $H(X)$ -- which is defined in (\ref{shannon_entropy_of_X_eqn}) -- to
\begin{equation}
\label{conditional_entropy_X|W=w_eqn_1}
H(X|W=w) \eqdef - \sum_{ x \in \mathcal{X}} p(x|w) \log_2 p(x|w). 
\end{equation}
Meanwhile, it follows from the definition of conditional PMF that \cite{DPJN08}
\begin{equation}
\label{cond_PMF_defn_eqn}
p(x|w) = \frac{p(x, w)}{p(w)}.
\end{equation}
Applying the properties of the logarithm to (\ref{cond_PMF_defn_eqn}) then produces
\begin{equation}
\label{log_2_cond_PMF_defn_eqn}
\log_2 p(x|w) =  \log_2 p(x, w) - \log_2 p(w) .
\end{equation}
Replacing (\ref{log_2_cond_PMF_defn_eqn}) and (\ref{cond_PMF_defn_eqn}) into the RHS of (\ref{conditional_entropy_X|W=w_eqn_1}) results in
\begin{equation}
\label{conditional_entropy_X|W=w_eqn_2}
H(X|W=w) \eqdef - \sum_{ x \in \mathcal{X}} \frac{p(x, w)}{p(w)}  \big[ \log_2 p(x, w) - \log_2 p(w) \big]. 
\end{equation}
Plugging (\ref{conditional_entropy_X|W=w_eqn_2}) into (\ref{conditional_entropy_X|W_eqn_1}) then leads to
\begin{multline}
\label{conditional_entropy_X|W_eqn_2}
H(X|W) = -\sum_{w \in \mathcal{W}} \sum_{ x \in \mathcal{X}} p(x, w) \log_2 p(x, w) + \\ \sum_{w \in \mathcal{W}} \Big[ \sum_{ x \in \mathcal{X}} p(x, w) \Big] \log_2 p(w) \stackrel{(a)}{=}  H(X, W)  +  \\ \sum_{w \in \mathcal{W}}  p(w) \log_2 p(w)
\end{multline}
where $(a)$ is because of the definition of joint entropy per Definition \ref{joint_entropy_defn_1} and the properties of joint PMF that lead to $\sum_{ x \in \mathcal{X}} p(x, w)=p(w)$ \cite{DPJN08}. To move on, we determine $p(w)$ using the PMF $p(x)$ -- which is equated in (\ref{message_distribution_eqn}) -- as  
\begin{equation}
\label{p_w_via_p_x_eqn_1}
\hspace{-3mm} p(w)=  \mu(w) \overbrace{p(w|w)}^{=1}  + \overbrace{\sum_{ \tilde{w} \neq w, \tilde{w} \in \mathcal{W}} \mu(\tilde{w}) p(w|\tilde{w})}^{=0} \stackrel{(a)}{=} \mu(w) , 
\end{equation}
where $(a)$ is because $p(w|w)=1$ and $p(w|\tilde{w})=0$ , $\forall$$\tilde{w} \neq w$.

Substituting (\ref{p_w_via_p_x_eqn_1}) into the RHS of (\ref{conditional_entropy_X|W_eqn_2}) produces
\begin{subequations}
\begin{align}
\label{conditional_entropy_X|W_eqn_3}
H(X|W) & =  H(X, W)  +   \sum_{w \in \mathcal{W}}  \mu(w) \log_2 \mu(w)   \\
\label{conditional_entropy_X|W_eqn_4}
 &\stackrel{(a)}{=} H(X, W) - H(W)    \\
\label{conditional_entropy_X|W_eqn_5}
&\stackrel{(b)}{=} H(X, W) - H(W|X),
\end{align}
\end{subequations}
where $(a)$ is due to (\ref{Shannon_entropy_of_W}) and $(b)$ is because of (\ref{conditional_entropy_W|X_eqn_6}). If we now employ the chain rule for entropy, it follows from (\ref{chain_rule}) that $H(X, W)=H(X) + H(W|X)$. Substituting this equality into the RHS of (\ref{conditional_entropy_X|W_eqn_5}) then gives
\begin{equation}
\label{conditional_entropy_X|W_eqn_6}
H(X|W)= H(X).
\end{equation}
If we subtract the equality in (\ref{conditional_entropy_W|X_eqn_6}) from the equality in (\ref{conditional_entropy_X|W_eqn_6}),
\begin{equation}
\label{equality_relation_1}
H(X)-H(W)=H(X|W)-H(W|X).
\end{equation}
Rearranging (\ref{equality_relation_1}) then results in the equality
\begin{equation}
\label{equality_relation_2}
H(X)=H(W)+H(X|W)-H(W|X).
\end{equation}
This is exactly (\ref{entropy_relationship_eqn}) and completes the proof of Theorem \ref{entropy_relationship_thm}.    \QEDclosed

\section{On Information Bottleneck (IB)}
\label{sec: on_IB}
Assume a source encoding of an information source is denoted by an RV $X$ and we wish to obtain its relevant quantization $\tilde{X}$ to compress $X$ as much as possible. Assume also that a relevance RV denoted by $Y$ (e.g., a classification label) that must not be independent from $X$ \cite{IB_method_Tishby'00}. Thus, $X$ and $Y$, have a positive mutual information $I(X; Y)$, and we presume that we have access to the joint PDF $p(x, y)$ \cite{DL_and_IB_Tishby'15,IB_method_Tishby'00}. Nonetheless, under these settings and contrary to the rate-distortion problem, we would like $\tilde{X}$ (the quantized information) to capture as much information about $Y$ (the relevance RV) as possible \cite{IB_method_Tishby'00}. The amount of information about $Y$ that is in $\tilde{X}$ is given by $I(\tilde{X}; Y)$ and defined as \cite[eq. (14)]{IB_method_Tishby'00} 
\begin{equation}
	\label{mutual_info_X_tilde_Y_definition}
	I(\tilde{X}; Y)= \sum_{y} \sum_{\tilde{x}} p(y, \tilde{x}) \log \frac{p(y, \tilde{x})}{p(y)p(\tilde{x})} \stackrel{(a)}{\leq}   I(X; Y), 
\end{equation}
where $(a)$ is for lossy compression cannot convey more information than the original signal, and hence, there is always a tradeoff between rate and distortion \cite{IB_method_Tishby'00}. Similarly to rate and distortion, there is a natural tradeoff between preserving meaningful information and compressing the original signal \cite{IB_method_Tishby'00}. Bearing in mind this tradeoff, the IB problem concerns maintaining a constant amount of meaningful information about the relevant signal $Y$ whilst minimizing the number of bits from the original information source $X$ (maximizing its compression) \cite{IB_method_Tishby'00}. This is equivalent to maximizing the meaningful information for a fixed compression of the original information signal \cite{IB_method_Tishby'00}. Accordingly this amounts to passing the information that $X$ provides about $Y$ through a \textquotedblleft bottleneck'' formed by the compact information content in $\tilde{X}$ \cite{IB_method_Tishby'00}.

On par with the aforementioned motivation, the IB problem boils down to solving the following optimization problem \cite[eq. (15)]{IB_method_Tishby'00}, \cite[eq. (2)]{Qiao_What_is_SemCom'21}:
\begin{equation}
	\label{IB_problem}
	\min_{p(\tilde{x}|x)}  I(\tilde{X}; X)-\beta I(\tilde{X}; Y), 
\end{equation}
where the conditional distribution $p(\tilde{x}|x)$ represents the considered source encoder and $\beta$ denotes the Lagrange multiplier connected to the constrained meaningful information \cite{IB_method_Tishby'00,Qiao_What_is_SemCom'21}. Meanwhile, the optimal solution for (\ref{IB_problem}) -- i.e., the optimal source encoder -- is task-dependent, and a generic algorithm computes the optimal solution by alternating iterations \cite{Qiao_What_is_SemCom'21}. In every iteration, minimization is performed by converging alternating iterations w.r.t. the PDFs $p(\tilde{x}|x)$, $p(\tilde{x})$, and $p(y|\tilde{x})$ \cite[Theorem 5]{DL_and_IB_Tishby'15}. This IB approach provides a unified framework for various information processing problems, including prediction, filtering and learning \cite{IB_method_Tishby'00}. Toward these ends, IB has many applications in DL \cite{DL_and_IB_Tishby'15}, ML \cite{IB_applications_Goldfeld'20}, SemCom \cite{Beck_SemCom_Info_Bottleneck_View'22}, and goal-oriented SemCom \cite{Shao_IB_for_Edge_Inference'22}.

\section{On Variants of IB}
\label{sec: on_variants_of_IB}
To inspire much more work on SemCom and goal-oriented SemCom theory, we highlight below the principles of graph IB (GIB) \cite{GIB_Wu'20}, robust IB (RIB), deterministic IB, and distributed IB (DIB), beginning with GIB.

\subsection{Graph IB (GIB)}
To formally define GIB, which is proposed by the authors of \cite{GIB_Wu'20}, let $Y$ be the target, $\mathcal{D} \eqdef \big\{(\bm{A}, X) \big\}$ be the input data for $\bm{A}$ being the graph structure and $X$ being the node features, and $Z$ be the representation. Concerning $Z$ being the representation, GIB is used to optimize $Z$ to capture the minimal sufficient information in input data $\mathcal{D}$ to predict the target $Y$ \cite{GIB_Wu'20}. To this end, the GIB problem reduces to solving the following optimization problem \cite{GIB_Wu'20}:
\begin{equation}
\label{GIB_optimization}
\min_{p(Z| \mathcal{D}) \in \Omega}  \textnormal{GIB}_{\beta} (\mathcal{D} , Y; Z) \eqdef   [ -I(Y; Z) + \beta  I(\mathcal{D}; Z)   ], 
\end{equation}
where $\Omega$ represents the search space of the optimal model $p(Z| \mathcal{D})$ \cite{GIB_Wu'20}.

\subsection{Robust IB (RIB)}
The authors of \cite{Xie_Robust_IB'22} propose to use a design criterion named RIB to design the goal-oriented SemCom system schematized in \cite[Fig. 1]{Xie_Robust_IB'22}. To define RIB formally, let the RVs $X$, $Y$, $Z$, and $\hat{Z}$ be the input datum, the target (label), the output of an encoder modeled as $p_{\phi}(z|x)$, and the output of a demodulator, respectively. From the vantage point of data compression, the optimal $Z$ can be approximated by optimizing the IB problem \cite{Xie_Robust_IB'22} such that $I(Y; \hat{Z})$ is maximized while being subjected to the constraint on the amount of preserved information $I(X; \hat{Z})$ \cite[eq. (5)]{Xie_Robust_IB'22}: 
\begin{equation}
\label{IB_optimization}
\max_{p_{\phi}(z|x)} I(Y; \hat{Z}) - \beta I(X; \hat{Z}).
\end{equation}
Apart from data compression, another crucial goal-oriented SemCom design criterion is the maximization of the transmission rate, and hence \cite[eq. (6)]{Xie_Robust_IB'22}
\begin{equation}
\label{MI_maximization_eqn}
\max_{p_{\phi}(z)} I(Z; \hat{Z}), 
\end{equation}
where $p_{\phi}(z)$ is the marginal distribution that depends on the parameters $\phi$ \cite{Xie_Robust_IB'22}. Meanwhile, combining (\ref{IB_optimization}) and (\ref{MI_maximization_eqn}) leads to the RIB design principle (or criterion) that is given by \cite[eq. (7)]{Xie_Robust_IB'22}
\begin{equation}
\label{RIB_principle}
\max_{p_{\phi}(z|x)} I(Y; \hat{Z}) + \beta [ I(Z; \hat{Z}) -I(X; \hat{Z}) ], 
\end{equation}
where $\beta$ is fixed and $\beta \geq 0$. 

We now move on to highlight deterministic IB \cite{Strouse_DIB'16}.

\subsection{Deterministic IB}
The authors of \cite{Strouse_DIB'16} introduce a modified IB criterion named deterministic IB, which they say better captures the essence of compression than an optimal tradeoff between discarding as many bits as possible and selectively keeping the ones that are most important \cite{Strouse_DIB'16}. Meanwhile, the deterministic IB problem boils down to solving the following optimization problem \cite[eq. (8)]{Strouse_DIB'16} (which is stated using the notation of Appendix \ref{sec: on_IB}):
\begin{equation}
\label{Deterministic_IB_problem} 
\min_{ p(\tilde{x} | x) } H(\tilde{X}) - \beta I(\tilde{X}; Y), 
\end{equation}
where the deterministic IB optimization of (\ref{Deterministic_IB_problem}) is subjected to the Markov constraint $\tilde{X} \leftrightarrow X \leftrightarrow Y$ \cite{Strouse_DIB'16}.

We now proceed to highlight DIB \cite{Strouse_DIB'16}.

\subsection{Distributed IB (DIB)}
To state and discuss the DIB framework \cite{Aguerri_Distributed_VRL'18}, we must first consider the distributed learning (e.g., multi-view learning) model depicted in \cite[Fig. 1]{Aguerri_Distributed_VRL'18}. Per \cite[Fig. 1]{Aguerri_Distributed_VRL'18}, $Y$ is the signal to be predicted and $(X_1, \ldots , X_K)$ are the relevant $K$ views of $Y$ that could each be useful to understand one or more aspects of it \cite{Aguerri_Distributed_VRL'18}. Accordingly, the relevant observations could be either distinct or redundant. This justifies the assumption $(X_1, \ldots , X_K)$ are independent given $Y$ \cite{Aguerri_Distributed_VRL'18}. This distributed learning problem's problem formulation \cite[Section 2]{Aguerri_Distributed_VRL'18} is highlighted below.

Let $K \in \mathbb{N}_{\geq 2}$ be given and $\mathcal{K} \eqdef [K]$. Let $(X_1, \ldots , X_K, Y)$ be a tuple of RVs that have a joint PMF $p_{X_{\mathcal{K}},Y}(x_{\mathcal{K}},y) \eqdef p_{X_1,\ldots, X_K, Y}(x_1, \ldots, x_K,y)$ for $ (x_1, \ldots, x_K) \in \mathcal{X}_1 \times \ldots \times \mathcal{X}_K$ and $y\in \mathcal{Y}$, given that $\mathcal{X}_k$ for all $k\in \mathcal{K}$ and $\mathcal{Y}$ represent the alphabet of $X_k$ and $Y$, respectively. Meanwhile, the Markov chain below is assumed to hold for all $k\in \mathcal{K}$ \cite[eq. (3)]{Aguerri_Distributed_VRL'18}:
\begin{equation}
\label{Markov_Chain_assumption}
X_k \leftrightarrow Y \leftrightarrow X_{\mathcal{K}/k}, 
\end{equation}
i.e., $p(x_{\mathcal{K}}, y) = p(y) \prod_{k=1}^K p(x_k|y)$ for $x_k \in \mathcal{X}_K$ and $y\in \mathcal{Y}$. The distributed learning problem aims to characterize how the goal variable $Y$ can be accurately estimated from the observations $(X_1, \ldots , X_K)$ when they are processed individually in different encoders \cite{Aguerri_Distributed_VRL'18}. 

Moreover, let a training dataset $\{(X_{1,i}, \ldots, X_{K,i}, Y_i)\}_{i=1}^n$ comprise $n$ i.i.d. random samples that are drawn from the joint PMF $p_{X_{\mathcal{K}},Y}$, which is assumed to be given \cite{Aguerri_Distributed_VRL'18}. The $k$-th encoder observes only the sequence $X_k^n$, which it would process to generate $J_k= \phi_k (X_k^n)$ per the following (possibly stochastic) mapping \cite[eq. (4)]{Aguerri_Distributed_VRL'18}:
\begin{equation}
\label{mapping_phi_eqn}
\phi_k : \mathcal{X}_k^n \to \mathcal{M}_k^n
\end{equation}
where $\mathcal{M}_k^n$ denotes an arbitrary set of descriptions \cite{Aguerri_Distributed_VRL'18}. Using $J_{\mathcal{K}} \eqdef (J_1, \ldots, J_K)$ as inputs, a (possibly stochastic) decoder $\psi(\cdot)$ processes all the inputs and returns $\hat{Y}^n$ (an estimate of $Y^n$) as \cite[eq. (5)]{Aguerri_Distributed_VRL'18}
\begin{equation}
\label{psi_defn_eqn}
\psi: \mathcal{M}_1^n \times \ldots \times \mathcal{M}_K^n \to \hat{\mathcal{Y}}^n.
\end{equation}
For the mapping in  (\ref{psi_defn_eqn}), the accuracy of $\hat{Y}^n$ is quantified in terms of \textit{relevance} \cite{Aguerri_Distributed_VRL'18}. Relevance is defined as  the information that the descriptions $\phi_1 (X_1^n), \ldots,  \phi_K (X_K^n)$ \textit{collectively preserve} about $Y^n$ and is given by \cite[eq. (6)]{Aguerri_Distributed_VRL'18}
\begin{equation}
\label{Relevance_defn}
\Delta^{(n)} (p_{X_{\mathcal{K}}, Y})  \eqdef \frac{1}{n} I_{p_{X_{\mathcal{K}}, Y}} (Y^n, \hat{Y}^n),
\end{equation}
where $\hat{Y}^n \eqdef \psi(\phi_1 (X_1^n), \ldots,  \phi_K (X_K^n))$ and the subscript $p_{X_{\mathcal{K}}, Y}$ implies that the mutual information is computed w.r.t. the joint distribution $p_{X_{\mathcal{K}}, Y}$ \cite{Aguerri_Distributed_VRL'18}. 

Should the encoder mappings $\{\phi_k\}_{k=1}^K$ be unconstrained, maximizing the RHS of (\ref{Relevance_defn}) would lead to overfitting \cite{Aguerri_Distributed_VRL'18}. Overfitting can be overcome by using better generalizability, which is usually obtained by constraining the \textit{complexity of the encoders} \cite{Aguerri_Distributed_VRL'18}. To this end, the encoding function $\phi_k(\cdot)$ of encoder $k \in \mathcal{K}$ needs to fulfill \cite[eq. (7)]{Aguerri_Distributed_VRL'18}
\begin{equation}
\label{R_k_constraint}
R_k \geq \frac{1}{n} \log |\phi_k(X_k^n)|, 
\end{equation}
where (\ref{R_k_constraint}) must be satisfied for all $X_k^n \in \mathcal{X}_k^n$ \cite{Aguerri_Distributed_VRL'18}. Meanwhile, optimal performance for distributed learning can be cast as finding the region of all simultaneously achievable \textit{relevance-complexity tuples} \cite{Aguerri_Distributed_VRL'18}, as defined below.
\begin{definition}[{\textbf{\cite[Definition 1]{Aguerri_Distributed_VRL'18}}}]
A tuple $(\Delta, R_1, \ldots, R_K)$ is termed achievable if there exists a training set of size $n$, encoders $\phi_k$ for $k\in[K]$, and a decoder $\psi$ such that \cite[eqs. (8) and (9)]{Aguerri_Distributed_VRL'18}
\begin{subequations}
\begin{align}
\label{Delta_upper_cound}
\Delta &\leq \frac{1}{n}   I_{p_{X_{\mathcal{K}}, Y}} \big(Y^n, \psi(\phi_1 (X_1^n), \ldots,  \phi_K (X_K^n)) \big)  \\
\label{R_K_lower_cound}
R_k & \geq \frac{1}{n} \log |\phi_k(X_k^n)|, \hspace{2mm} \forall k\in \mathcal{K}.
\end{align}
\end{subequations}
The relevance-complexity region $\mathcal{RI}_{\textnormal{DIB}}$ is expressed by the closure of all attainable tuples $(\Delta, R_1, \ldots, R_K)$ \cite{Aguerri_Distributed_VRL'18}.
\end{definition}
The region $\mathcal{RI}_{\textnormal{DIB}}$ is characterized by the following theorem.  
\begin{theorem}[{\textbf{\cite[Theorem 1]{Aguerri_Distributed_VRL'18}}}]
\label{Thm: DIB}
The relevance-complexity region $\mathcal{RI}_{\textnormal{DIB}}$ of a distributed learning problem with a joint PMF $p_{X_{\mathcal{K}}, Y}$ -- for which the Markov chain of (\ref{Markov_Chain_assumption}) holds -- is expressed by the union of all tuples $(\Delta, R_1, \ldots, R_K) \in \mathbb{R}_+^{K+1}$ fulfilling, for all $\mathcal{S} \subseteq \mathcal{K}$, \cite[eq. (14)]{Aguerri_Distributed_VRL'18}:
\begin{equation}
\label{Delta_union_eqn}
\Delta \leq \sum_{k \in \mathcal{S}}  \big[ R_k - I(X_k; U_k | Y, T) \big]  + I(Y; U_{\mathcal{S}^c} | T),
\end{equation}
for some of the PMFs $\{  p_{U_1|X_1, T}, \ldots, p_{U_K|X_K, T}, p_T\}$ with a joint distribution of the form \cite[eq. (15)]{Aguerri_Distributed_VRL'18}:
\begin{equation}
\label{Prod_distribution_eqn}
p_T(t) p_Y(y)  \sum_{k=1}^K p_{X_k|Y}(x_k|y)  \sum_{k=1}^K p_{U_k|X_k, T (u_k|x_k, t)}.
\end{equation}
\proof The proof is provided in \cite[Section 7.1]{Aguerri_Distributed_VRL'18}.
\end{theorem}
Theorem \ref{Thm: DIB} extends the single encoder IB principle to the distributed learning model with K encoders, which is dubbed the DIB problem \cite{Aguerri_Distributed_VRL'18}.

\section*{Acknowledgment}
The first author acknowledges Dr. Hamid Gharavi (\textit{Life Fellow, IEEE}) of NIST, MD, USA for funding and leadership support.  

\section*{Disclaimer} 
The identification of any commercial product or trade name does not imply endorsement or recommendation by the National Institute of Standards and Technology, nor is it intended to imply that the materials or equipment identified are necessarily the best available for the purpose.


\balance
\end{document}